\documentclass[a4paper,12pt]{article}
\usepackage[utf8]{inputenc}
\usepackage[english]{babel}
\usepackage{amsmath, amssymb,graphicx}
\usepackage{braket}
\usepackage{caption}
\usepackage{subcaption}
\usepackage{ulem}
\usepackage{multicol}
\usepackage{textcomp}
\usepackage{multirow}

\pdfoutput=1
\usepackage{jheppub}
\newcommand{\be}{\begin{equation}}
\newcommand{\ee}{\end{equation}}
\newcommand{\bea}{\begin{eqnarray}}
\newcommand{\eea}{\end{eqnarray}}
\newcommand{\nn}{\nonumber}

\newcommand{\cA}{\mathcal{A}}

\newcommand{\fb}{\mathfrak{b}}

\newcommand{\fg}{\mathfrak{g}}

\numberwithin{equation}{section}

\title{Holographic Anisotropic Model for Heavy Quarks in Anisotropic
  Hot Dense QGP with External Magnetic Field}

\author{Irina Ya. Aref'eva$^a$, Kristina Rannu$^a$ and Pavel Slepov$^a$}

\affiliation{$^a$Steklov Mathematical Institute, Russian Academy of
  Sciences,\\ Gubkina str. 8, 119991, Moscow, Russia
}

\emailAdd{arefeva@mi-ras.ru}
\emailAdd{rannu-ka@rudn.ru}
\emailAdd{slepov@mi-ras.ru}

\abstract{We present a five-dimensional fully anisotropic holographic model
  for heavy quarks supported by Einstein-dilaton-three-Maxwell action;
  one of the Maxwell fields is related to an external magnetic
  field. Influence of the external magnetic field on the 5-dim black
  hole solution and the confinement/deconfinement phase diagram is
  considered. The effect of the inverse magnetic catalyses is revealed
  and positions of critical end points are found.
}

\keywords{AdS/QCD, holography, phase transition, Wilson loops, heavy
  quarks, magnetic field}

\begin{document}

\maketitle


\newpage

\section{Introduction}

Isotropic  holographic QCD (HQCD) models, see for review \cite{Solana,
  IA, DeWolf}, were considered in numerous papers
\cite{0611304}-\cite{K-dubna}. These models have been considered to
study the confinement/deconfinement phase transition \cite{0611304,
  0812.0792, 0903.2859, 1006.5461, 1201.0820, 1301.0385, 1303.6929,
  1406.1865, 1411.5332, 1505.07894,1506.05930, 1512.04062, 1803.06764,
  1808.05596, ARS-2019qfthep, 1906.12316,
  2010.04578,IA-NICA,S-dubna,K-dubna}, as well as the chiral phase
transition \cite{1206.2824, 1511.02721, 1512.06493, 1810.12525,
  1810.07019, 1910.02269, 2002.00075, 2005.00500, 2009.05694,
  2010.04578, 2010.06762} and also to search for quarkyonic phase
transition \cite{1908.02000}. One of the popular methods for
constructing an HQCD is a potential reconstruction method for models
with Einstein-dilaton-Maxwell action. It consists in starting with a
given form of metric, which is a deformed form $AdS_5$ \cite{DeWolf,
  1506.05930}, and then finding the dilaton potential that supports
the given form of the metric. Metric deformation is realized by the
warp factor, which in turn determines the blackening function and,
consequently, the thermodynamics of the model. Choice of the warp
factor in the metrics strongly influences the phase transition
structure  of  HQCD. As a guiding principle in the choice of the warp
factor one usually uses the agreement with the main features of
lattice results. Unfortunately lattice cannot provide us a full
picture of the QCD phase diagram and shows only its individual
parts. The absence of general scheme of study different regimes in
QCD is the origin to develop  holographic approach to QCD. It occurred
that the phase diagram  describing heavy quarks can be obtained via
deformation of $AdS_5$ by wrap factor that is  the exponential of a
polynomial on holographic coordinate. For the simplest quadratic
polynomial \cite{0611304, 1506.05930} the model reproduces some
features of phase diagram describing heavy quarks, meanwhile to
reproduce phase diagram of light quarks one has to use rational
functions \cite{1006.5461, 1703.09184}.\\

It was recognized that \cite{1202.4436} it is important to add
anisotropy in the holographic theory as QGP is an anisotropic media
just after the HIC, and an estimation for isotropisation time is about
$1$--$5$ fm/c $\sim 10^{-24}$ s \cite{Strickland:2013uga}. To deal
with anisotropic HQCD one considers Einstein-dilaton-two Maxwell model
with additional Maxwell field to support the anisotropy in
metrics. Such anisotropic model was considered in \cite{1802.05652}
for heavy quarks and in \cite{2009.05562} for light quarks. In these
models anisotropy is defined by a parameter $\nu$, and it's value of
about $\nu=4.5$ gives the dependence of the produced entropy on energy
in accordance with the experimental data for the energy dependence of
the total multiplicity of particles created in heavy ion collisions
\cite{Alice}. Isotropic holographic models had not been able to
reproduce the experimental multiplicity dependence on energy
(\cite{AG} and refs therein). As shown in \cite{1808.05596}, the
model \cite{1802.05652} describes smeared confinement/deconfinement
phase transitions. This happens since the location on phase diagram of
confinement/deconfinement transition line depends on orientation of
quark pair in respect to the collision line. That model also indicates
the relations of the fluctuations of the multiplicity, i.e. the
entanglement entropy, with the background phase transitions
\cite{APS}. More precisely, the fluctuations of entropy are directly
related to the background phase transition, and points to  the
locations of chiral symmetry breaking. Only in particular cases, when
the decay of small black holes to large ones throws out the stable
phase to the zone with no dynamical wall, the background phase
transition is also the confinement/deconfinement transition. In more
general cases, as it has stressed in \cite{2010.09392}, entanglement
entropy is  sensitive to mass gap, i.e. the correlation length. Note
that anisotropy of the background metric also influences on
corresponding jet quenching \cite{1202.4436, 1606.03995,1710.06417}.\\

There also is another source of anisotropy -- an external magnetic
field. Strong magnetic field in the physical 4-dimensional space-time
appears in non-central HIC \cite{0907.1396, 1103.4239, 1111.1949,
  1201.5108, 1604.06231}. This external magnetic field presents
another type of anisotropy and the phenomenon of magnetic catalysis
(MC)/inverse magnetic catalysis (IMC) is associated with it. Strong
magnetic fields are present in neutron stars and magnetars
\cite{1503.06313, 1507.02924}, as well as in the early cosmology
\cite{9308270, 0009061}. So it is important to take into account this
anisotropy in the holographic approach. This anisotropy for improved
holographic QCD with Maxwell action has been considered in
\cite{1505.07894, 1610.04618, 1907.01852, 2004.01965} and for the
model with a stack of branes in the Veneziano limit in
\cite{1611.06339, 1707.00872, 1811.11724}.\\

In this paper we present a fully anisotropic holographic model for
heavy quarks. It is set up by Einstein-dilaton-three-Maxwell
action. This model describes two different types of anisotropy:
anisotropy for producing the multiplicity dependence on energy (this
type of anisotropy is supported by the 2nd Maxwell field in action)
and anisotropy connected  with the magnetic field (this type is
supported by 3rd Maxwell field). The 1st Maxwell field is introduced
to describe chemical potential. We start from a deformed anisotropic
metric and recover the dilaton potential and some of kinetic gauge 
functions from the Einstein-dilaton Maxwell equations. The form of
deformation is chosen to reproduce a typical for heavy quarks phase
structure, supported by lattice calculations. Deformation of the
metric for light quarks is different \cite{1301.0385, 1703.09184,
  2009.05562}, and this agrees with so-called Columbia plot
\cite{Brown:1990ev, 1602.06129, 1912.04827}.\\

This paper is organised as follows. In Sect.\ref{model} the 5-dim
holographic model of hot dense anisotropic QCD in magnetic field is
presented. In Sect.\ref{metricEOM} action and metric to set up the
model are introduced and in Sect.\ref{solution} the corresponding
5-dim BH solution reconstructing heavy quarks model is
discussed. Sect.\ref{thermo} contains thermodynamic properties of the
model in external magnetic field: temperature and entropy behavior
(Sect.\ref{Ts}), magnetic field influence on the solution properties
following from the free energy -- the lifetime of the unstable state
and BH-BH phase transition picture (BH-BH phase transition curve and
critical end points positions) (Sect.\ref{FEBB}), temporal Wilson
loops and crossover regions on the phase diagram with magnetic field
(Sect.\ref{WL}). The main conclusions of this investigation and
subjects for further research are given in
Sect.\ref{conclusions}. Details on magnetic field setting and
following EOM handling are given in Appendix \ref{appendixA}. A review
of holographic heavy quarks models with different polynomial
warp-factors for zero and non zero  magnetic field are presented in
appendices \ref{appendixB1} and \ref{appendixB2} respectively.

\section{Model}\label{model}

\subsection{Metric and EOM}\label{metricEOM}

We take the action in Einstein frame
\begin{gather}
  \begin{split}
    S &= \cfrac{1}{16\pi G_5} \int d^5x \ \sqrt{-g} \ \times \\
    &\qquad \quad \times \left[ R - \cfrac{f_1(\phi)}{4} \ F^{_{(1)}2} 
      - \cfrac{f_2(\phi)}{4} \ F^{_{(2)}2}
      - \cfrac{f_B(\phi)}{4} \ F^{_{(B)}2}
      - \cfrac{1}{2} \ \partial_{\mu} \phi \partial^{\mu} \phi
      - V(\phi) \right],
  \end{split}\label{eq:2.01} \\
  \begin{split}
    F_{\mu\nu}^{(1)} = \partial_{\mu} A_{\nu} - \partial_{\nu} A_{\mu},\
&  {\mbox{i.e.}} \quad A_{\mu}^{(1)} = A_t (z) \delta_\mu^0, \\
    F_{\mu\nu}^{(2)} = q \ dy^1 \wedge dy^2, \ &  {\mbox{i.e.}} \quad
    F_{y_1 y_2}^{(2)} = q,\\
    F_{\mu\nu}^{(B)} = q_B \ dx \wedge dy^1, \ &  {\mbox{i.e.}} \quad
   F_{x y_1}^{(B)} = q_B.
  \end{split}\label{eq:2.02} 
\end{gather}
where $\phi = \phi(z)$ is the scalar field, $f_1(\phi)$, $f_2(\phi)$
and $f_B(\phi)$ are the coupling functions associated with the Maxwell
fields $A_{\mu}$, $F_{\mu\nu}^{(2)}$ and $F_{\mu\nu}^{(B)}$
correspondingly, $q$ and $q_B$ are constants and $V(\phi)$ is the
scalar field potential. Thus \eqref{eq:2.01} is the extended version
of the action used in \cite{1802.05652,2009.05562}, where we add an
external magnetic field $F_{\mu\nu}^{(B)}$\footnote{There also exists
  another choice of $F_{\mu\nu}^{(B)}$-field different from
  $F_{\mu\nu}^{(2)}$. The corresponding EOM can be found in
  Appendix \ref{appendixA}.}.

To consider action \eqref{eq:2.01} let us take the ansatz in the
following view:
\begin{gather}
  ds^2 = \cfrac{L^2}{z^2} \ \fb(z) \left[
    - \, g(z) dt^2 + dx^2 + \left(
      \cfrac{z}{L} \right)^{2-\frac{2}{\nu}} \hspace{-5pt} dy_1^2
    + e^{c_B z^2} \left( \cfrac{z}{L} \right)^{2-\frac{2}{\nu}}
    \hspace{-5pt} dy_2^2
    + \cfrac{dz^2}{g(z)} \right], \label{eq:2.03} \\
  \fb(z) = e^{2{\cA}(z)}, \label{eq:2.04}
\end{gather}
where $L$ is the AdS-radius, $\fb(z)$ is the warp-factor, ${\cA}(z)$
is  related with $ \fb(z)$ according \eqref{eq:2.04}, $g(z)$ is the
blackening function, $\nu$ is the parameter of primary anisotropy,
caused by non-symmetry of heavy-ion collision (HIC), and $c_B$ is the
coefficient of secondary anisotropy related to the external magnetic
field $F_{\mu\nu}^{(B)}$. Choice of ${\cA}(z)$ determines the
heavy/light quarks description of the model, so we follow previous
works and consider ${\cA}(z) = - \, c z^2/4$ for heavy quarks
\cite{1802.05652} and ${\cA}(z) = - \, a \ \ln (b z^2 + 1)$ for
light-quarks \cite{2009.05562}.

The corresponding EOM have the form \footnote{Note, that this is
  transformed EOM, for more details see Appendix \ref{appendixA}.}
\begin{gather}
  \begin{split}
    \phi'' + \phi' \left( \cfrac{g'}{g} + \cfrac{3 \fb'}{2 \fb} -
      \cfrac{\nu + 2}{\nu z} + c_B z \right)
    &+ \left( \cfrac{z}{L} \right)^2 \cfrac{\partial f_1}{\partial
      \phi} \ \cfrac{(A_t')^2}{2 \fb g} \
    - \left( \cfrac{L}{z} \right)^{2-\frac{4}{\nu}} \cfrac{\partial
      f_2}{\partial \phi} \ \cfrac{q^2 \ e^{-c_Bz^2}}{2 \fb g} \ - \\
    &- \left( \cfrac{z}{L} \right)^{\frac{2}{\nu}} \cfrac{\partial
      f_B}{\partial \phi} \ \cfrac{q_B^2}{2 \fb g} \
    - \left( \cfrac{L}{z} \right)^2 \cfrac{\fb}{g} \ \cfrac{\partial
      V}{\partial \phi} = 0,
  \end{split} \label{eq:2.05} \\
  A_t'' + A_t' \left( \cfrac{\fb'}{2 \fb} + \cfrac{f_1'}{f_1} +
    \cfrac{\nu - 2}{\nu z} + c_B z \right) = 0, \label{eq:2.06} \\
   g'' + g' \left(\cfrac{3 \fb'}{2 \fb} - \cfrac{\nu + 2}{\nu z} - c_B
     z \right) 
   - 2 g \left(\cfrac{3 \fb'}{2 \fb} - \cfrac{2}{\nu z} - c_B z
   \right) c_B z
   - \left( \cfrac{z}{L} \right)^2 \cfrac{f_1 (A_t')^2}{\fb} = 0,
   \label{eq:2.07} \\
  \fb'' - \cfrac{3 (\fb')^2}{2 \fb} + \cfrac{2 \fb'}{z}
  - \cfrac{4 \fb}{3 \nu z^2} \left( 1 - \cfrac{1}{\nu}
    + \left( 1 - \cfrac{3 \nu}{2} \right) c_B z^2
    - \cfrac{\nu c_B^2 z^4}{2} \right)
  + \cfrac{\fb \, (\phi')^2}{3} = 0, \label{eq:2.08} \\
  2  g' \ \cfrac{\nu - 1}{\nu}
  + 3  g \ \cfrac{\nu - 1}{\nu} \left(
    \cfrac{\fb'}{\fb} - \cfrac{4 \left( \nu + 1 \right)}{3 \nu z}
    + \cfrac{2 c_B z}{3} \right)
  + \left( \cfrac{L}{z} \right)^{1-\frac{4}{\nu}} \cfrac{L \, q^2 \,
    e^{-c_Bz^2} f_2}{\fb} = 0, \label{eq:2.09} \\
  \begin{split}
    \cfrac{\fb''}{\fb} + \cfrac{(\fb')^2}{2 \fb^2}
    + \cfrac{3 \fb'}{\fb} \left( \cfrac{g'}{2 g}
      - \cfrac{\nu + 1}{\nu z}
      + \cfrac{2 c_B z}{3} \right)
    &- \cfrac{g'}{3 z g}  \left( 5 + \cfrac{4}{\nu} - 3 c_B z^2
    \right)
    + \cfrac{8}{3 z^2} \left( 1 + \cfrac{3}{2 \nu} + \cfrac{1}{2
        \nu^2} \right) - \\
    &- \cfrac{4 c_B}{3} \left( 1 + \cfrac{3}{2 \nu} - \cfrac{c_B
        z^2}{2} \right)
    + \cfrac{g''}{3 g} + \cfrac{2}{3} \left( \cfrac{L}{z} \right)^2
    \cfrac{\fb V}{g} = 0,
  \end{split} \label{eq:2.10}
\end{gather}
where $'= \partial/\partial z$ and
\begin{gather}
 f_{B} = 2 \left( \cfrac{z}{L} \right)^{-\frac{2}{\nu}} \fb g \
  \cfrac{c_B z}{q_B^2} \left( \cfrac{3 \fb'}{2 \fb} -
    \cfrac{2}{\nu z} + c_B z + \cfrac{g'}{g} \right). \label{eq:2.11}
\end{gather}

We use the boundary conditions \cite{2009.05562}:
\begin{gather}
  A_t(0) = \mu, \quad A_t(z_h) = 0, \label{eq:2.17} \\
  g(0) = 1, \quad g(z_h) = 0, \label{eq:2.18} \\
  \phi(z_0) = 0. \label{eq:2.19}
\end{gather}
where $z_h$ is a size of horizon and $z_0$ is the integration
boundary, $0\leq z_0\leq z_h$. Taking $z_0 = 0$ repeats consideration
in \cite{1703.09184} and $z_0 = z_h$ was used in \cite{1802.05652}. The
form of the integration boundary $z_0 = z_0(z_h)$ determines the
string tension behavior in the model. This aspect was discussed in
more details in \cite{2009.05562, S-dubna}.

Excluding anisotropy and normalizing to the AdS-radius, i.e. putting
$L = 1$, $\nu = 1$ and $f_2 = f_B = c_B = 0$ into
\eqref{eq:2.05}--\eqref{eq:2.10}, we get the expressions that fully
coincide with the EOM (2.12)--(2.16) from \cite{1802.05652}. 


\subsection{Solution for heavy quarks model}\label{solution}



To solve EOM (\ref{eq:2.05}--\ref{eq:2.10}) we first need to determine
the form of the coupling function $f_1$. To do this we base on our
previous experience. In anisotropic heavy quarks \cite{1802.05652} and
light quarks models \cite{2009.05562} we used the following
expressions:
\begin{gather}
  f_{1\,HQ} = z^{-2+\frac{2}{\nu}}, \qquad
  f_{1\,LQ} = e^{-cz^2-{\cA}(z)} \ z^{-2+\frac{2}{\nu}} \label{eq:3.01}.
\end{gather}

To describe heavy quarks' behavior holographically, let us take $\fb =
e^{-\frac{cz^2}{2}}$and $f_1 = z^{-2+\frac{2}{\nu}}$, therefore system
(\ref{eq:2.05}--\ref{eq:2.10}) has the solution \footnote{More
  general forms  of $P(z)$ do not admit explicit express for $A_t$ and
  $g$, and will be considered separately.}
\begin{gather}
  A_t = \mu \ \cfrac{e^{\frac{1}{4}(c-2c_B)z^2} -
    e^{\frac{1}{4}(c-2c_B)z_h^2}}{1 -
    e^{\frac{1}{4}(c-2c_B)z_h^2}} \, , \label{eq:3.06} \\
  \begin{split}
    g &= e^{c_B z^2} \left\{ 1
      - \cfrac{\Gamma\left(1 + \frac{1}{\nu} \ ; 0\right) -
        \Gamma\left(1 + \frac{1}{\nu} \ ; \frac{3}{4} (2 c_B - c)
          z^2\right)}{\Gamma\left(1 + \frac{1}{\nu} \ ; 0\right) - 
        \Gamma\left(1 + \frac{1}{\nu} \ ; \frac{3}{4} (2 c_B - c)
          z_h^2 \right)} \right. - \\ 
    &\qquad \qquad \quad - \cfrac{\mu^2 \ (2 c_B -
      c)^{-\frac{1}{\nu}}}{4 L^2 \left( 1 -
        e^{(c-2c_B)\frac{z_h^2}{4}} \right)^2} \
    \left(\Gamma\left(1 + \frac{1}{\nu} \ ; 0\right) -
      \Gamma\left(1 + \frac{1}{\nu} \ ; \frac{3}{4} \ (2 c_B - c) \
        z^2\right)\right) \times \\
    &\left. \times \left[ 1
        - \cfrac{\Gamma\left(1 + \frac{1}{\nu} \ ; 0\right) -
          \Gamma\left(1 + \frac{1}{\nu} \ ; \frac{3}{4} \ (2 c_B - c)
            z^2\right)}{\Gamma\left(1 + \frac{1}{\nu} \ ; 0\right) -  
          \Gamma\left(1 + \frac{1}{\nu} \ ; \frac{3}{4} \ (2 c_B - c)
            z_h^2 \right)} \ 
        \cfrac{\Gamma\left(1 + \frac{1}{\nu} \ ; 0\right) -
          \Gamma\left(1 + \frac{1}{\nu} \ ; (2 c_B - c)
            z_h^2\right)}{\Gamma\left(1 + \frac{1}{\nu} \ ; 0\right) -
          \Gamma\left(1 + \frac{1}{\nu} \ ; (2 c_B - c) z^2
          \right)}\right] \right\}, \label{eq:3.07}
  \end{split} \\
  f_B = - \, 2 \left( \cfrac{z}{L} \right)^{-\frac{2}{\nu}}
  e^{-\frac{1}{2}cz^2} \ \cfrac{c_B z}{q_B^2} \ g \left( \cfrac{3 c
      z}{2} + \cfrac{2}{\nu z} - c_B z - \cfrac{g'}{g}
  \right), \label{eq:3.08} \\
  f_2 = 4 \left( \cfrac{z}{L} \right)^{2-\frac{4}{\nu}}
  e^{-\frac{1}{2}(c-2c_B)z^2} \ \cfrac{\nu - 1}{q^2 \nu z} \ g \left(
    \cfrac{\nu + 1}{\nu z} + \cfrac{3 c - 2 c_B}{4} \ z -
    \cfrac{g'}{2g} \right), \label{eq:3.09} \\
  \phi = \int\limits_{z_0}^z \cfrac{1}{\nu \xi} \
  \sqrt{4 \nu - 4 
    + \left( 4 \nu c_B + 3 (3 c - 2 c_B) \nu^2 \right) \xi^2 
    + \left( \cfrac{3}{2} \ \nu^2 c^2 - 2 c_B^2 \right) \xi^4}
  \ d \xi, \label{eq:3.10} 
\end{gather}
\begin{gather}
  \begin{split}
    V = - \, \cfrac{e^{\frac{1}{2}cz^2}}{4 L^2 \nu^2} \ &\biggl\{
    \bigl[ 
    8 (1 + 2 \nu) (1 + \nu) + 2 (3 + 2 \nu) (3 c - 2 c_B) \nu z^2 + (3
    c - 2 c_B)^2 \nu^2 z^4 \bigr] g \ - \\
    &\quad - \left[ 2 (4 + 5 \nu) + 3 (3 c - 2 c_B) \nu z^2 \right] g'
    + 2 g'' \nu^2 z^2 \biggr\}. \label{eq:3.11}
  \end{split}
\end{gather}
Note that the solution (\ref{eq:3.06}--\ref{eq:3.11}) was obtained for
the external magnetic field set as $F_{x y_1}^{(B)} = q_B$. 

To keep in touch not only with our previous model for heavy quarks
\cite{1802.05652}, but also with our holographic description for light
quarks \cite{2009.05562} we assume $c = 0.227$.

\begin{figure}[t!]
  \centering
  \includegraphics[scale=0.2]{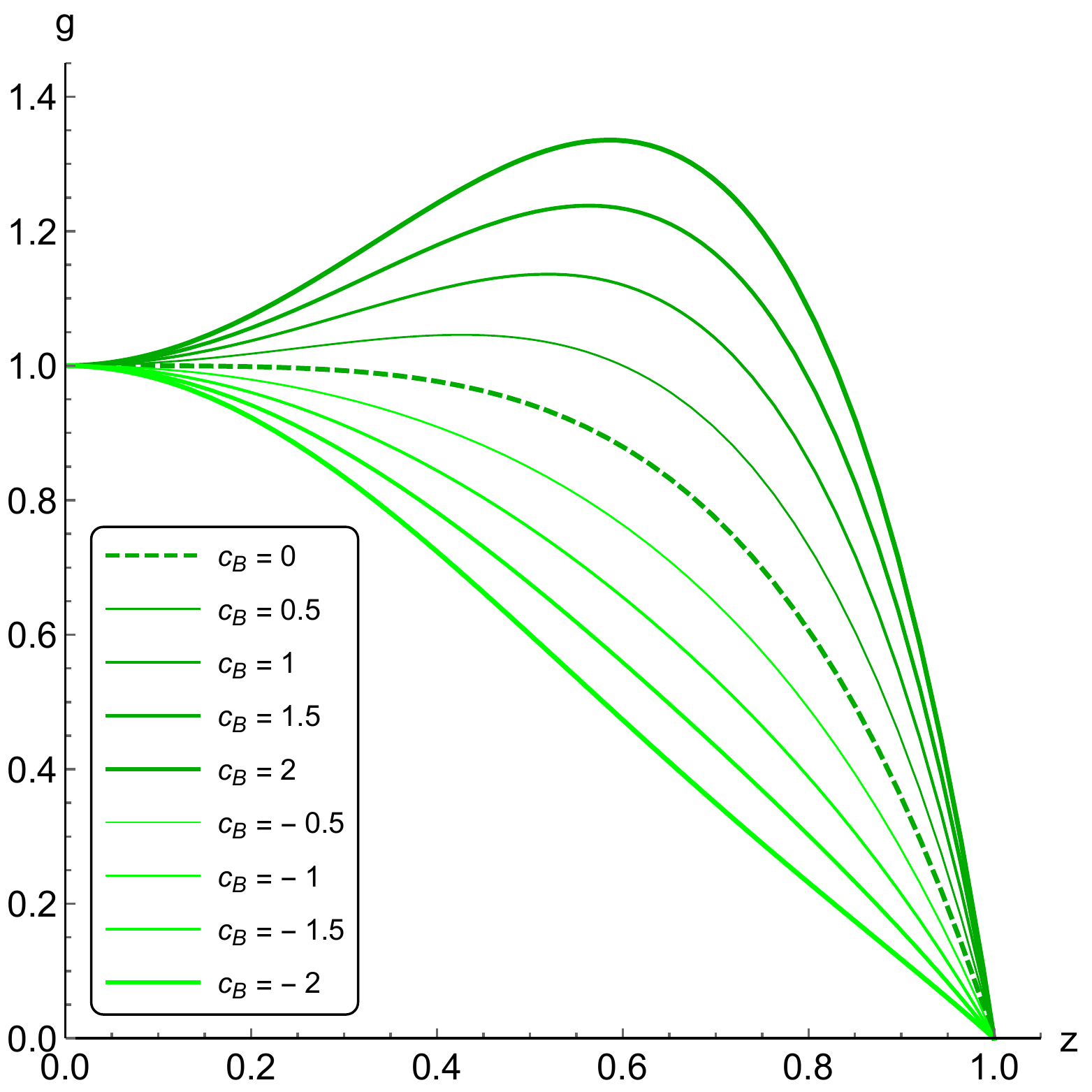} \quad
  \includegraphics[scale=0.2]{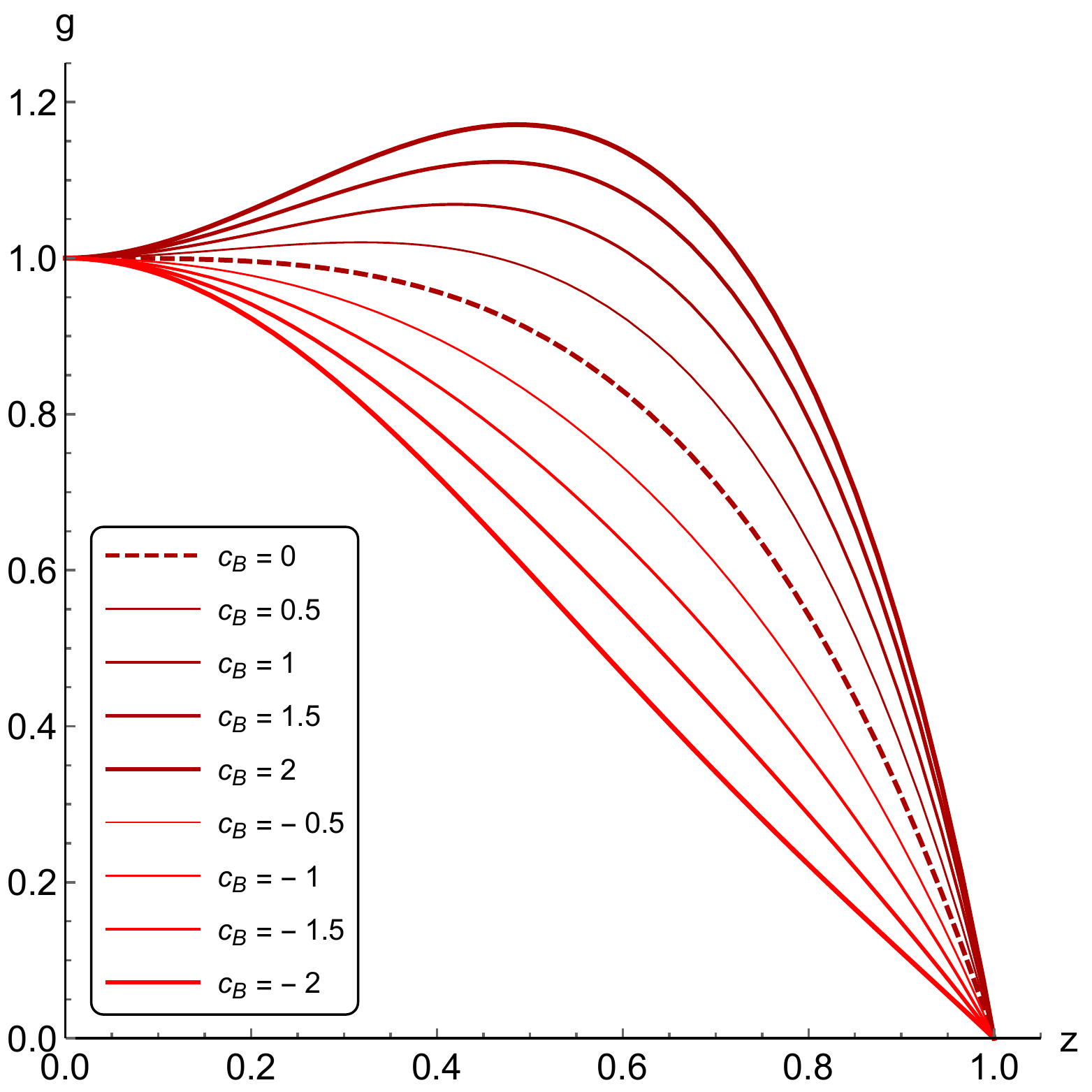} \quad  
  \includegraphics[scale=0.2]{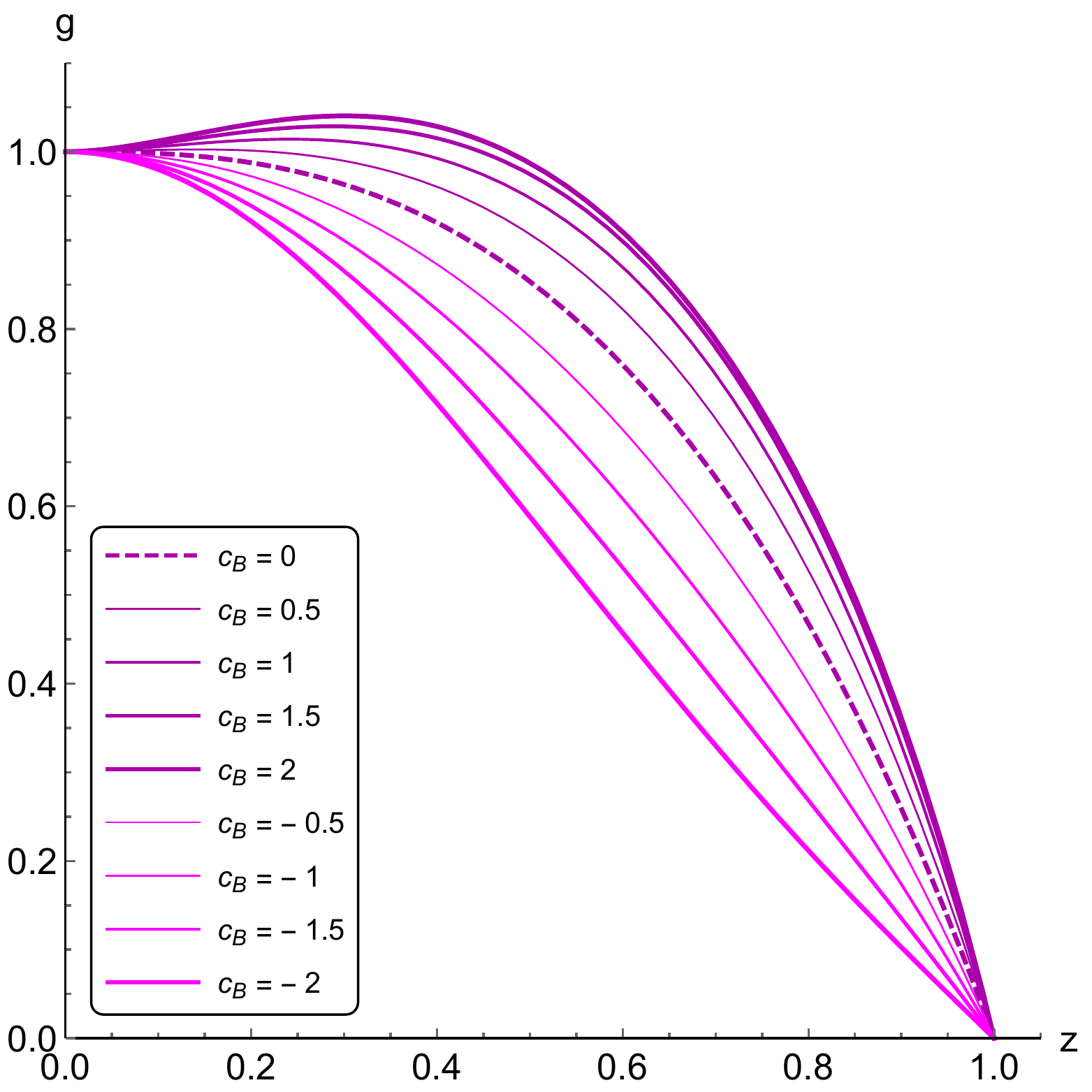} \quad
  \includegraphics[scale=0.2]{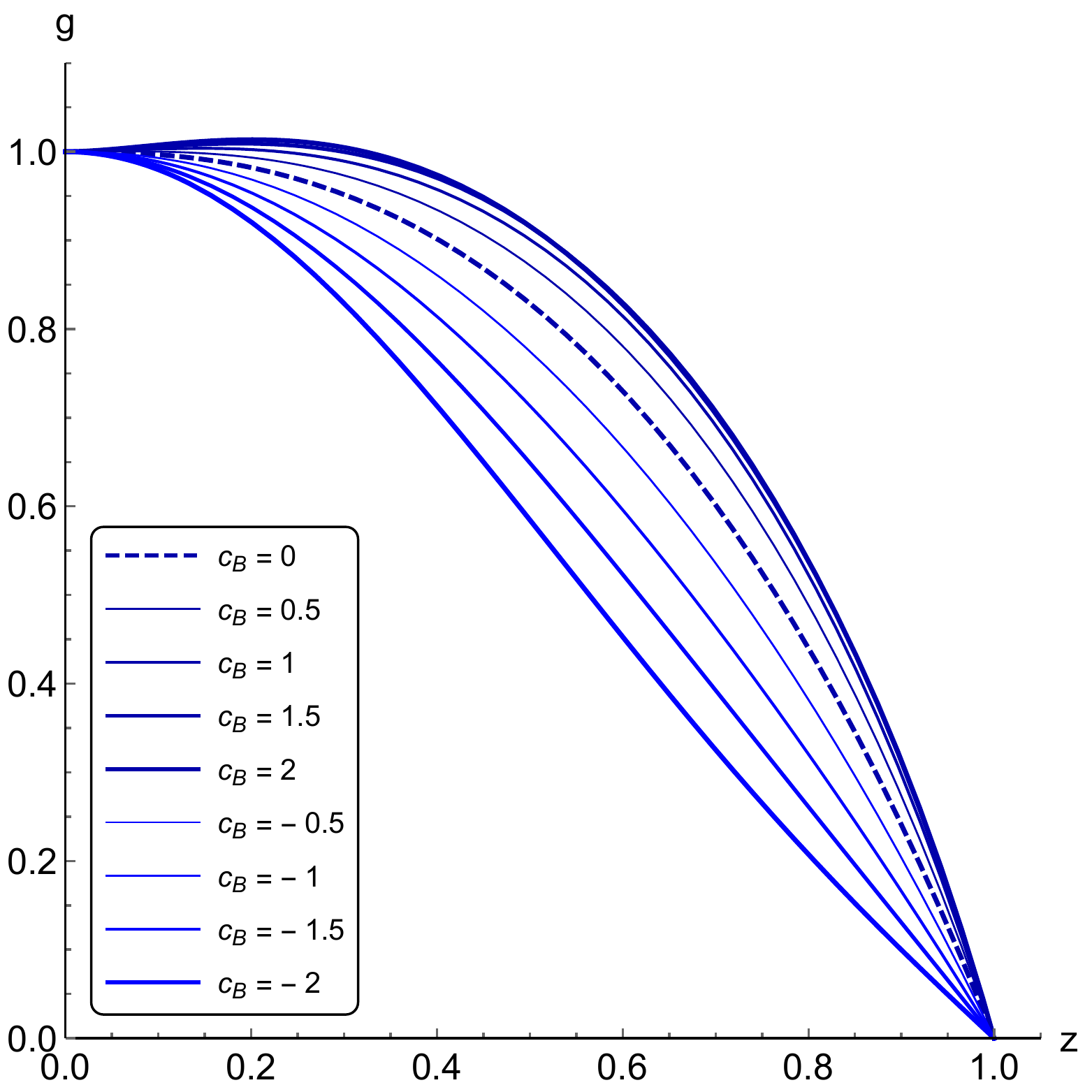} \\
  A \hspace{90pt} B \hspace{90pt}
  C \hspace{90pt} D
  \caption{Blackening function $g(z)$ for different $c_B$ for $\nu =
    1$ (A), $\nu = 1.5$ (B), $\nu = 3$ (C), $\nu = 4.5$ (D); $c =
    0.227$, $z_h = 1$, $\mu = 0$.}
  \label{Fig:gzcb}
\end{figure}
\begin{figure}[t!]
  \centering
  \includegraphics[scale=0.26]{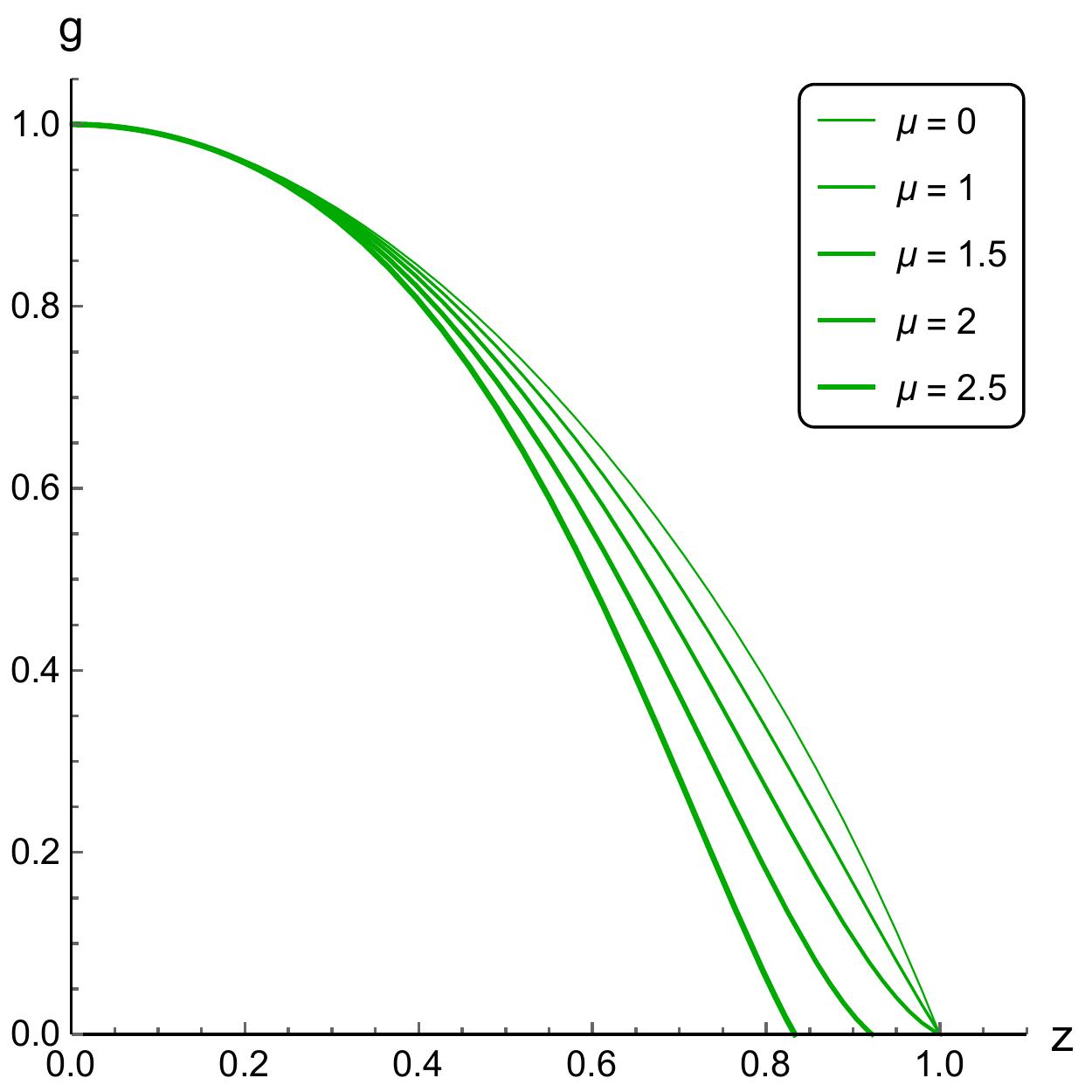} \quad
  \includegraphics[scale=0.26]{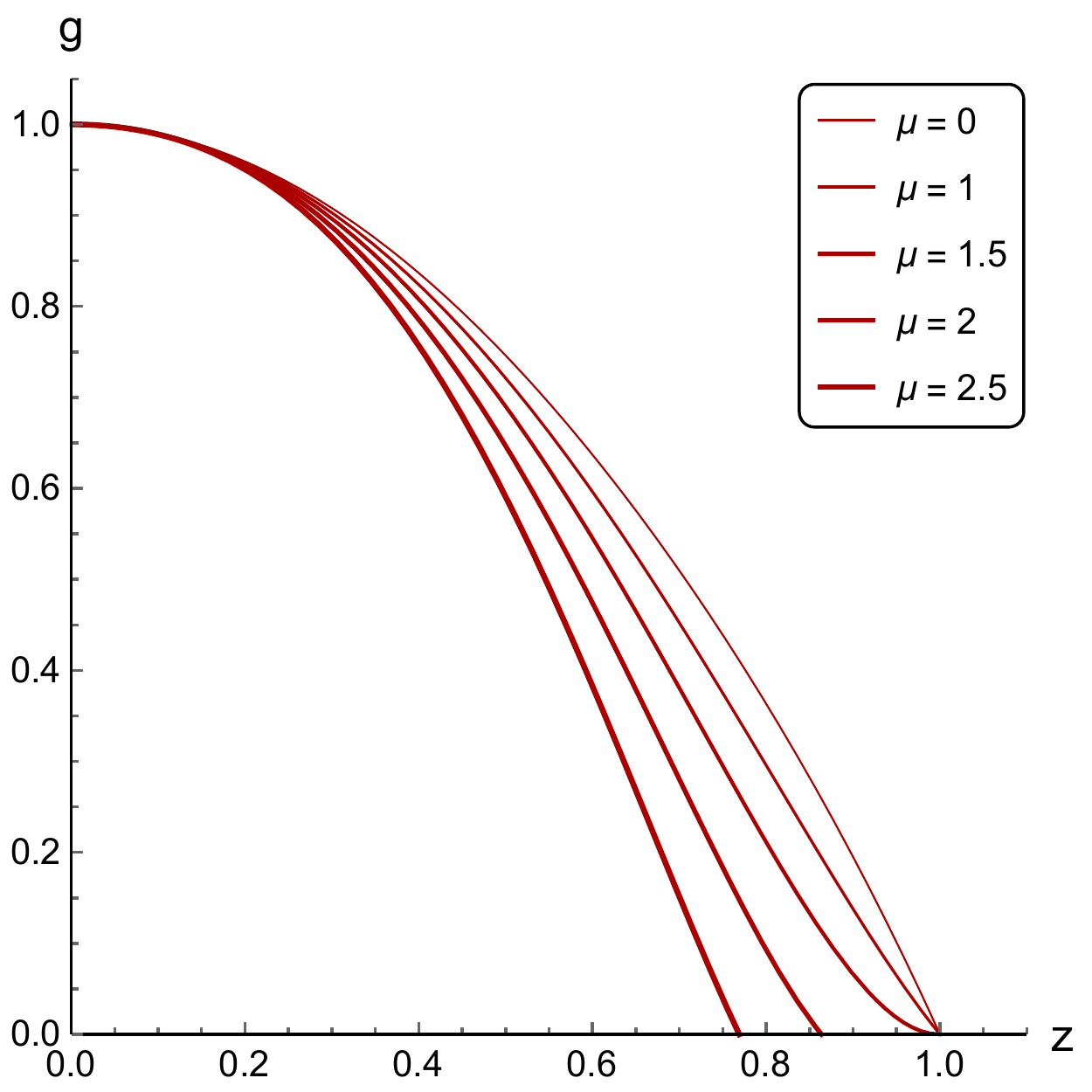} \quad
  \includegraphics[scale=0.26]{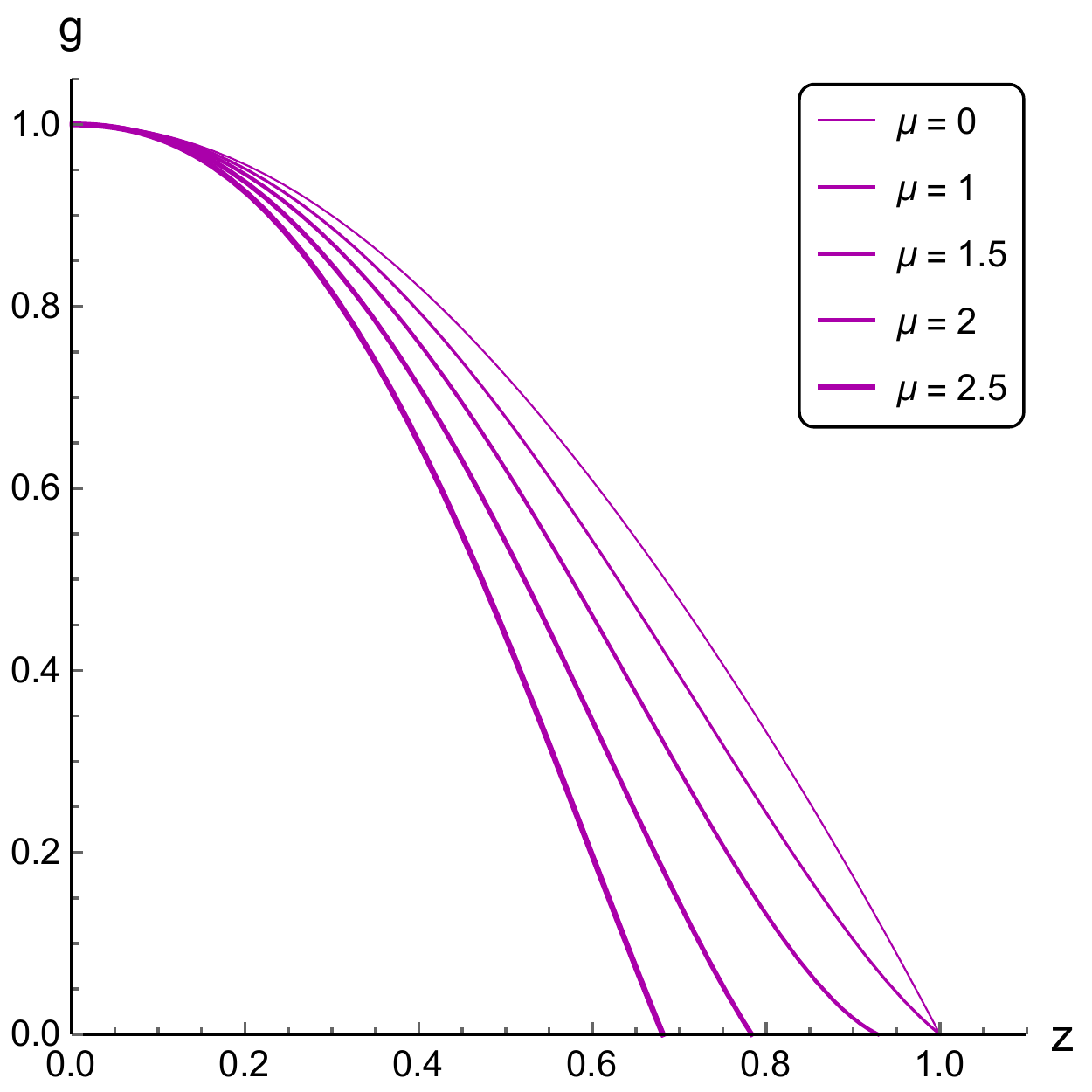} \quad
  \includegraphics[scale=0.26]{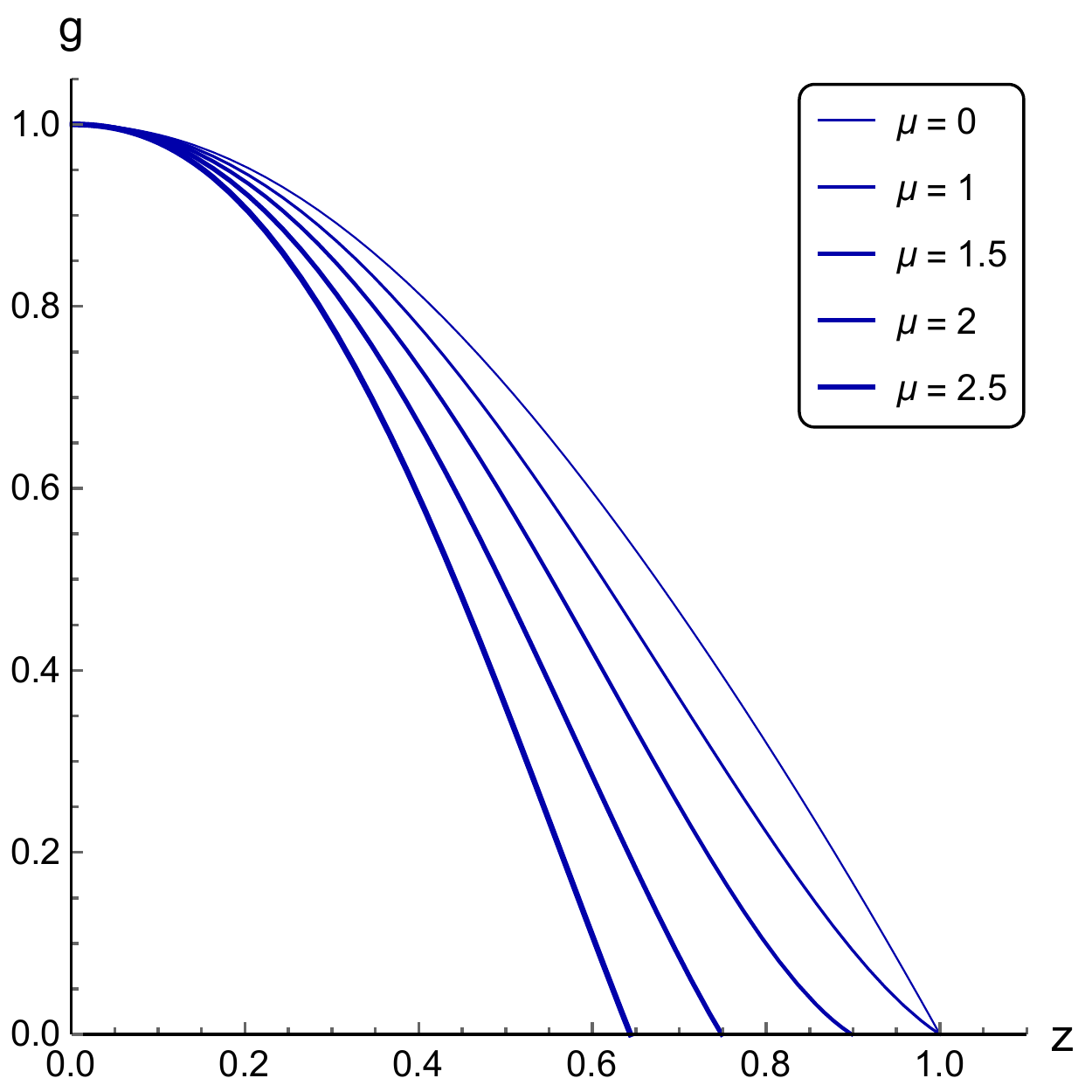} \\
  A \hspace{90pt} B \hspace{90pt}
  C \hspace{90pt} D
  \caption{Blackening function $g(z)$ for different $\mu$ for $\nu =
    1$ (A), $\nu = 1.5$ (B), $\nu = 3$ (C), $\nu = 4.5$ (D); $c =
    0.227$, $z_h = 1$, $c_B = - \, 1$.}
  \label{Fig:gzmu}
\end{figure}
\begin{figure}[t!]
  \centering
  \includegraphics[scale=0.22]{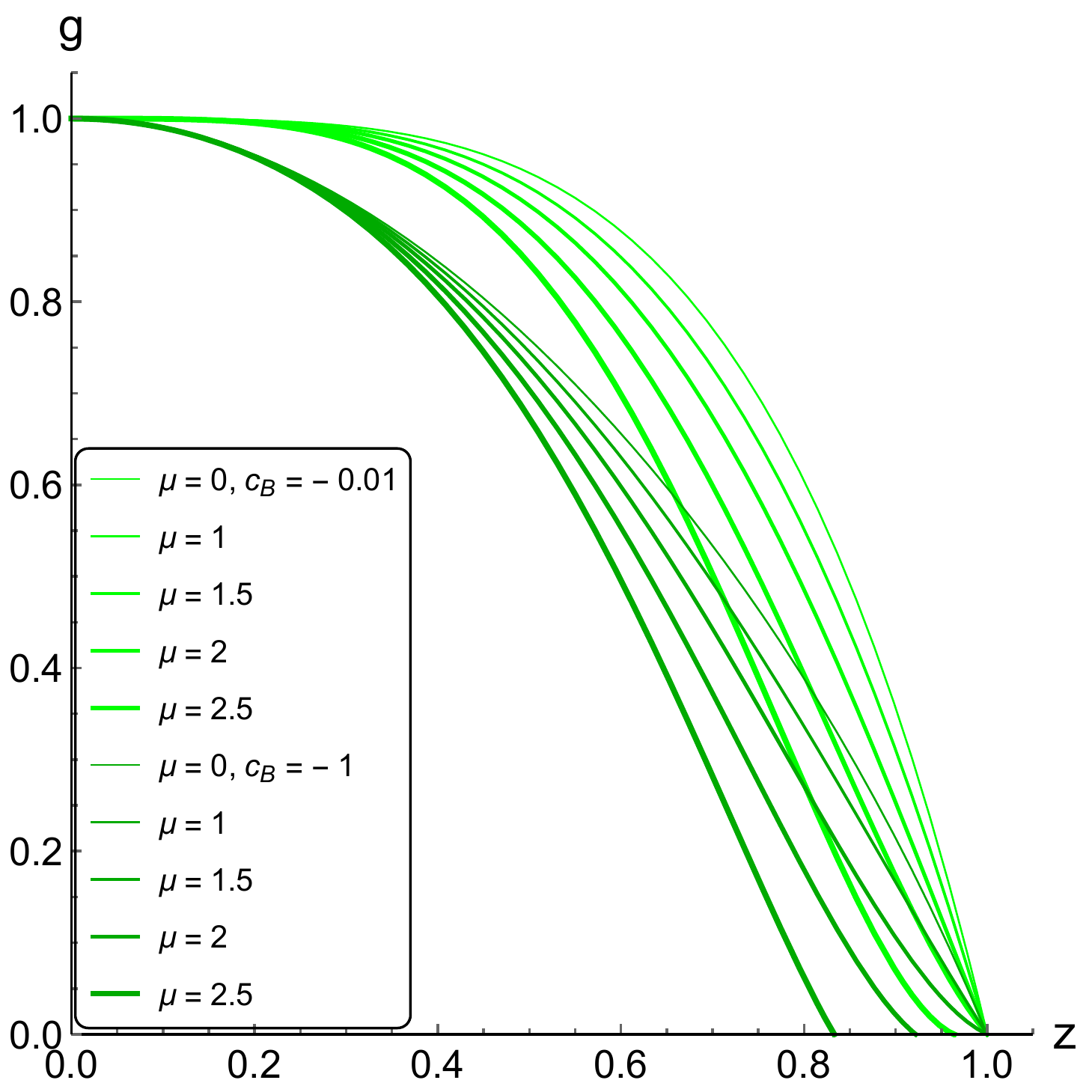} \quad
  \includegraphics[scale=0.22]{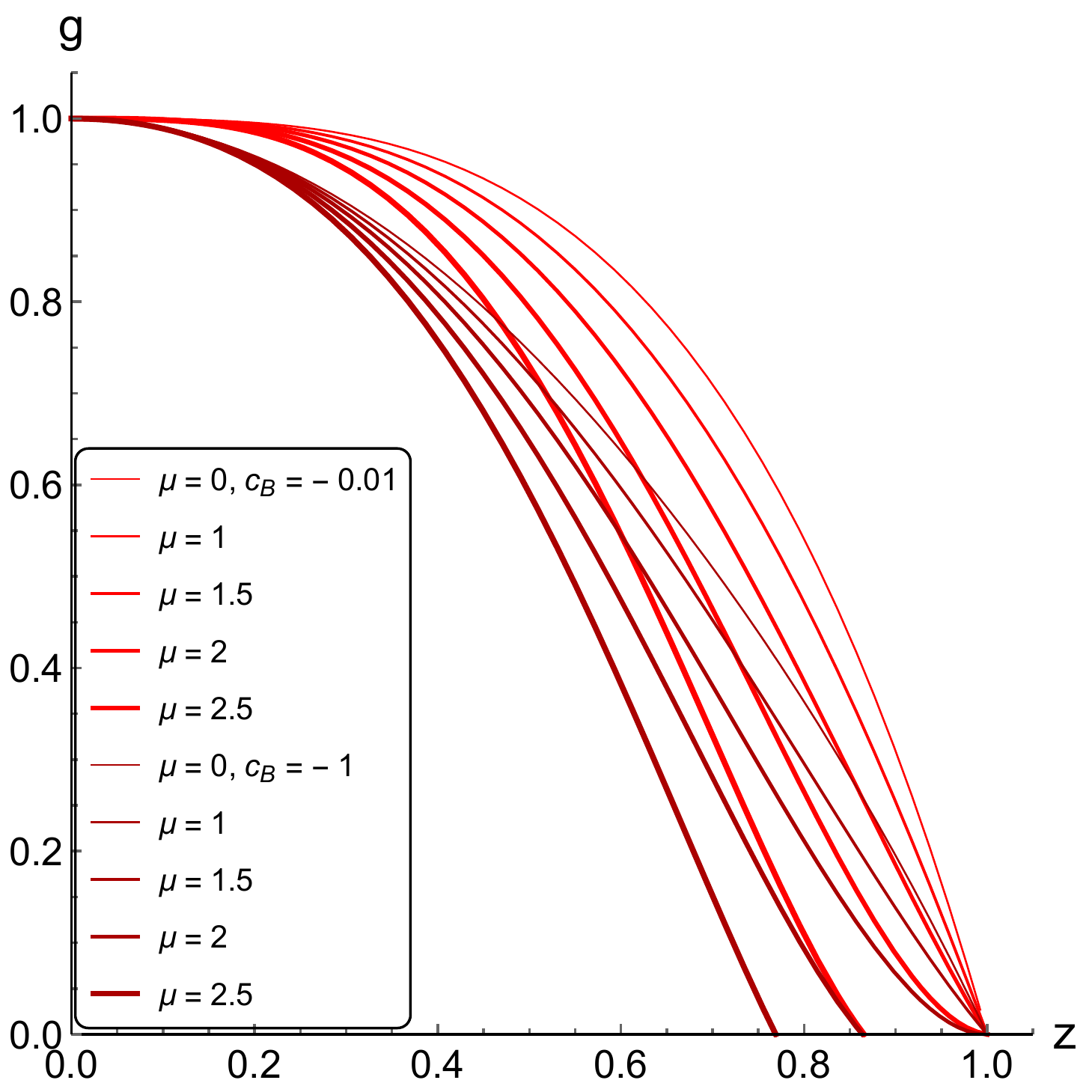} \quad
  \includegraphics[scale=0.22]{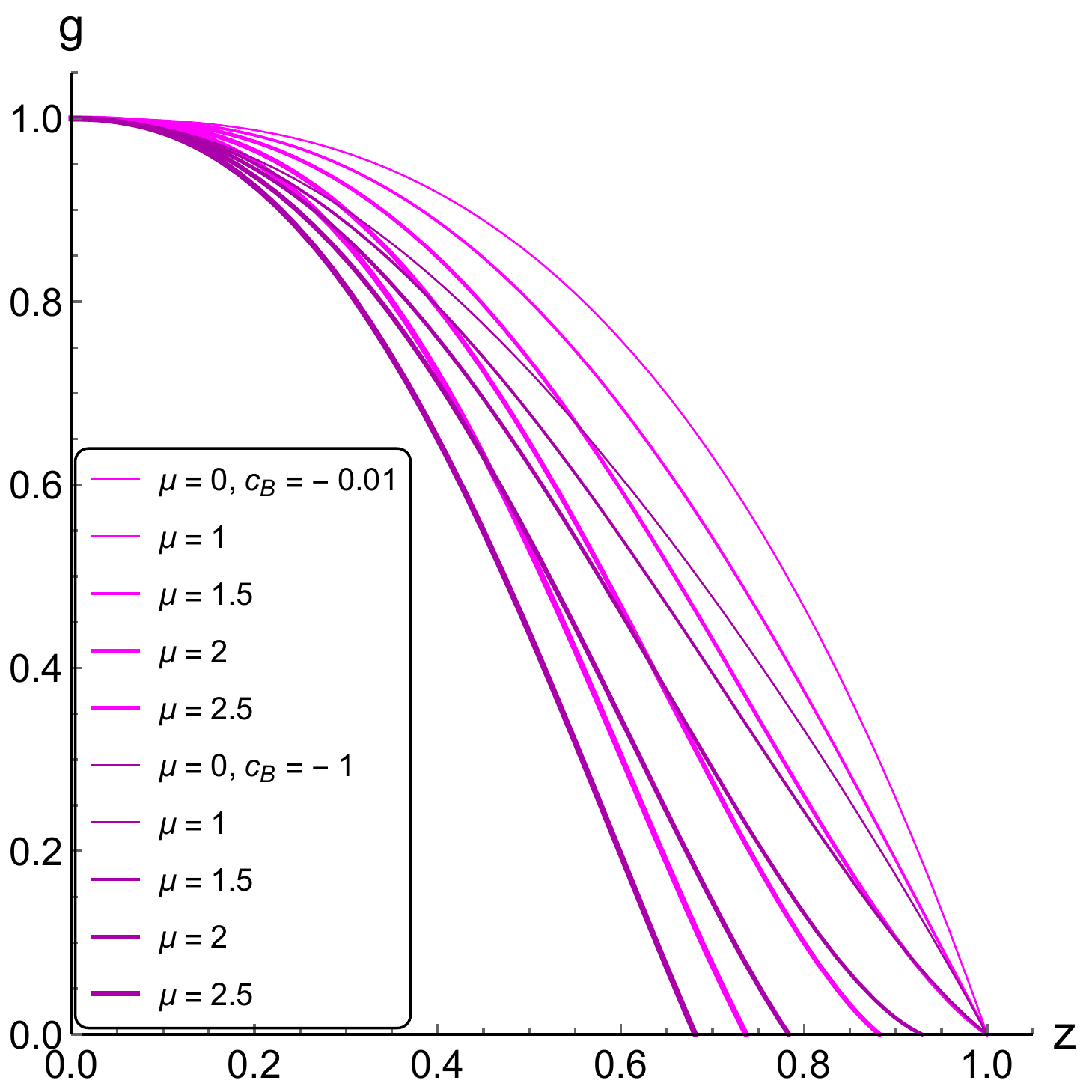} \quad
  \includegraphics[scale=0.22]{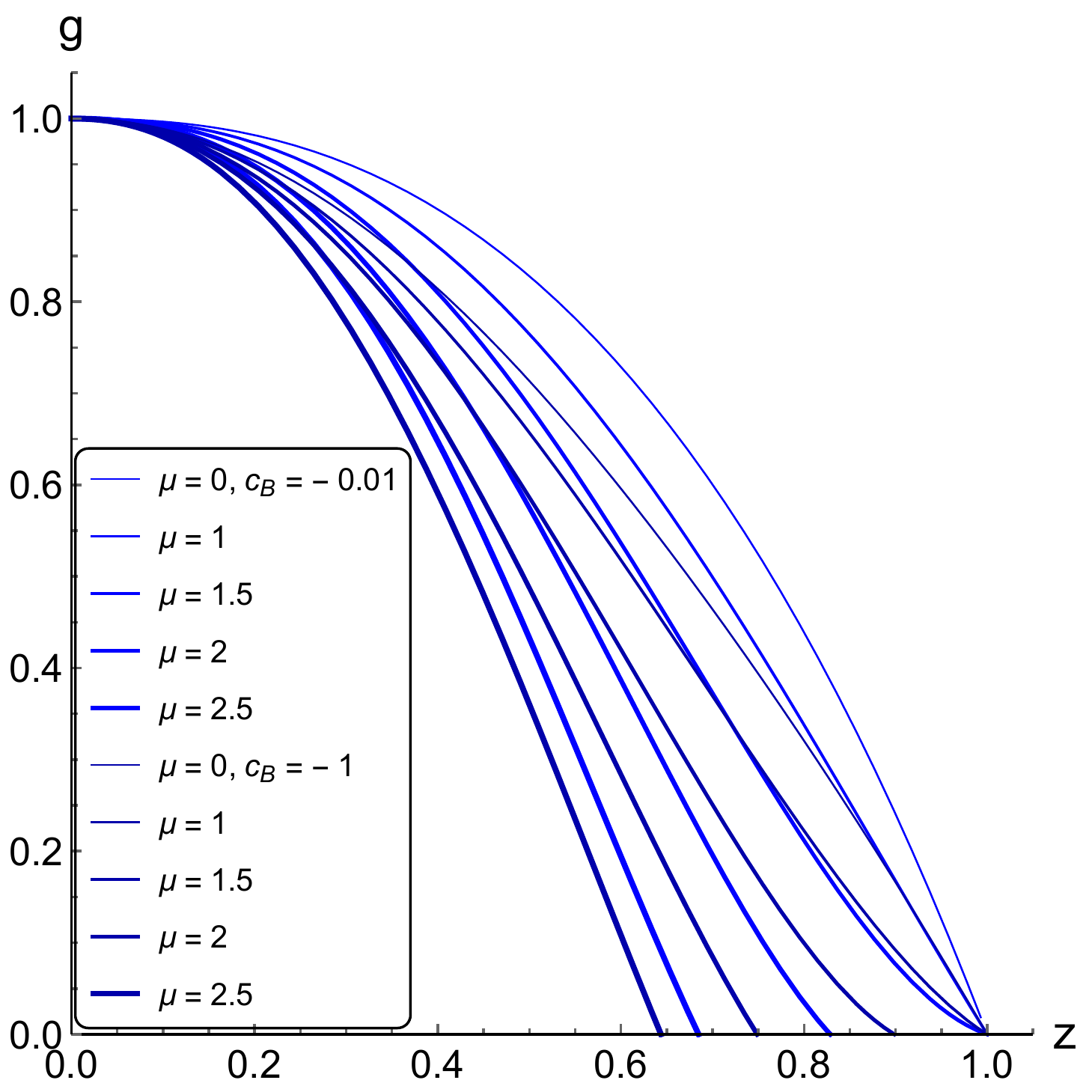} \\
  A \hspace{90pt} B \hspace{90pt}
  C \hspace{90pt} D
  \caption{Blackening function $g(z)$ for different $\mu$ for $\nu =
    1$ (A), $\nu = 1.5$ (B), $\nu = 3$ (C), $\nu = 4.5$ (D); $c =
    0.227$, $z_h = 1$, $c_B = - \, 0.01$ (light curves) and $c_B = - \
    1$ (dark curves).}
  \label{Fig:gzmucb}
\end{figure}

As we can see from (\ref{eq:3.07}), blackening function doesn't
actually depend on magnetic ``charge'' $q_B$, but it depends on
coupling coefficient $c_B$, that characterises influence of the
external magnetic field on metric along the $y_2$-direction
(Fig.\ref{Fig:gzcb}). Positive $c_B$ causes the appearence of local
maximum of the blackening function under the horizon. This maximum is
expressed the brighter the higher the $c_B$ value is. Negative $c_B$
decreases the blackening function values till the horizon. Putting
$c_B = 0$ brings us back to the previous model without the external
magnetic field \cite{1802.05652}. The only restriction on parameters'
values that we obtain here is $c \ne 2 c_B$, since for $c = 2 c_B$
leads to singular behavior of curvature invariants.

We set $q_b = 1$ in most of the following calculations. 

Comparing Fig.\ref{Fig:gzcb}.A-D one can notice, that the primary QGP
anisotropy, paramet\-riz\-ed by $\nu$, suppresses the external
magnetic field influence on the blackening function. Moreover, this
suppression isn't equal for positive and negative values of coupling
coefficient $c_B$. For $\nu = 4.5$ (Fig.\ref{Fig:gzcb}.D) difference
between $c_B = 0$ and $c_B = 0.5$ is the smallest comparing to the
lesser anisotropy cases (Fig.\ref{Fig:gzcb}.A-C), while curves for
$c_B = 1.5$ and $c_B = 2$ can be hardly distunguished from each
other.

On Fig.\ref{Fig:gzmu} blackening function depending on chemical
potential for $c_B = - \, 1$ and primary anisotropy ($1 \le \nu \le
4.5$) is plotted. One can see that the presence of magnetic field
doesn't change the picture in general: increasing chemical potential
still leads to decreasing of the blackening function values and to the
shrinkage of the black hole horizon value as it was in
\cite{1802.05652}. Increase of primary anisotropy (larger $\nu$
values) enhances the effect of the chemical potential on the
blackening function behavior.

On Fig.\ref{Fig:gzmucb} blackening function dependence on $\mu$ for
different $c_B$ is compared. For larger absolute value of $c_B$
chemical potential demonstrates weaker effect of blackening function
for small $z$, but stronger influence on the horizon position. In the
absence of primary anisotropy ($\nu = 1$ on Fig.\ref{Fig:gzmucb}.A)
this influence on the horizon position is the most obvious.




\begin{figure}[t!]
  \centering
  \includegraphics[scale=0.21]{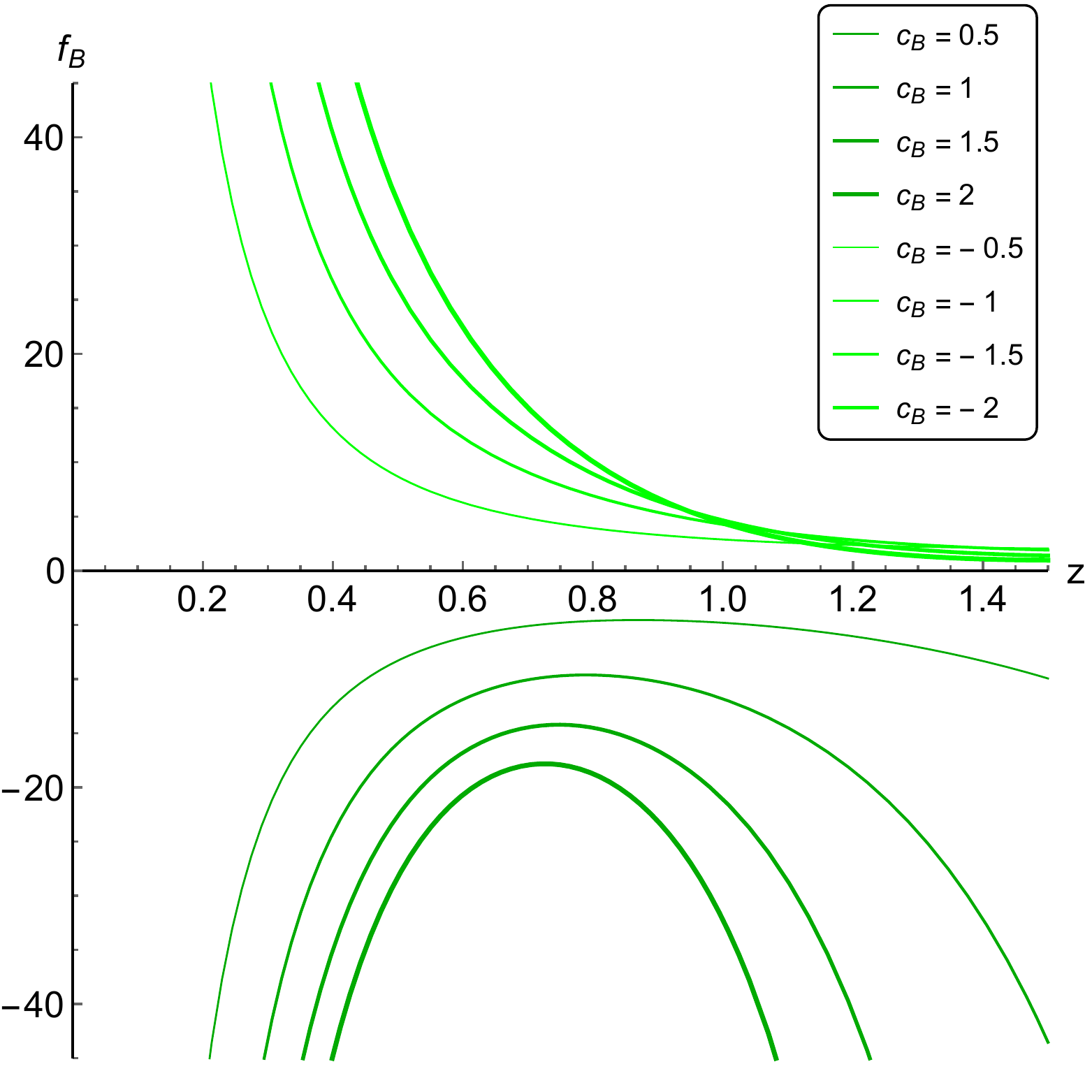} \quad
  \includegraphics[scale=0.21]{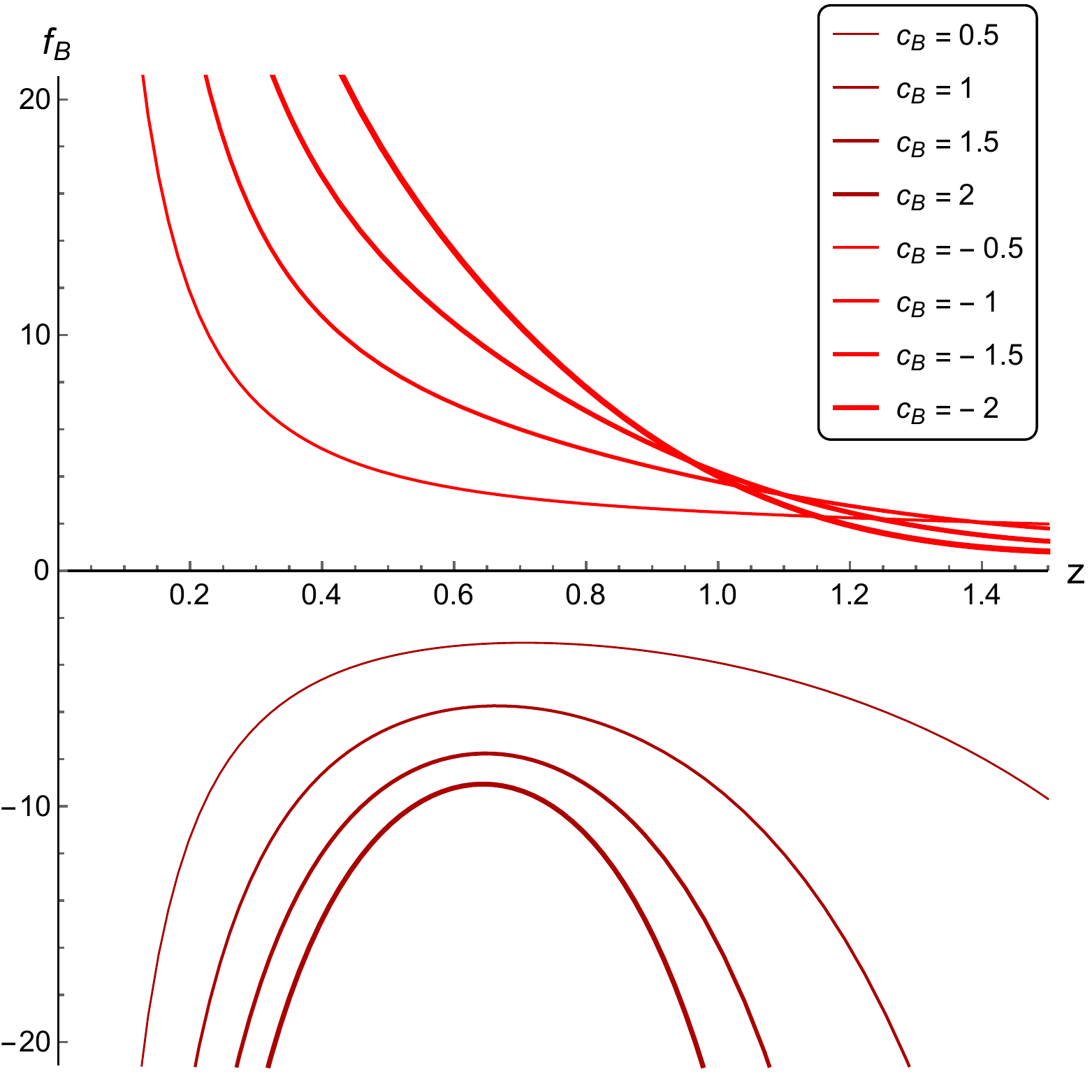} \quad
  \includegraphics[scale=0.21]{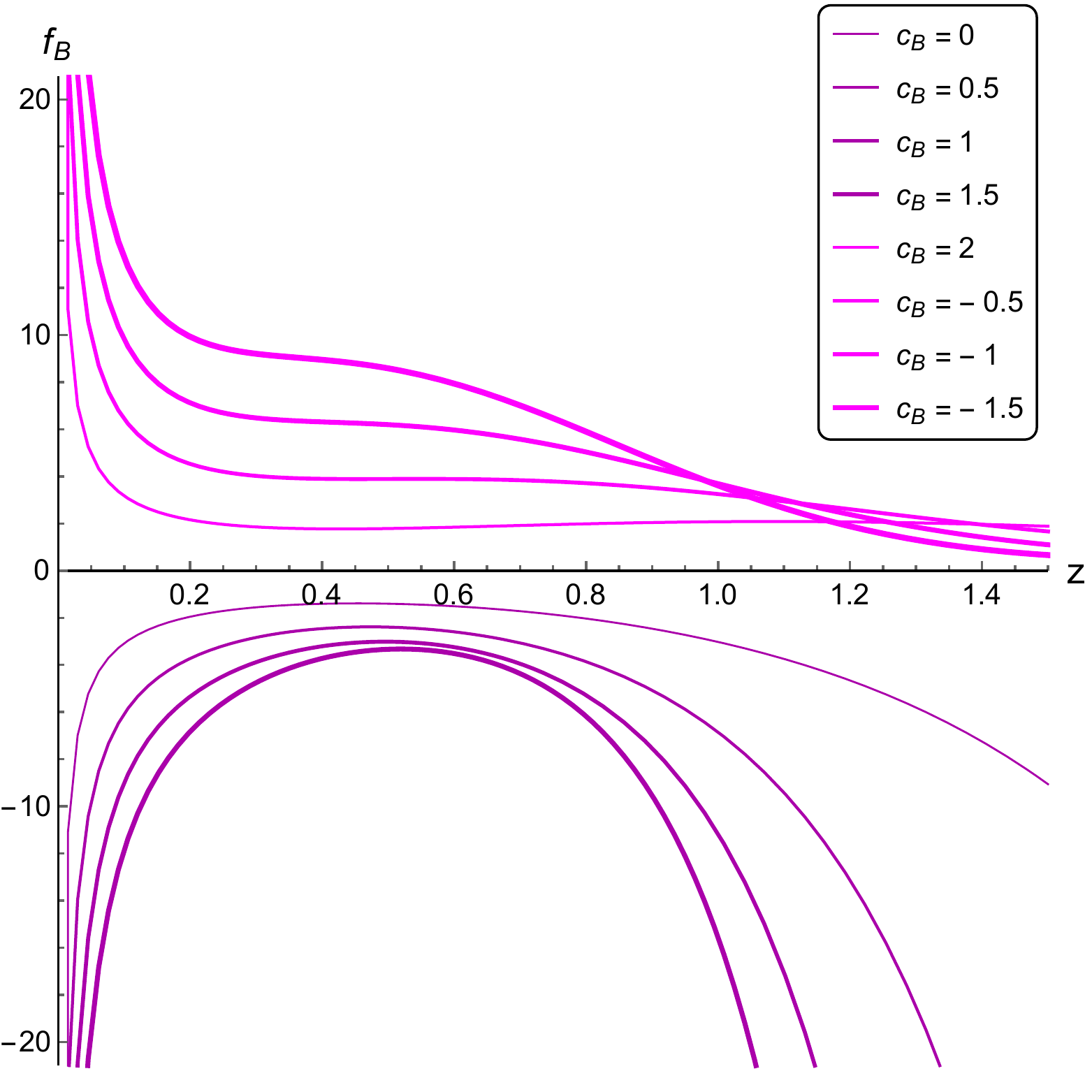} \quad
  \includegraphics[scale=0.21]{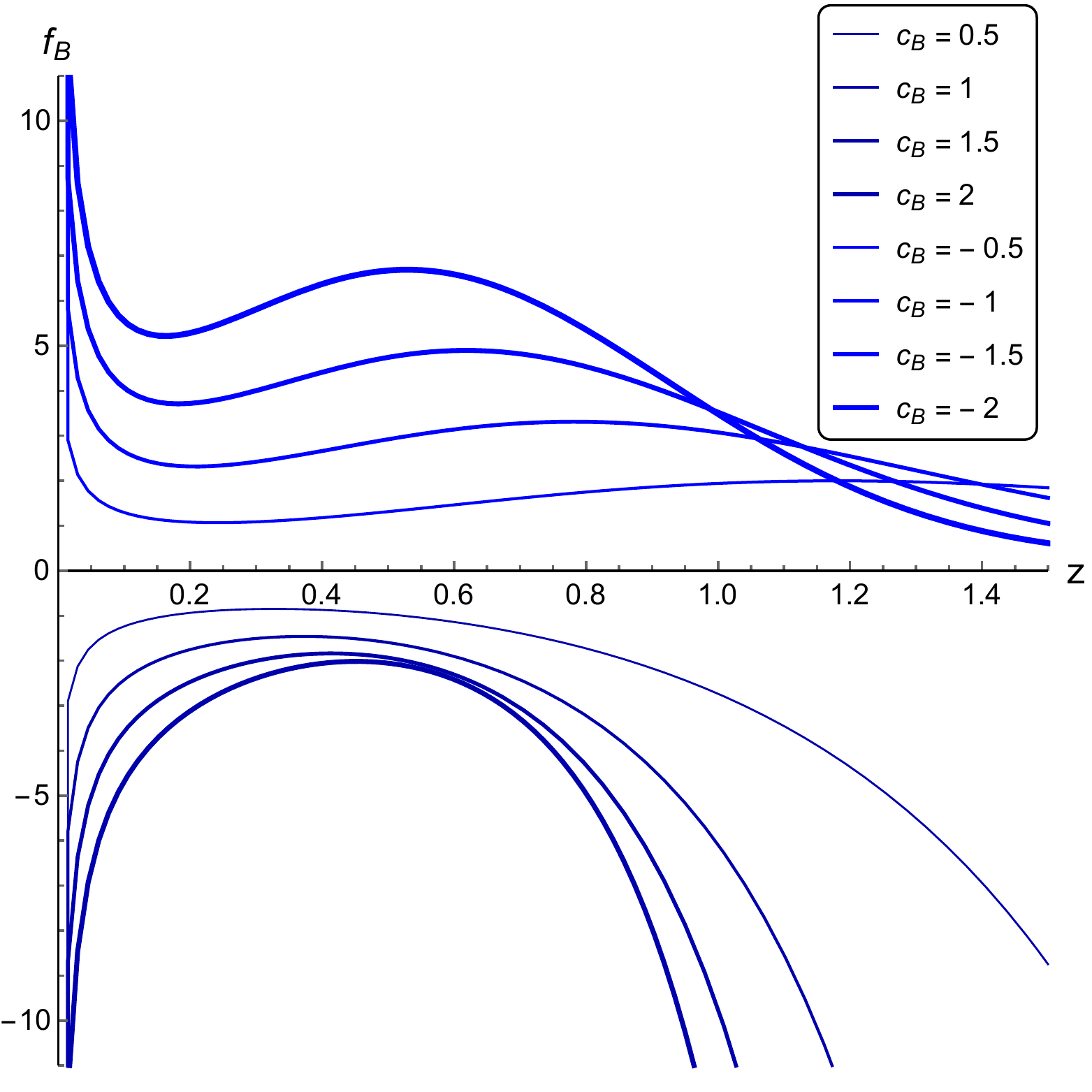} \\
  A \hspace{90pt} B \hspace{90pt}
  C \hspace{90pt} D
  \caption{Coupling function $f_B(z)$ for different $c_B$ for $\nu =
    1$ (A), $\nu = 1.5$ (B), $\nu = 3$ (C), $\nu = 4.5$ (D); $c =
    0.227$, $z_h = 1$, $\mu = 0$, $q_B = 1$.}
  \label{Fig:fBzcb}
\end{figure}
\begin{figure}[t!]
  \centering
  \includegraphics[scale=0.23]{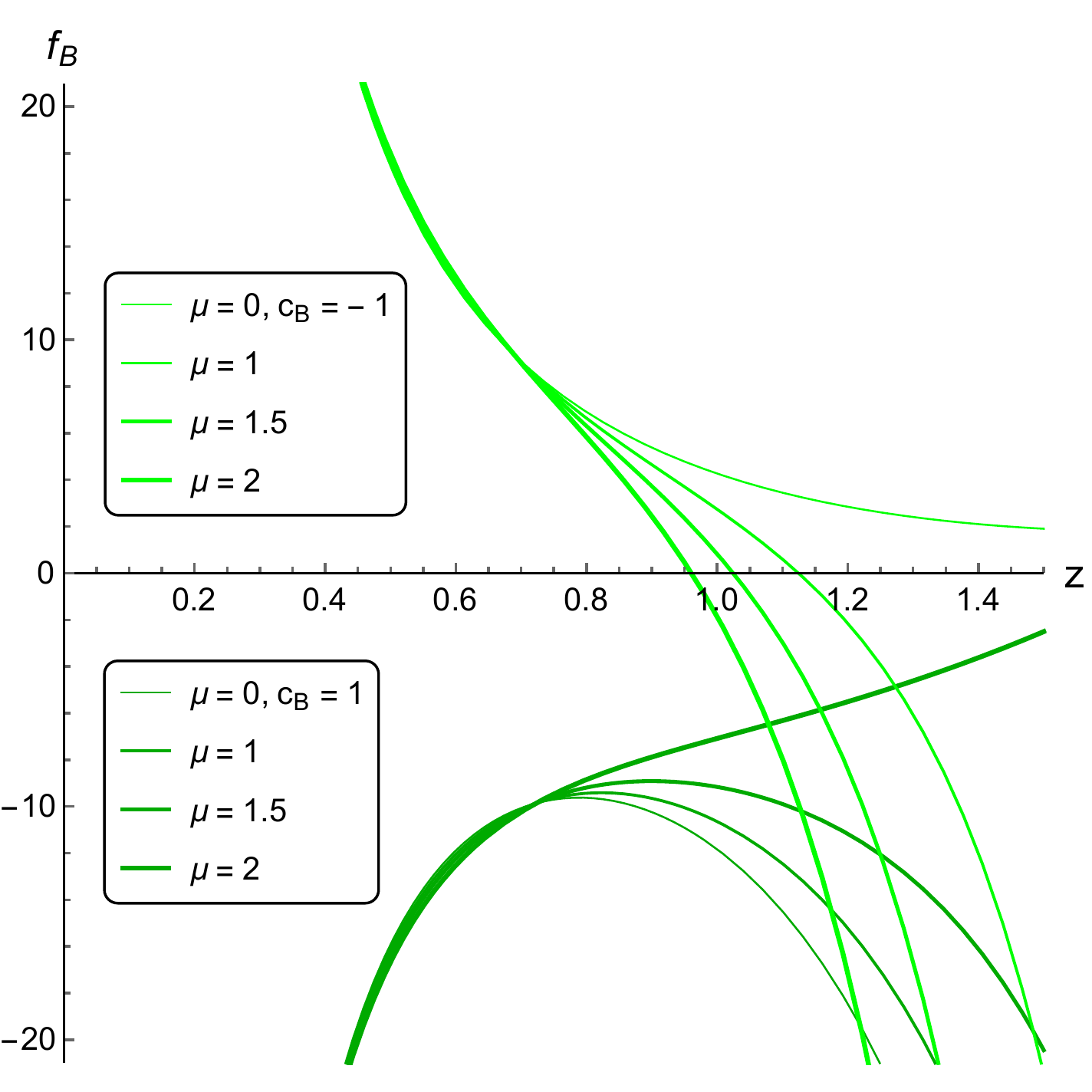} \quad
  \includegraphics[scale=0.23]{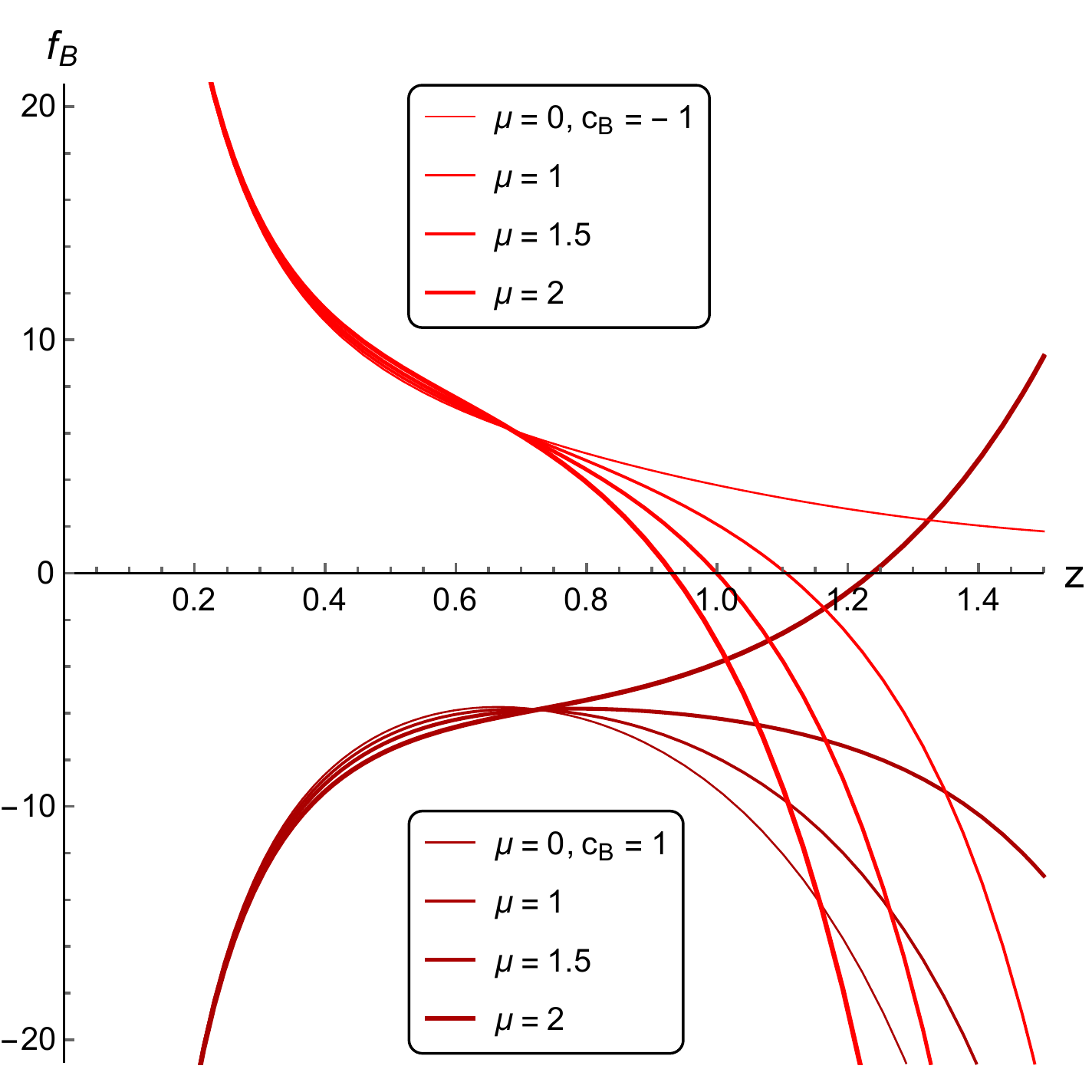} \quad
  \includegraphics[scale=0.23]{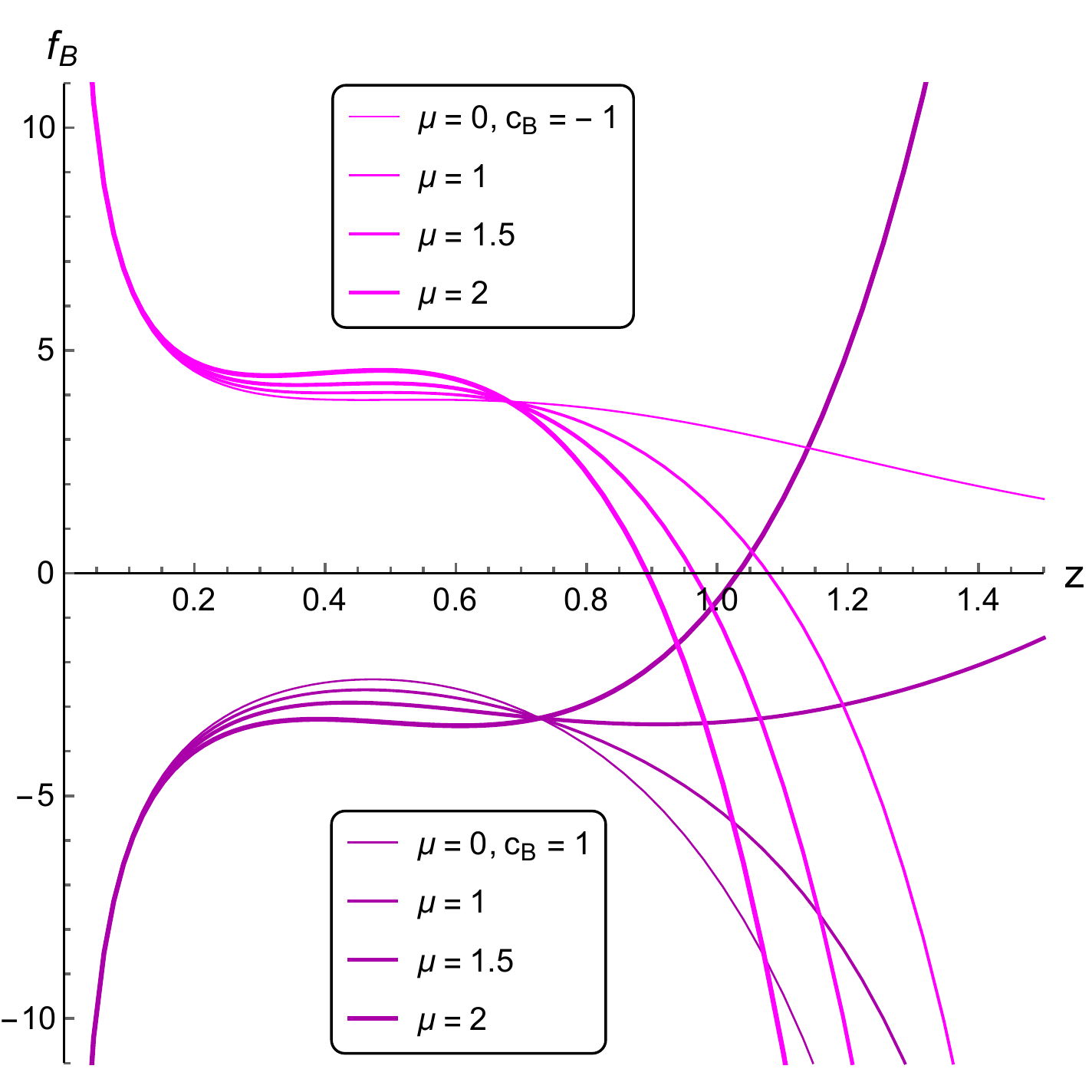} \quad
  \includegraphics[scale=0.23]{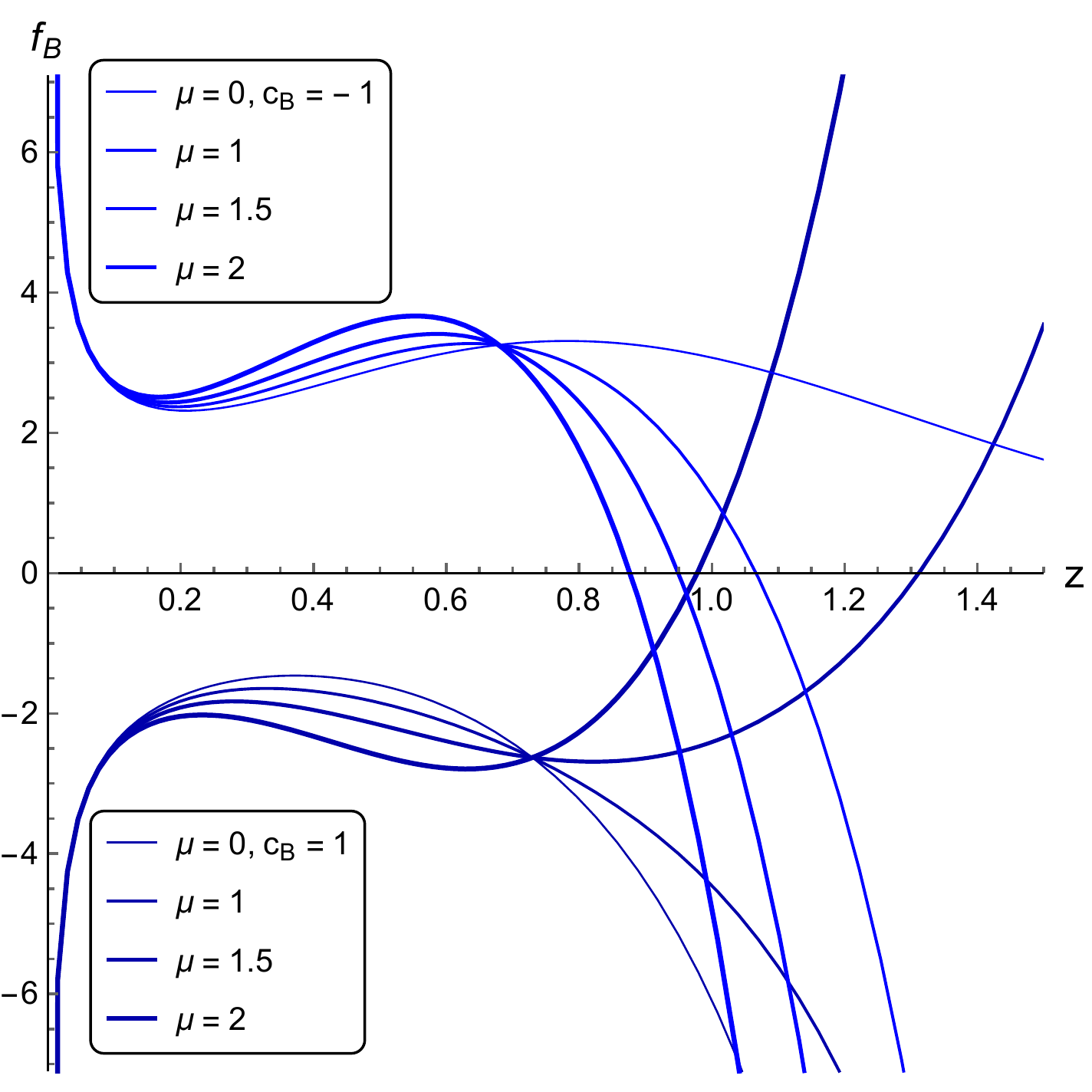} \\
  A \hspace{90pt} B \hspace{90pt}
  C \hspace{90pt} D
  \caption{Coupling function $f_B(z)$ for different $\mu$ for $\nu =
    1$ (A), $\nu = 1.5$ (B), $\nu = 3$ (C), $\nu = 4.5$ (D); $c =
    0.227$, $z_h = 1$, $c_B = \pm \ 1$, $q_B = 1$.}
  \label{Fig:fBzmu}
\end{figure}
\begin{figure}[t!]
  \centering
  \includegraphics[scale=0.23]{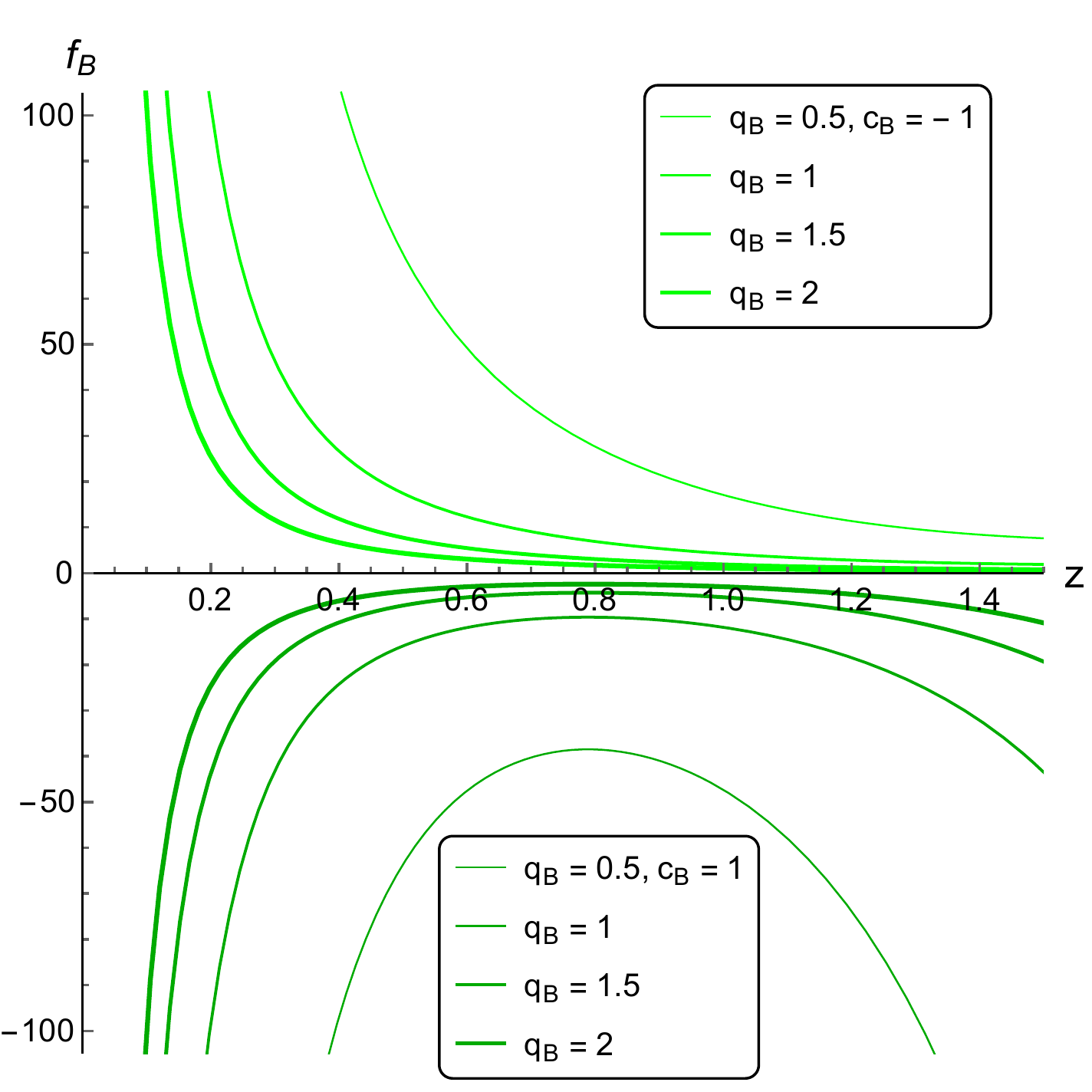} \quad
  \includegraphics[scale=0.23]{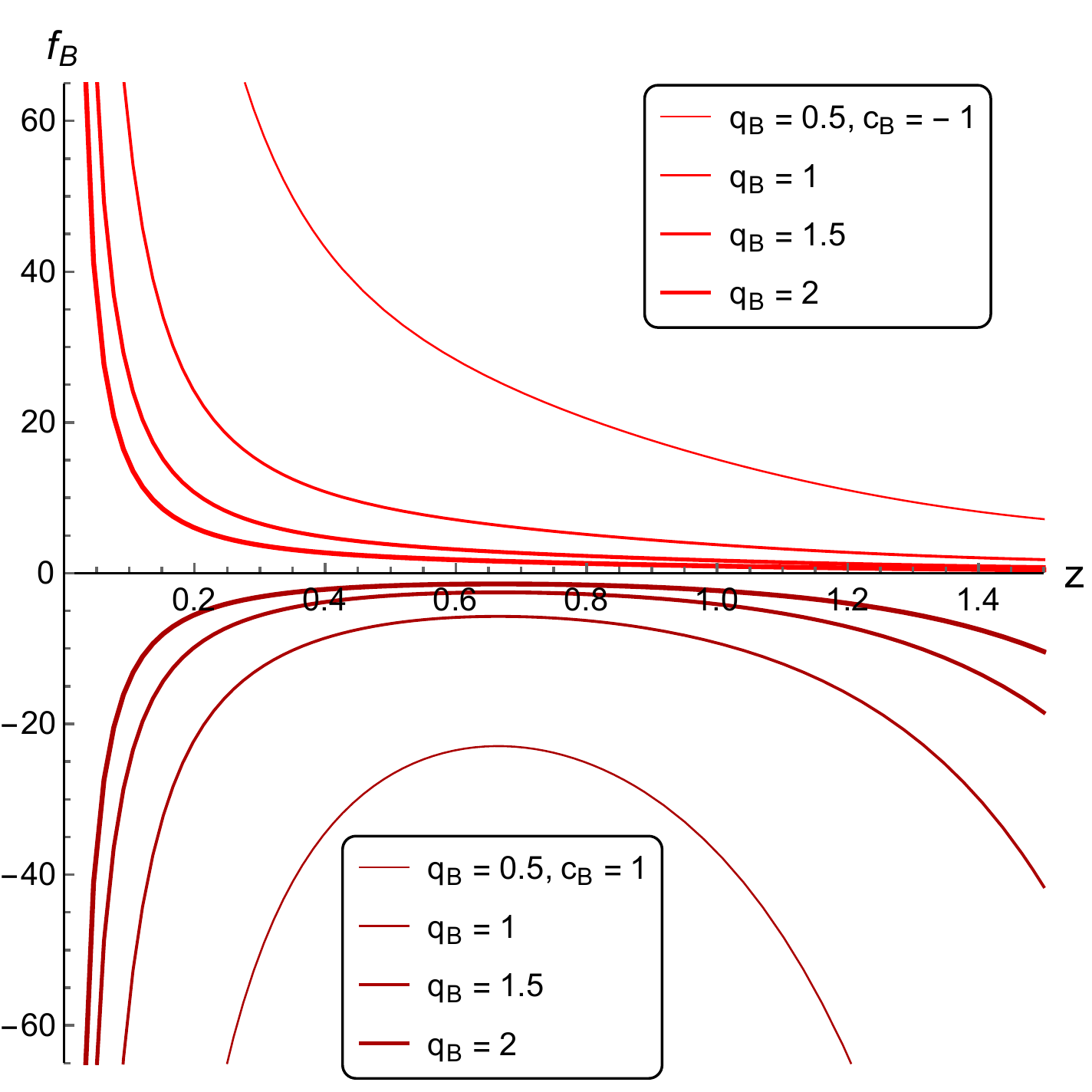} \quad
  \includegraphics[scale=0.23]{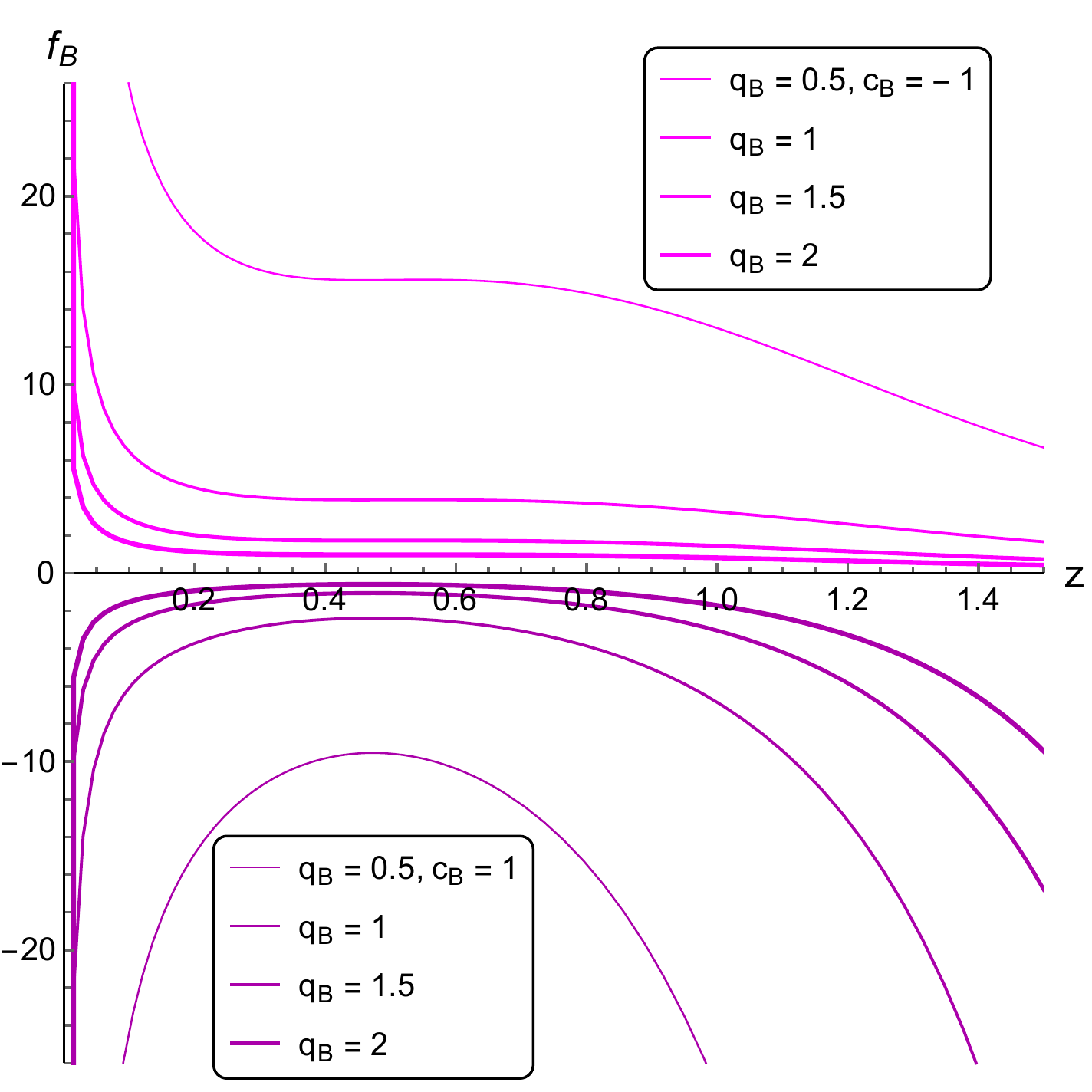} \quad
  \includegraphics[scale=0.23]{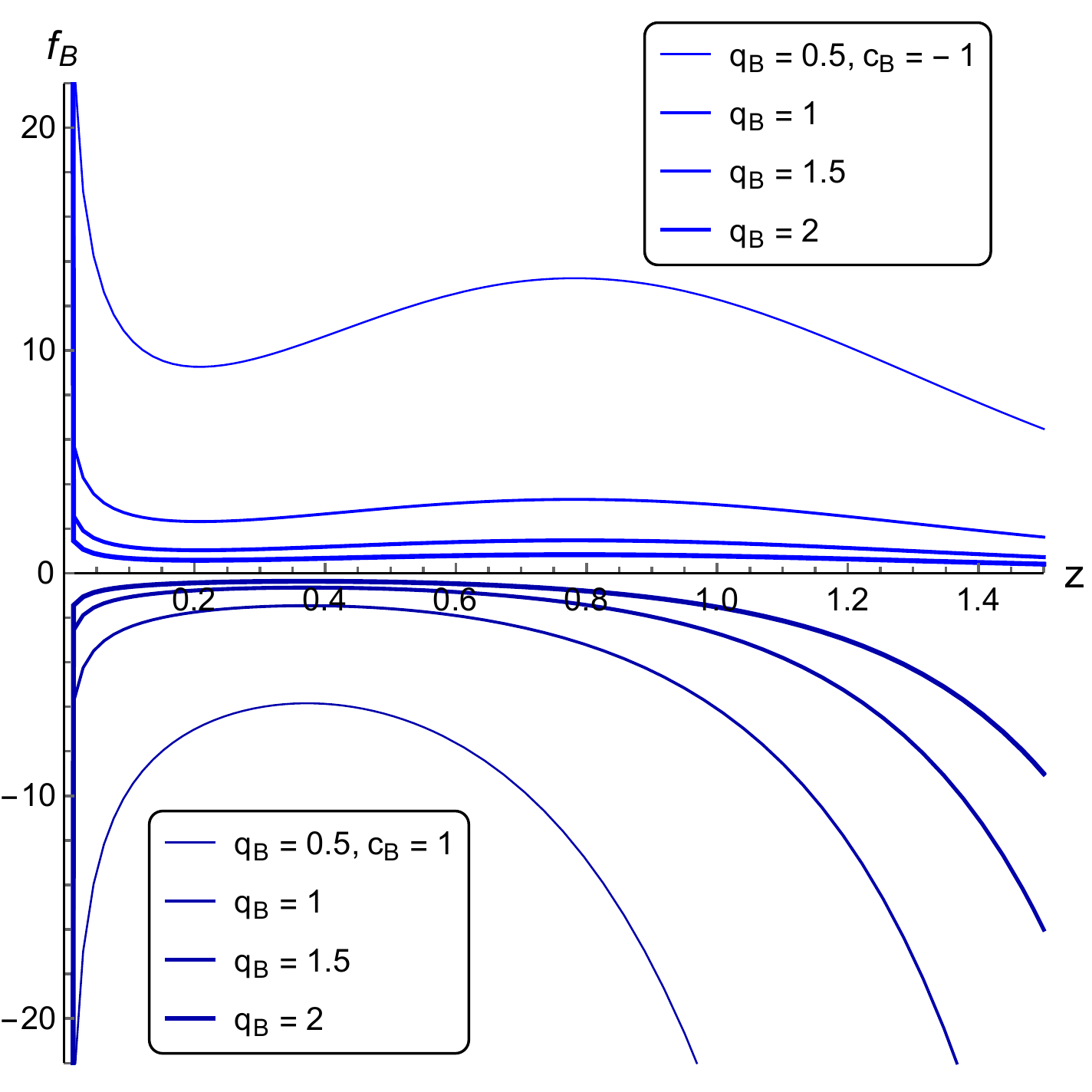} \\
  A \hspace{90pt} B \hspace{90pt}
  C \hspace{90pt} D
  \caption{Coupling function $f_B(z)$ for different $q_B$ for $\nu =
    1$ (A), $\nu = 1.5$ (B), $\nu = 3$ (C), $\nu = 4.5$ (D); $c =
    0.227$, $z_h = 1$, $\mu = 0$, $c_B = \pm \ 1$.}
  \label{Fig:fBzqb}
\end{figure}

Coupling function $f_B$ (\ref{eq:3.08}) depends both on the
coefficient $c_B$ and the magnetic ``charge'' $q_B$. First of all,
$f_B$ behavior forces us to limit ourselves to $c_B < 0$ only, as $c_B
> 0$ makes $f_B$ take negative values above the horizon ($z < z_h$)
thus violencing the NEC (Fig.\ref{Fig:fBzcb}). For $\nu = 1, \ 1.5$
(Fig.\ref{Fig:fBzcb}.A,B) $f_B$ decreases monotonically. Larger
primary anisotropy (Fig.\ref{Fig:fBzcb}.C,D for $\nu = 3, \ 4.5$)
leads to more complex function behavior and even to the appearence of
local maximum (Fig.\ref{Fig:fBzcb}.D). This effect is more significant
for the larger absolute values of $c_B$.

Increasing chemical potential makes the coupling function $f_B$ with
$c_B < 0$ decrease faster (Fig.\ref{Fig:fBzmu}) and reach negative
values earlier. Does it mean that large chemical potential forces
$f_B$ to break NEC at the outer vicinity of horizon? Not really. We
should not forget that the horizon itself shifts to smaller $z_h$
values because of large chemical potentials. Let us make sure that
point, where $f_B$ reaches zero, lies inside of horizon anyway as
$f_B(z_h)$ is still positive.

According to our boundary condition (\ref{eq:2.18}) $g(z_h) =
0$. Therefore expression (\ref{eq:3.08}) can be simplified as
\begin{gather}
  f_B(z_h) = 2 \left( \cfrac{z}{L} \right)^{-\frac{2}{\nu}}
  e^{-\frac{1}{2}cz^2} \ \cfrac{c_B z}{q_B^2} \
  g'(z_h) > 0, \label{eq:3.12}
\end{gather}
At first horizon (the one that really matters) blackening function is
decreasing, so $g'(z_h) < 0$. If we also take $c_B < 0$, their product
is positive, all the other multipliers in (\ref{eq:3.12}) are positive
as well, therefore $f_B(z_h) > 0$ for any negative $c_B$ in the $z_h$
interval we need.

Until now we only spoke about dependences on $c_B$, that characterizes
the connection between metric and the external magnetic field, but
told nothing about the strength of that magnetic field itself. This
quality is described by ``charge'' $q_B$.

From (\ref{eq:3.08}) we see that the coupling function $f_B$ depends
on the inverse square of the ``charge'' $q_B$. So the larger $q_B$ is,
the faster $f_B$ tends to zero and and the longer is in its vicinity
(Fig.\ref{Fig:fBzqb}).



  
\begin{figure}[t!]
  \centering
  \includegraphics[scale=0.36]{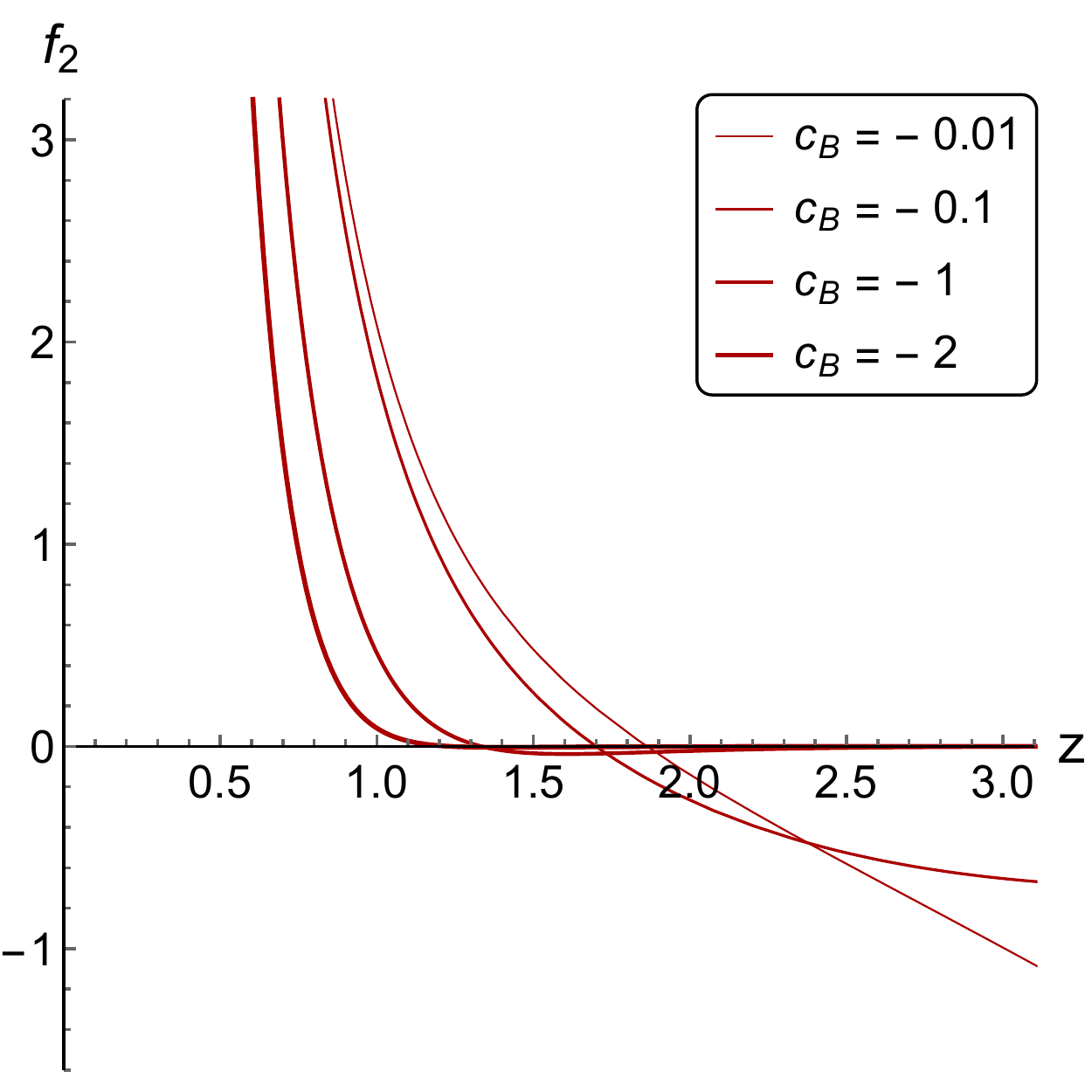} \quad
  \includegraphics[scale=0.36]{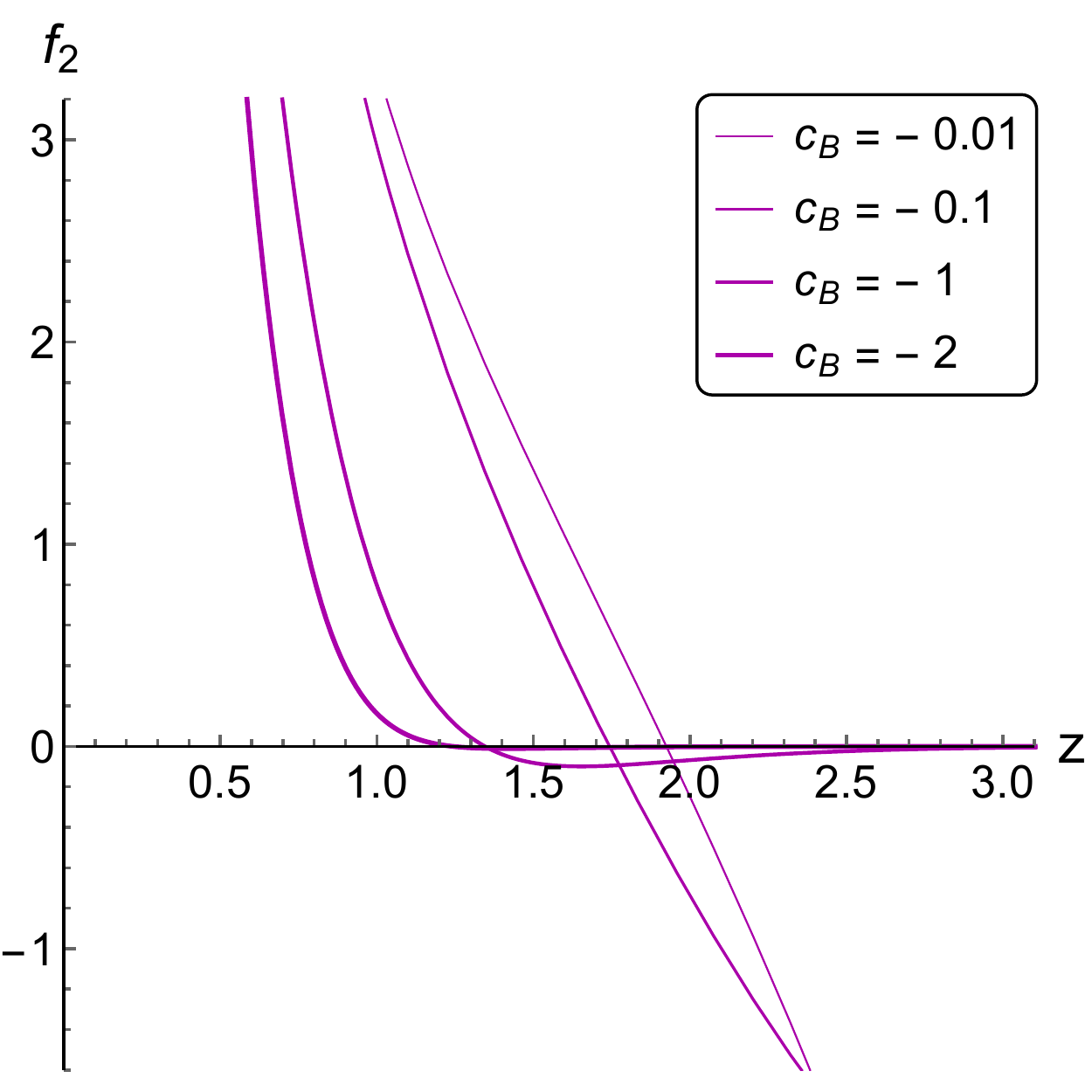} \quad
  \includegraphics[scale=0.36]{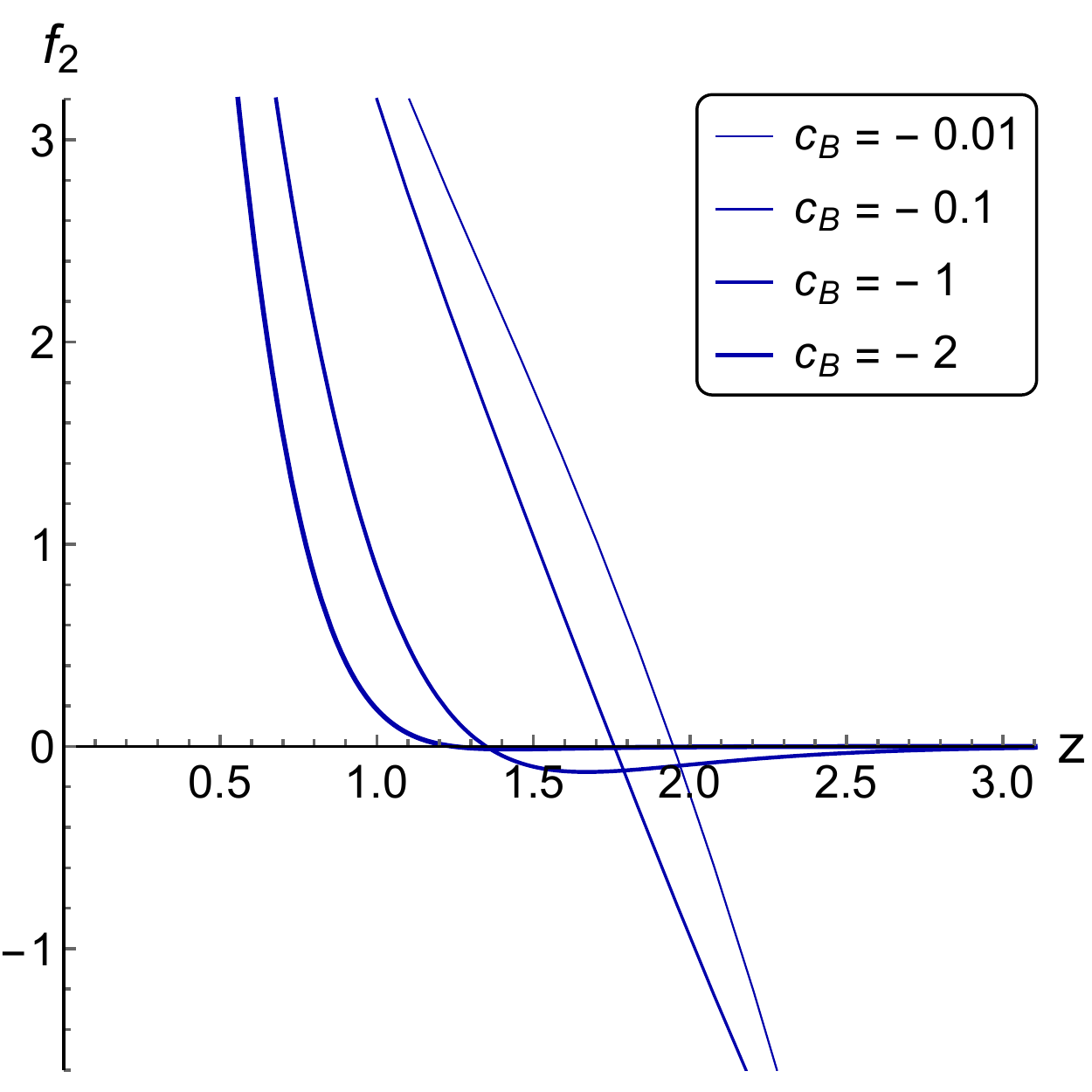}\\
  \includegraphics[scale=0.36]{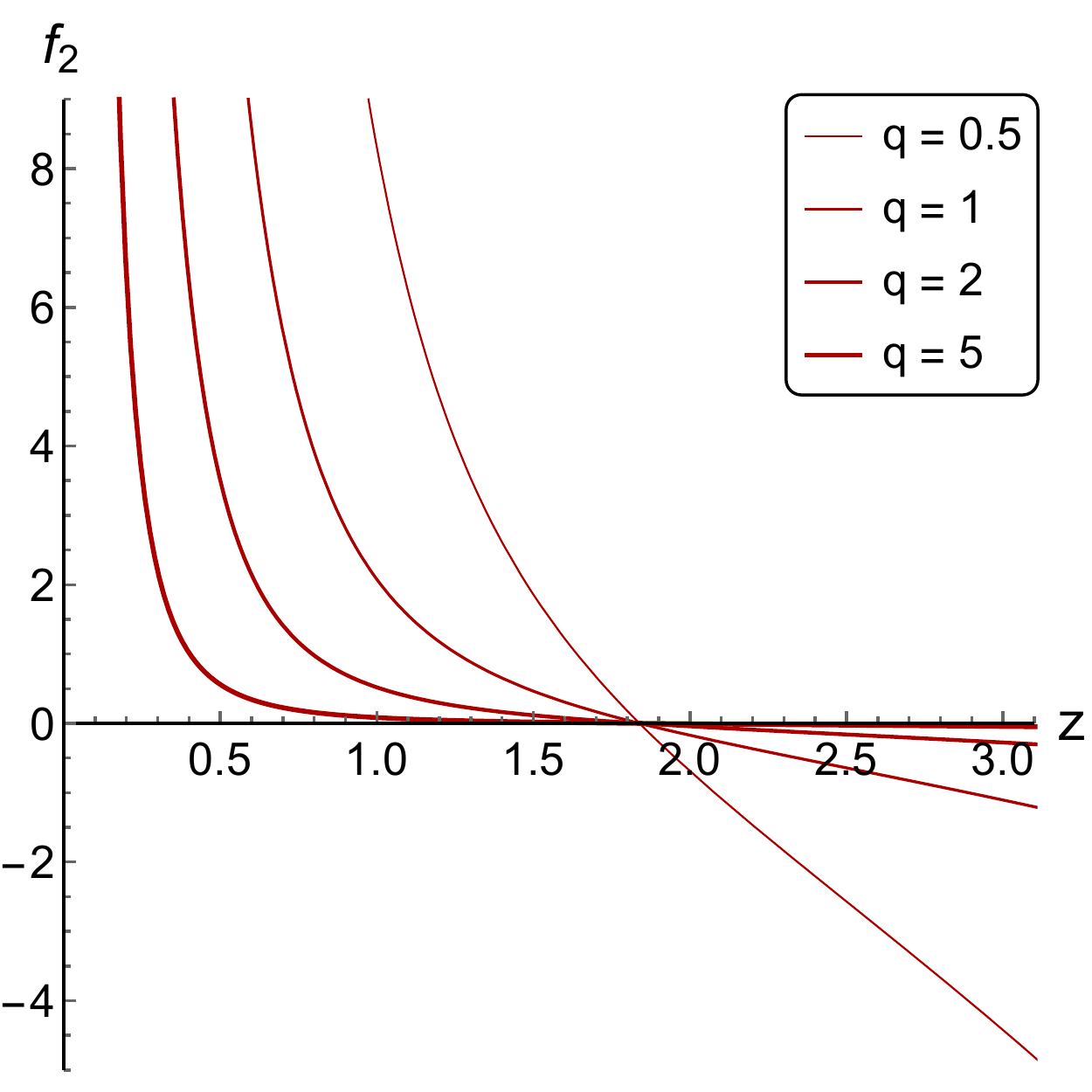} \quad
  \includegraphics[scale=0.36]{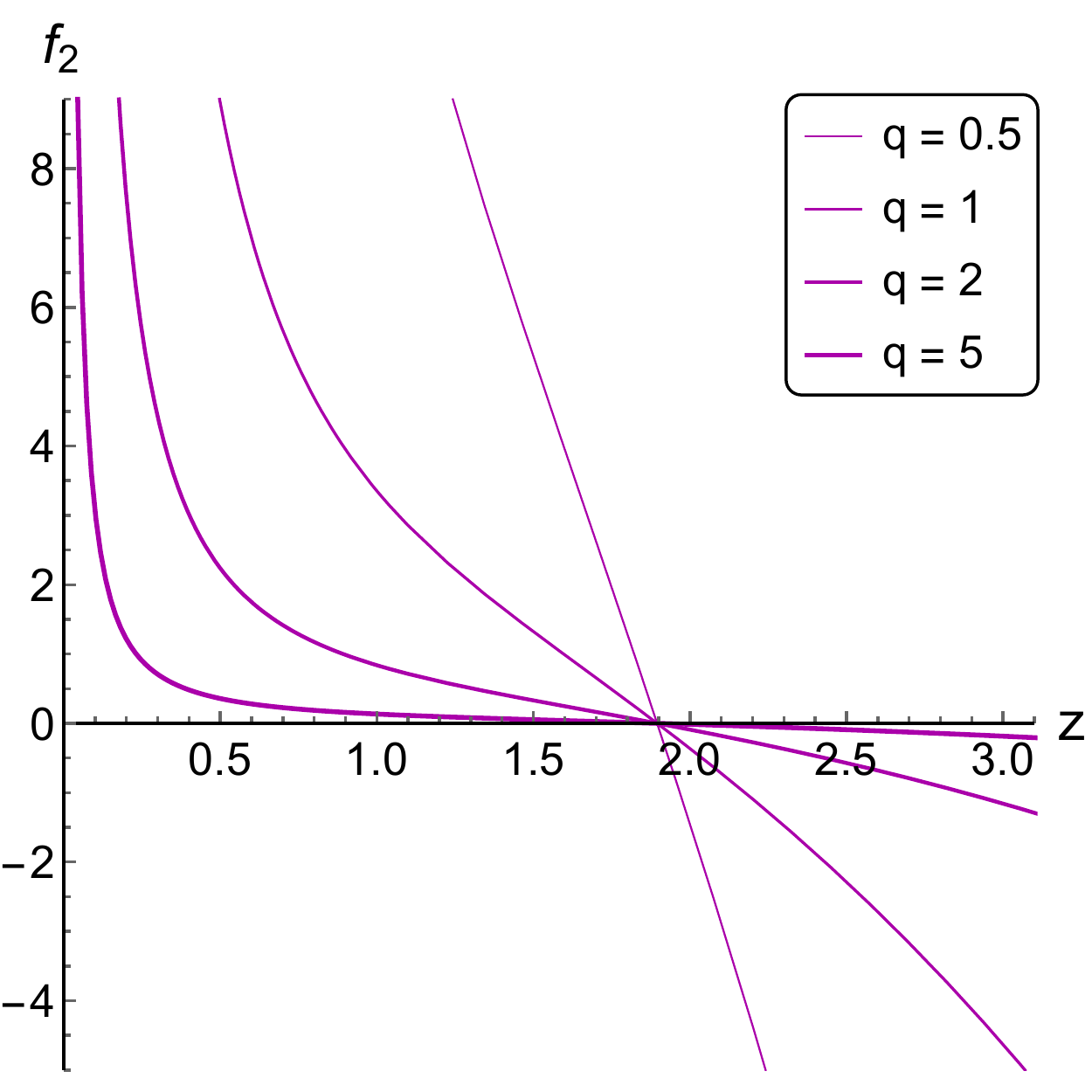} \quad
  \includegraphics[scale=0.36]{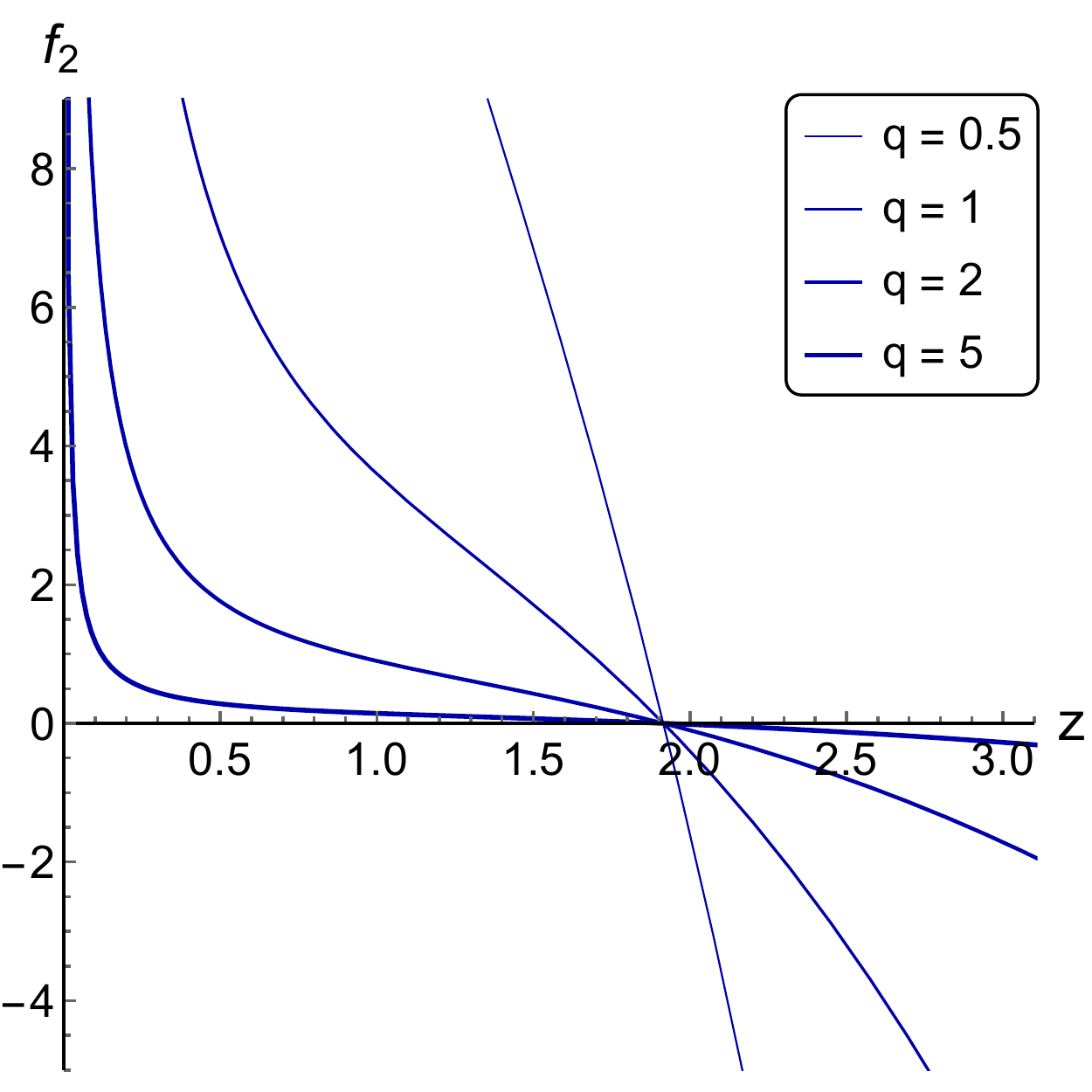}\\
  \includegraphics[scale=0.36]{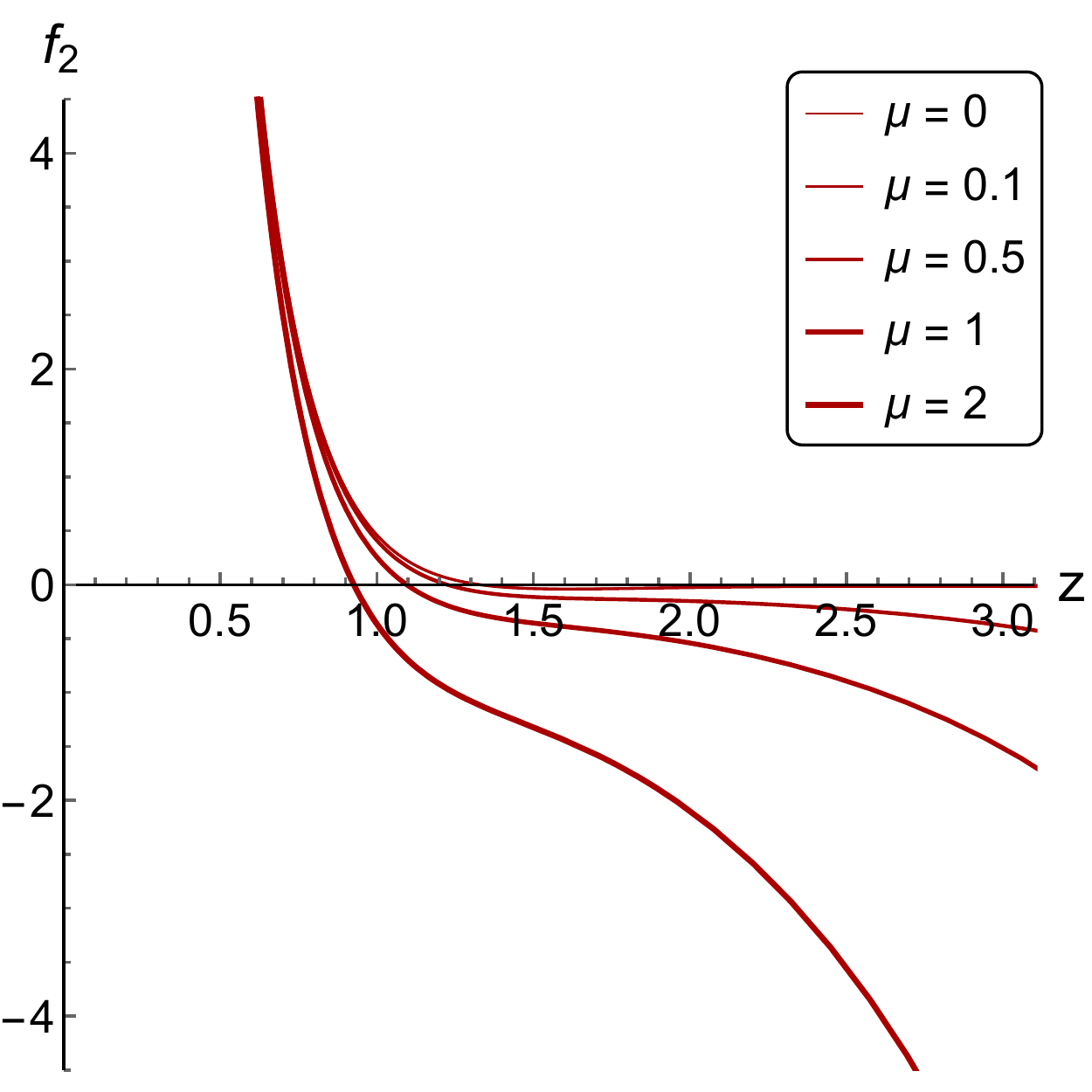} \quad
  \includegraphics[scale=0.36]{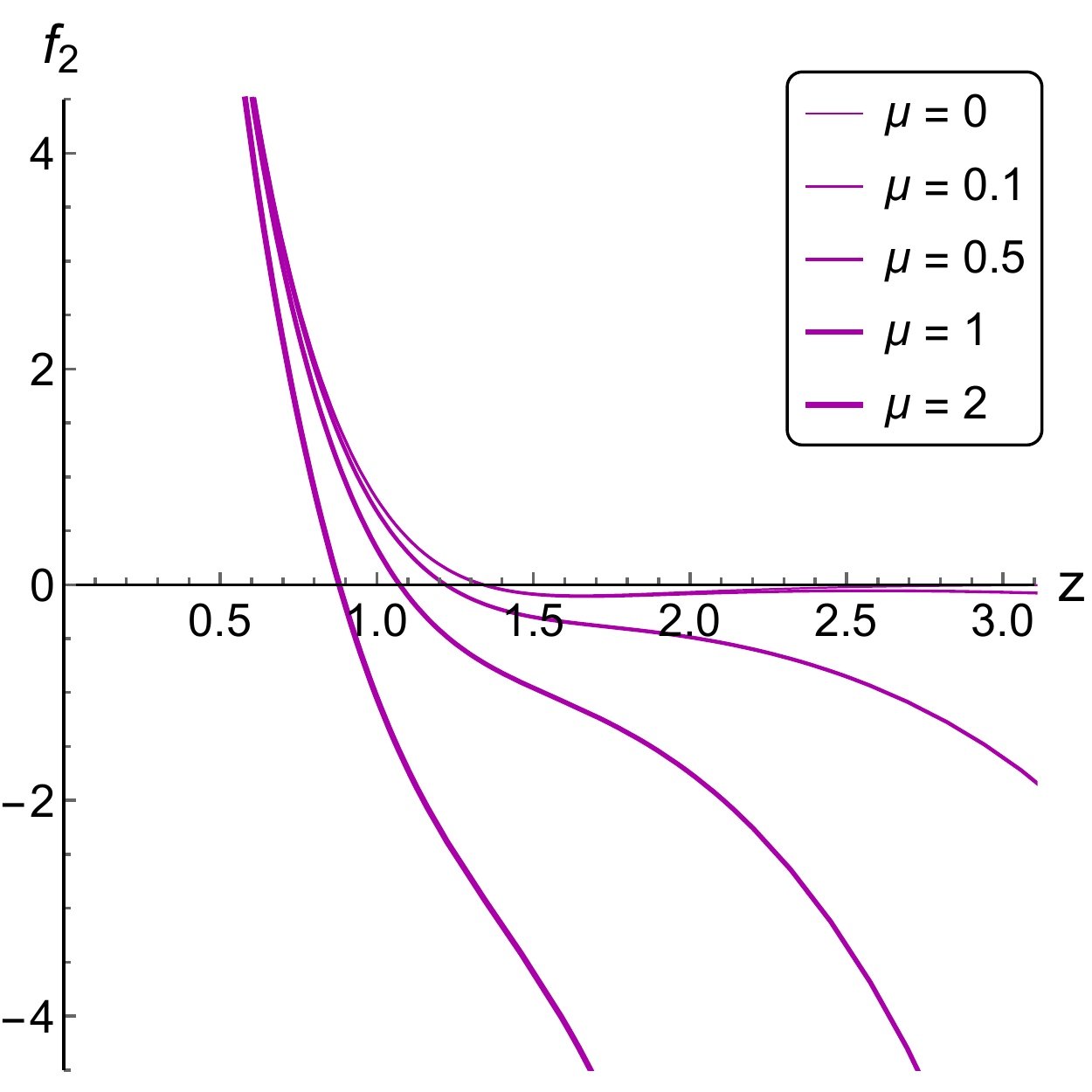} \quad
  \includegraphics[scale=0.36]{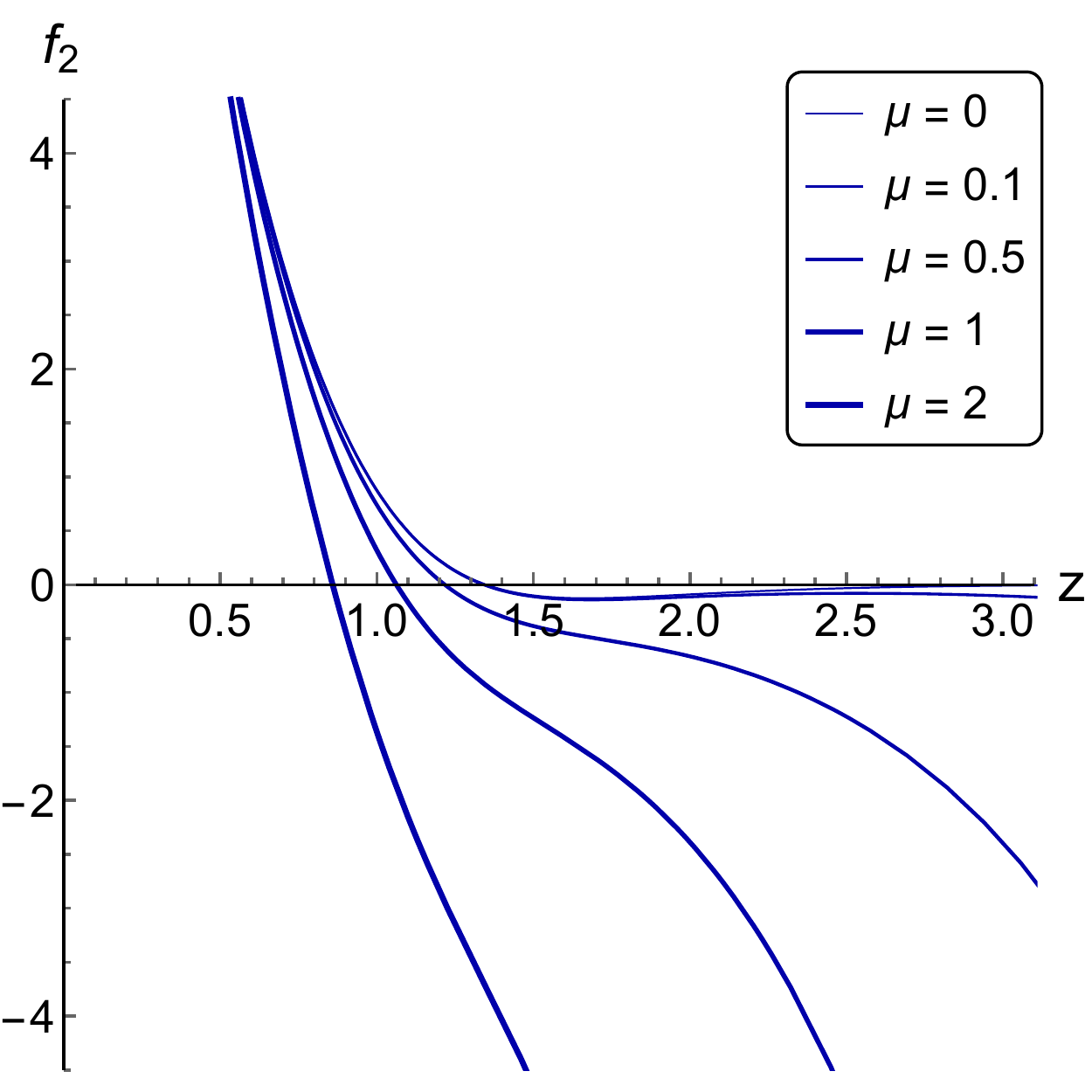}\\
  A \hspace{130pt} B \hspace{130pt} C
  \caption{Coupling function $f_2(z)$ for different $c_B$, $\mu = 0$,
    $q = 1$ (first line), for different $q$ $\mu = 0$, $c_B = - \, 1$
    (second line), for different $\mu$, $c_B = - \, 1$, $q = 1$ (third
    line) in anisotropic cases for $\nu = 1.5$ (A), $\nu = 3$ (B),
    $\nu = 4.5$ (C); $c = 0.227$, $z_h = 1$.}
  \label{Fig:f2z}
\end{figure}


Let us remind that, unlike the third Maxwell $F_{\mu\nu}^{(B)}$, the
second Maxwell field $F_{\mu\nu}^{(2)}$ isn't physical and serves to
maintain primary anisotropy, set by parameter $\nu$, of the black hole
solution within the holographic approach.

Behavior of the coupling function $f_2$ (\ref{eq:3.09}) in this model
is quite similar to its behavior in the absence of external magnetic
field \cite{1802.05652} (Fig.\ref{Fig:f2z}). Mainly this function
descreases monotonically, larger chemical potentials and larger $q$
-- ``charges'' of $F_{\mu\nu}^{(2)}$ -- provide faster
descrease. These effects are more obvious for larger $\nu$. Case $\nu
= 1$ doesn't make sense, as $F_{\mu\nu}^{(2)} \equiv 0$ and its
coupling function $f_2$ do not exist.

For rather large absolute values of coefficient $c_B$ (for example
$c_B = - 1$) function $f_2$ has local minimum, that is more obvious
for larger primary anisotropy (Fig.\ref{Fig:f2z}.C). But this effect
is suppressed by chemical potential.

\begin{figure}[t!]
  \centering
  \includegraphics[scale=0.29]{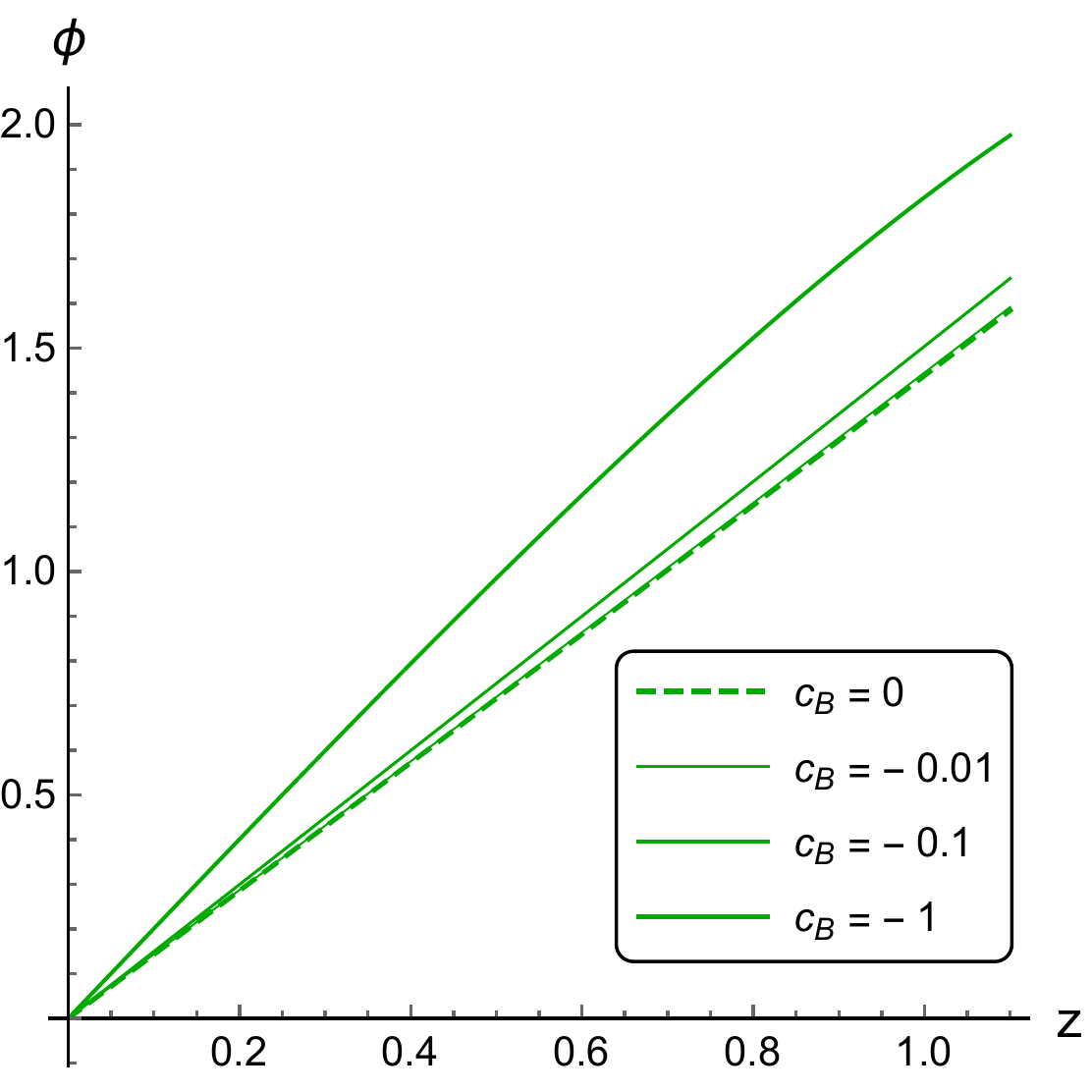} \quad
  \includegraphics[scale=0.29]{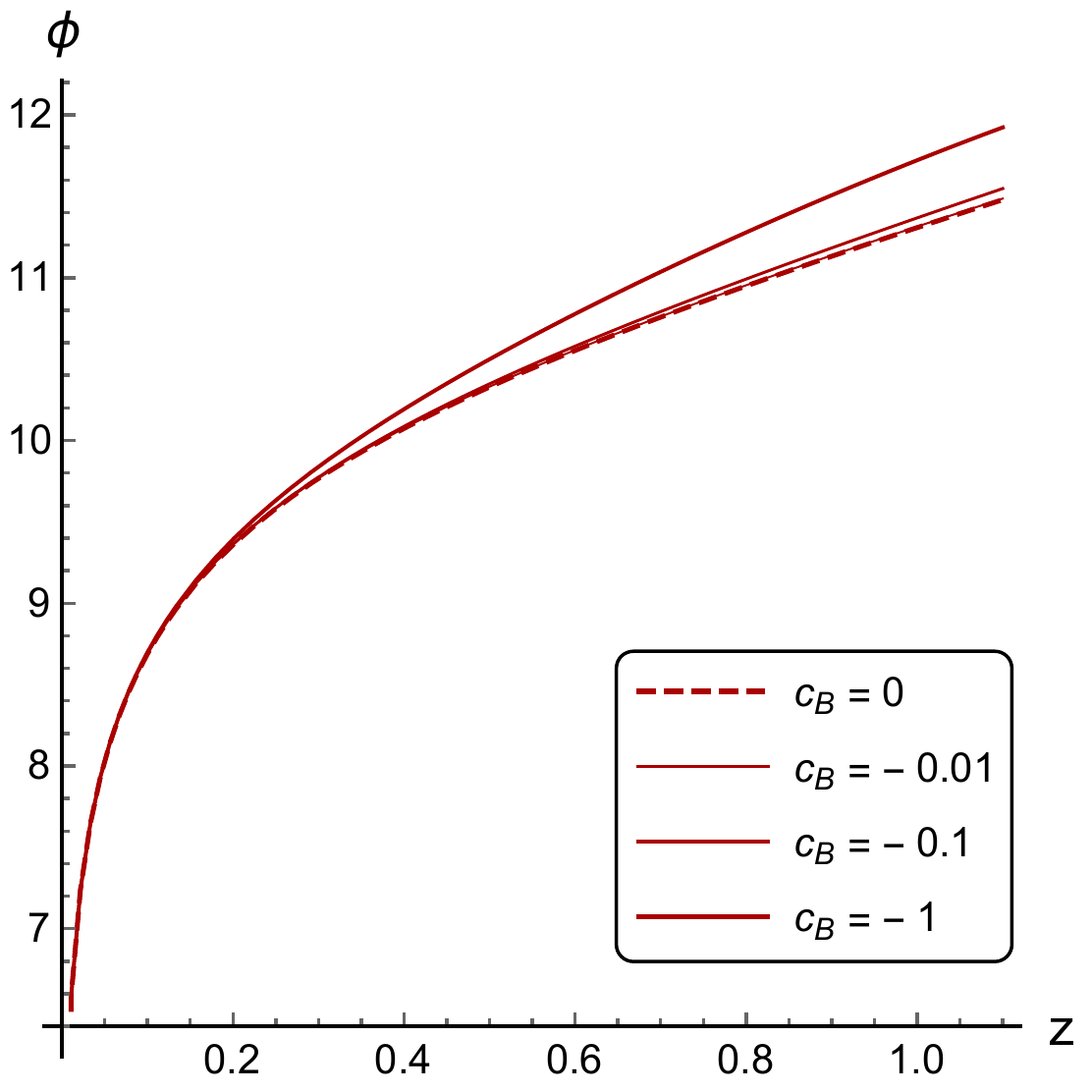} \quad
  \includegraphics[scale=0.29]{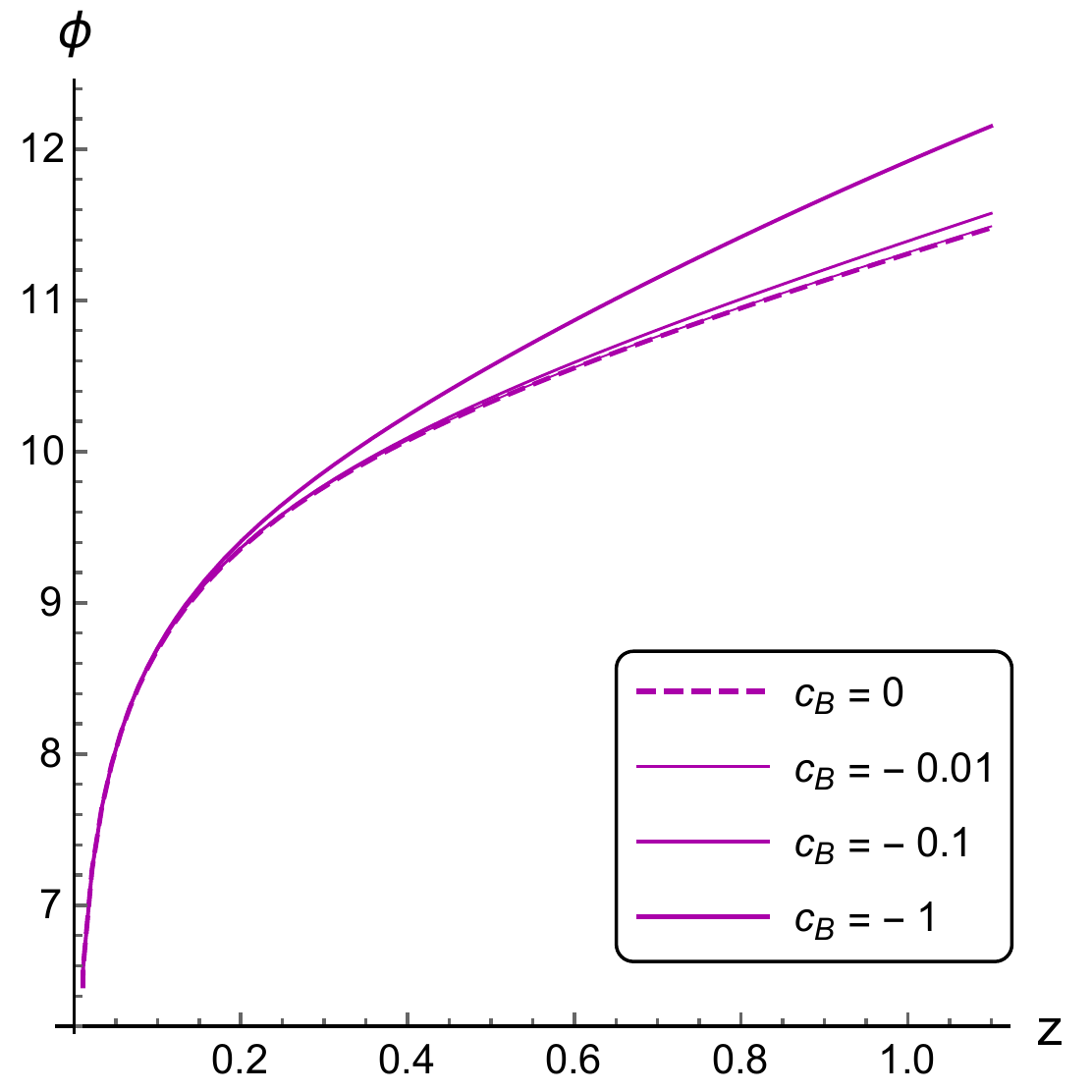} \quad
  \includegraphics[scale=0.29]{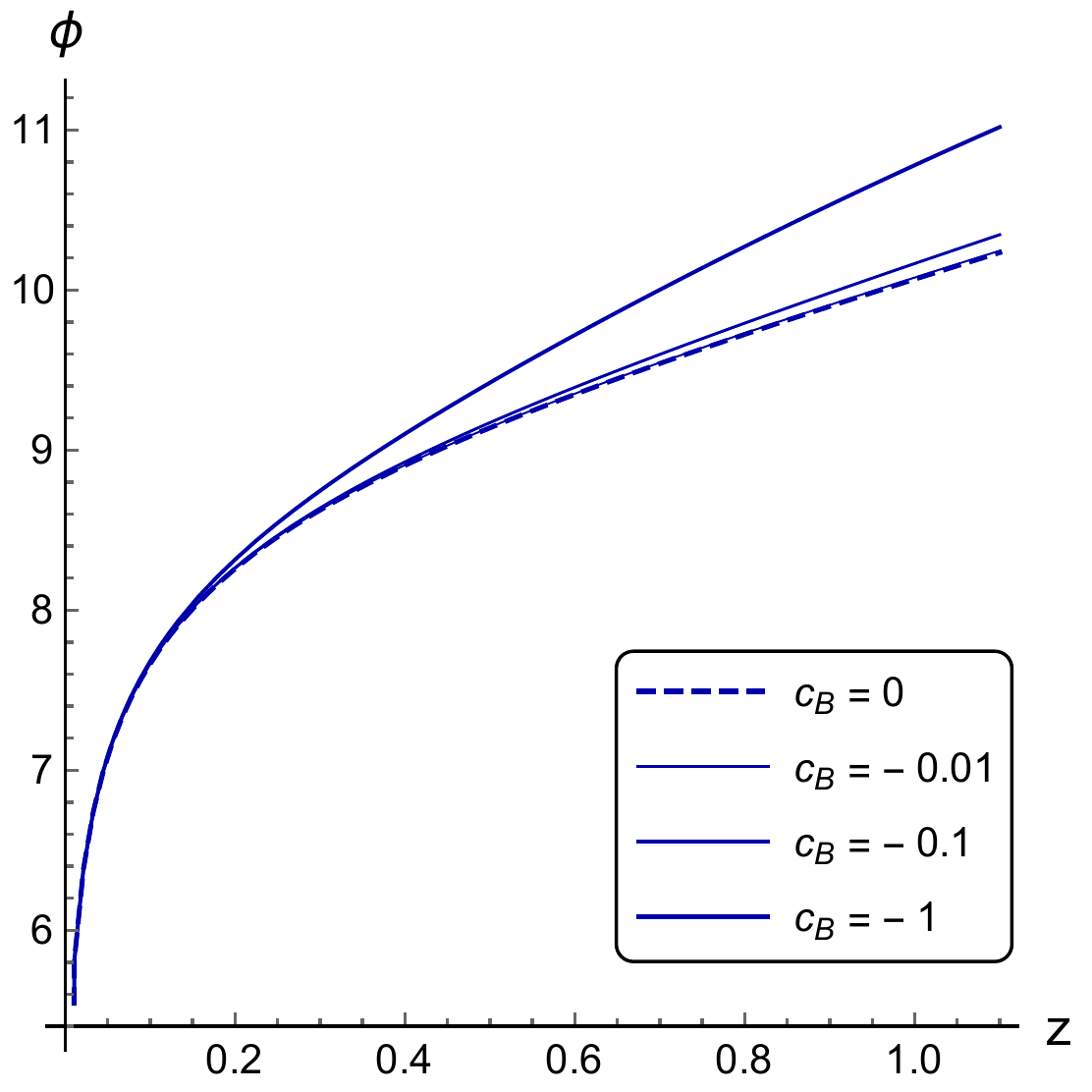} \\
  A \hspace{90pt} B \hspace{90pt}
  C \hspace{90pt} D
  \caption{Coupling function $f_B(z)$ for different  $c_B$ for $\nu =
    1$ (A), $\nu = 1.5$ (B), $\nu = 3$ (C), $\nu = 4.5$ (D); $c =
    0.227$, $z_h = 1$, $\mu = 0$.}
  \label{Fig:phiz}
\end{figure}

We can show that $f_2$ doesn't break NEC above the horizon like it was
done for $f_B$ earlier:
\begin{gather}
  f_2(z_h) = 4 \left( \cfrac{z}{L} \right)^{2-\frac{4}{\nu}}
  e^{-\frac{1}{2}(c-2c_B)z^2} \ \cfrac{\nu - 1}{q^2 \nu z} \ \left( -
    \, g' \right). \label{eq:3.13}
\end{gather}
At first horizon $g(z_h) = 0$ and $g'(z_h) < 0$. As $f_2$ has sense
for $\nu > 1$ only, all the multipliers in (\ref{eq:3.13}) are
positive, so $f_2(z_h) > 0$ and NEC isn't broken so far.

\begin{figure}[t!]
  \centering
  \includegraphics[scale=0.26]{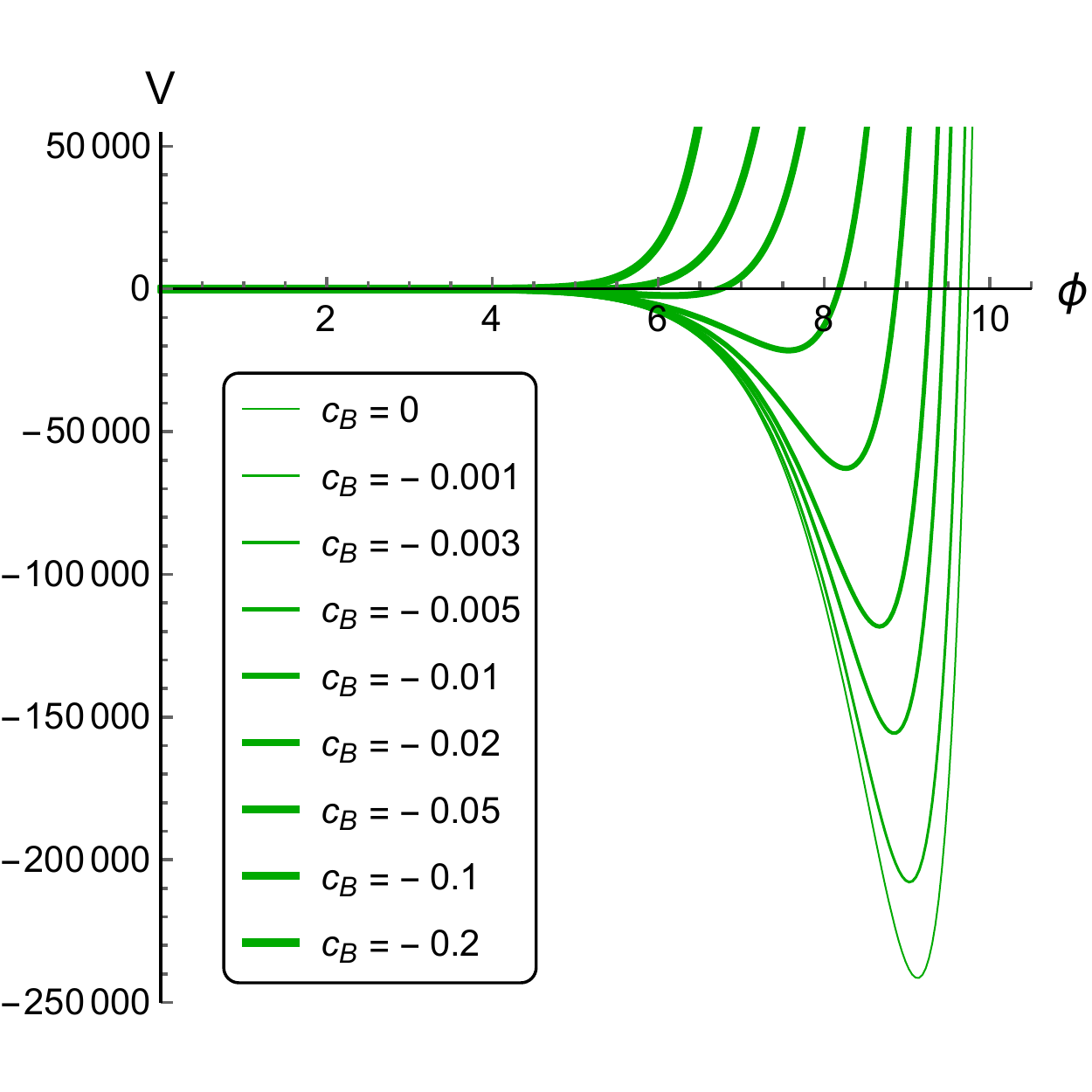} \quad
  \includegraphics[scale=0.26]{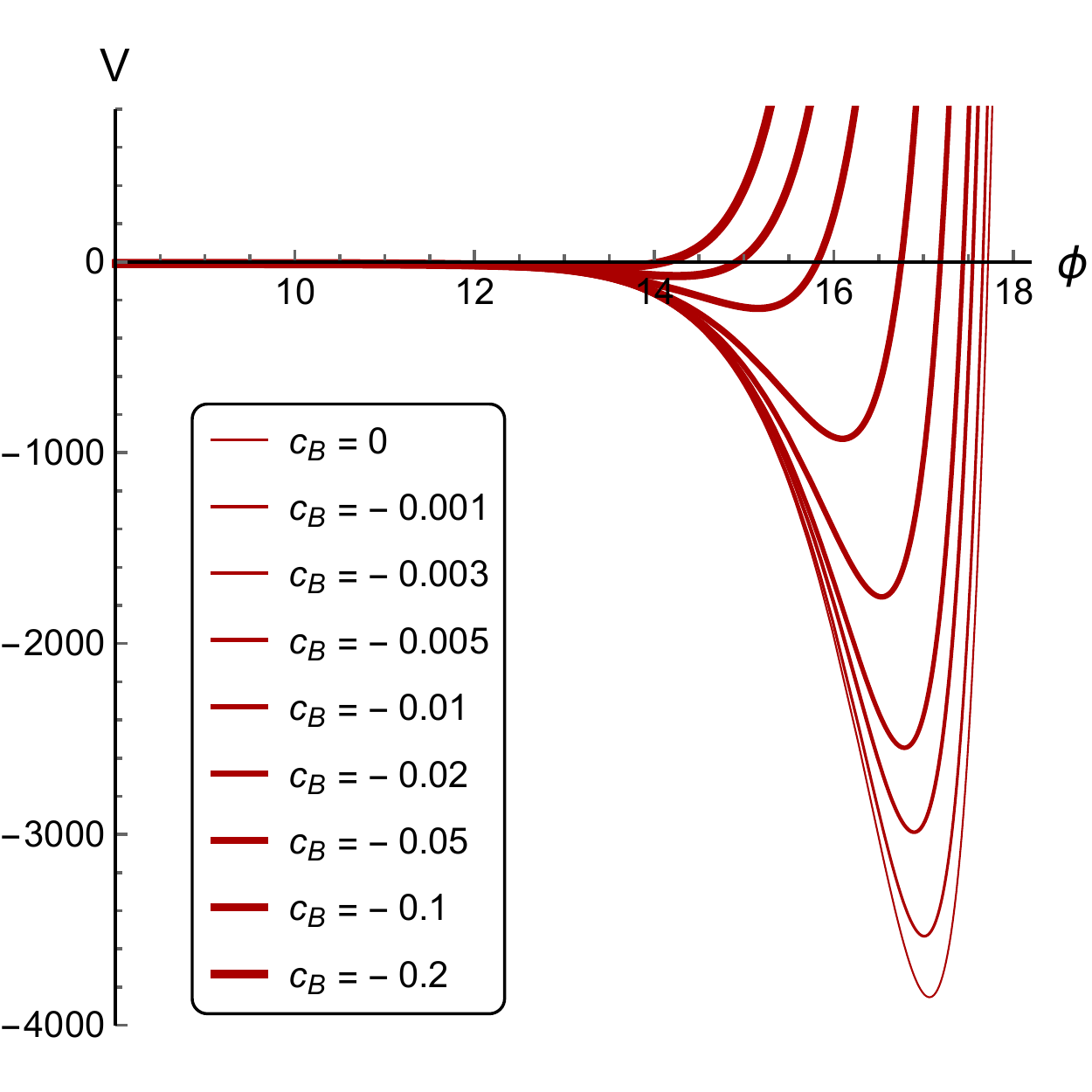} \quad
  \includegraphics[scale=0.26]{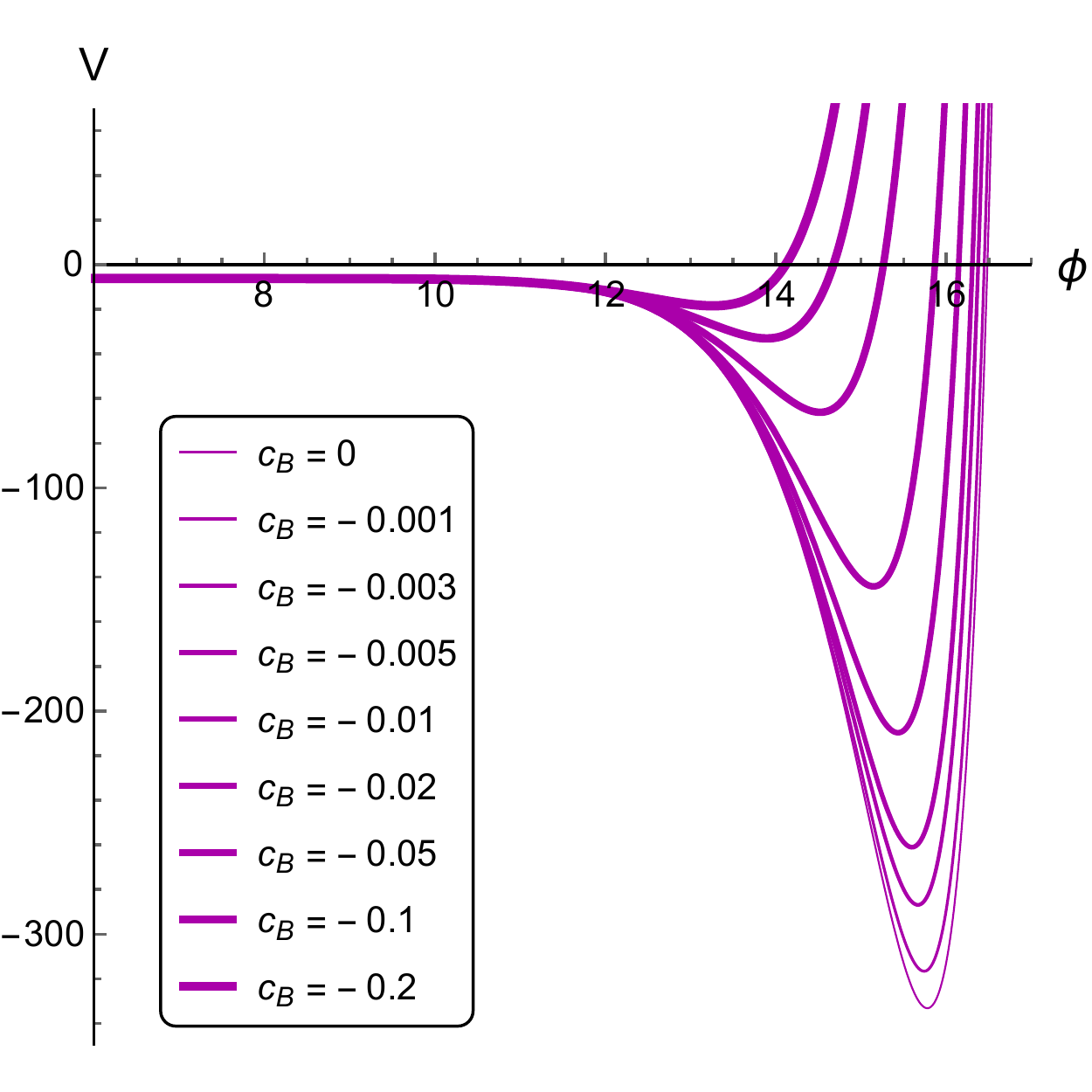} \quad
  \includegraphics[scale=0.26]{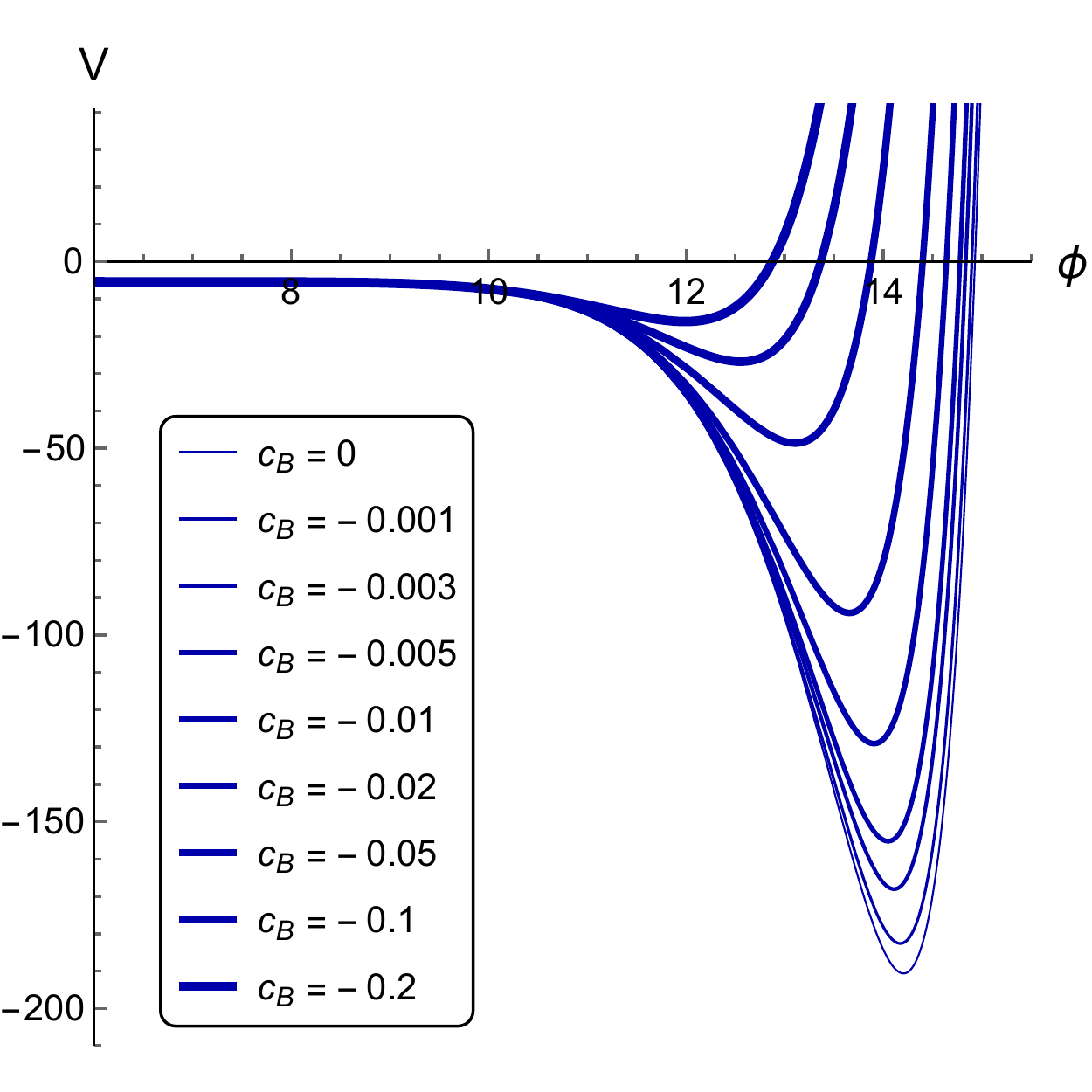} \\
  A \hspace{90pt} B \hspace{90pt}
  C \hspace{90pt} D
  \caption{Scalar potential $V(\phi)$ for different $c_B$ for $\nu =
    1$ (A), $\nu = 1.5$ (B), $\nu = 3$ (C), $\nu = 4.5$ (D); $c =
    0.227$, $z_h = 1$, $\mu = 0$.}
  \label{Fig:VphicB}
\end{figure}

\begin{figure}[t!]
  \centering
  \includegraphics[scale=0.26]{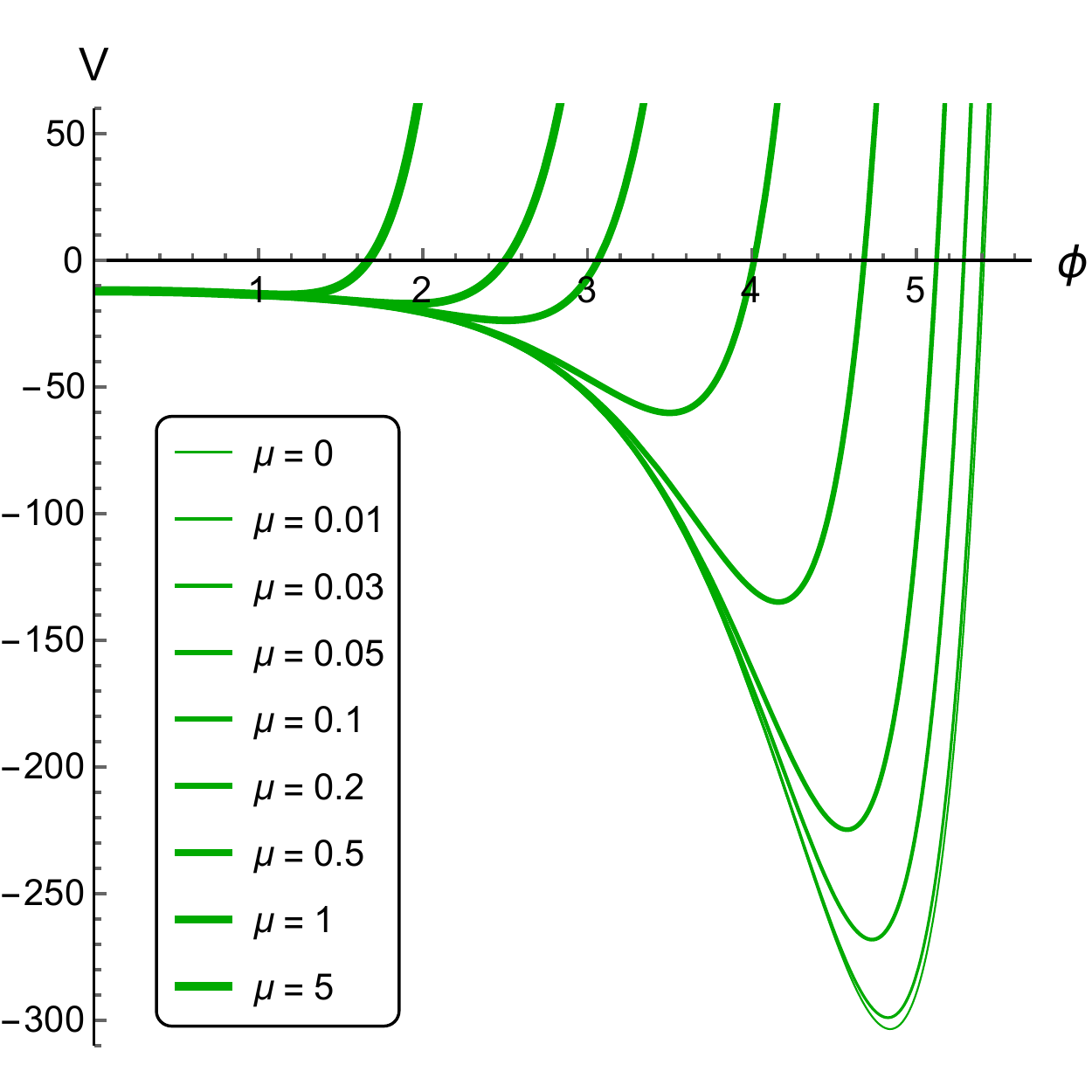} \quad
  \includegraphics[scale=0.26]{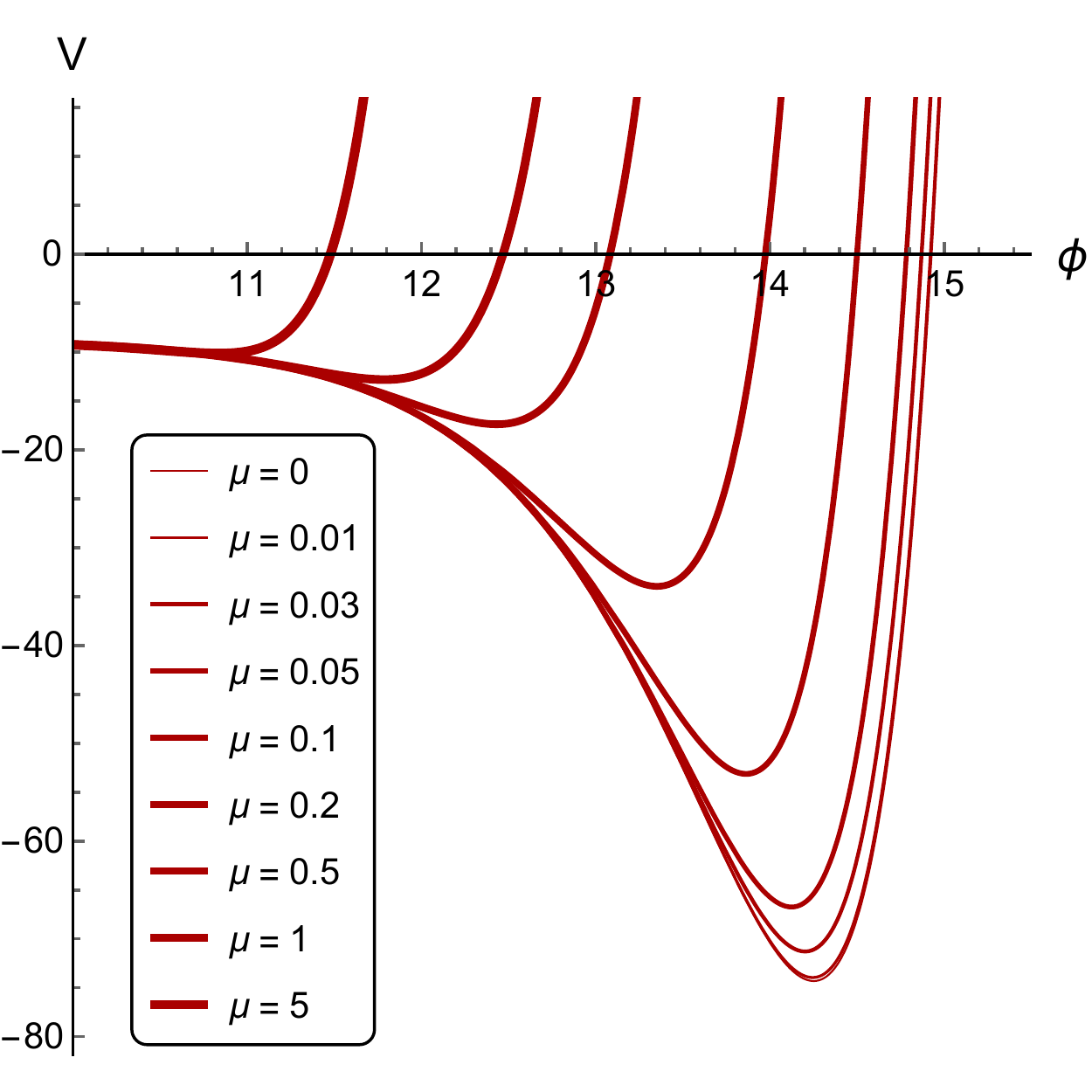} \quad
  \includegraphics[scale=0.26]{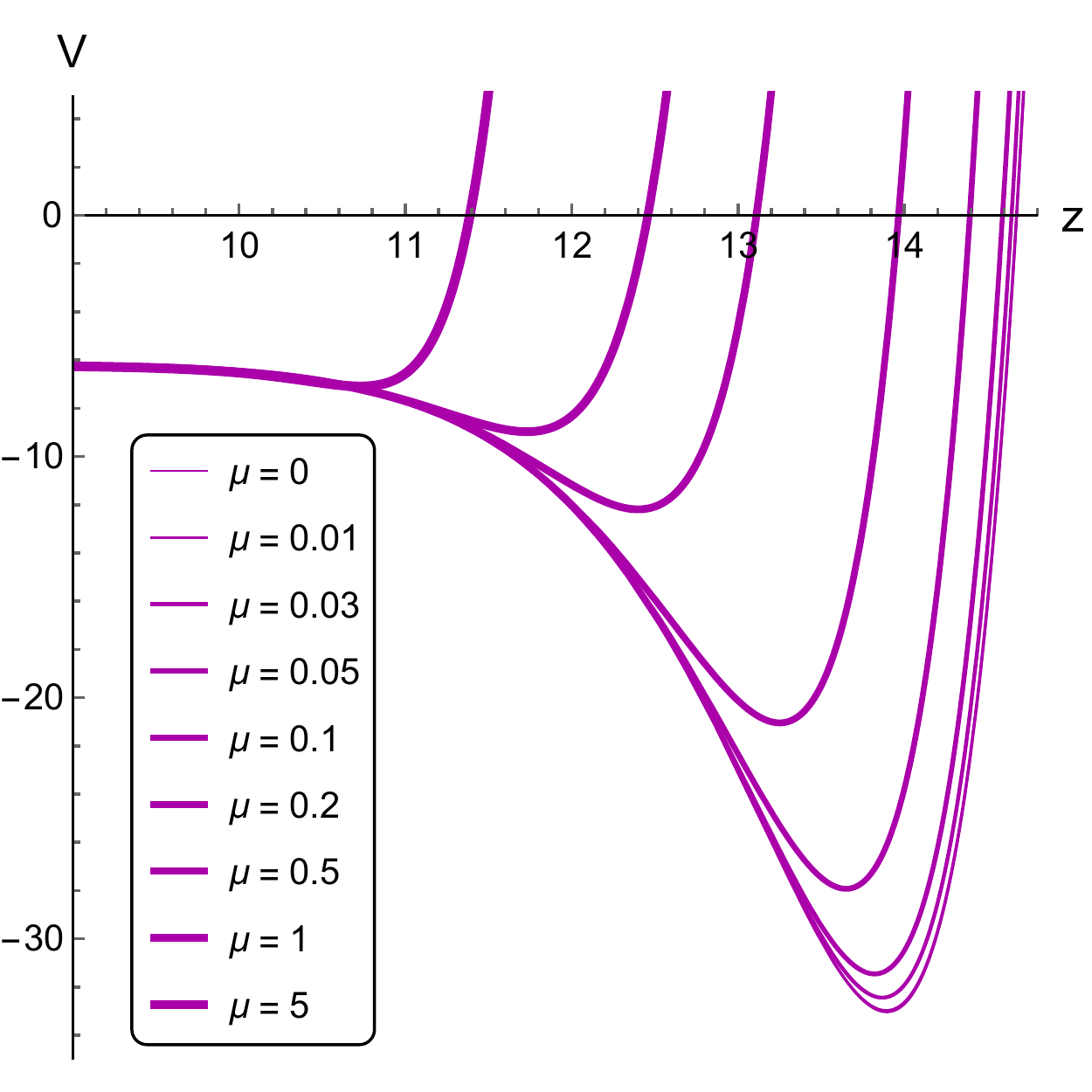} \quad
  \includegraphics[scale=0.26]{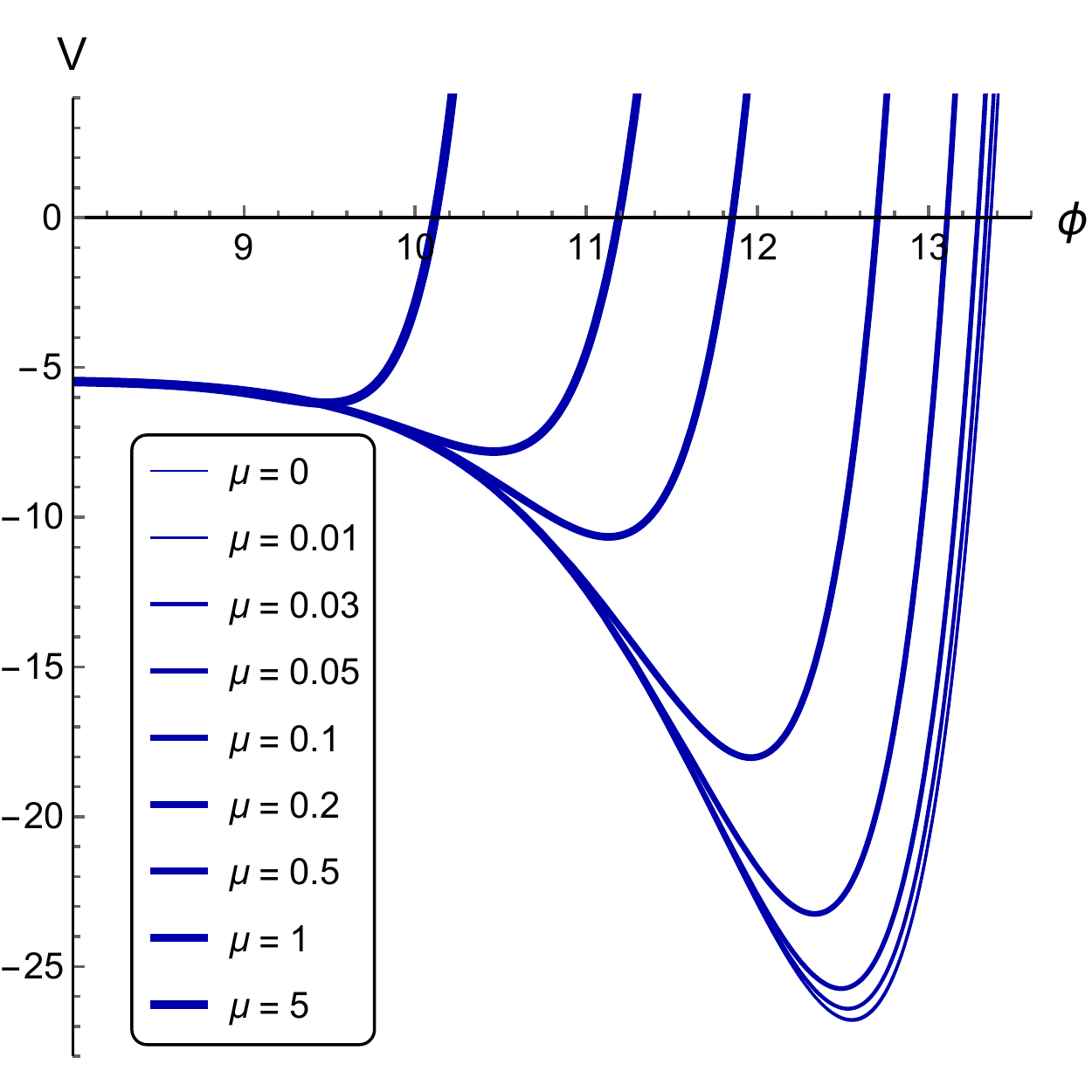} \\
  A \hspace{90pt} B \hspace{90pt}
  C \hspace{90pt} D
  \caption{Scalar potential $V(\phi)$ for different $\mu$ for $\nu =
    1$ (A), $\nu = 1.5$ (B), $\nu = 3$ (C), $\nu = 4.5$ (D); $c =
    0.227$, $z_h = 1$, $c_B = - \, 0.1$.}
  \label{Fig:Vphimu}
\end{figure}

Unlike our previous light quarks model \cite{2009.05562}, this
time $z_0 = 0$ is restricted even for $\nu = 1$. But we can take $z_0$
small enough to maintain the connection to the isotropic limit
\cite{1506.05930}.

Scalar field doesn't depend on $q_B$ that characterizes the external
magnetic field $F_{\mu\nu}^{(B)}$. Dependence on the coupling
coefficient $c_B$ is rather slight and weakens with the rise of
primary anisotropy, parametrized by $\nu$ (Fig.\ref{Fig:phiz}).

  

The effect of increasing $c_B$ absolute value on $V(\phi)$
(Fig.\ref{Fig:VphicB}) is similar to increasing chemical potential 
(Fig.\ref{Fig:Vphimu}): local minimum becomes larger and its position
shifts to the left. Case $\nu = 1$ still differs from the others, as 
$V(\phi_{min} = 0) = 0$. For larger $\nu$ we can say that if $\nu_1 <
\nu_2$, therefore $V(0)|_{\nu_1} < V(0)|_{\nu_2}$, $V_{min}|_{\nu_1} <
V_{min}|_{\nu_2}$ and for $\phi_*: \ V(\phi) = V_{min}$ we have
$\phi_*|_{\nu_1} > \phi_*|_{\nu_2}$, but $\phi_*|_{\nu\ne1} >
\phi_*|_{\nu=1}$. For larger primary anisotropy $\nu$ influence of
both $c_B$ and $\mu$ weakens.



\section{Thermodynamics}\label{thermo}

\subsection{Temperature and entropy}\label{Ts}

For the metric (\ref{eq:2.03}) and the warp-factor $f_{1\,HQ} =
z^{-2+\frac{2}{\nu}}$ (\ref{eq:3.01}) temperature and entropy can be
written as:
\begin{gather}
  \begin{split}
    T &= \cfrac{|g'|}{4 \pi} \, \Bigl|_{z=z_h}
    = \cfrac{1}{2 \pi} \left| - \, e^{\frac{1}{4}(3c-2c_B)z_h^2} \ (2
      c_B - c) \ z_h^{1+\frac{2}{\nu}} \left\{
        \cfrac{\mu^2 \ e^{\frac{1}{4}(c-2c_B)z_h^2}}{4 L^2 \left( 1 -
            e^{\frac{1}{4}(c-2c_B)z_h^2} \right)^2} \right. \right. +
    \\
    &+ \ \left( \cfrac{3}{4} \right)^{1+\frac{1}{\nu}}
    \cfrac{(2 c_B - c)^{\frac{1}{\nu}}}{\Gamma\left(1 + \frac{1}{\nu}
        \ ; 0\right) - \Gamma\left(1 + \frac{1}{\nu} \ ; \frac{3}{4}
        (2 c_B - c) z_h^2 \right)} \ \times \label{eq:4.01} \\
  \end{split}
\end{gather}
\begin{gather}
    \left. \times \left. \left[
          1 - \cfrac{\mu^2 \ (2 c_B - c)^{-\frac{1}{\nu}}}{4 L^2
            \left( 1 - e^{\frac{1}{4}(c-2c_B)z_h^2} \right)^2} \left(
            \Gamma\left(1 + \frac{1}{\nu} \ ; 0\right) - \Gamma\left(1
              + \frac{1}{\nu} \ ; (2 c_B - c) z_h^2 \right) \right)
        \right] \right\} \right|, \nn \\
  s = \cfrac{1}{4} \left( \cfrac{L}{z_h} \right)^{1+\frac{2}{\nu}}
  e^{-\frac{1}{4}(3c-2c_B)z_h^2}. \label{eq:4.02}
\end{gather}

\begin{figure}[b!]
  \centering
  \includegraphics[scale=0.22]{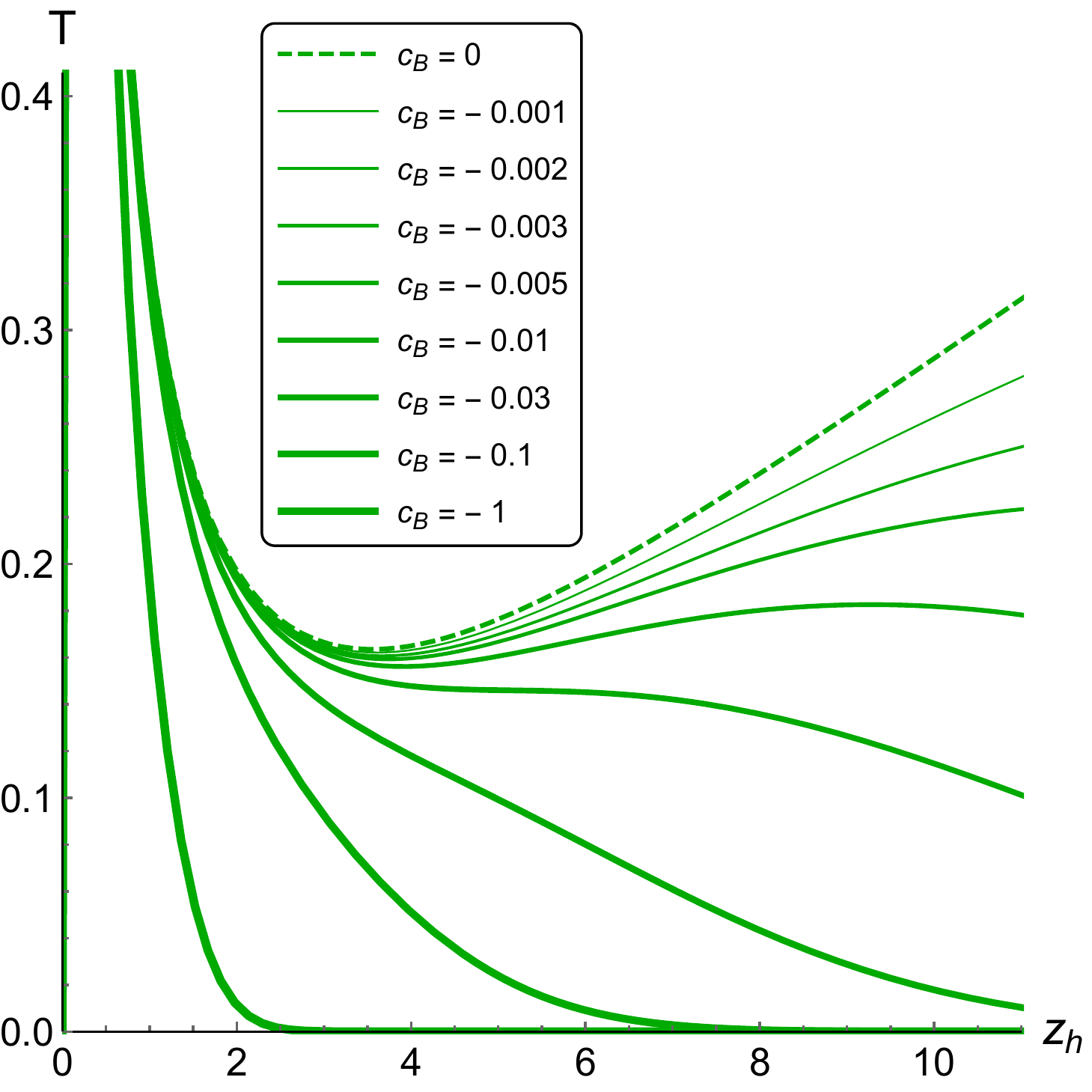} \quad
  \includegraphics[scale=0.22]{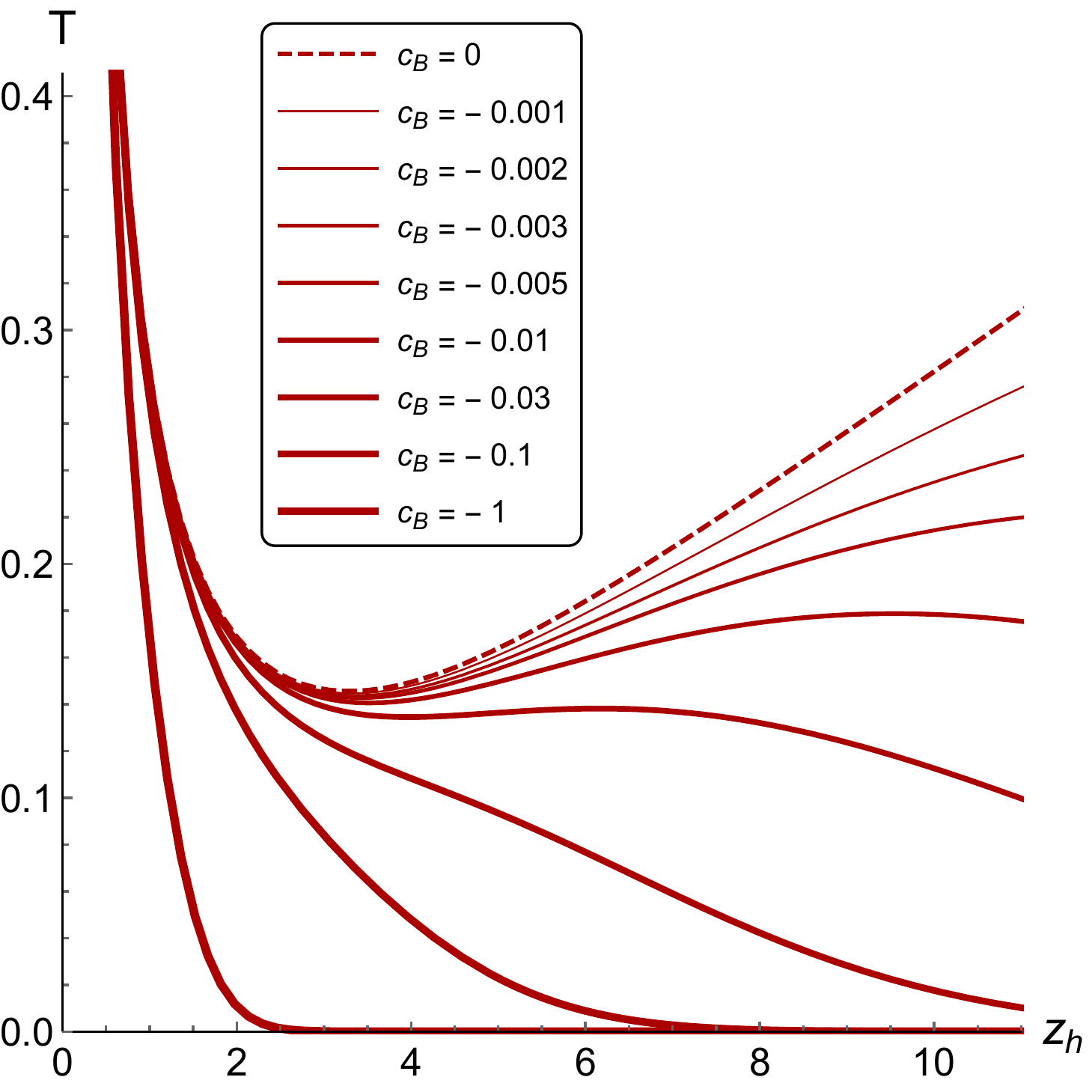} \quad
  \includegraphics[scale=0.22]{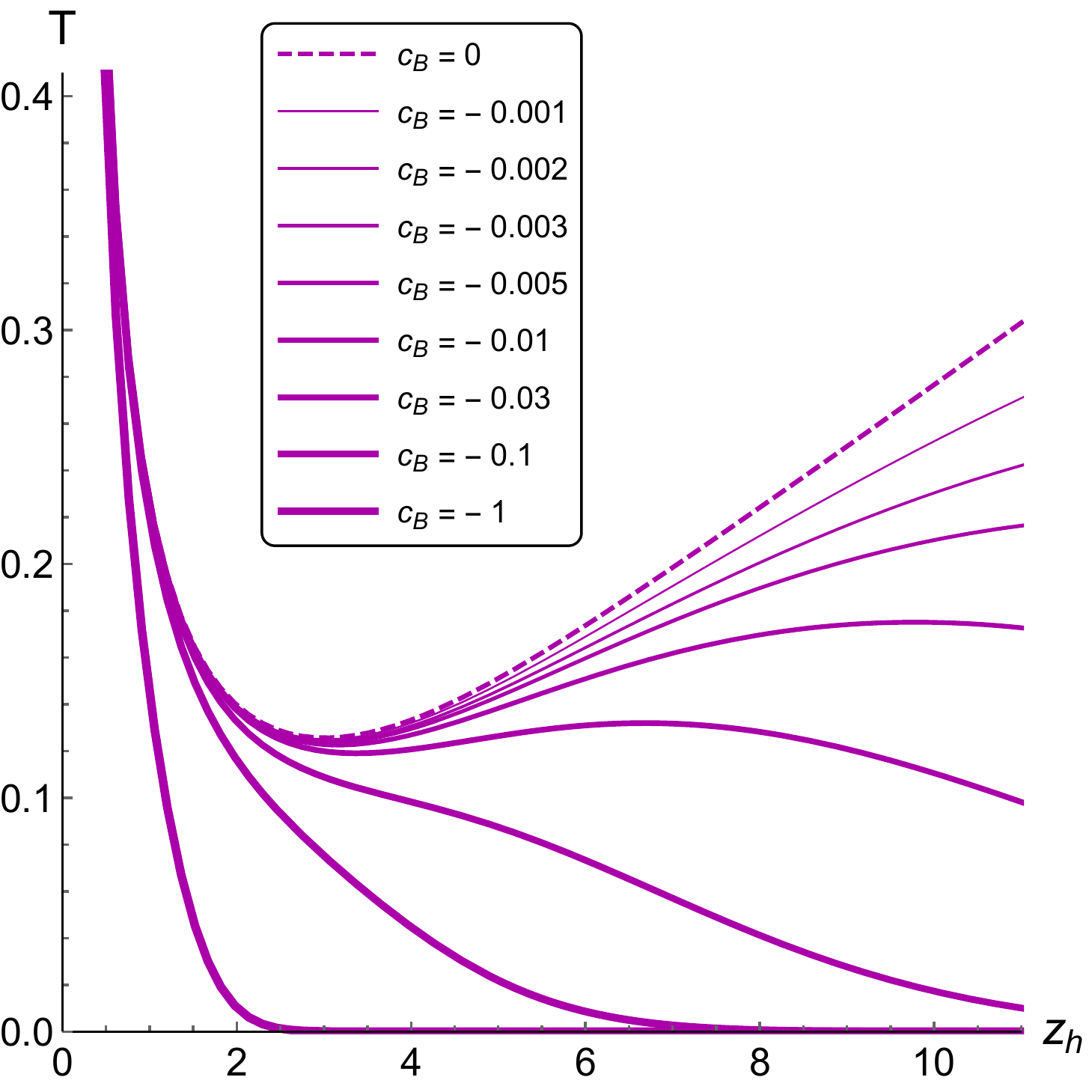} \quad
  \includegraphics[scale=0.22]{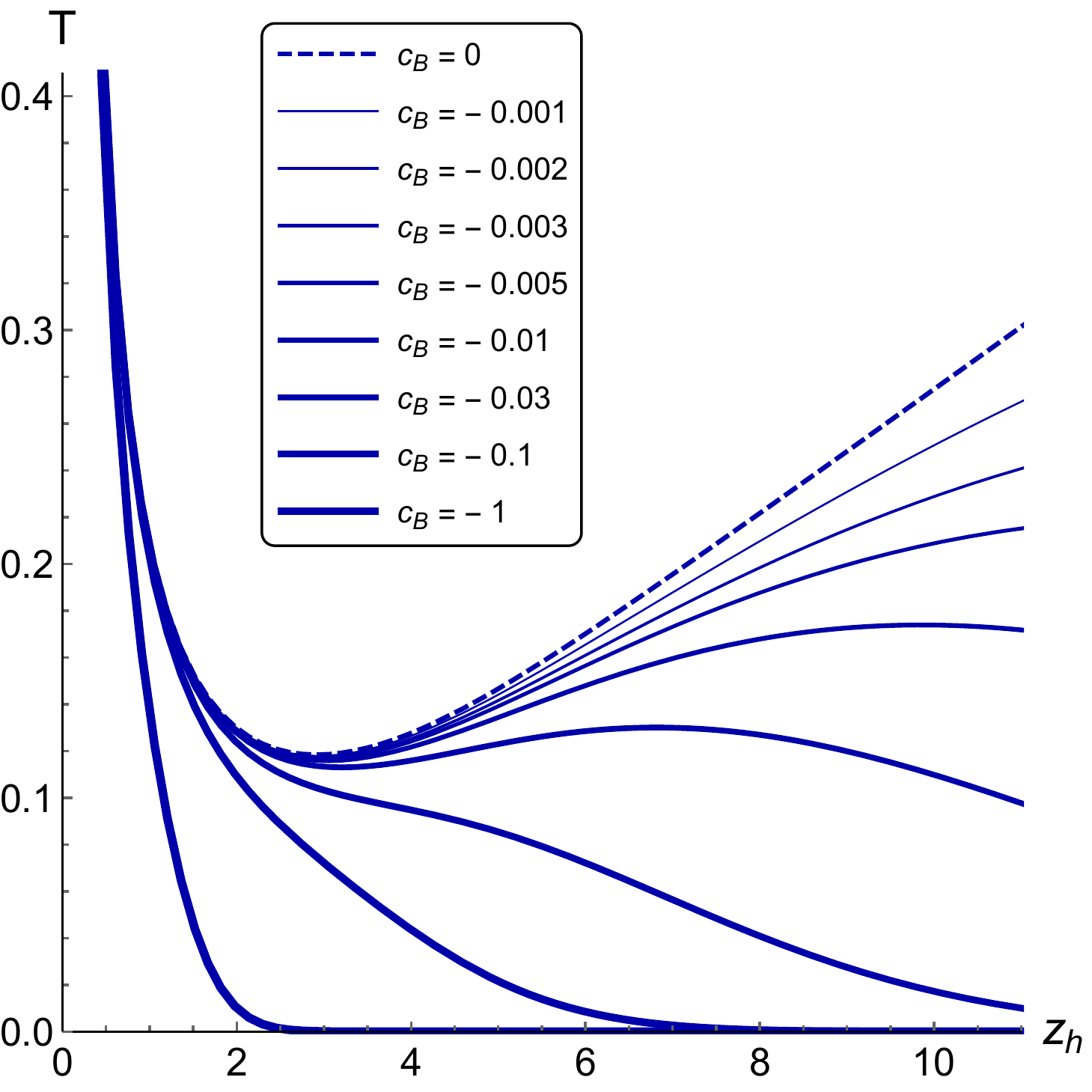} \\
  A \hspace{90pt} B \hspace{90pt}
  C \hspace{90pt} D
  \caption{Temperature $T(z_h)$ for different $c_B$ for $\nu =
    1$ (A), $\nu = 1.5$ (B), $\nu = 3$ (C), $\nu = 4.5$ (D); $c =
    0.227$, $\mu = 0$.}
  \label{Fig:TzhcB}
\end{figure}

On Fig.\ref{Fig:TzhcB} we see, that temperature is quite sensible to
$c_B$ even for zero chemical potential. Function $T(z_h)$ is
multivalued for $c_B > - \, 0.02 \div - \, 0.01$ (for differen
$\nu$). For larger anisotropy the ambiguity of temperature is
preserved till larger absolute value of coupling coefficient $c_B$.

\begin{figure}[t!]
  \centering
  \includegraphics[scale=0.26]{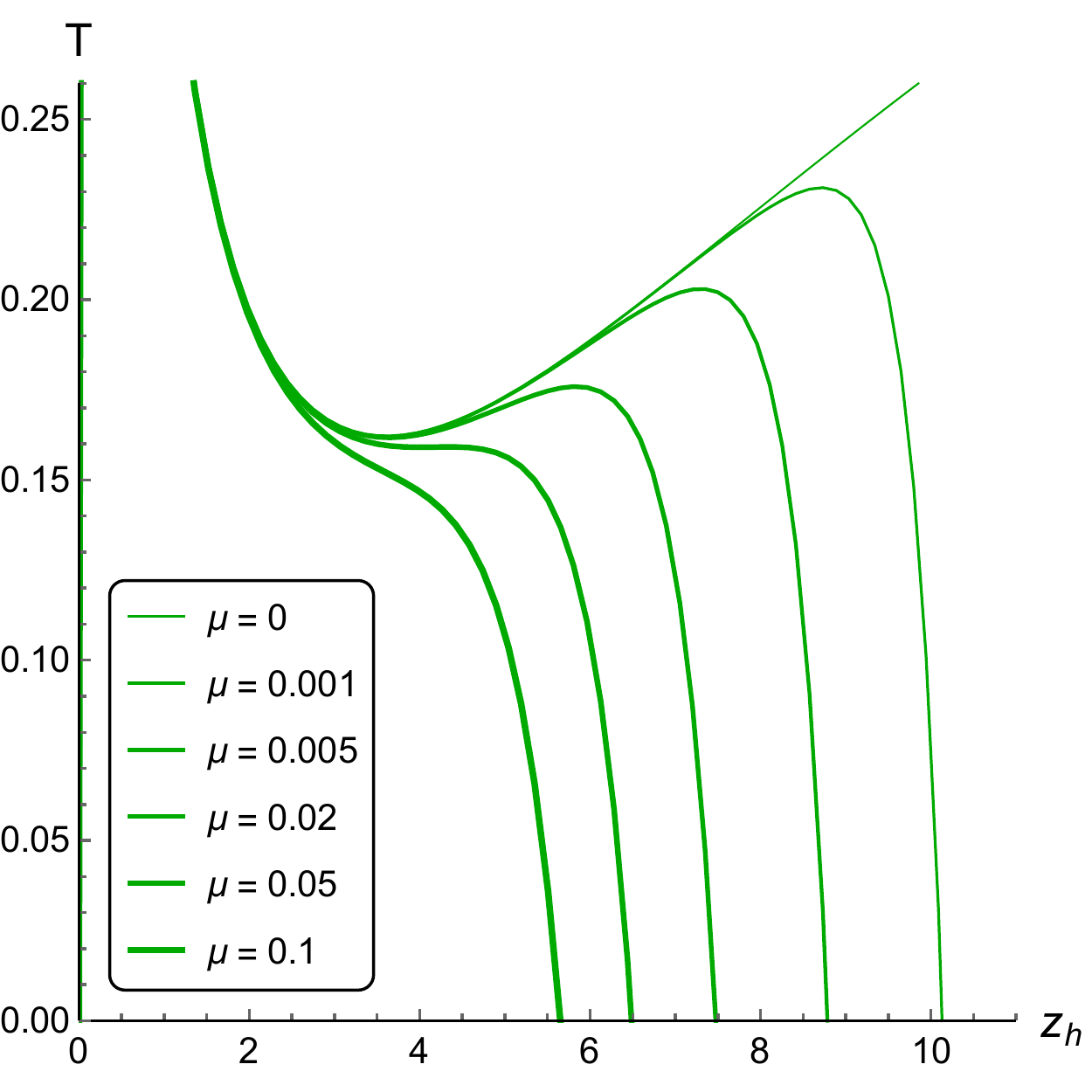} \quad
  \includegraphics[scale=0.26]{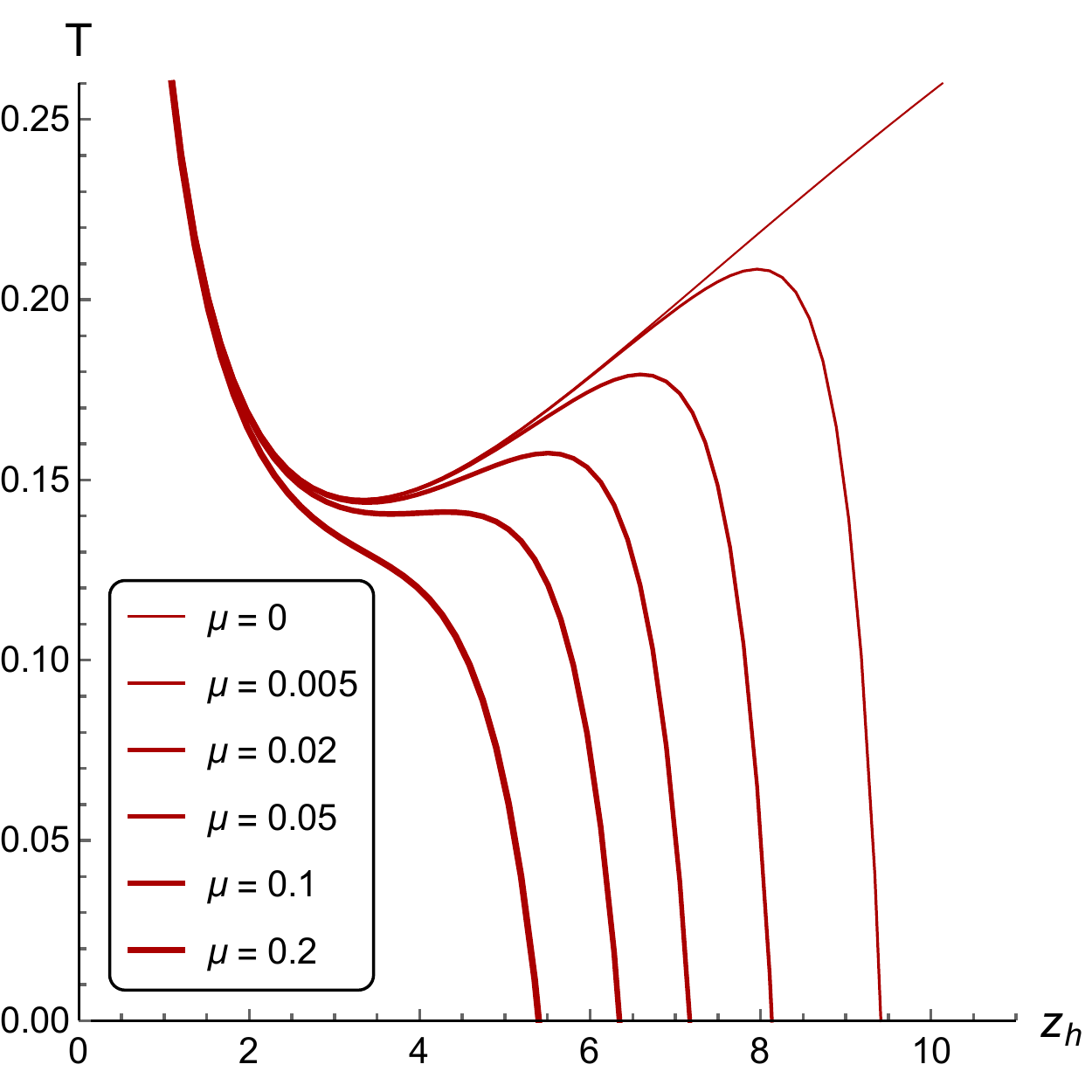} \quad
  \includegraphics[scale=0.26]{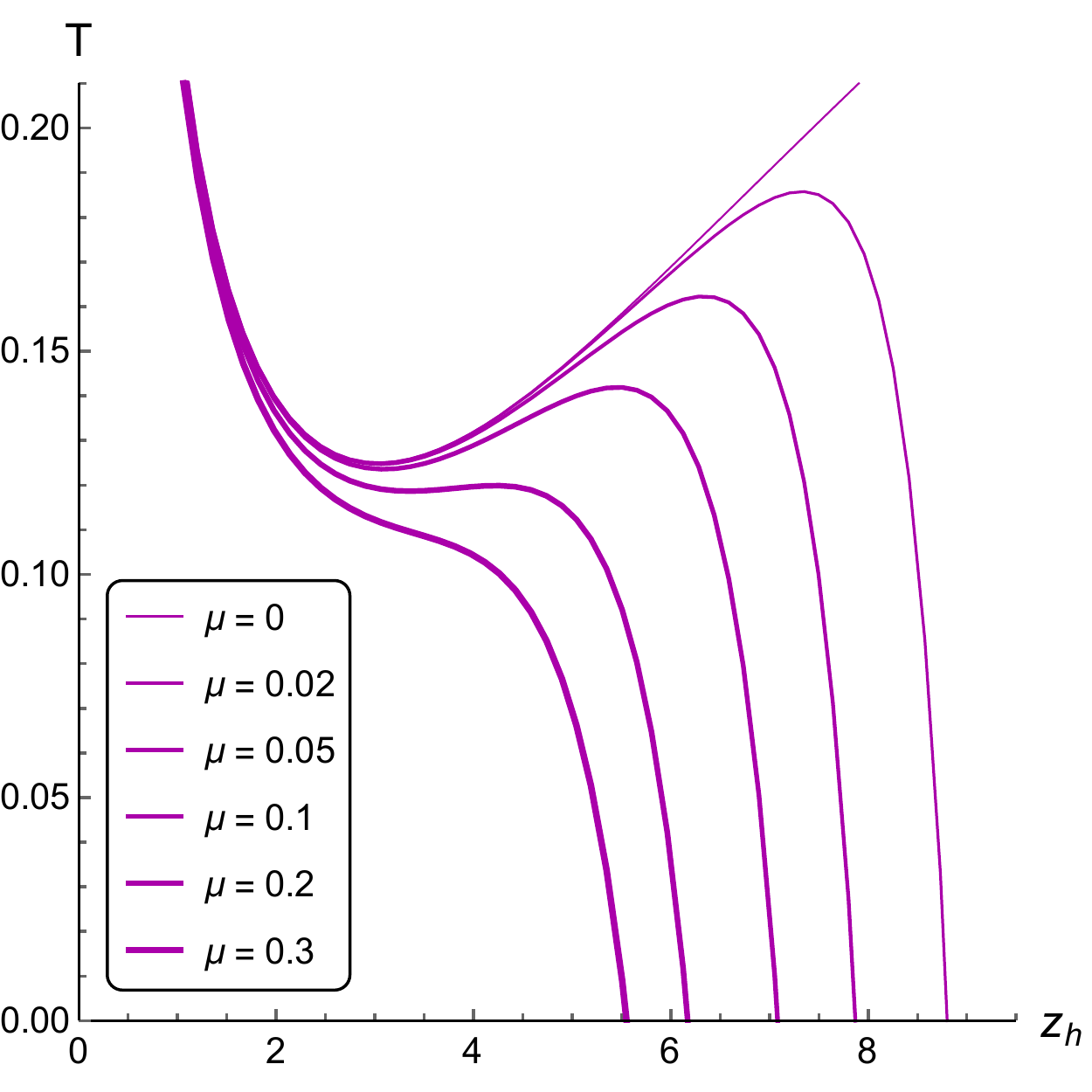} \quad
  \includegraphics[scale=0.26]{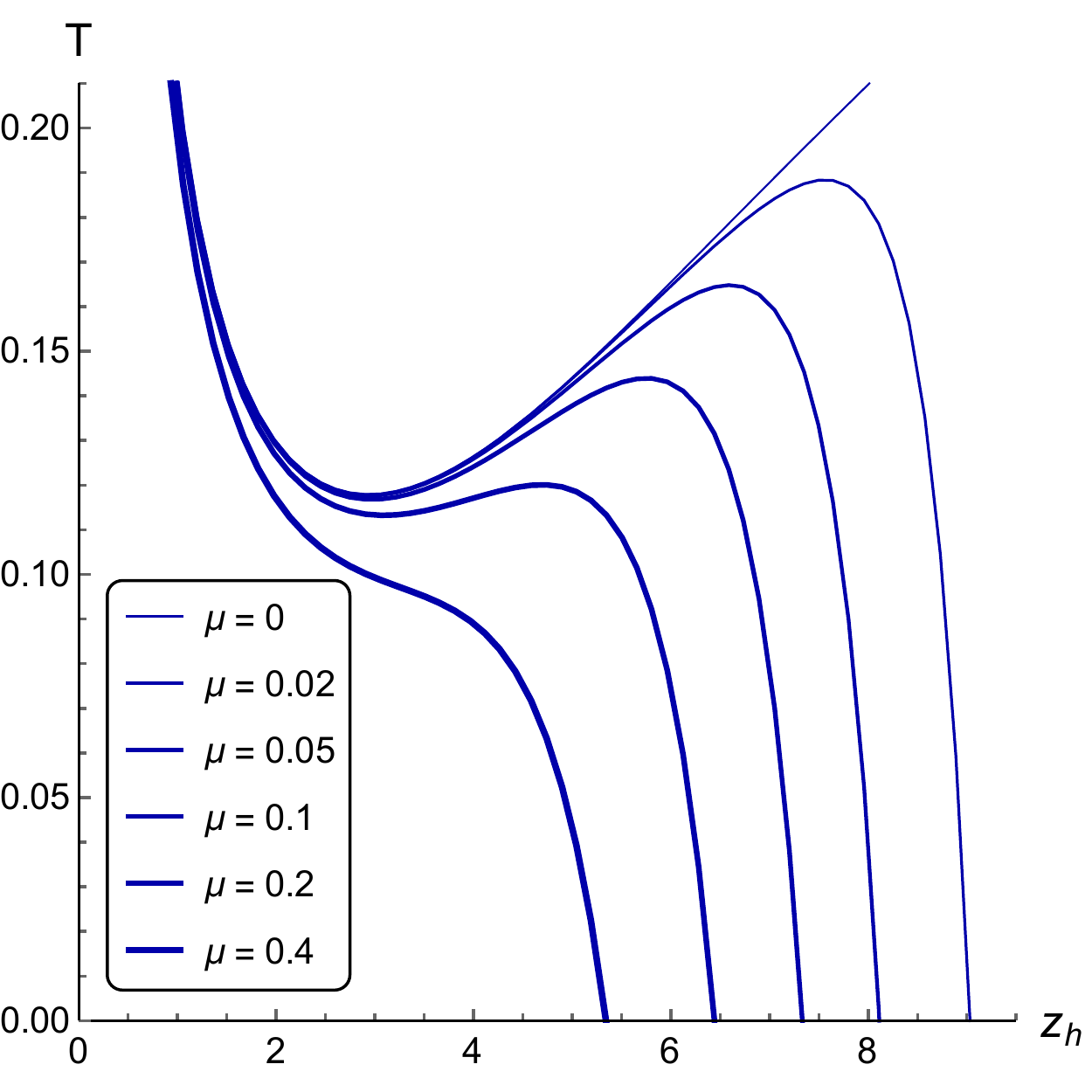} \\
  A \hspace{90pt} B \hspace{90pt}
  C \hspace{90pt} D
  \caption{Temperature $T(z_h)$ for different $\mu$ for $\nu =
    1$ (A), $\nu = 1.5$ (B), $\nu = 3$ (C), $\nu = 4.5$ (D); $c =
    0.227$, $c_B = - \, 0.001$.}
  \label{Fig:Tzhmu}
\end{figure}

Temperature dependence on chemical potential in the presence of
external magnetic field is similar to the previous model
\cite{1802.05652} (Fig.\ref{Fig:Tzhmu}). For larger anisotropy the
ambiguity of temperatue is preserved for larger $\mu$ values: till
$\mu \approx 0.05$ for $\nu = 1$ (Fig.\ref{Fig:Tzhmu}.A) vs $\mu
\approx 0.3$ for $\nu = 4.5$ (Fig.\ref{Fig:Tzhmu}.D).

\begin{figure}[t!]
  \centering
  \includegraphics[scale=0.26]{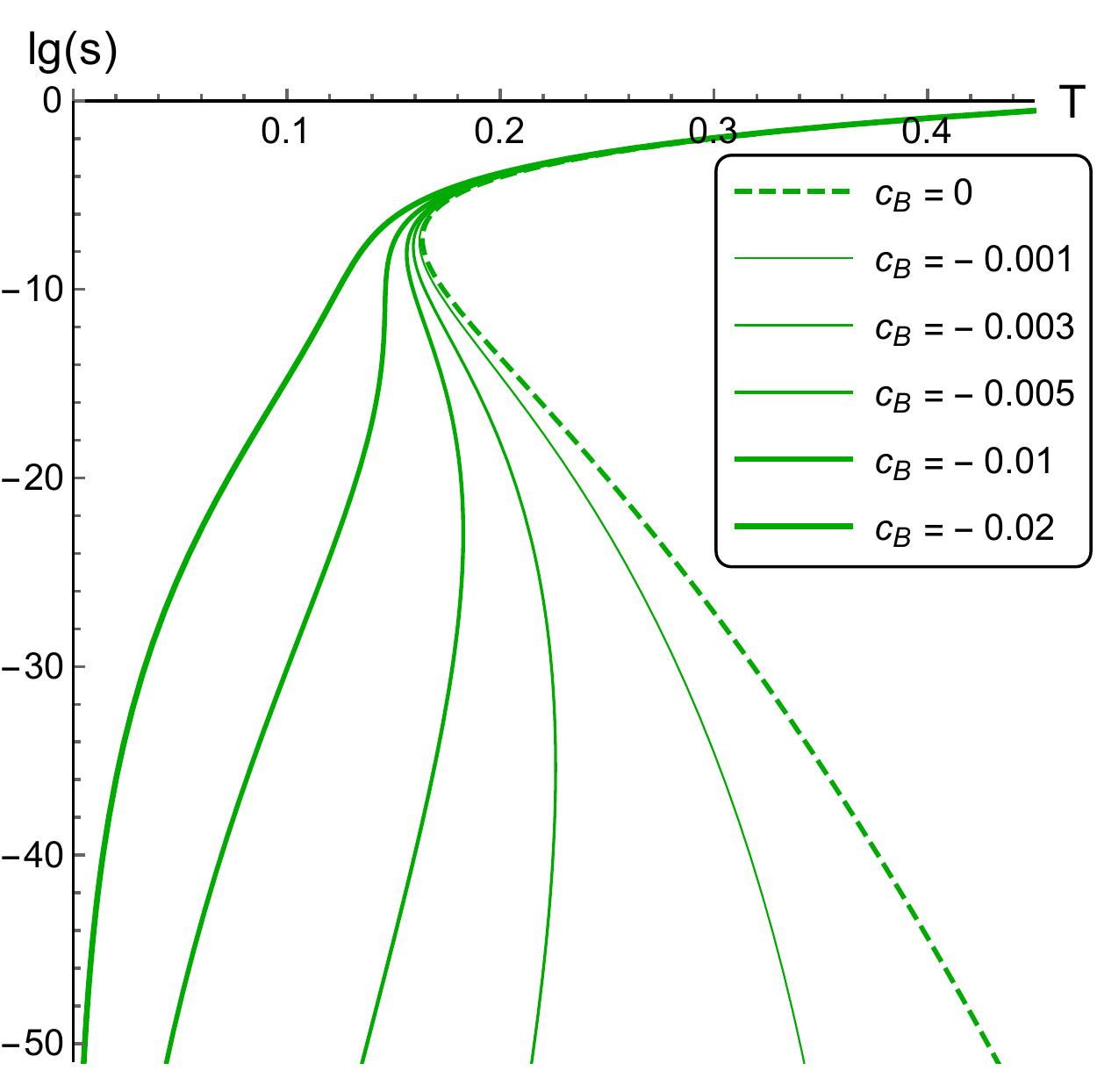} \quad
  \includegraphics[scale=0.26]{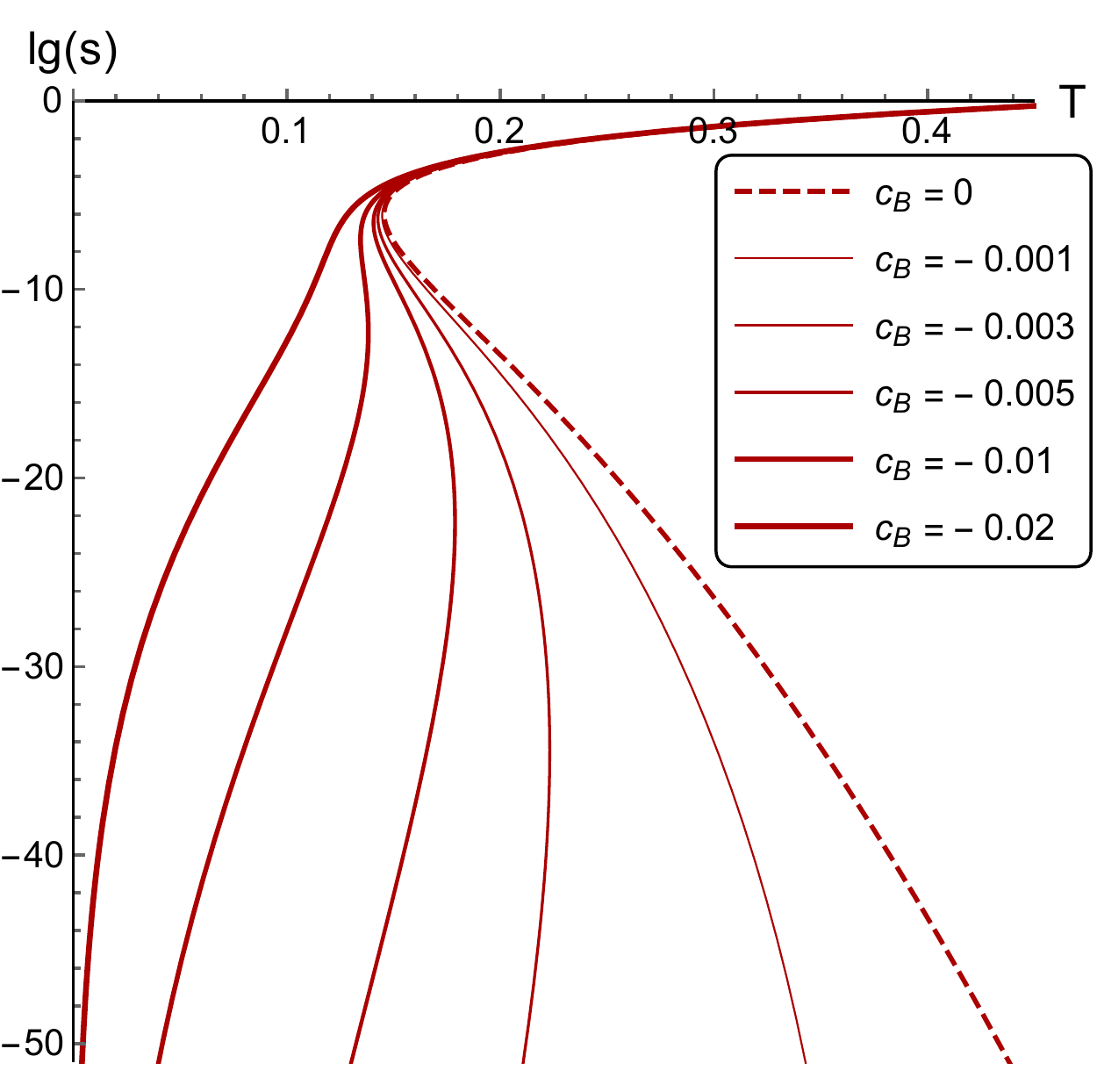} \quad
  \includegraphics[scale=0.26]{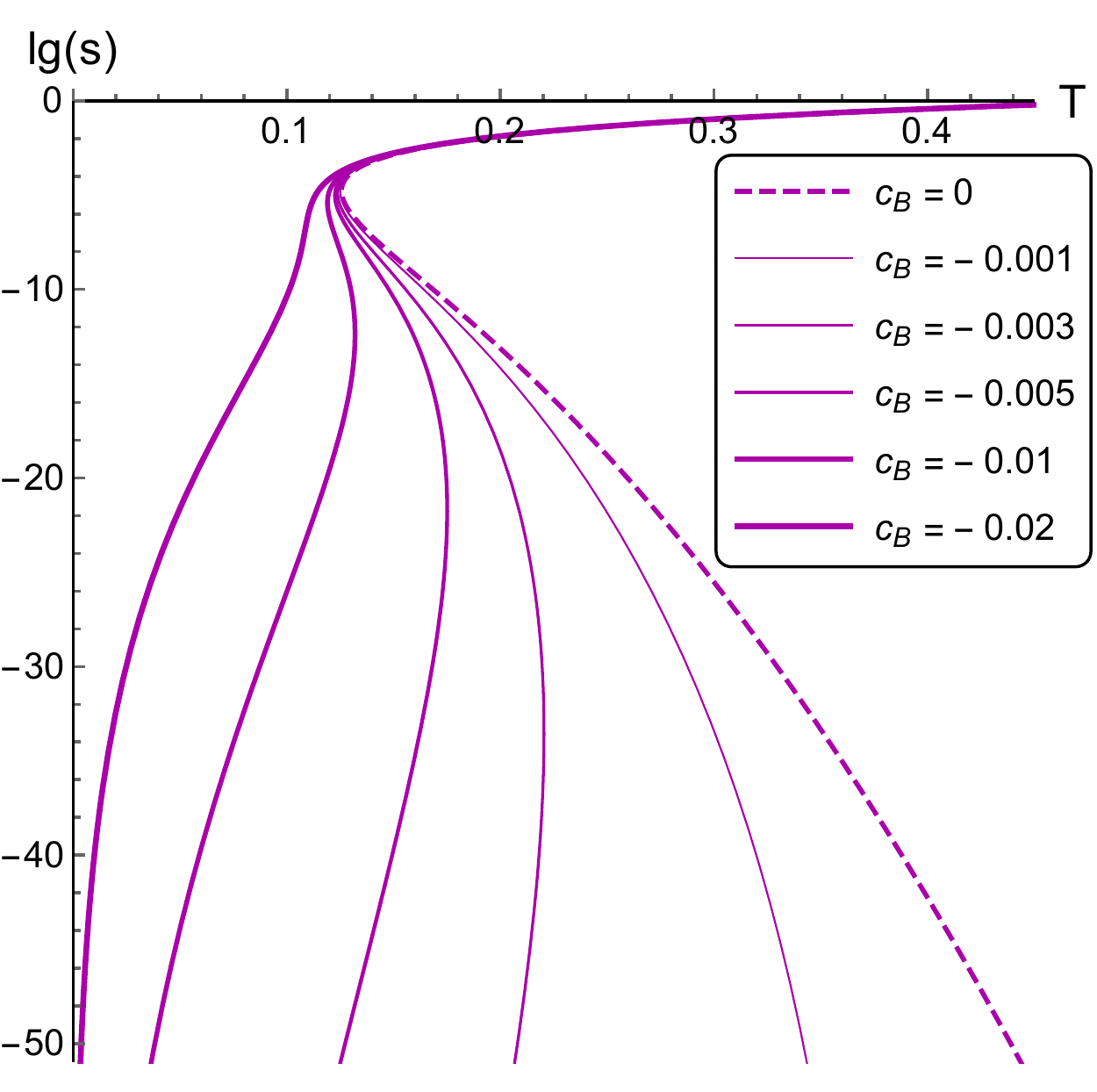} \quad
  \includegraphics[scale=0.26]{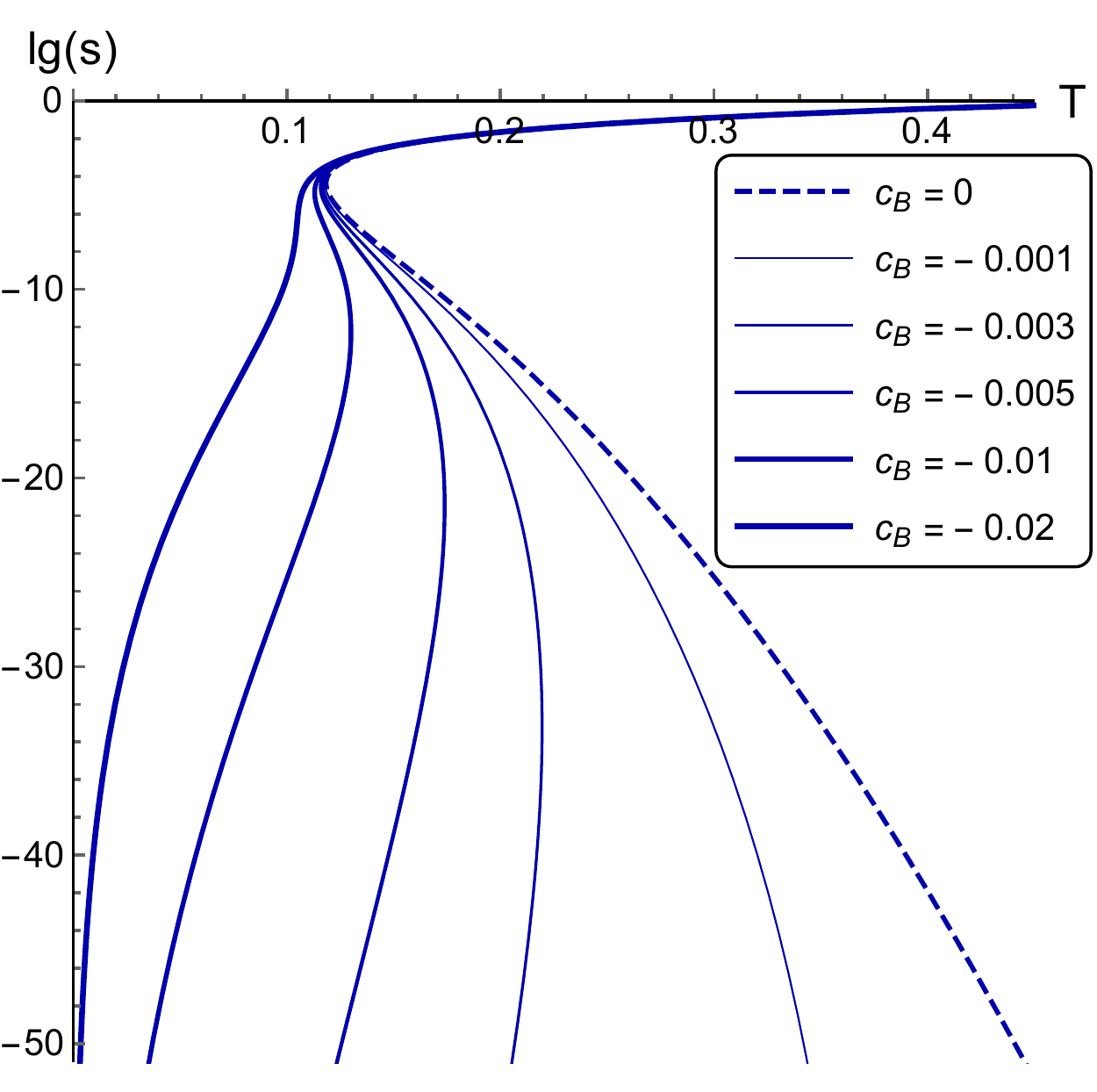} \\
  A \hspace{90pt} B \hspace{90pt}
  C \hspace{90pt} D
  \caption{Entropy $s(T)$ (in logarithmic scale) for different $c_B$
    for $\nu = 1$ (A), $\nu = 1.5$ (B), $\nu = 3$  (C), $\nu = 4.5$
    (D); $c = 0.227$, $\mu = 0$.} 
  \label{Fig:sTcB}
\end{figure}

\begin{figure}[t!]
  \centering
  \includegraphics[scale=0.26]{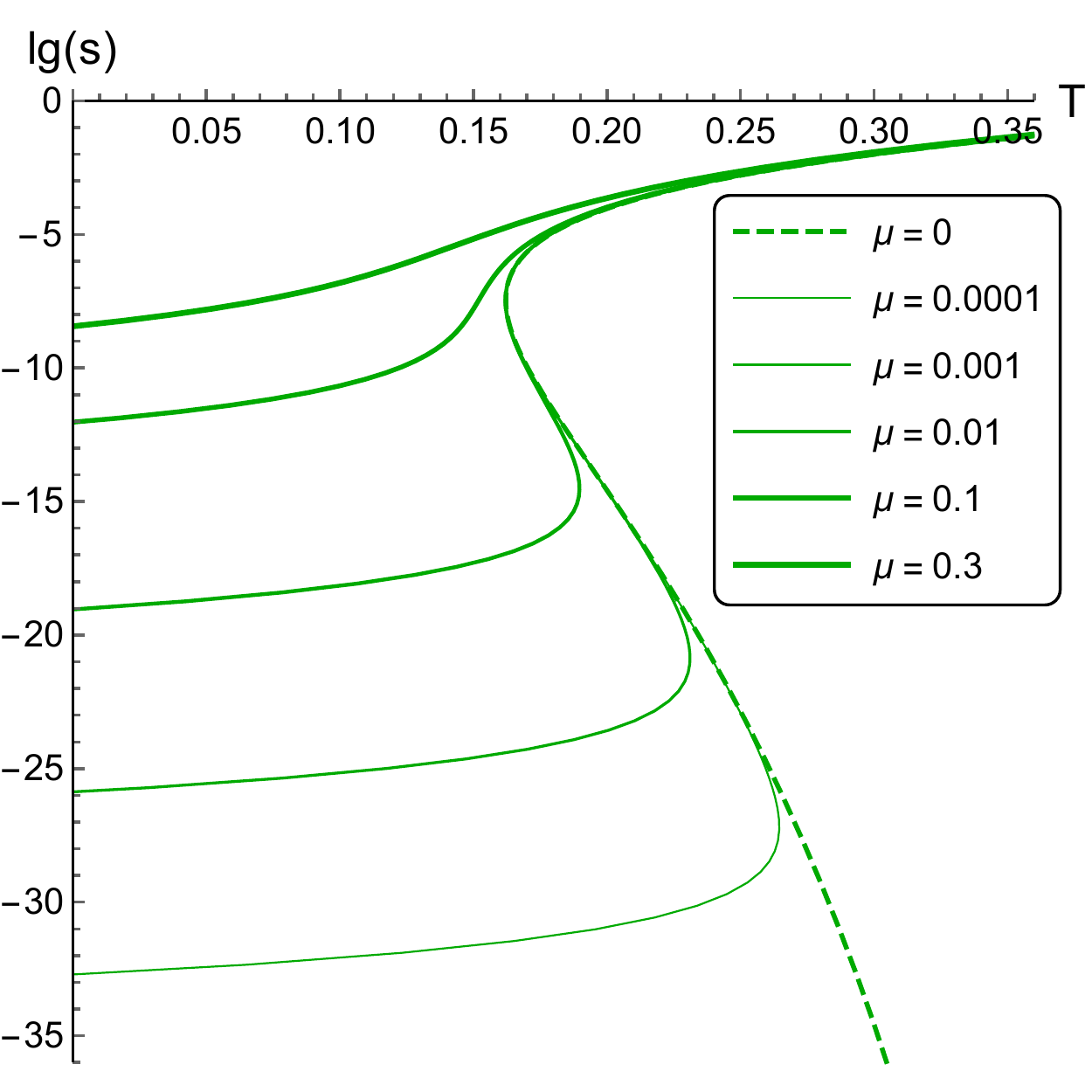} \quad
  \includegraphics[scale=0.26]{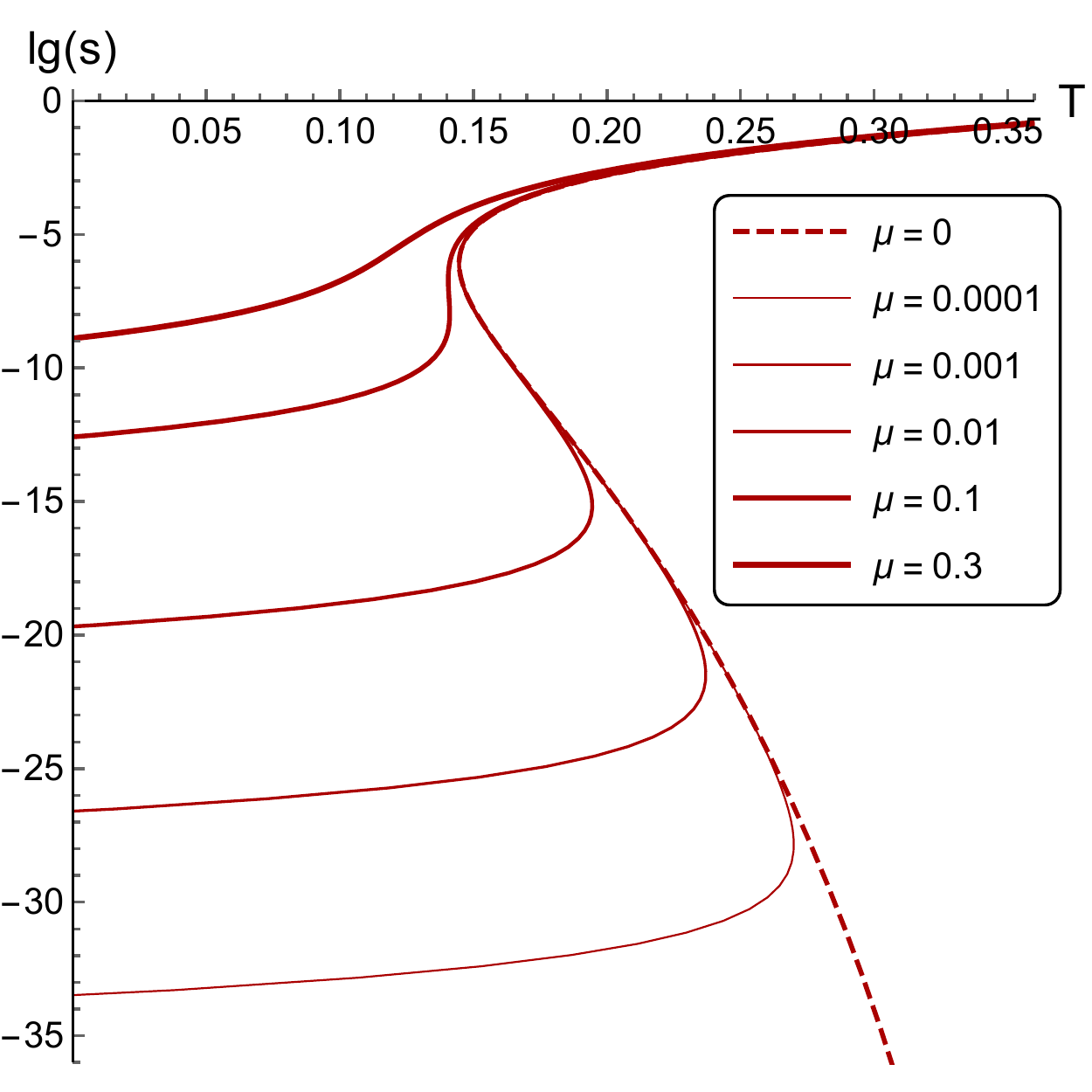} \quad
  \includegraphics[scale=0.26]{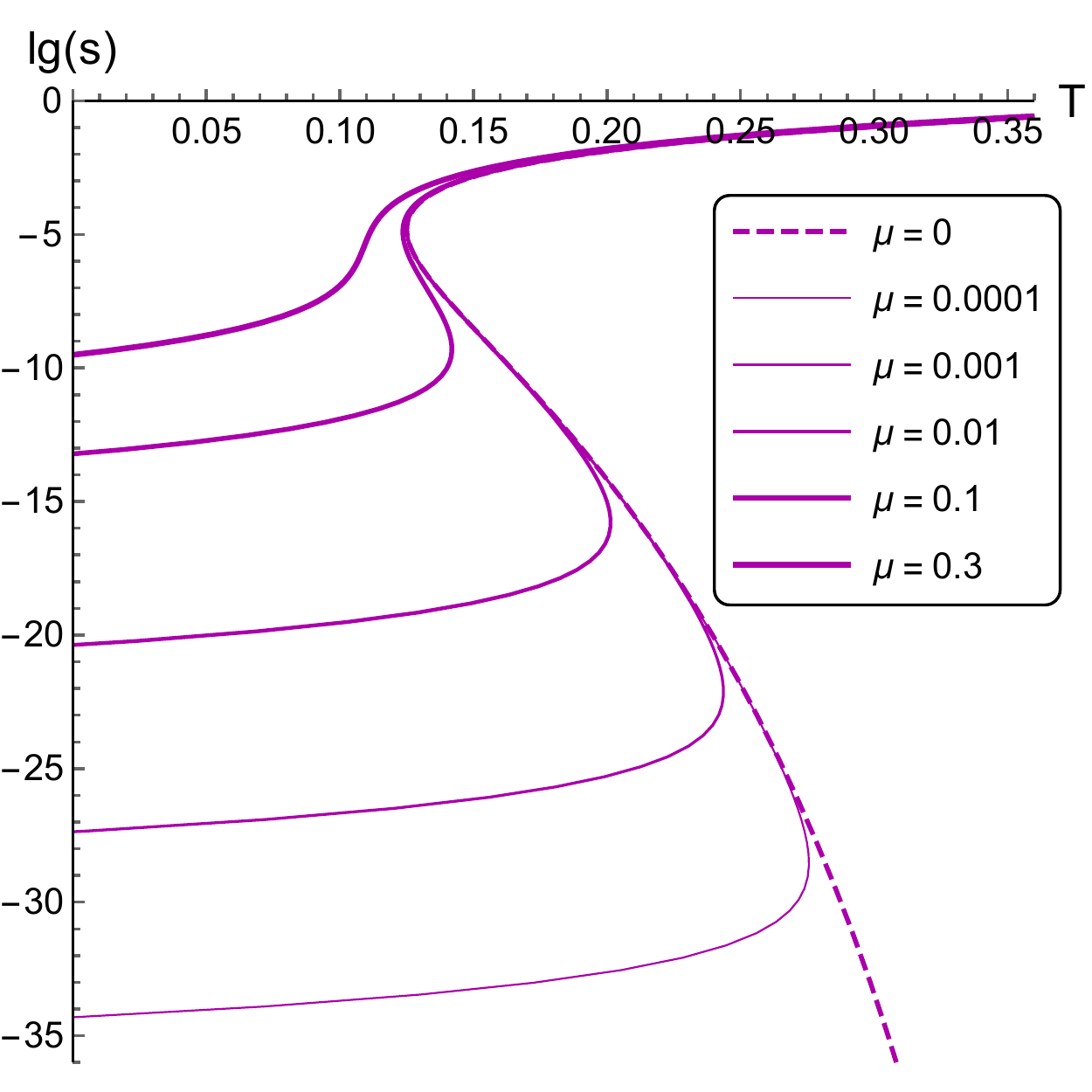} \quad
  \includegraphics[scale=0.26]{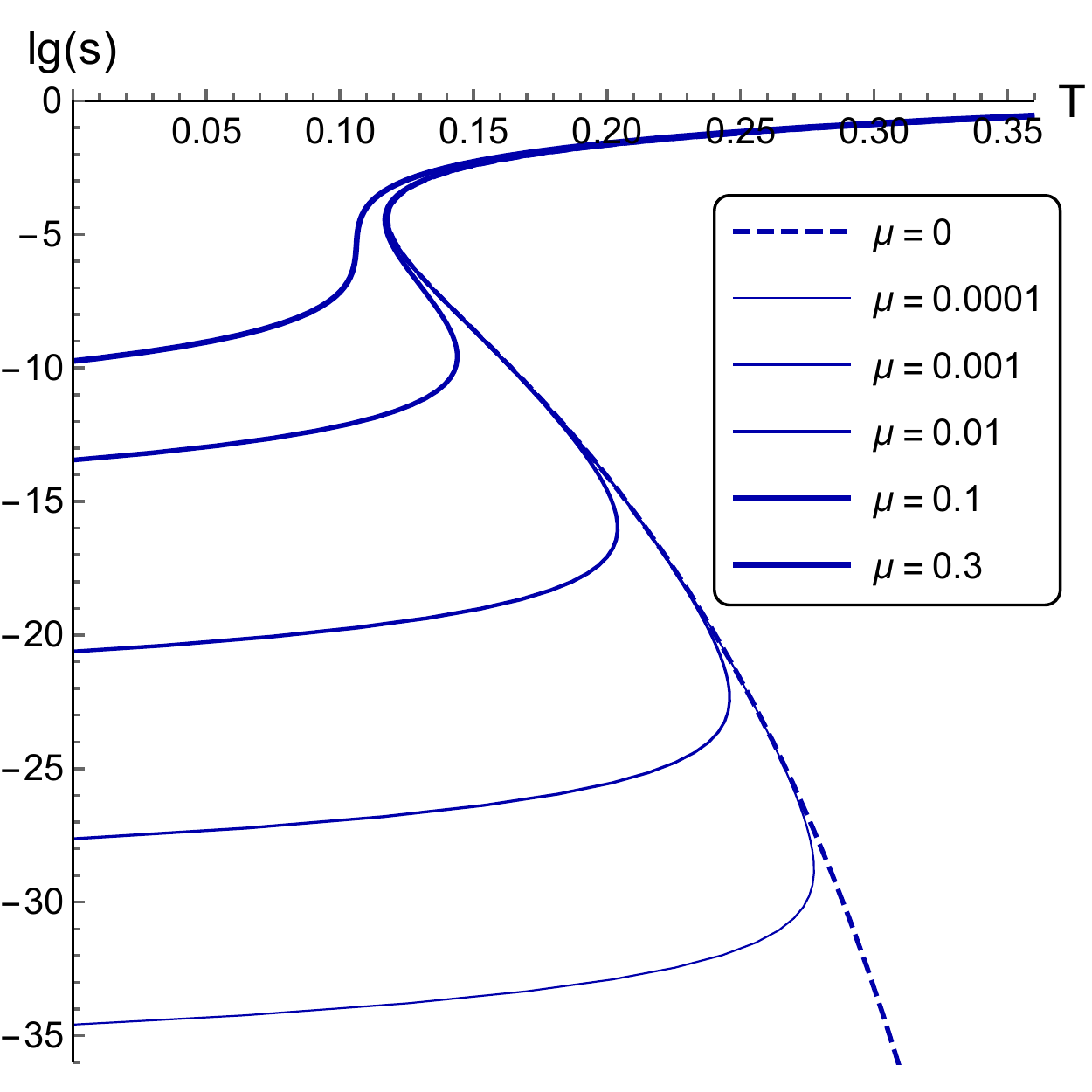} \\
  A \hspace{90pt} B \hspace{90pt}
  C \hspace{90pt} D
  \caption{Entropy $s(T)$ (in logarithmic scale) for different $\mu$
    for $\nu = 1$ (A), $\nu = 1.5$ (B), $\nu = 3$ (C), $\nu = 4.5$
    (D); $c = 0.227$, $z_h = 1$, $c_B = - \, 0.001$.}
  \label{Fig:sTmu}
\end{figure}

Behavior of entropy as function of temperature for different values of
the coupling coefficient $c_B$ (Fig.\ref{Fig:sTcB}) and chemical
potential $\mu$ (Fig.\ref{Fig:sTmu}) convinces us of the conclusions
drawn above. Therefore the confinement/deconfinement phase transition
provided by the background (that we called a Hawking-Page-like phase
transition previously \cite{1802.05652, 2009.05562,
  1808.05596,K-dubna}) in the presence of magnetic field significantly
depends on the coupling coefficient $c_B$. If the connection of metric
with the external magnetic field is too strong ($c_B < - \, 0.02 \div
- \, 0.01$ for different $\nu$), the background phase transition
shouldn't happen at all. Note, that this process doesn't generally
depend on the strength of the field $F_{\mu\nu}^{(B)}$ itself, as both
temperature and entropy functions do not contain ``charge'' $q_B$
explicitly. So the coupling coefficient $c_B$ turns out to be the
parameter that can be fitted via background phase transition.

\begin{figure}[t!]
  \centering
  \includegraphics[scale=0.4]{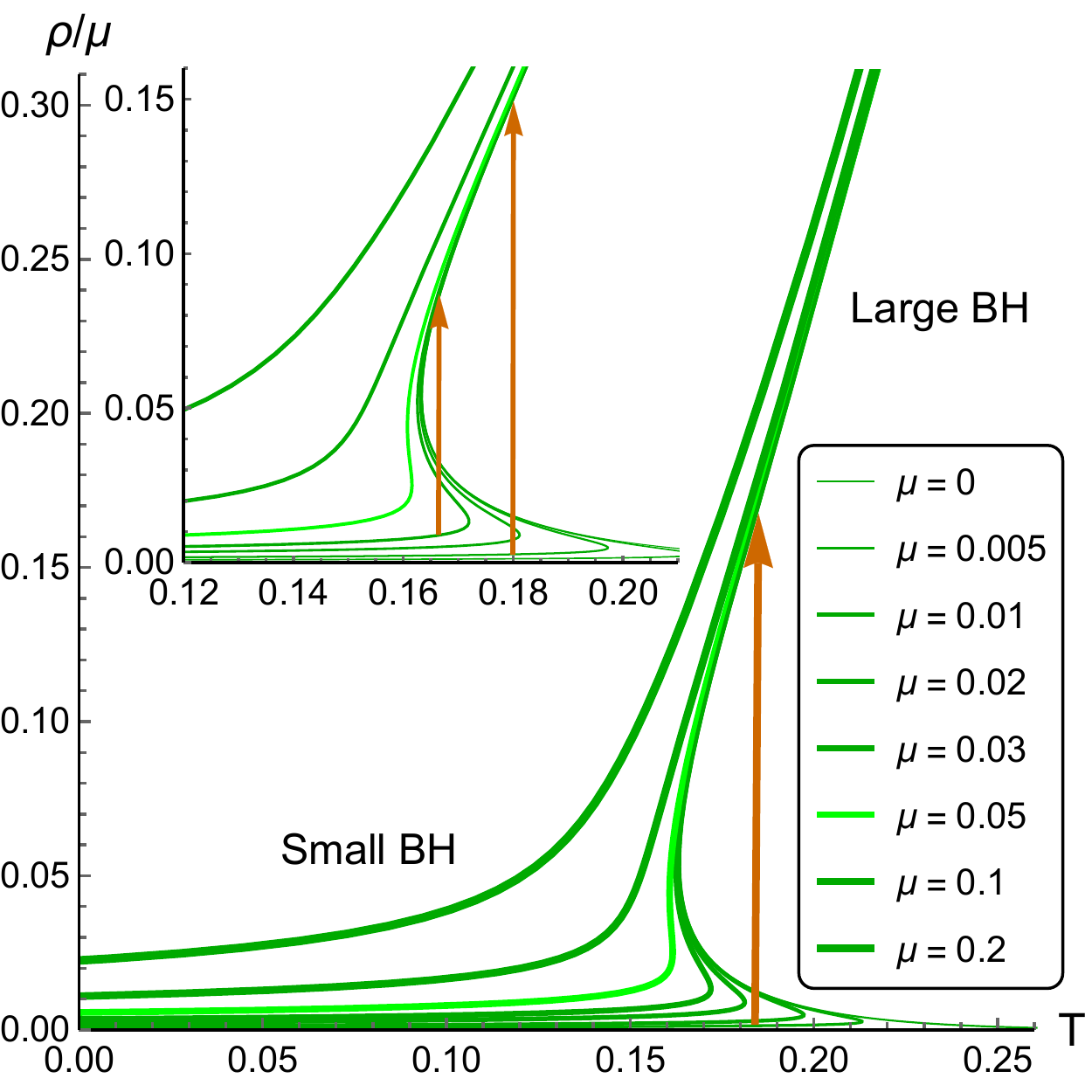} \qquad\qquad
  \includegraphics[scale=0.4]{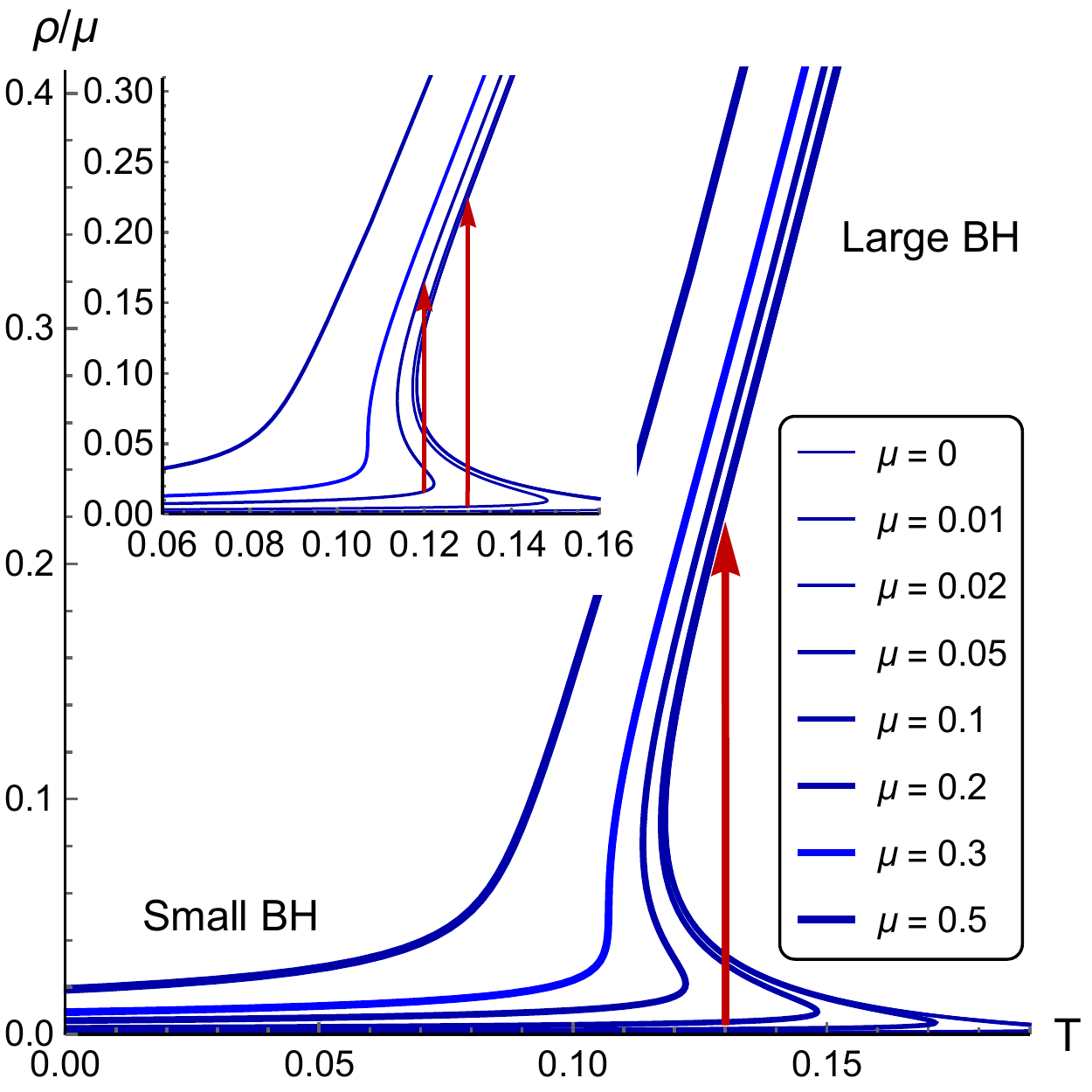} \\
  A \hspace{160pt} B
  \caption{Density $\rho/\mu(T)$ for different $\mu$ for $\nu = 1$ (A)
    and $\nu = 4.5$ (B); $c = 0.227$, $c_B = 0$.}
  \label{Fig:rhoT}
\end{figure}

To ensure that BH-BH transition caused by three-valued temperature
function is a phase transition let us consider density $\rho$, that is
a coefficient in $A_t$ expansion:
\begin{gather}
  A_t = \mu - \rho z^2 + \dots 
  = \mu + \cfrac{\left(c - 2 c_B \right) \mu z^2}{4 \left( 1 -
      e^{\frac14(c-2c_B)z_h^2} \right)} + \dots, \label{eq:4.03} \\
  \rho = - \, \cfrac{\left(c - 2 c_B \right) \mu}{4 \left( 1 -
      e^{\frac14(c-2c_B)z_h^2} \right)} \, . \label{eq:4.04}
\end{gather}
On Fig.\ref{Fig:rhoT} $\rho/\mu$ ratio as a function of temperature
for primary isotropic solution ($\nu = 1$, Fig.\ref{Fig:rhoT}.A) and
anisotropic solution ($\nu = 4.5$, Fig.\ref{Fig:rhoT}.B) are
plotted. Vertical arrows show the BH-BH transition direction. Function
$\rho/\mu$ is a three-digit function of $T$ as expected, and we can
see that collapse from small black holes (larger $z_h$) to the large
ones (smaller $z_h$) is accompanied by a sharp rise of the density for
any appropriate chemical potential. On Fig.\ref{Fig:rhoT} $c_B = 0$,
but for $c_B$ close to zero enough to preserve ambiguity of
$T(z_h)$-function density curves will differ from the zero magnetic
field case negligibly.

\subsection{Free energy and background phase transition}\label{FEBB}

To get BH-BH phase transition, that originates from the model
background, we need to consider the free energy as a function of
temperature:
\begin{gather}
  F =  \int_{z_h}^{\infty} s \, T' dz,
  \qquad
  dF = s \, T'. \label{eq:4.05}
\end{gather}

\begin{figure}[t!]
  \centering
  \includegraphics[scale=0.5]{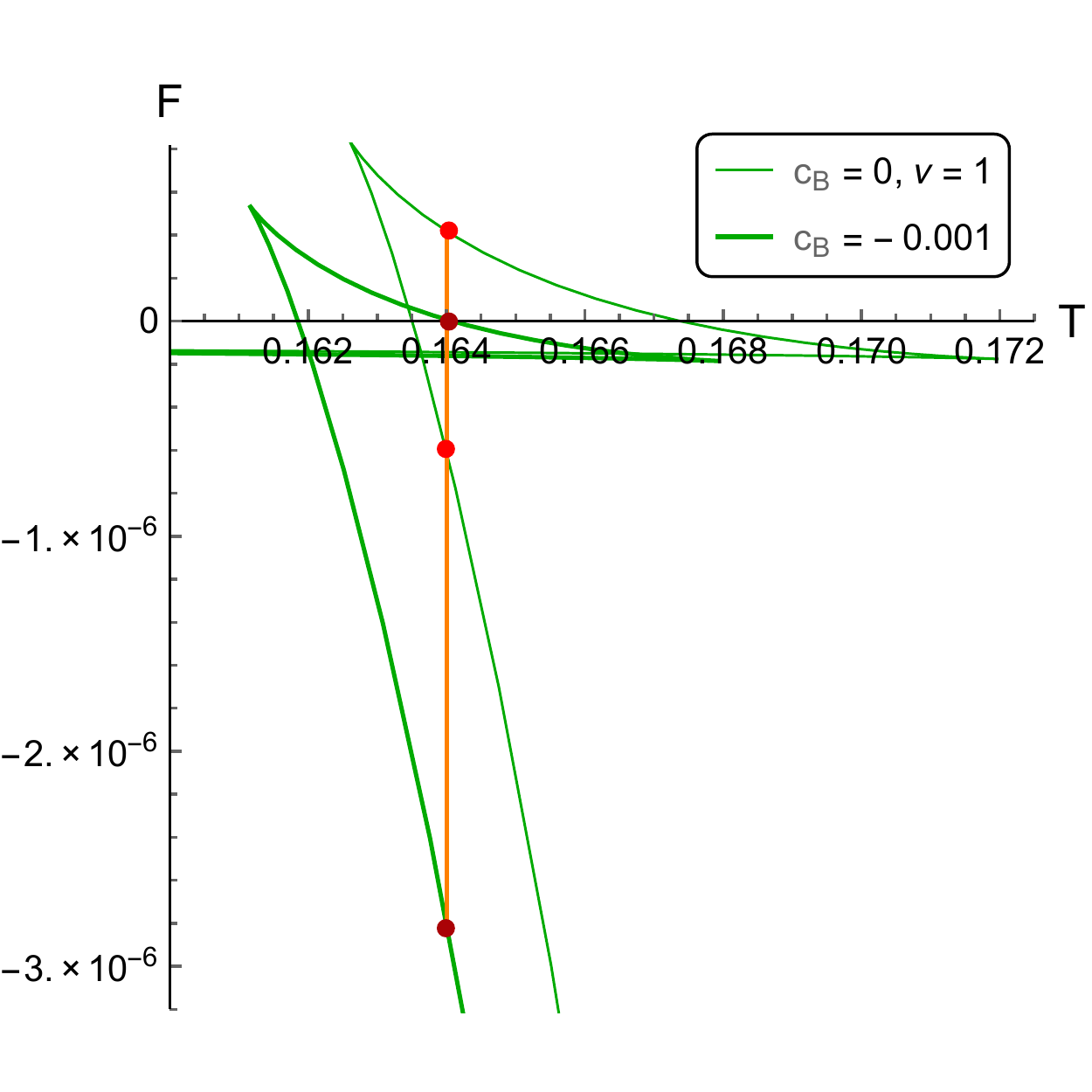} \qquad
  \includegraphics[scale=0.5]{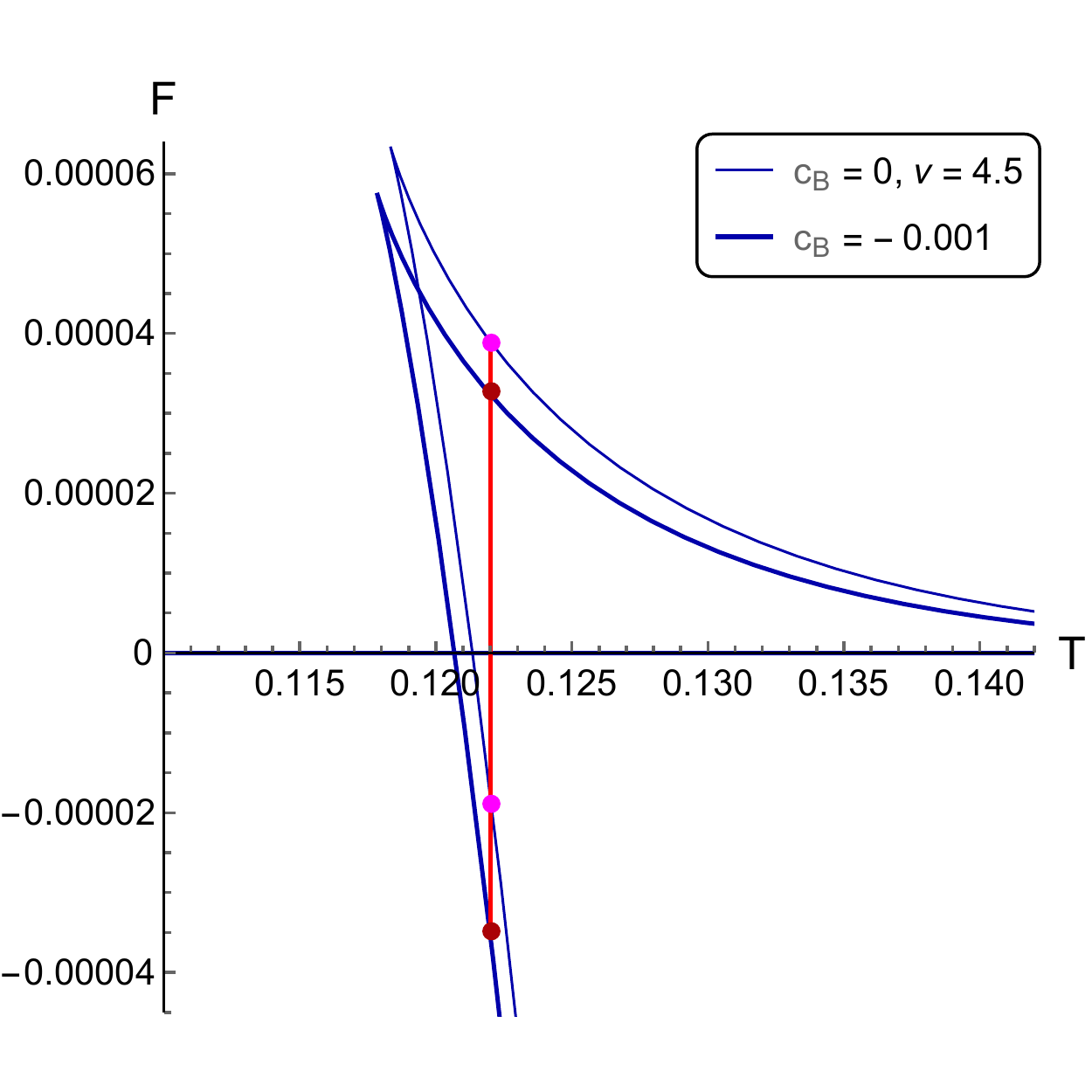} \\
  A \hspace{200pt} B
  \caption{Free energy difference $\Delta F(T)$ for $c_B = 0$ and
    $c_B = - \, 0.001$, $\nu = 1$ (A) and $\nu = 4.5$ (B); $c = 0.227$,
    $\mu = 0.03$ and $T = 0.164$.}
  \label{Fig:dFT}
\end{figure}

Using free energy we can estimate the lifetime of the unstable
state. For the general case we should include the free energy of the
black hole $F_{BH}$ and the free energy of the quarks pair
$F_{q\bar{q}}$ into consideration and also take both connected and
disconnected string configurations into account. Therefore the total
free energy characterizing our system is $F_{tot} = F_{q\bar{q}} +
F_{BH}$. However for small distance $\ell$ between quarks the
difference of free energy values between connected and disconnected
string configurations $\Delta F_{q\bar{q}}$ is small \cite{1506.05930,
  1907.01852}. For small $\ell$ we assume $\Delta F_{q\bar{q}} \ll
\Delta F_{BH}$. Therefore the total free energy difference between
configurations is $\Delta F_{tot} = \Delta F_{q\bar{q}} + \Delta
F_{BH} \approx \Delta F_{BH}$. In this case the lifetime of the
unstable state can be estimated as:
\begin{gather}
  \tau = \cfrac{\hbar}{\Delta F_{tot}(T, \mu)}= \cfrac{\hbar}{\Delta
    F_{BH}(T, \mu)} \, .
  \label{eq:4.06}
\end{gather}
On Fig.\ref{Fig:dFT} an example of such an estimation is
illustrated. Vertical lines show the chosen temperature $T = 0.164$
(Fig.\ref{Fig:dFT}.A,B) or $T = 0.122$ (Fig.\ref{Fig:dFT}.C). It is
orange for $\nu = 1$ (Fig.\ref{Fig:dFT}.A) and red for $\nu = 4.5$
(Fig.\ref{Fig:dFT}.B,C).

In the absence of magnetic field the free energy difference between
two states of the transition is the distance between two red points
located on the orange temperature vertical for $\nu = 1$
(Fig.\ref{Fig:dFT}.A) and between two magenta points located on the
red temperature vertical for $\nu = 4.5$ (Fig.\ref{Fig:dFT}.B,C). For
$c_B = - \, 0.001$ the free energy difference is the distance between
two dark red points on the same temperature verticals. But on scale of
Fig.\ref{Fig:dFT}.B curves for $c_B = 0$ and $c_B = - \, 0.001$ almost
coincide, so red and magenta points merge on this plot.

Taking the coordinates of all these points we have
\begin{gather}
  \begin{split}
    \tau \Big|_{\nu = 1, \, \mu = 0.03, \, T = 0.164 \, c_B = 0}
    \qquad \quad
    &= \cfrac{\hbar}{4.2 \cdot 10^{-7} - (- \, 6.131 \cdot 10^{-7})}
    \approx 6.39 \cdot 10^{-19} \ \mbox{s}, \\
    \tau \Big|_{\nu = 1, \, \mu = 0.03, \, T = 0.164 \, c_B = - 0.001}
    \quad \,
    &= \cfrac{\hbar}{1.314 \cdot 10^{-8} - (- \, 2.807 \cdot 10^{-6})}
    \approx 2.34 \cdot 10^{-19} \ \mbox{s}, \\
    \tau \Big|_{\nu = 4.5, \, \mu = 0.03, \, T = 0.164 \, c_B = 0, -0.001}
    &= \cfrac{\hbar}{3.469 \cdot 10^{-6} - (- \, 3.132 \cdot 10^{-3})}
    \approx 2.10 \cdot 10^{-22} \ \mbox{s}, \\
    \tau \Big|_{\nu = 4.5, \, \mu = 0.03, \, T = 0.122 \, c_B = 0}
    \qquad \ \,
    &= \cfrac{\hbar}{3.92 \cdot 10^{-5} - (- \, 1.81 \cdot 10^{-5})}
    \approx 1.15 \cdot 10^{-20} \ \mbox{s}, \\
    \tau \Big|_{\nu = 4.5, \, \mu = 0.03, \, T = 0.122 \, c_B = -
      0.001} 
    \ \
    &= \cfrac{\hbar}{3.264 \cdot 10^{-5} - (- \, 3.413 \cdot 10^{-5})}
    \approx 9.88 \cdot 10^{-21} \ \mbox{s}.
  \end{split} \label{eq:4.07}
\end{gather}
Therefore the lifetime of the unstable state is about
$10^{-22}$--$10^{-19}$ s, that is larger than the lifetime of QGP. For
$\nu = 1$ presence of magnetic field decreases this value in about
$2.7$ times, while for $\nu = 4.5$ -- in less than $1.2$ times. So the
external magnetic field generally reduces the lifetime of an unstable
state, but primary anisotropy $\nu > 1$ weakens this effect.

\begin{figure}[t!]
  \centering
  \includegraphics[scale=0.57]{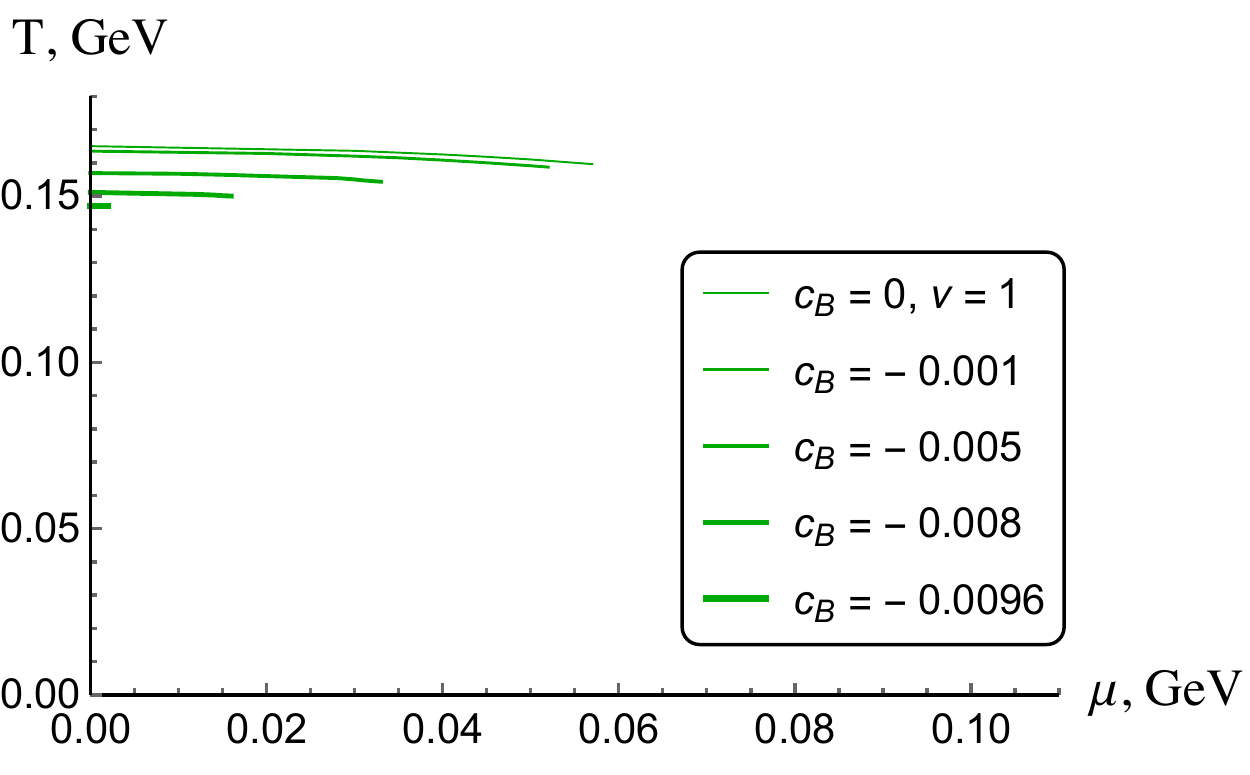} \quad
  \includegraphics[scale=0.57]{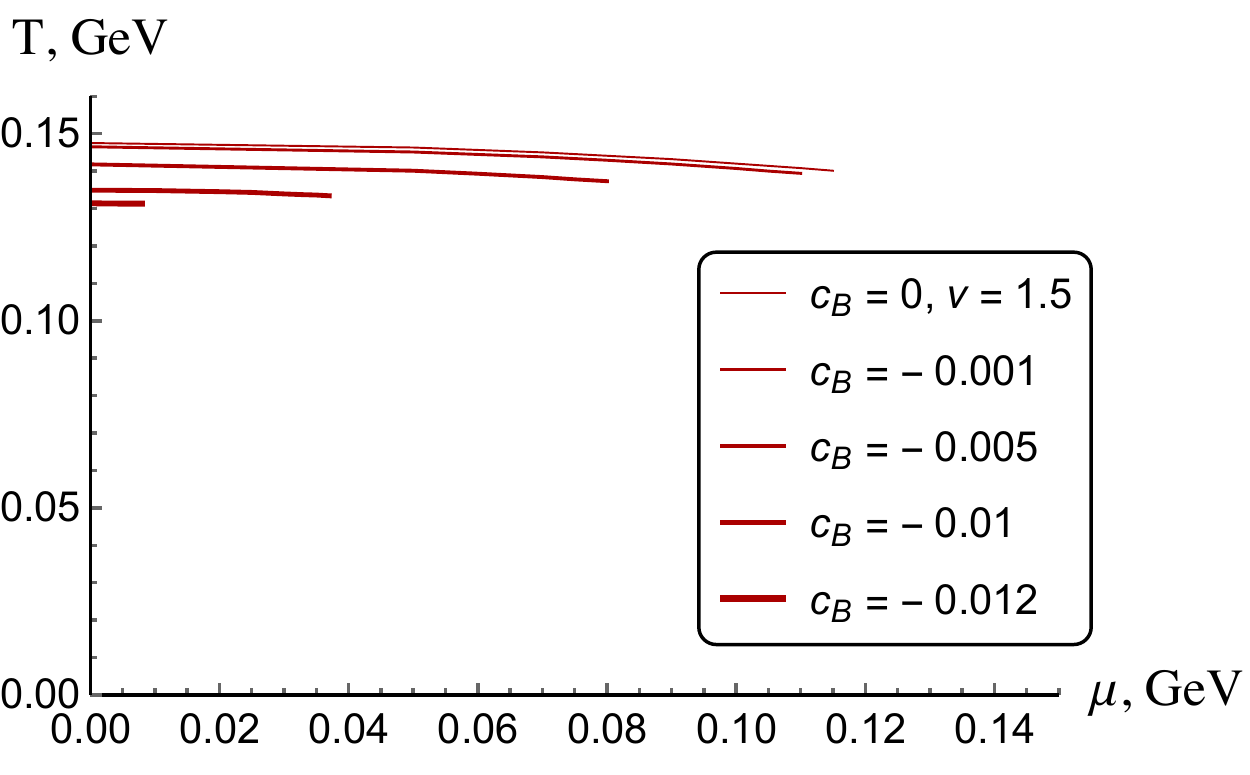} \\
  A \hspace{220pt} B \\
  \includegraphics[scale=0.57]{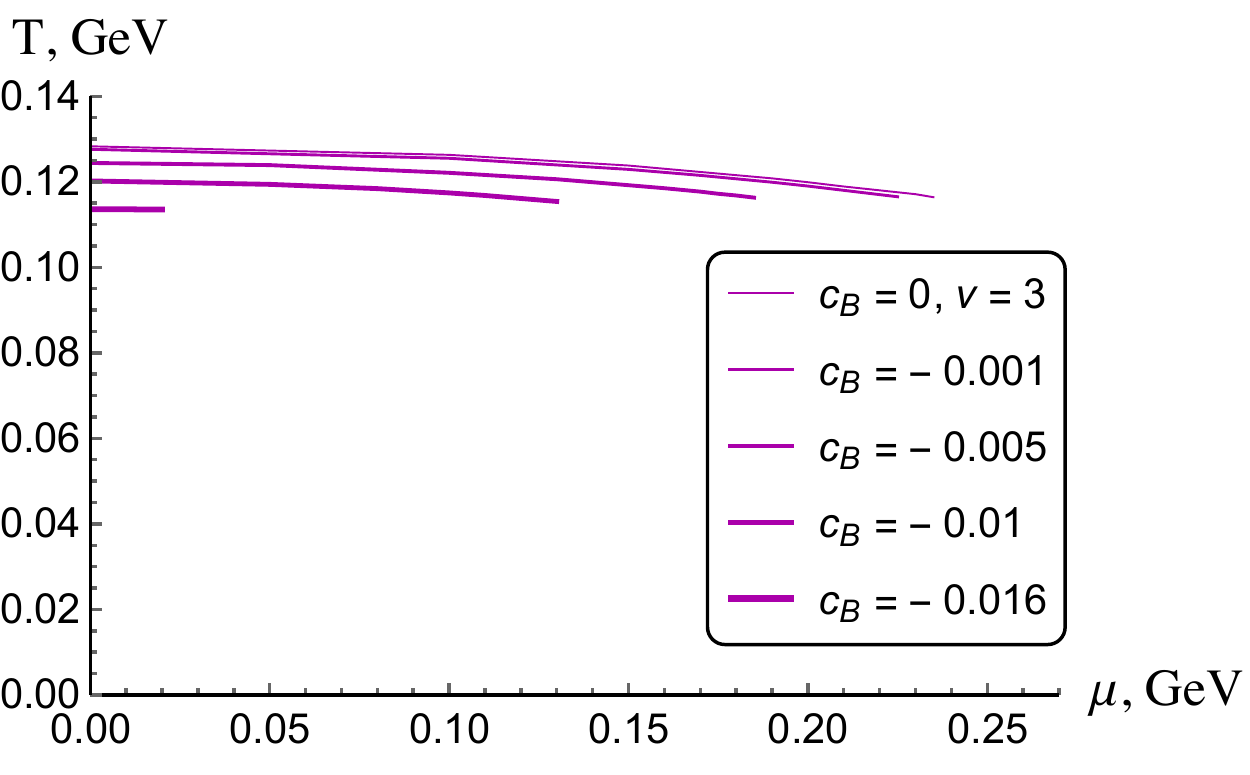} \quad
  \includegraphics[scale=0.57]{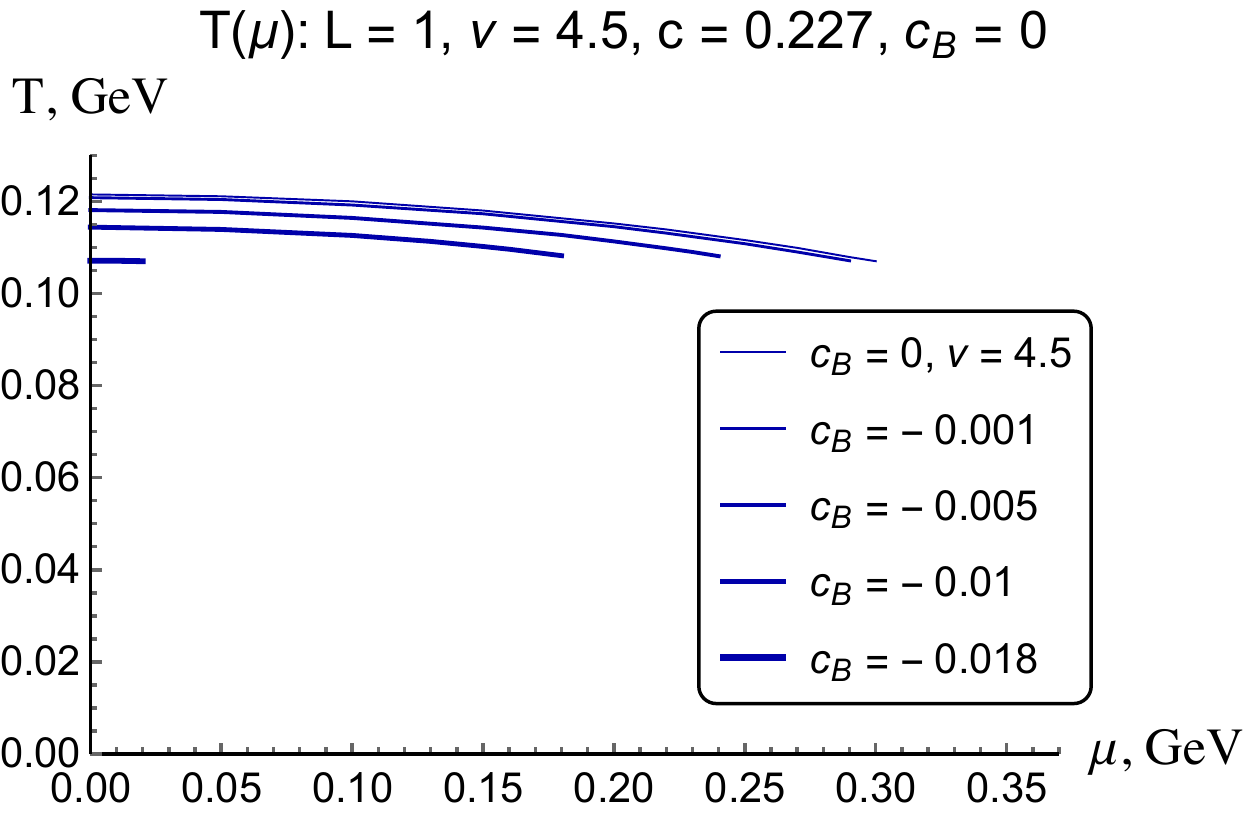} \\
  C \hspace{220pt} D \\
  \includegraphics[scale=0.68]{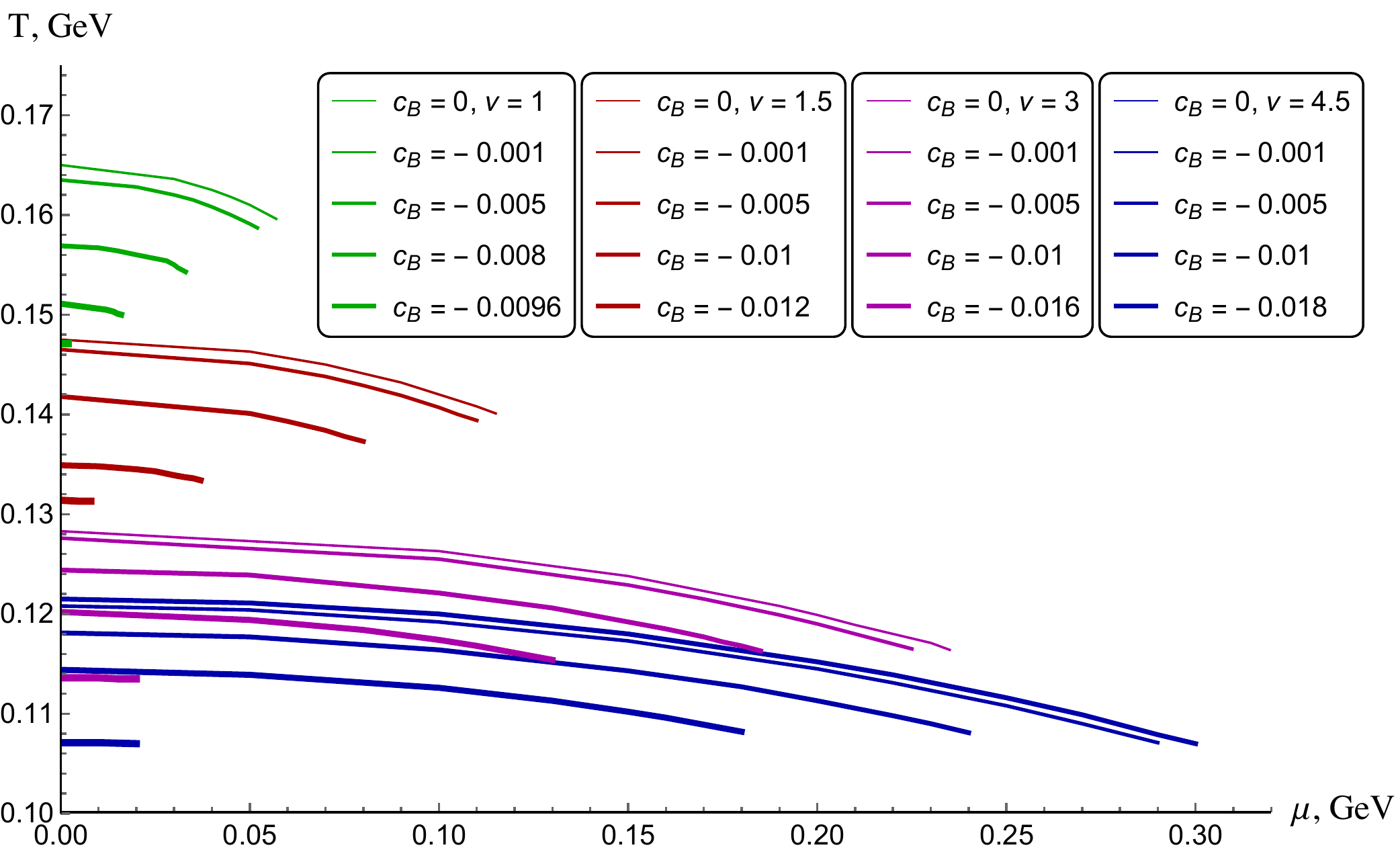} \\
  E
  \caption{Background phase transition lines depending on
      $c_B$ for $\nu = 1$ (A), $\nu = 1.5$ (B), $\nu = 3$ (C), $\nu =
      4.5$ (D) and all of them together (E); $c = 0.227$.}
  \label{Fig:BBcB}
\end{figure}

On Fig.\ref{Fig:BBcB} BH-BH transition lines for different
values of primary and secondary anisotropy are
presented. Fig.\ref{Fig:BBcB}.A-D display families of curves
representing BH-BH phase transition lines for different
values of primary anisotropy parameter, i.e. Fig.\ref{Fig:BBcB}.A for
$\nu = 1$, Fig.\ref{Fig:BBcB}.B for $\nu = 1.5$, Fig.\ref{Fig:BBcB}.C
for $\nu = 3$ and Fig.\ref{Fig:BBcB}.D for $\nu = 4.5$.

First of all, larger absolute values of the coupling coefficient $c_B$
lead to decrease of the transition temperature, so the effect of the
inverse magnetic catalysis takes place. Besides, the length of the
Hawking-Page line also decreases and eventually degenerates. However
primary anisotropy weakens these effects, so BH-BH phase
transition is preserved for larger absolute values of the coupling
coefficient. For example, background phase transition still exists for
$c_B = - \, 0.018$ and $\nu =~4.5$ (Fig.\ref{Fig:BBcB}.D), whereas for
$c_B = - \, 0.001$ and $\nu = 1$ no such curve can be plotted
(Fig.\ref{Fig:BBcB}.A).

On Fig.\ref{Fig:BBcB}.E all the BH-BH phase transition
lines are combined into one plot. One can see, that the distance
between lines for the same pairs of $c_B$ values shrinks for larger
$\nu$, while their length along the $\mu$-axis, on the contrary,
grows. Both anisotropy parameters $\nu$ and $c_B$ lower the transition
temperature, but other their effects on the background phase transition
picture can be considered the opposite.

\begin{figure}[t!]
  \centering
  \includegraphics[scale=0.39]{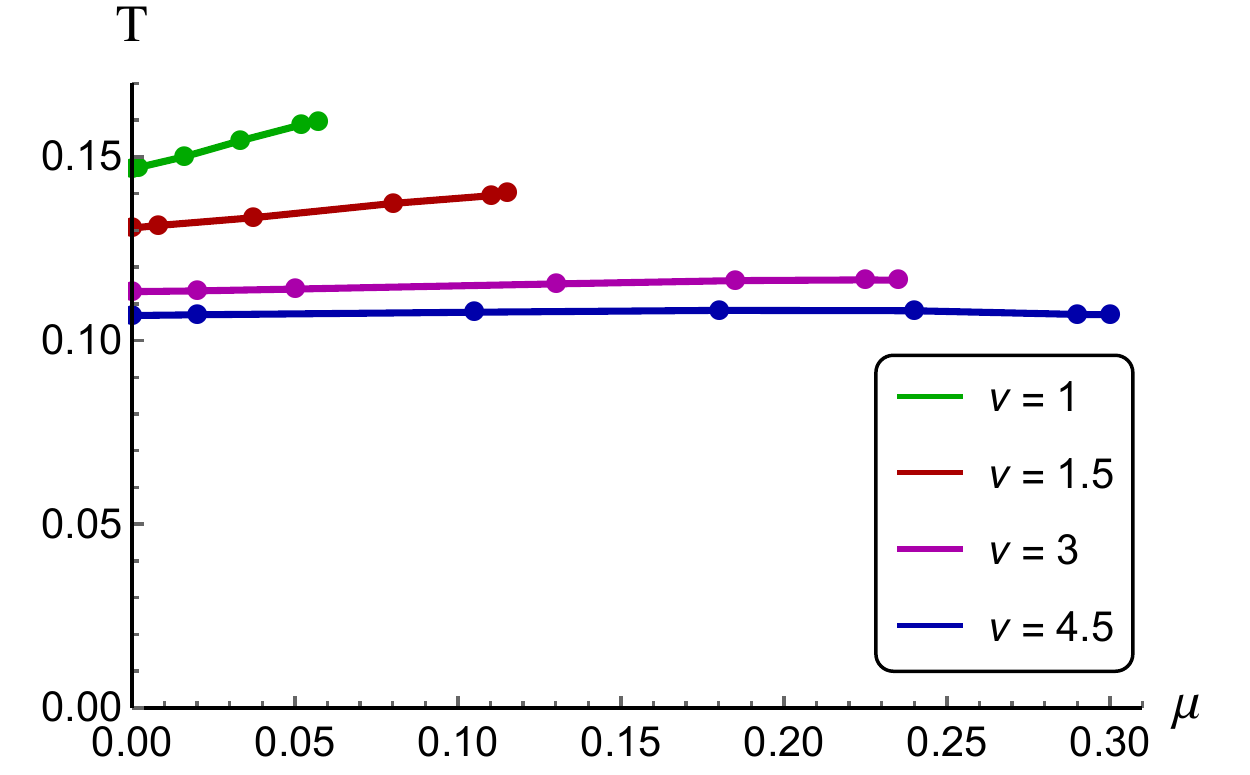} 
  \includegraphics[scale=0.39]{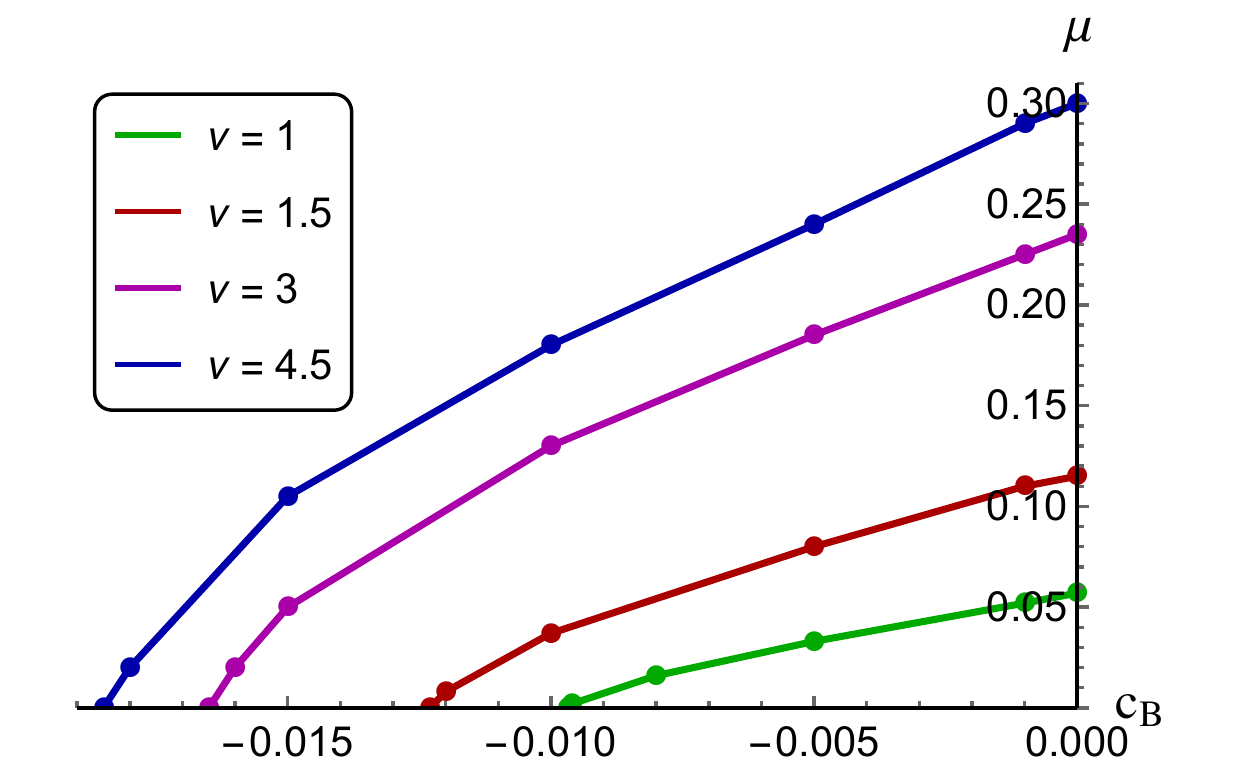}
  \includegraphics[scale=0.39]{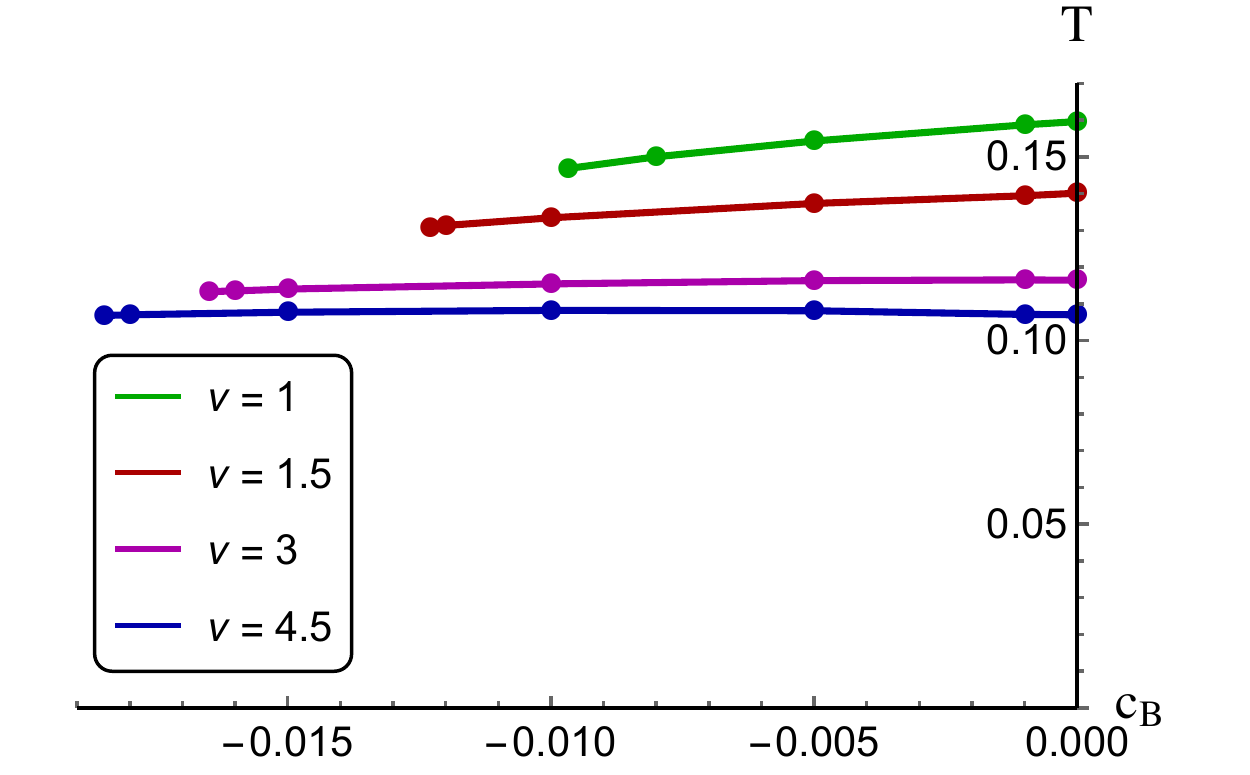} \\
  A \hspace{130pt} B \hspace{130pt} C
  \caption{Position of critical end-points CEP$_{HQ}$ for heavy quarks
    model depending on $\mu$ and $T$ (A), $c_B$ and $\mu$ (B), $c_B$
    and $T$ (C) for $\nu = 1, \ 1.5, \ 3, \ 4.5$, $c = 0.227$.}
  \label{Fig:CEP}
\end{figure}

\begin{figure}[t!]
  \centering
  \includegraphics[scale=0.25]{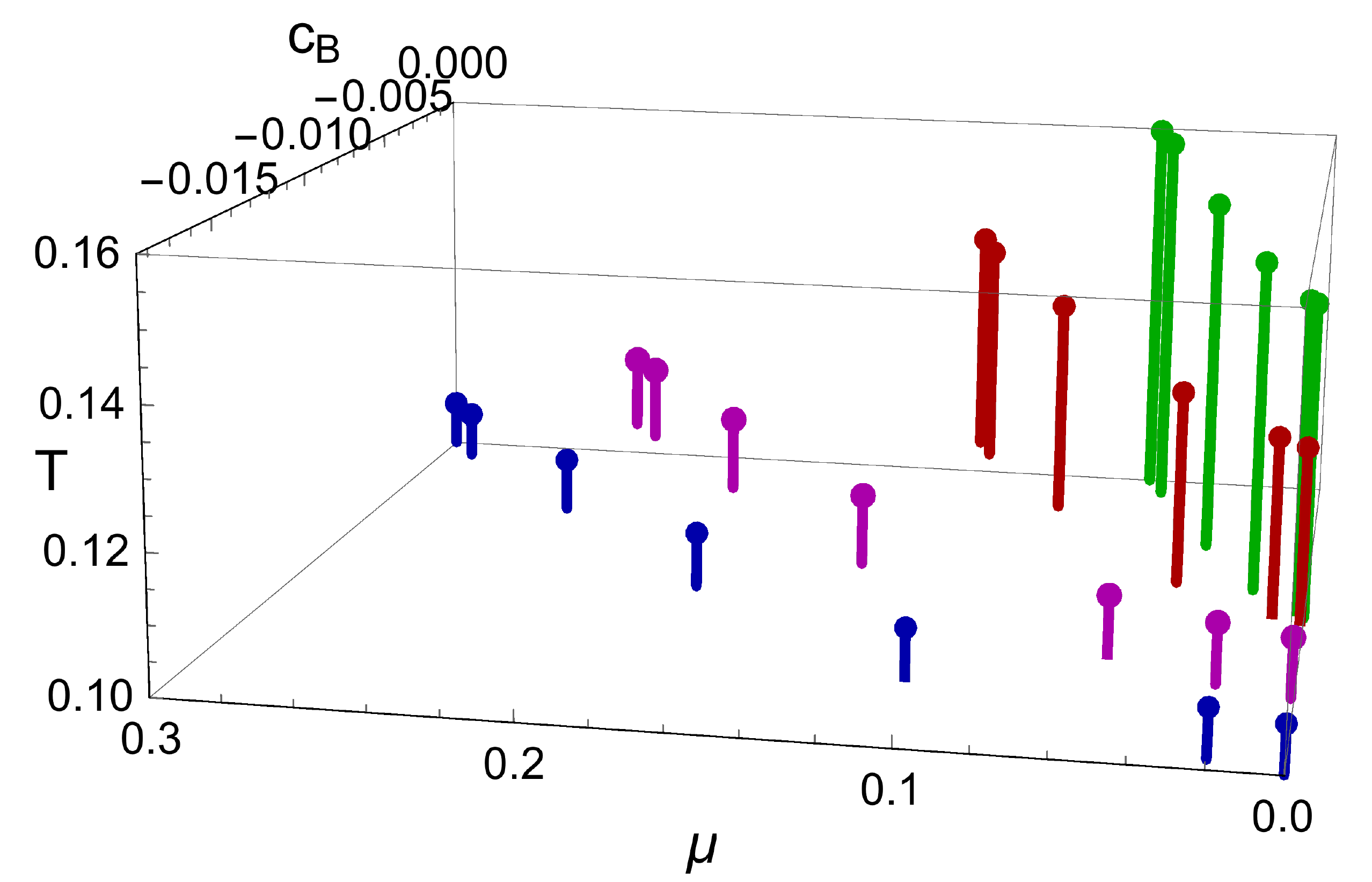} \qquad
  \includegraphics[scale=0.26]{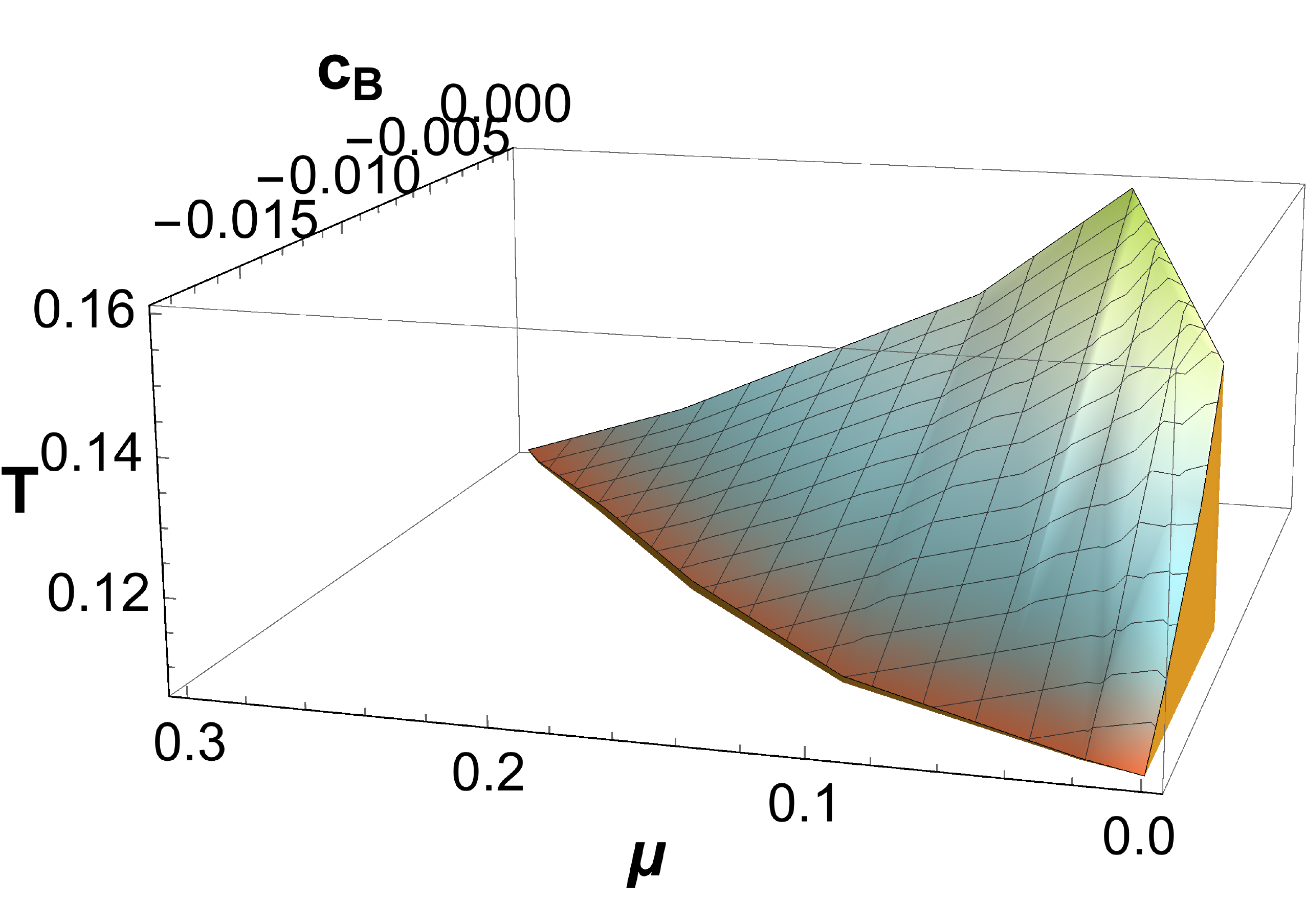} \\
  A \hspace{220pt} B
  \caption{Temperature of critical end-points CEP$_{HQ}$ for heavy
    quarks model depending on $c_B$ and $\mu$ for $1 \le \nu \le 4.5$
    as a set of point (A) and as a 3D-surface (B), $c = 0.227$.}
  \label{Fig:CEP3D}
\end{figure}

We can also consider positions of the ``free ends'' of the
BH-BH phase transition lines (let us call it ``critical end-points for
heavy quarks'', CEP$_{HQ}$). On Fig.\ref{Fig:CEP} these positions are
depicted on $T$--$\mu$ (Fig.\ref{Fig:CEP}.A), $\mu$--$c_B$
(Fig.\ref{Fig:CEP}.B) and $T$--$c_B$ (Fig.\ref{Fig:CEP}.C) planes. For
$\nu = 1$ temperature of the CEP$_{HQ}$ falls relatively quickly with
the $c_B$ absolute value growth, while for $\nu = 4.5$ this
temperature is almost constant (Fig.\ref{Fig:CEP}.A,C). On the
contrary, chemical potential of the CEP$_{HQ}$-points falls more for
larger primary anisotropy, especially quickly decreasing near the $\mu
= 0$. On Fig.\ref{Fig:CEP3D} the 3D-plots for the CEP$_{HQ}$ positions
depending on coupling coefficient $c_B$, chemical potential and
temperaure for different values of primary anisotropy are displayed.

\subsection{Temporal Wilson loops}\label{WL}

\begin{figure}[t!]
  \centering
  \includegraphics[scale=0.6]{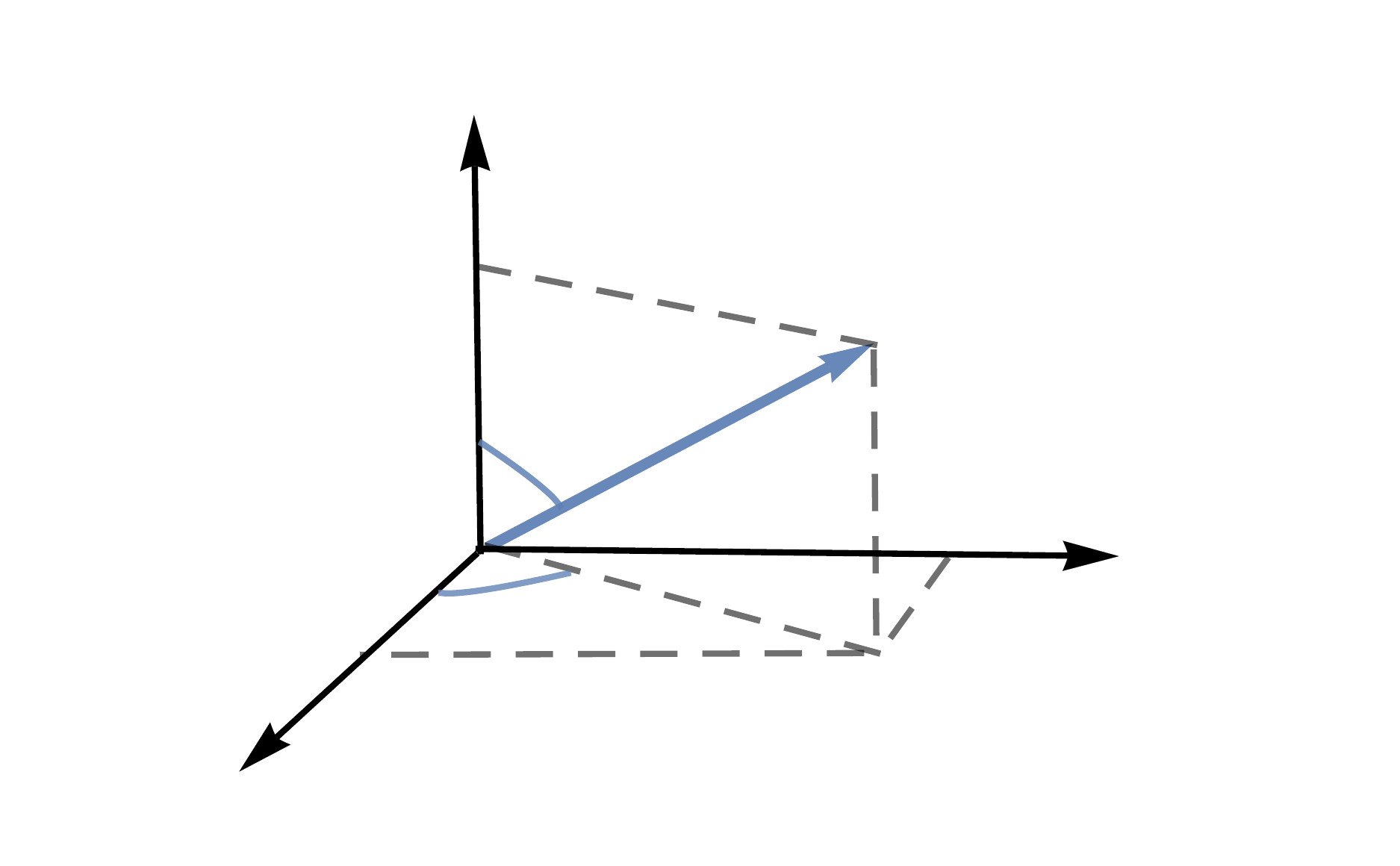} 
\begin{picture}(0,0)\put(-285,20){$\Large{x}$}
\put(-85,60){$\Large{y_1}$}
\put(-210,160){$\Large{y_2}$}
\put(-220,55){$\Large{\vartheta}$}
\put(-210,95){$\Large{\alpha}$}
\put(-190,95){ $\Large{\vec n}$}
\end{picture}
  \caption{World sheet parametrization for the Wilson loop orientation
  vector $\vec n$.}
  \label{WLparam}
\end{figure}
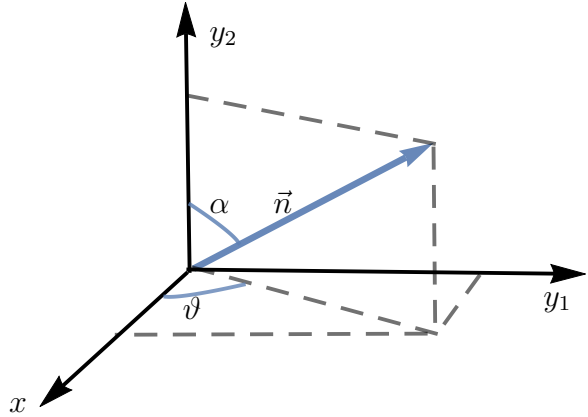

To calculate the expectation value of the temporal Wilson loop
\begin{gather}
  W[C_\vartheta] = e^{-S_{\vartheta,t}}, \label{eq:4.08}
\end{gather}
oriented along vector $\vec n$:
\begin{gather}
  n_{x1} = \cos \vartheta \sin \alpha, \quad
  n_{x2} = \sin \vartheta \sin \alpha, \quad 
  n_{x3}= \cos \alpha \label{eq:4.09}
\end{gather}
we use our metric (\ref{eq:2.03}) as a background:
\begin{gather}
  ds^2 = G_{\mu\nu}dx^{\mu}dx^{\nu}
  = \cfrac{L^2}{z^2} \ \fb_s(z) \left[
    - \, g(z) dt^2 + \fg_1 dx^2 + \fg_2 dy_1^2 + \fg_3 dy_2^2
    + \cfrac{dz^2}{g(z)} \right], \label{eq:4.10}
\end{gather}
Following the holographic approach we calculate the value of the
Nambu-Goto action for test string in our background in string frame:
\begin{gather}
  S = \frac{1}{2 \pi \alpha'} \int d\xi^{0} \, d\xi^{1} \sqrt{- \det
    h_{\alpha\beta}}, \qquad
  h_{\alpha\beta} = G_{\mu\nu} \, \partial_{\alpha}
  X^{\mu} \, \partial_{\beta} X^{\nu}. \label{eq:4.11}
\end{gather}
The world sheet is parameterized as (Fig.\ref{WLparam})
\begin{gather}
  \begin{split}
    X^{0} \equiv t = \xi^0, \quad
    X^{1} &\equiv x = \xi^1 \cos\vartheta \sin \alpha, \quad
    X^{2} \equiv y_{1} = \xi^1\sin\vartheta \sin \alpha, \\
    X^{3} &\equiv y_{2} = \xi^1 \cos \alpha, \qquad \ \ \,
    X^{4} \equiv z = z(\xi^1).
  \end{split}
    \label{eq:4.12}
\end{gather}


Let us denote $\xi \equiv \xi^1$, rewrite Nambu-Goto action
(\ref{eq:4.11})
\begin{gather}
  \begin{split}
    &S = - \, \cfrac{\tau}{2 \pi \alpha'} \int d\xi \
    M\left(z(\xi)\right) \ \sqrt{{\cal F}(z\left(\xi)\right) +
      \left(z'(\xi)\right)^2}, \qquad 
    \tau = \int d \xi^0, \\
    &M\left(z(\xi)\right) = \cfrac{\fb_s\left(z(\xi)\right)}{z^2(\xi)}
    \, , \\
    &{\cal F}\left(z(\xi)\right) = g\left(z(\xi^1)\right) \left(\fg_1
      \cos^2\vartheta \sin^2 \alpha + \fg_2 \sin^2\vartheta \sin^2
      \alpha + \fg_3 \cos^2 \alpha \right)
  \end{split}\label{eq:4.13}
\end{gather}
and introduce the effective ``potential''
\begin{gather}
  \begin{split}
    {\cal V}\left(z(\xi)\right) 
    &\equiv M\left(z(\xi)\right) \sqrt{{\cal F}\left(z(\xi)\right)} =
    \\
    &= \cfrac{\fb_s\left(z(\xi)\right)}{(z^2(\xi))}
    \sqrt{g\left(z(\xi)\right) \left(\fg_1 \cos^2\vartheta \sin^2
        \alpha + \fg_2 \sin^2\vartheta \sin^2 \alpha + \fg_3 \cos^2
        \alpha \right)}.
  \end{split}\label{eq:4.14}
\end{gather}

Thus for the current heavy quarks model Wilson loops, oriented along
$x$, $y_1$ and $y_2$ axes, are defined:
\begin{gather}
  \begin{split}
  {\cal DW}_x \ &\equiv - \, cz + \sqrt{\cfrac23} \ \phi'(z) +
  \cfrac{g'}{2 g} - \cfrac{2}{z} \ \Big|_{z = z_{DWx}} \hspace{-15pt}
  = 0, \\
  {\cal DW}_{y_1} &\equiv - \, cz + \sqrt{\cfrac23} \ \phi'(z) +
  \cfrac{g'}{2 g} - \cfrac{\nu + 1}{\nu z} \ \Big|_{z = z_{DWy_1}}
  \hspace{-15pt} = 0, \\
  {\cal DW}_{y_2} &\equiv - \, cz + \sqrt{\cfrac23} \ \phi'(z) +
  \cfrac{g'}{2 g} - \cfrac{\nu + 1}{\nu z} + c_B z \ \Big|_{z =
    z_{DWy_2}} \hspace{-15pt} = 0. \\
  \end{split}\label{eq:4.15}
\end{gather}
It is obvious, that for $c_B = 0$ expressions in (\ref{eq:4.15}) for
$y_1$ and $y_2$ become equal, and putting $\nu = 1$ leaves the
$x$-expression only. Behavoir of WLx and WLy$_1$ was already
considered \cite{1802.05652, 1808.05596}.

\begin{figure}[t!]
  \centering
  \includegraphics[scale=0.58]{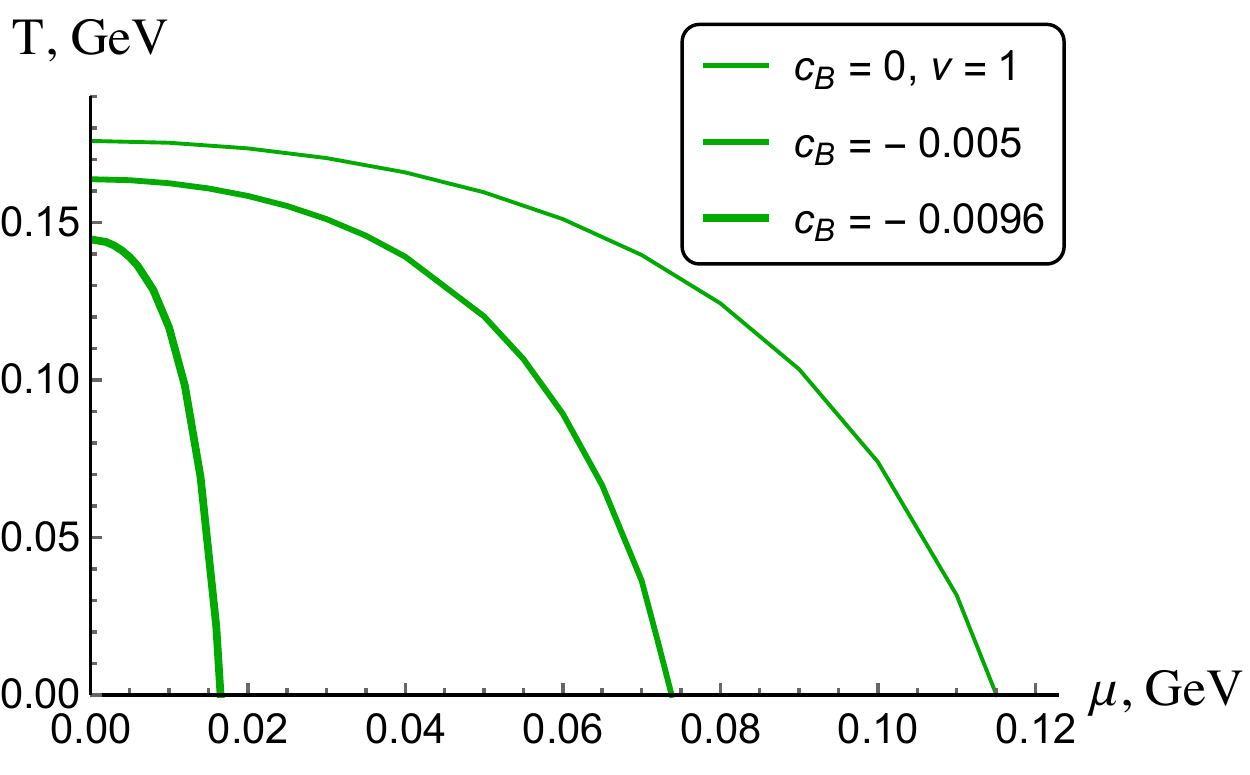} \
  \includegraphics[scale=0.58]{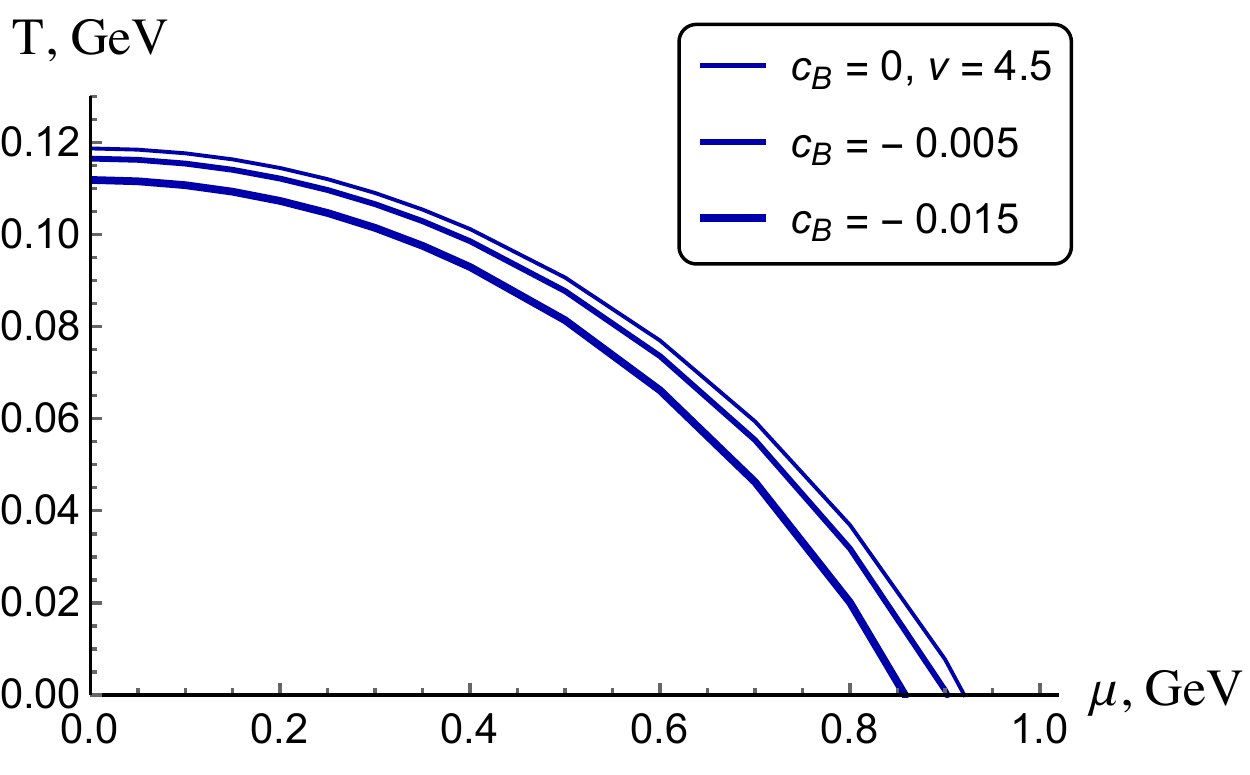} \\
  A \hspace{220pt} B
  \caption{Wilson loop transition lines for different $c_B$ for $\nu =
    1$ (A) and $\nu = 4.5$ (B); $c = 0.227$.}
  \label{Fig:WLy2}
\end{figure}

\begin{figure}[t!]
  \centering
  \includegraphics[scale=0.58]{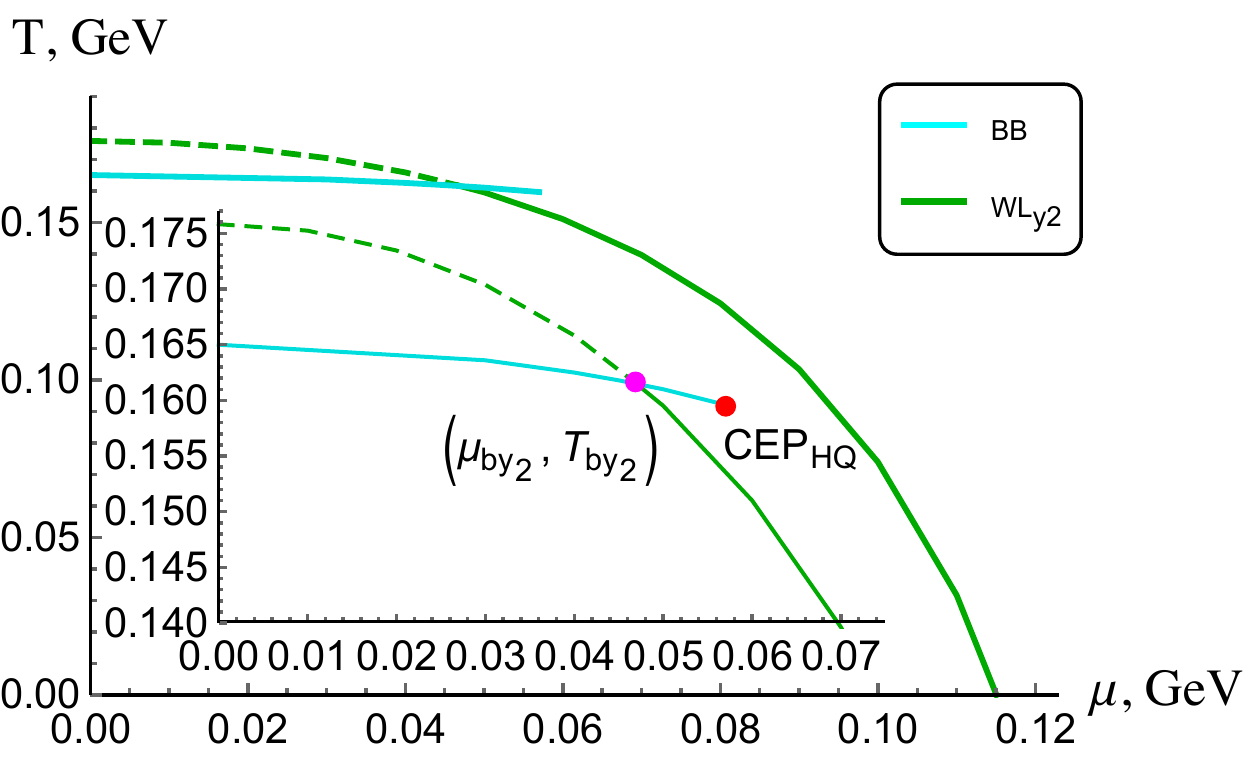} \
  \includegraphics[scale=0.58]{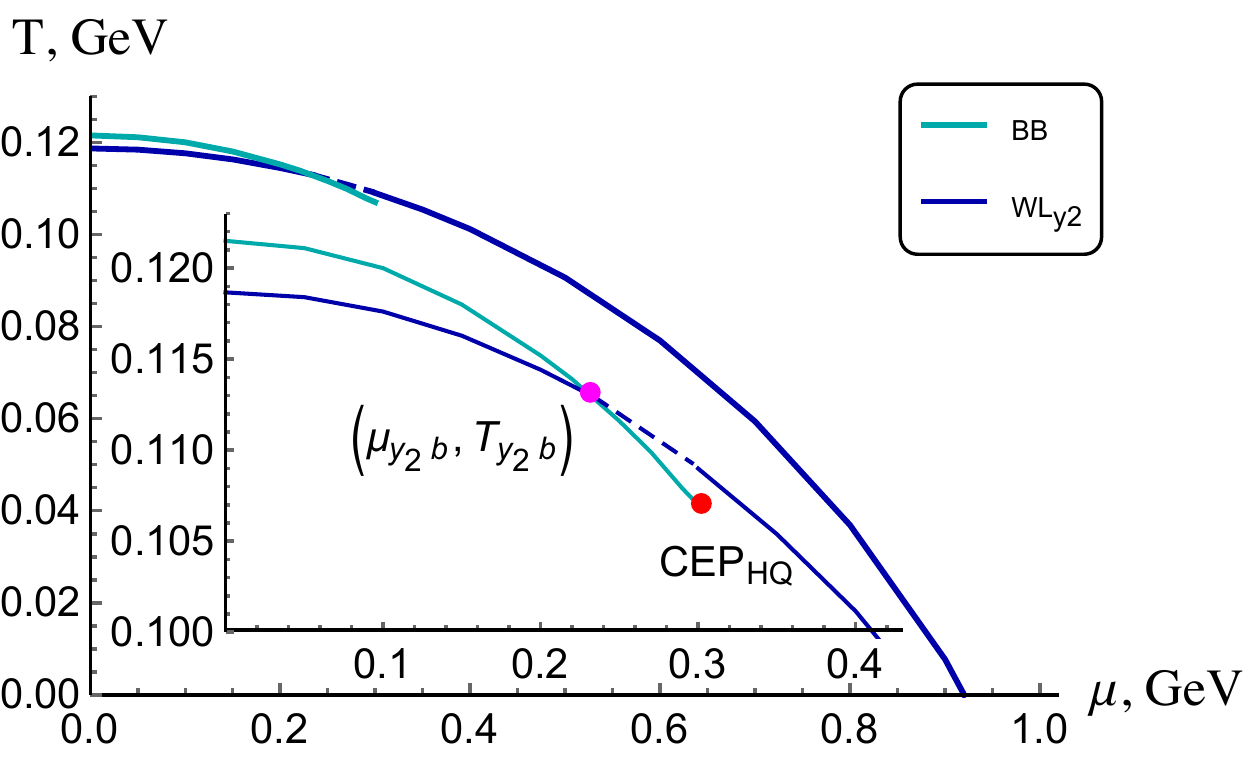} \\
  A \hspace{220pt} B \\
  \includegraphics[scale=0.58]{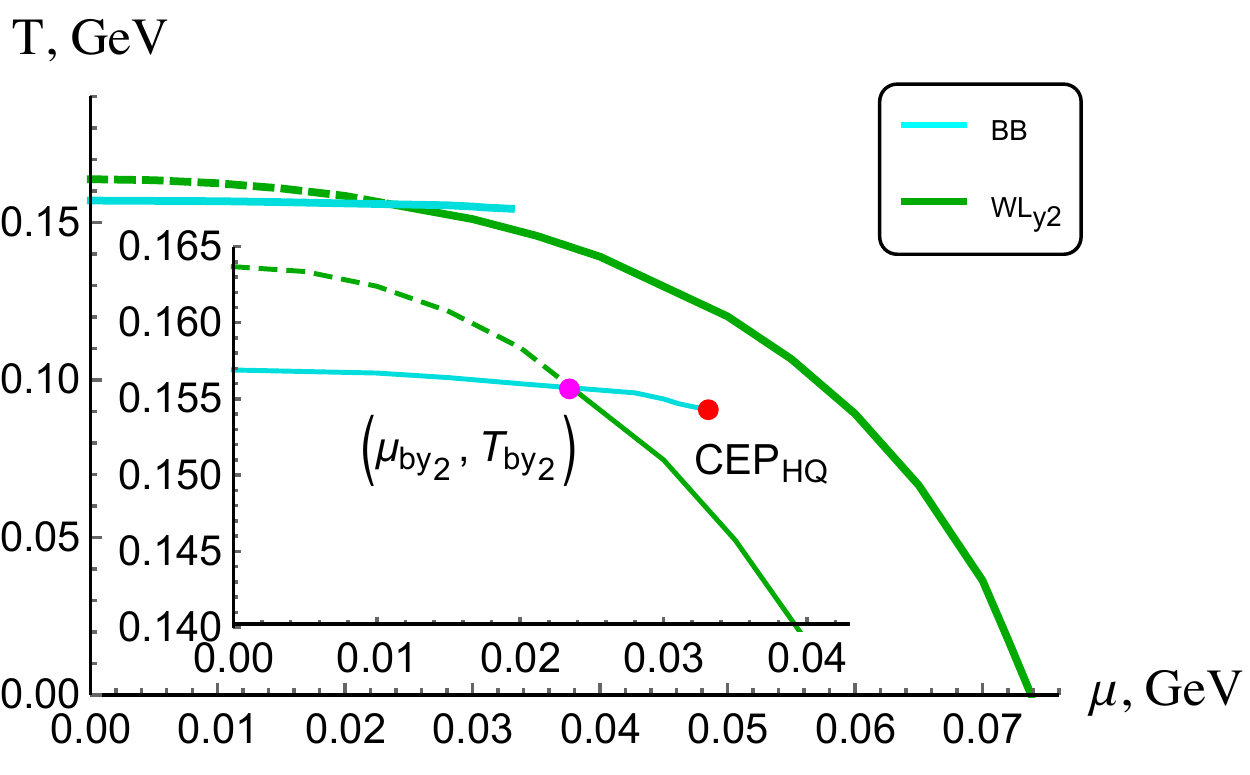} \
  \includegraphics[scale=0.58]{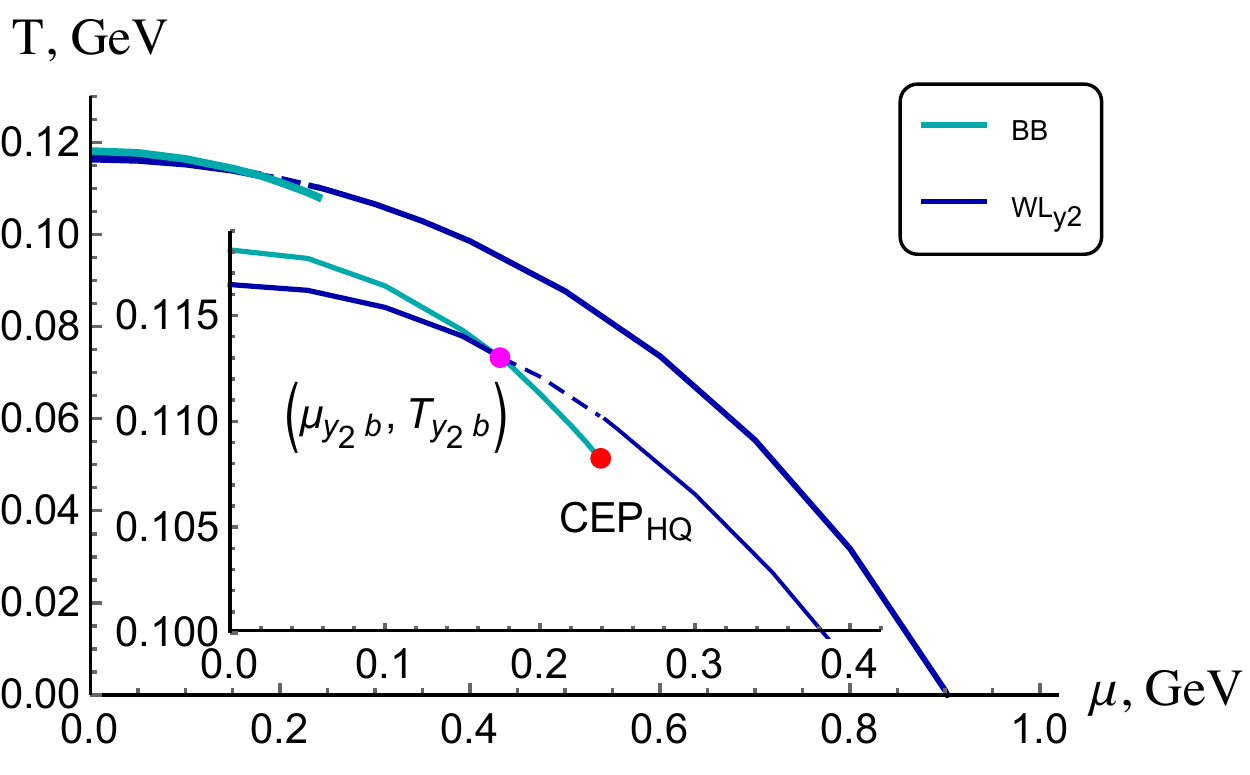} \\
  C \hspace{220pt} D \\
  \includegraphics[scale=0.58]{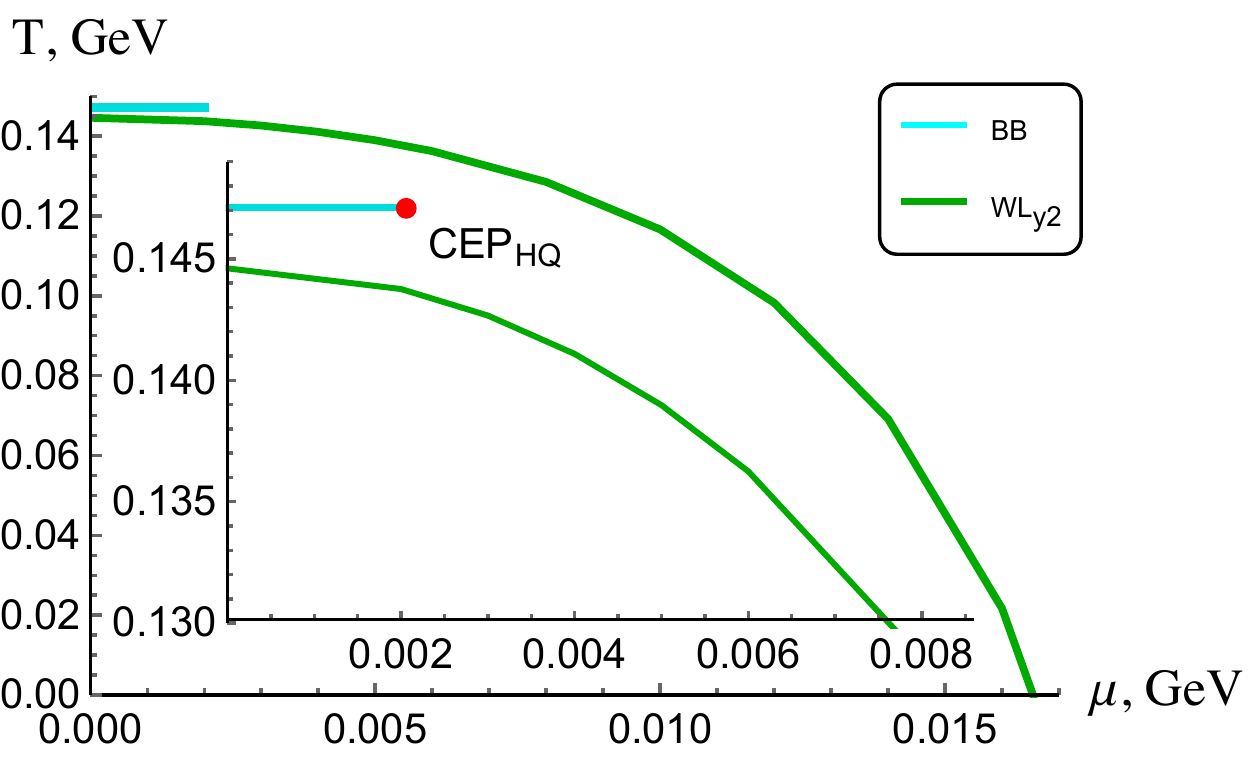} \
  \includegraphics[scale=0.58]{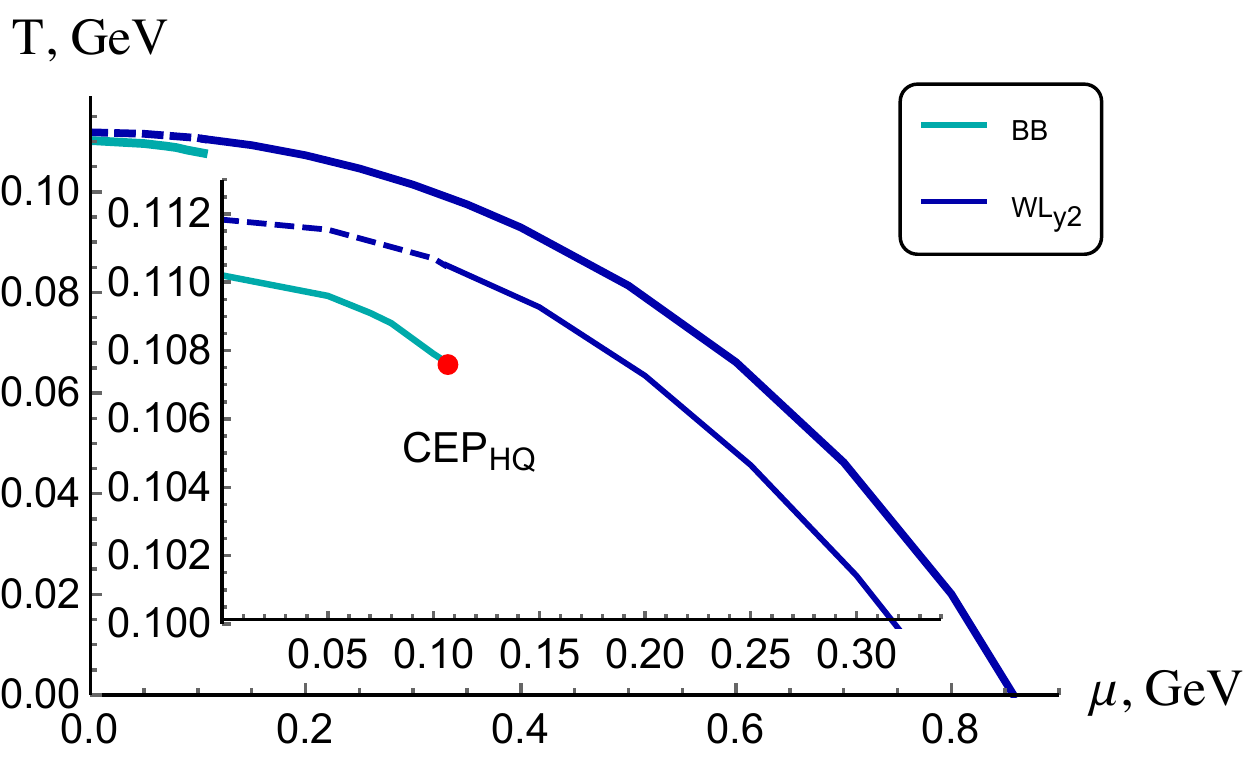} \\
  E \hspace{220pt} F \\
  \caption{Phase diagram for $y_2$-orientation for $c_B = 0$, $\nu =
    1$ (A) and $\nu = 4.5$ (B), $c_B = - \, 0.005$, $\nu = 1$ (C) and
    $\nu = 4.5$ (D), $c_B = - \, 0.0096$, $\nu = 1$ (E) and $c_B = - \
    0.015$, $\nu = 4.5$ (F); $c = 0.227$.}
  \label{Fig:Tmuy2}
\end{figure}

On Fig.\ref{Fig:WLy2} curves for $c_B = 0$ can also be considered to
represent WLy$_1$ transition lines. We see, that the appearence of
external magnetic field ($c_B \ne 0$) makes the Wilson loop phase
transition line to shrink. For $\nu = 1$ maximum chemical potential at
$T = 0$ desreases faster than maximum temperature at $\mu = 0$. For
$\nu = 4.5$ the shrinkage of the phase transition line corresponding
to $y_2$ Wilson loop is much weaker and looks almost cocentric.

On Fig.\ref{Fig:Tmuy2} phase transition diagrams for $y_2$-oriented
Wilson loops are displayed. We see the picture quite similar to our
previous results for heavy quarks.

For $\nu = 1$ without magnetic field (Fig.\ref{Fig:Tmuy2}.A) BH-BH
line determines the confinement/deconfinement phase transition for $0
\le \mu \le \mu_{by_2} = 0.04666$ till the intersection with the $y_2$
Wilson loop line at point $(0.04666; 0.1617)$. 

If we turn on the external magnetic field, the intersection between
BH-BH line and Wilson loop line shifts to smaller temperature and
chemical potential, $(0.0235; 0.1558)$ for $c_B = - \, 0.005$
(Fig.\ref{Fig:Tmuy2}.C), and the total area of confinement phase
predictably decreases. Aslo phase transition lines, corresponding to
the background (BH-BH) and string action (WLy$_2$), become closer to
each other.

When the coupling with the $F_{\mu\nu}^{(B)}$ is almost maximum ($c_B
= - \, 0.0096$), BH-BH transition line turns out to be
above the $y_2$ Wilson loop line without any intersection
(Fig.\ref{Fig:Tmuy2}.E).

For $\nu = 4.5$ the picture can be called the opposite from some point
of view. Without the external magnetic field (Fig.\ref{Fig:Tmuy2}.B)
$y_2$ Wilson loop line playes the main role for $\mu \le \mu_{y_2b} =
0.2448$ till the intersection with the BH-BH line at $(0.2448;
0.1121)$. BH-BH phase transition line lasts till the
CEP$_{HQ}$ at $(0.3; 0.107)$, where the main role comes back to Wilson
loop line via the jump to higher temperature.

For $c_B = - \, 0.005$ the picture remains the same qualitatively
(Fig.\ref{Fig:Tmuy2}.D). Transition lines intersect at point ($0.1753;
0.113$), i.e. at lesser chemical potential, but a little higher
temperature.

Eventually BH-BH transition line turns out to lie entirely
under the WL-line till ($0.24, 0.1081$), where it's CEP is located
(Fig.\ref{Fig:Tmuy2}.F). Critical value ofthe coupling function, for
which $T_{BH-BH}|_{\mu = 0} = T_{WLy_2}|_{\mu = 0}$ is about $c_{B\,
  crit} \approx - \, 0.11$ in this case.

In the backgrounds with instability one has also to check that the
extended surface lies in the stability zone \cite{IA-NICA}. This
allows to determine roles of different transition lines in the general
confinement/deconfinement picture.

\begin{figure}[t!]
  \centering
  \includegraphics[scale=0.13]{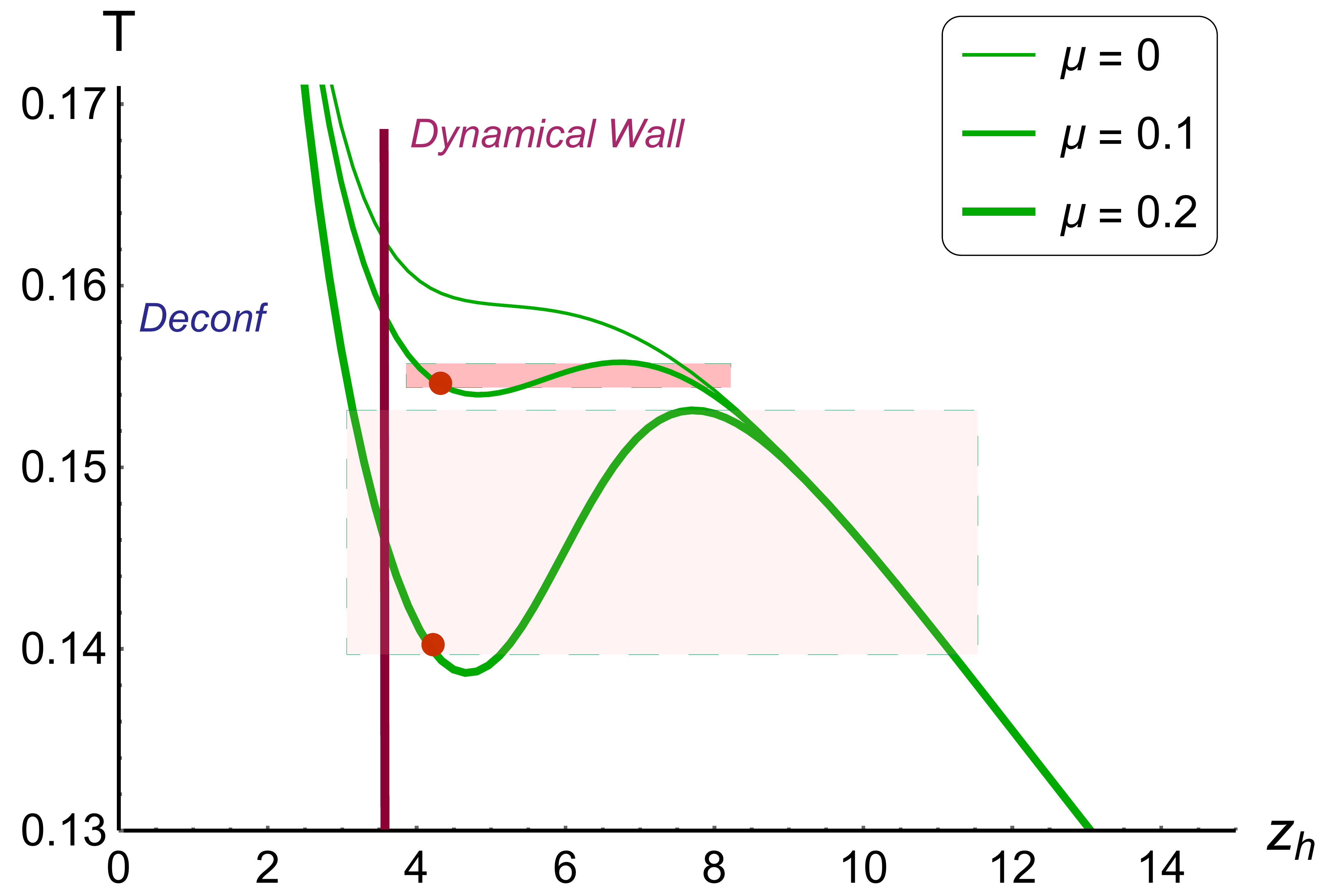} 
  \includegraphics[scale=0.13]{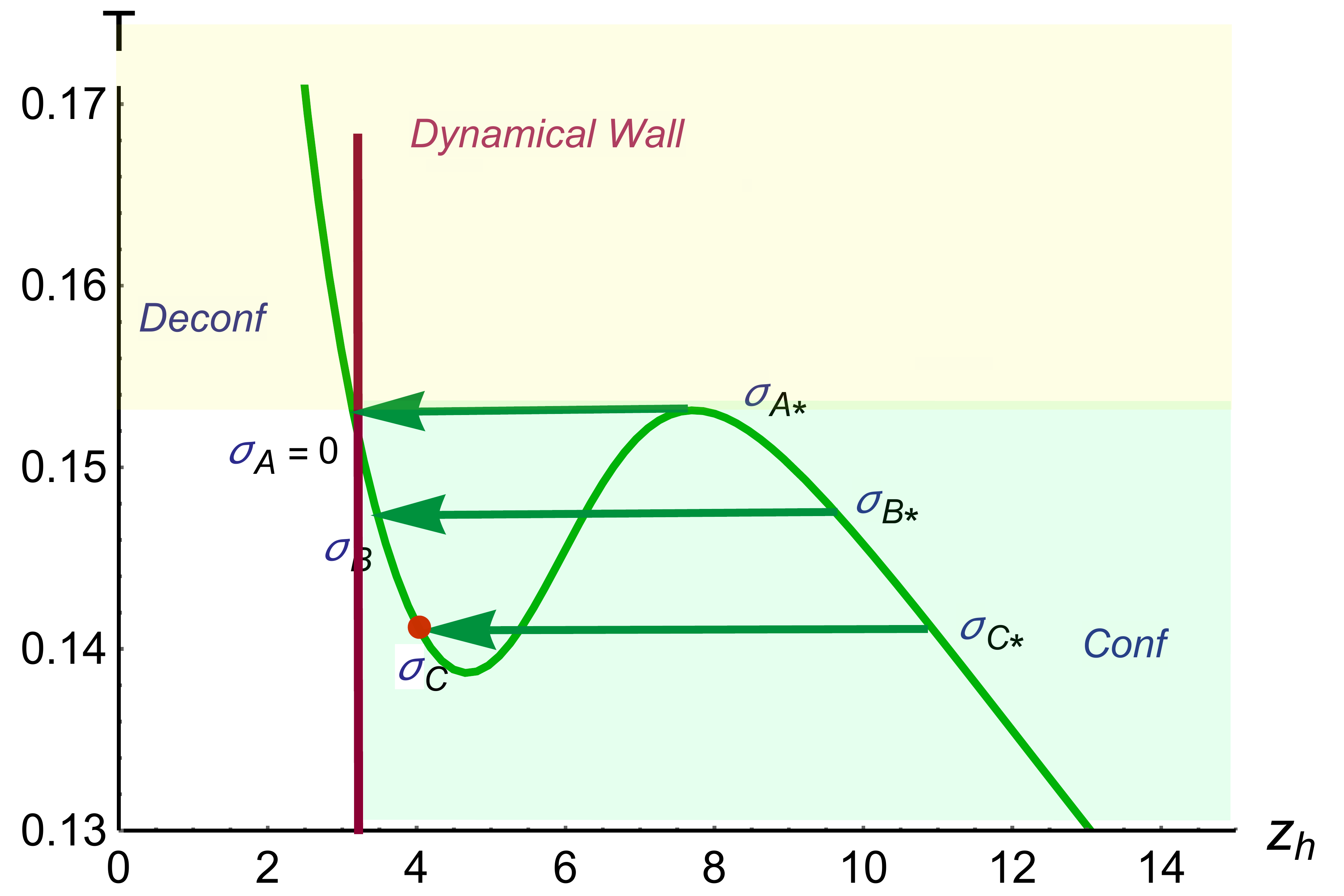}
  \includegraphics[scale=0.13]{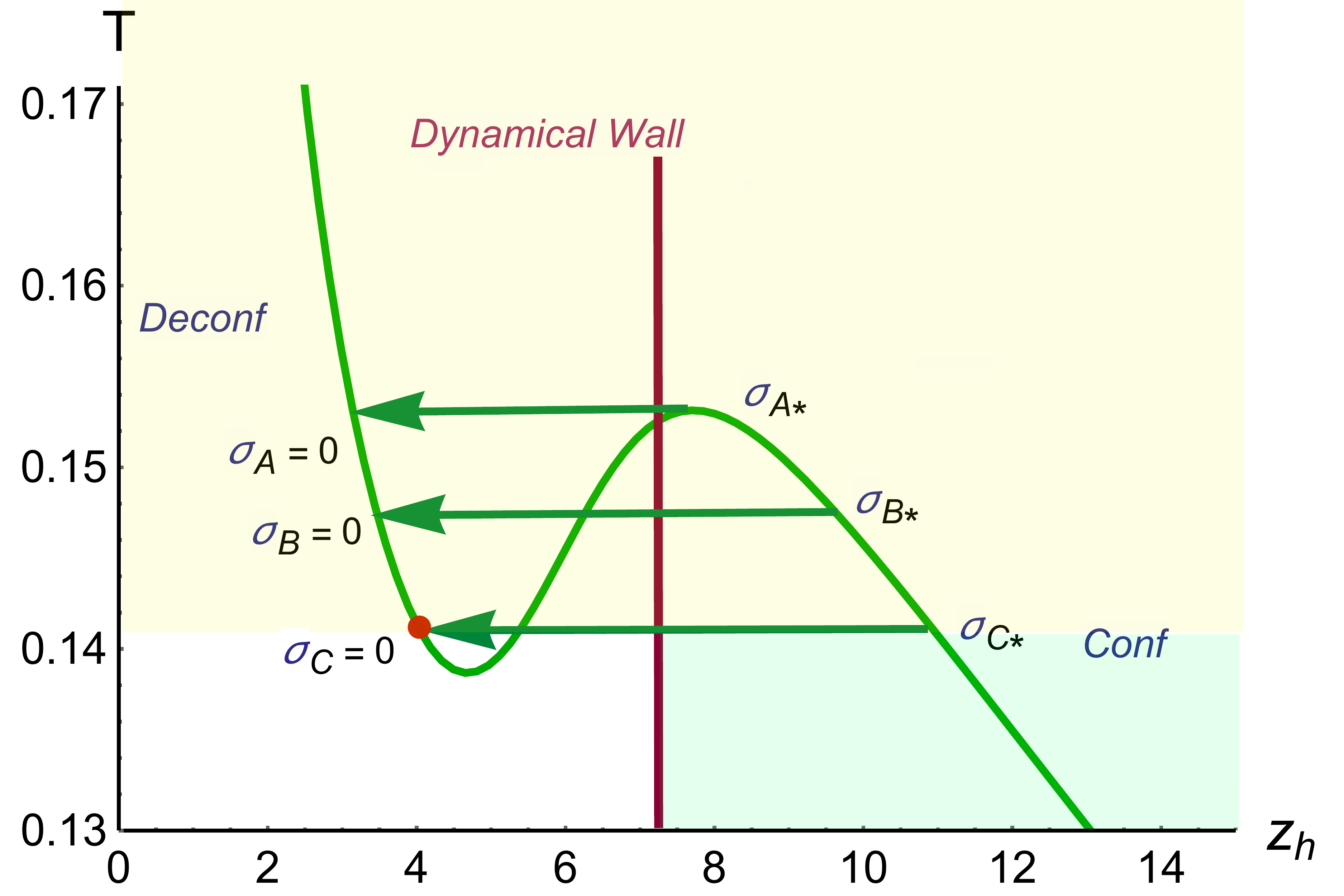} \\
  A \hspace{130pt} B \hspace{130pt} C
  \caption{Scheme of confinement/deconfuinement depending on location
    of the BH-BH phase transition points relative to the dynamic
    wall.}
  \label{Fig:DWvsBB}
\end{figure}

There are different situations in this matter. For given $\mu$ the
dynamic wall (DW) can be out of instability zone (small rectangular
pink region on Fig.\ref{Fig:DWvsBB}.A), or cross the instability zone
starting from some temperature (large rectangular pink region on
Fig.\ref{Fig:DWvsBB}.A). In the first case the jump to the stable zone
keeps confinement, the string tension decreases from $\sigma_*$ to
$\sigma$, $\sigma_* > \sigma$ and we land in the stable phase, that
can be interpreted as quarkyonic (Fig.\ref{Fig:DWvsBB}.B). If the DW
lies in unstable zone we lose the linear interaction between quarks in
the quarkyonic phase (Fig.\ref{Fig:DWvsBB}.C).

\begin{figure}[t!]
  \centering
  \includegraphics[scale=0.27]{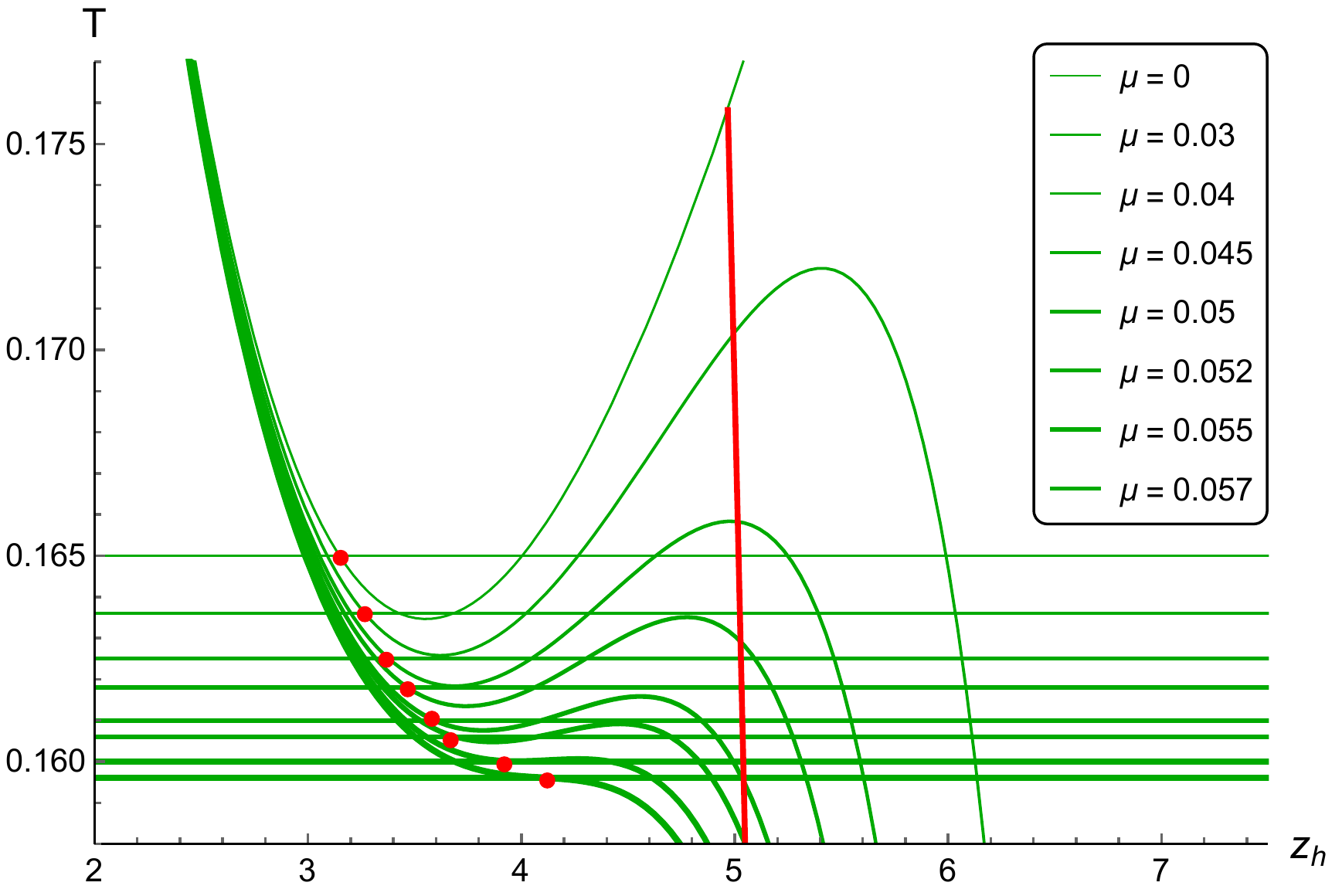} \
  \includegraphics[scale=0.27]{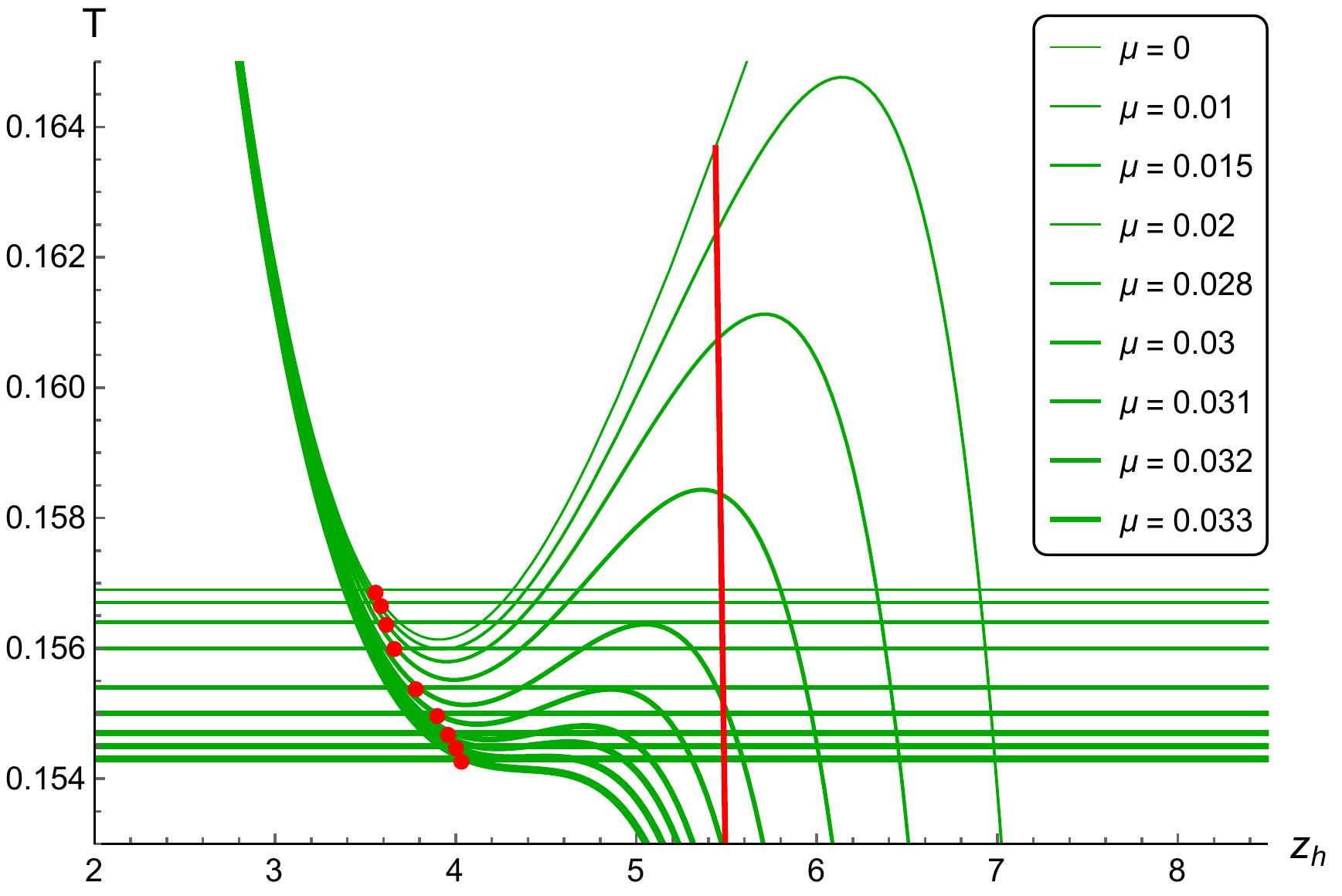} \
  \includegraphics[scale=0.27]{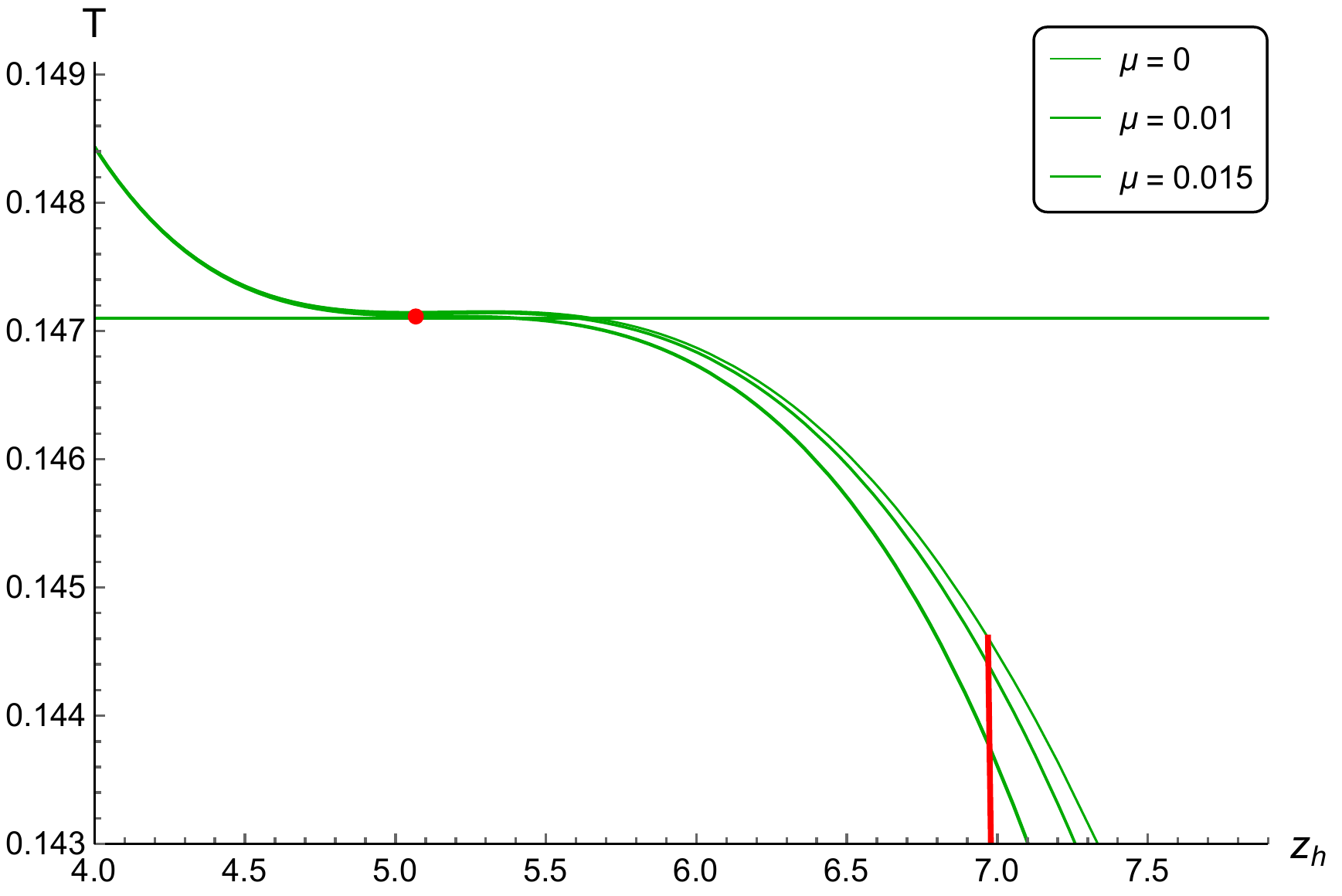} \\ \ \\
  \includegraphics[scale=0.27]{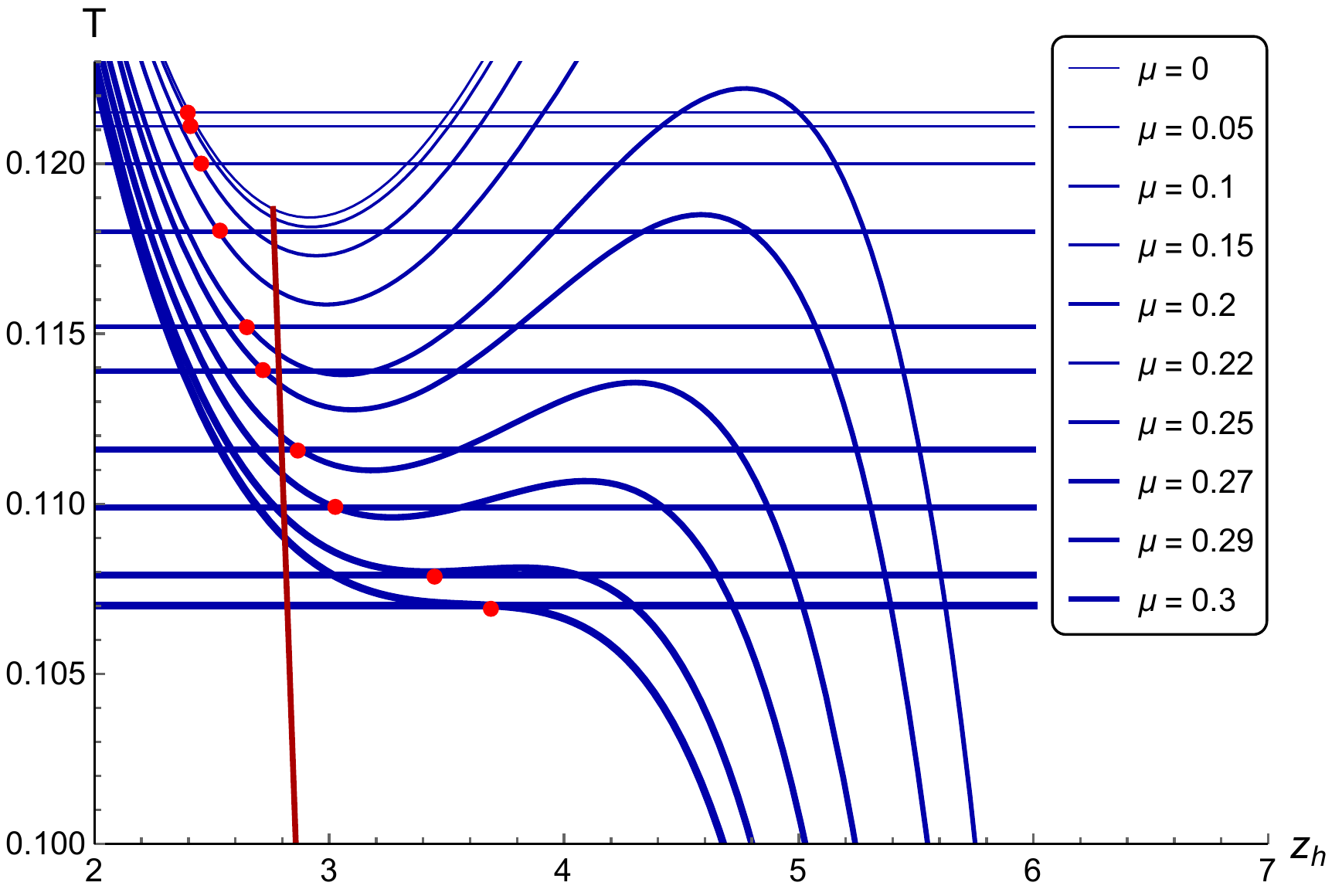} \
  \includegraphics[scale=0.27]{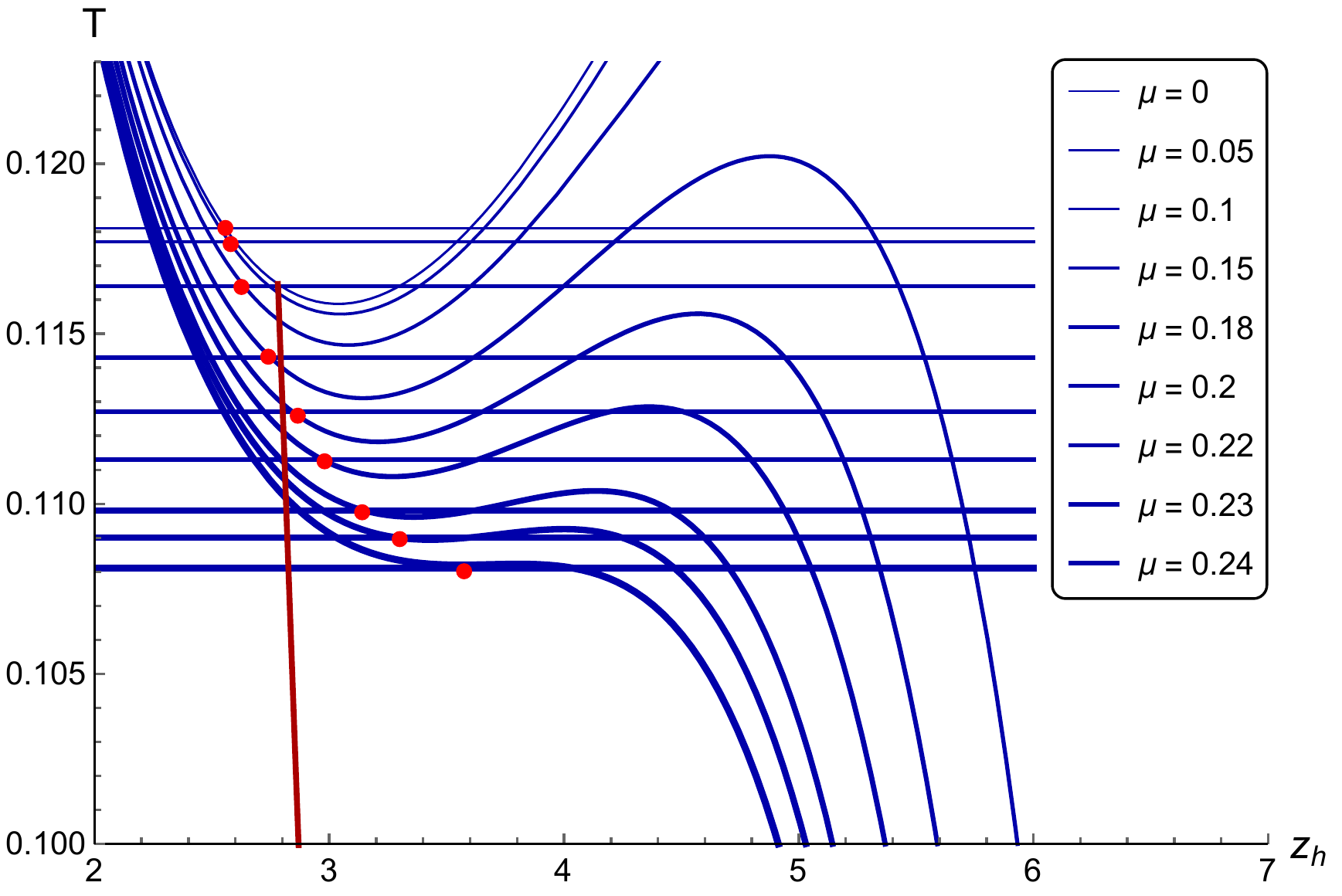} \
  \includegraphics[scale=0.27]{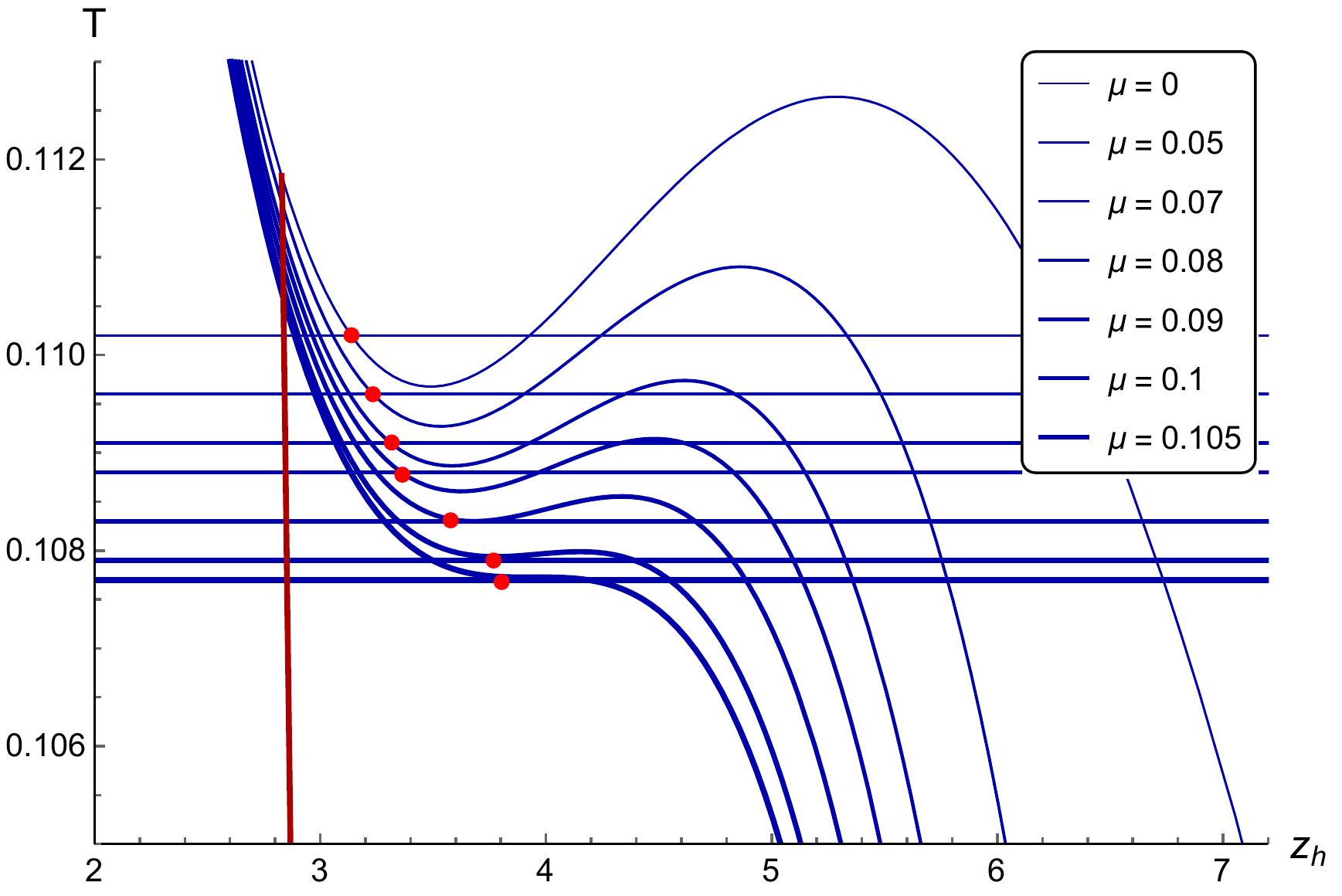} \\ \ \\
  \includegraphics[scale=0.27]{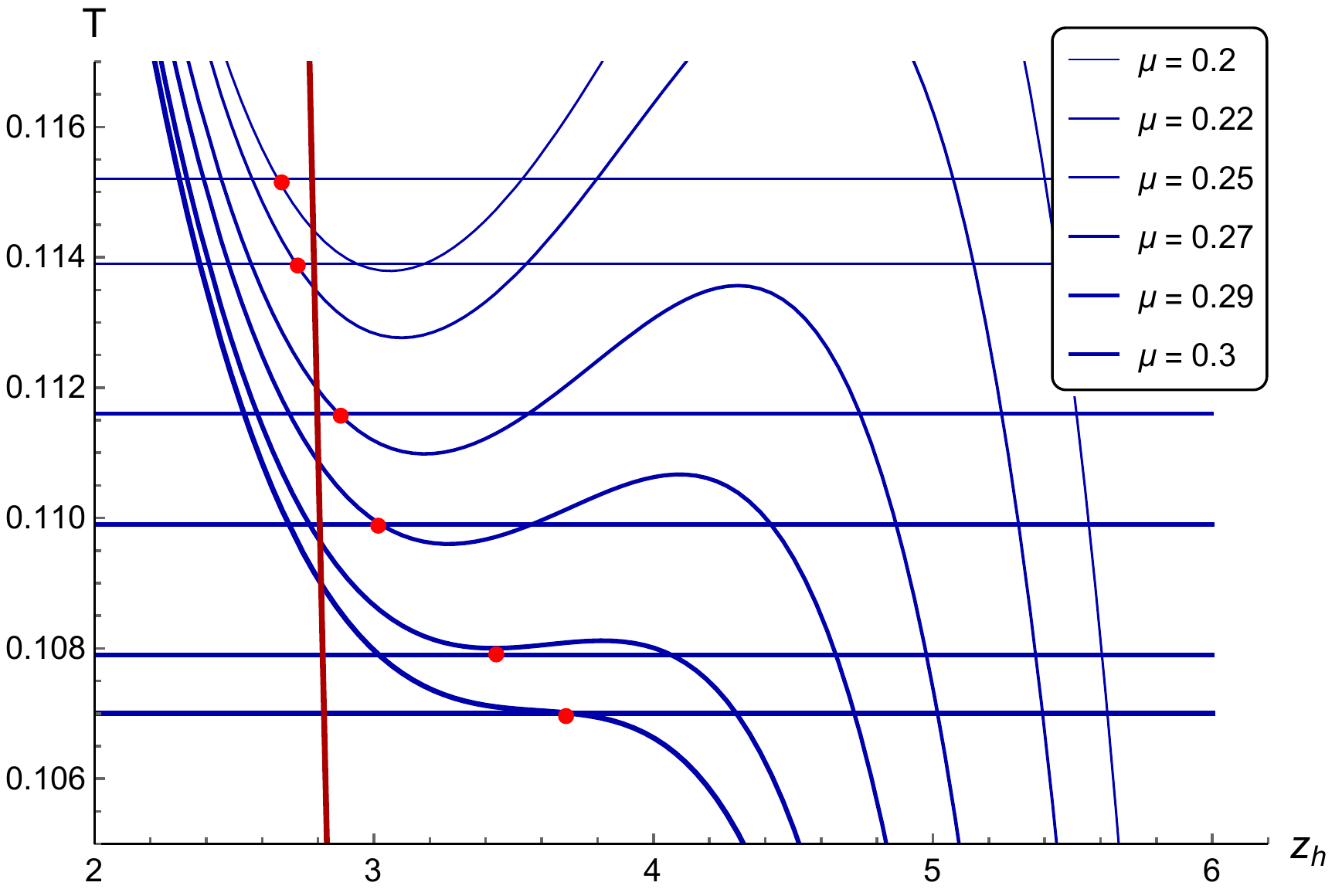} \
  \includegraphics[scale=0.27]{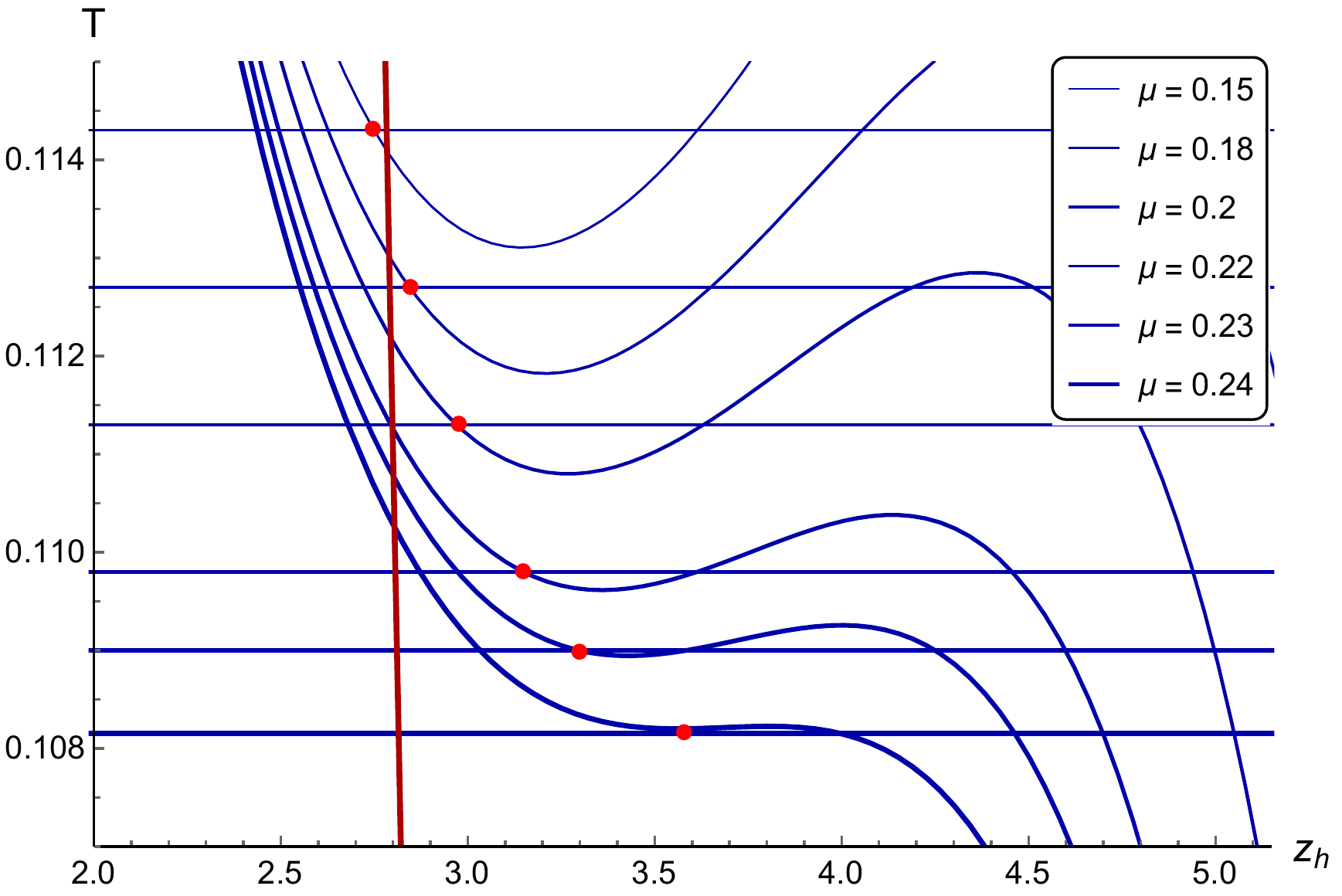} \
  \includegraphics[scale=0.27]{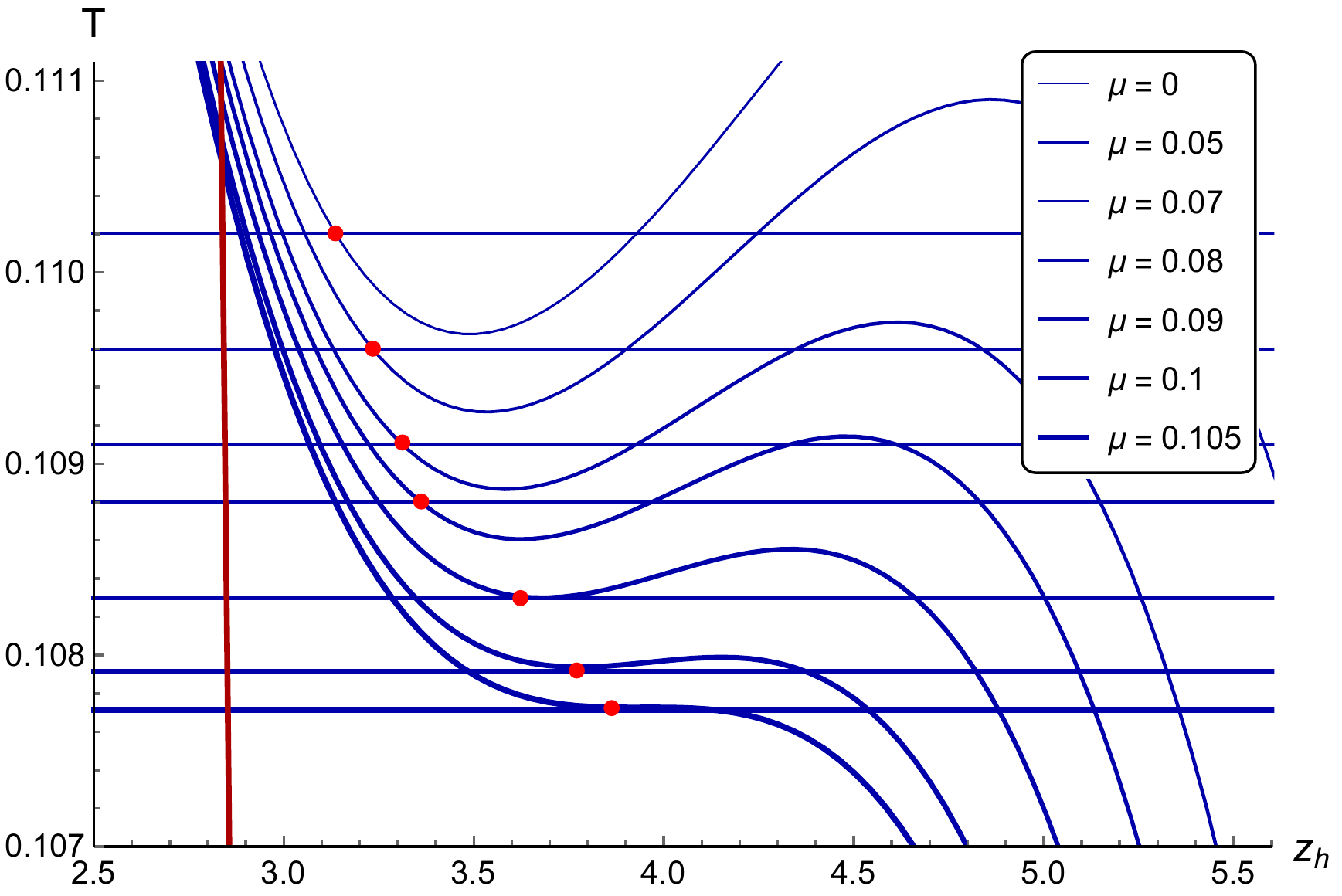} \\
  A \hspace{130pt} B \hspace{130pt} C
  \caption{Location of the BH-BH phase transition points (red point)
    relative to the dynamic wall (red lines) for $\nu = 1$ (first
    line) and $\nu = 4.5$ (second and third line, third line is the
    zoom of the second) for $c_B = 0$ (A), $c_B = - 0.005$ (B), $c_B =
    - 0.0096$ for $\nu = 1$ and $c_B = - 0.015$ for $\nu = 4.5$, $c =
    0.227$.}
  \label{Fig:DWy2}
\end{figure}

This is exactly the situation that takes place for $\nu = 1$ in our
model (first line in Fig.\ref{Fig:DWy2}). The dynamic wall is located
in the instability region, and the BH-BH phase transition
make us land in the area of deconfinement (Fig.\ref{Fig:DWy2}.A,B,
first line). So in the absence of the primary anisotropy
confinement/deconfinement phase transition is detemined by the first
order phase transition (BH-BH line) till the intersection with the
WL-line, providing us with the crossover. The part of the BH-BH line,
that turns out to be in the deconfinement area to the right of the
crossover (Fig.\ref{Fig:Tmuy2}.A,C), seems to cause density jump in
the QGP. For large $c_B$ BH-BH phase transition lies above the DW,
that is still in the instability region (Fig.\ref{Fig:DWy2}.C, first
line). So the entire BH-BH line serves for density jumps in QGP, not
separating confinement and deconfinement areas
(Fig.\ref{Fig:Tmuy2}.E).

For $\nu = 4.5$ the situation is quite different, as the dynamic wall
is located in the stable region. In the absence of external magnetic
field first order phase transition lands into the deconfinement region
for $0 \le \mu \le 0.2448$, closing the crossover from above till the
intersection with the WL-line. For $\mu > 0.2448$ the
Hawling-Page-like phase transition lands in the confinement area
(Fig.\ref{Fig:DWy2}.A, second line). This time the BH-BH line produces
the density jump of the confined matter, while
confinement/deconfinement phase transition is determined by the
$y$-Wilson loop. (Let us to remind that for $c_B = 0$ there is no
difference between $y_1$ and $y_2$ orientations.) On
Fig.\ref{Fig:DWy2}.A in the third line show the same picture more
clearly, zooming the BH-BH transition to the confinement area from
Fig.\ref{Fig:DWy2}.A, second line.

Adding the external magnetic field makes BH-BH line go under the
WL-line into the confinement area earlier, i.e. for lesser $\mu$
values (Fig.\ref{Fig:DWy2}.B, second line and zoom in
Fig.\ref{Fig:DWy2}.B, third line). The curved triangle limited by the
intercection point from the left, CEP temperature from the right,
BH-BH line from below and WL from above should represent the area of
higher density in the confinement area (Fig.\ref{Fig:Tmuy2}.B,D).

When the coupling coefficient reaches critical value $c_{B\, crit}
\approx - \, 0.11$ the entire BH-BH line sinks into the confinement
area. Therefore the confinement/deconfinement phase transition goes
along the Wilson loop line only (Fig.\ref{Fig:DWy2}.C, second line and
zoom in Fig.\ref{Fig:DWy2}.C, third line), and we have a higher
density sector between the BH-BH and WL-lines (Fig.\ref{Fig:Tmuy2}.F).


\section{Conclusions}\label{conclusions}

In this work holographic description of hot dense anisotropic QGP with
heavy quarks in magnetic field for heavy quarks model was
constructed. This model is the extension of the previous solution
\cite{1802.05652} that serves as a zero magnetic field limit.\\

The magnetic field $F^{(B)}$ is characterized by two parameters --
``charge'' $F_{x y_1}^{(B)} = q_B$ and metric coupling coefficient
$c_B$. First of them characterises the Maxwell field itself, while the
second one describes the metric deformation due to the presence of
external magnetic field. In our model kinetic coupling function has
simplest gaussian form \cite{1612.06248, 1907.01852, 2004.01965} and
depends of the coefficient $c_B$ and the external ``charge'' $q_B$
which not connected to each other. Such a parameterization was
introduced for more generality of the model. As a result, it turns out
that the magnetic ``charge'' is almost not involved into the main
properties of our holographic description for QGP, so we actually
consider $c_B$ as an efficient factor of the Maxwell field $F_{x
  y_1}^{(B)}$ and put $q_B=1$. But this doesn't mean at all that the
phase transition structure is independent from the strength of
external magnetic field. The fact is that we can always link these
parameters determining coupling coefficient as a function of
``charge'', $c_B = f(q_B)$. The specific shape of this function is an
important aspect that should allow to fit lattice results,
experimental data as well as reach agreement with other theoretical
considerations. \\

\begin{figure}[t!]
\centering
\includegraphics[scale=0.17]{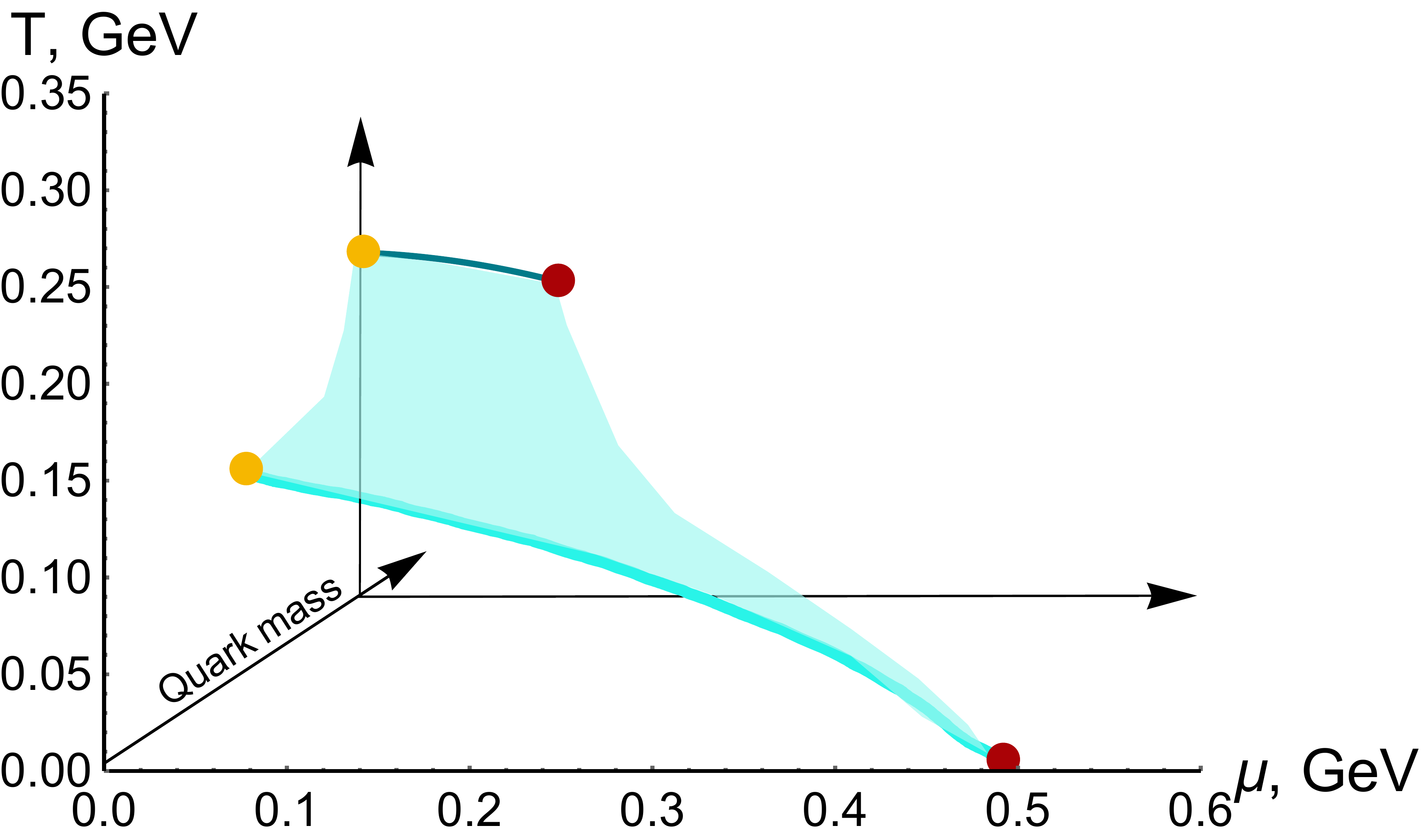}\quad
\includegraphics[scale=0.17]{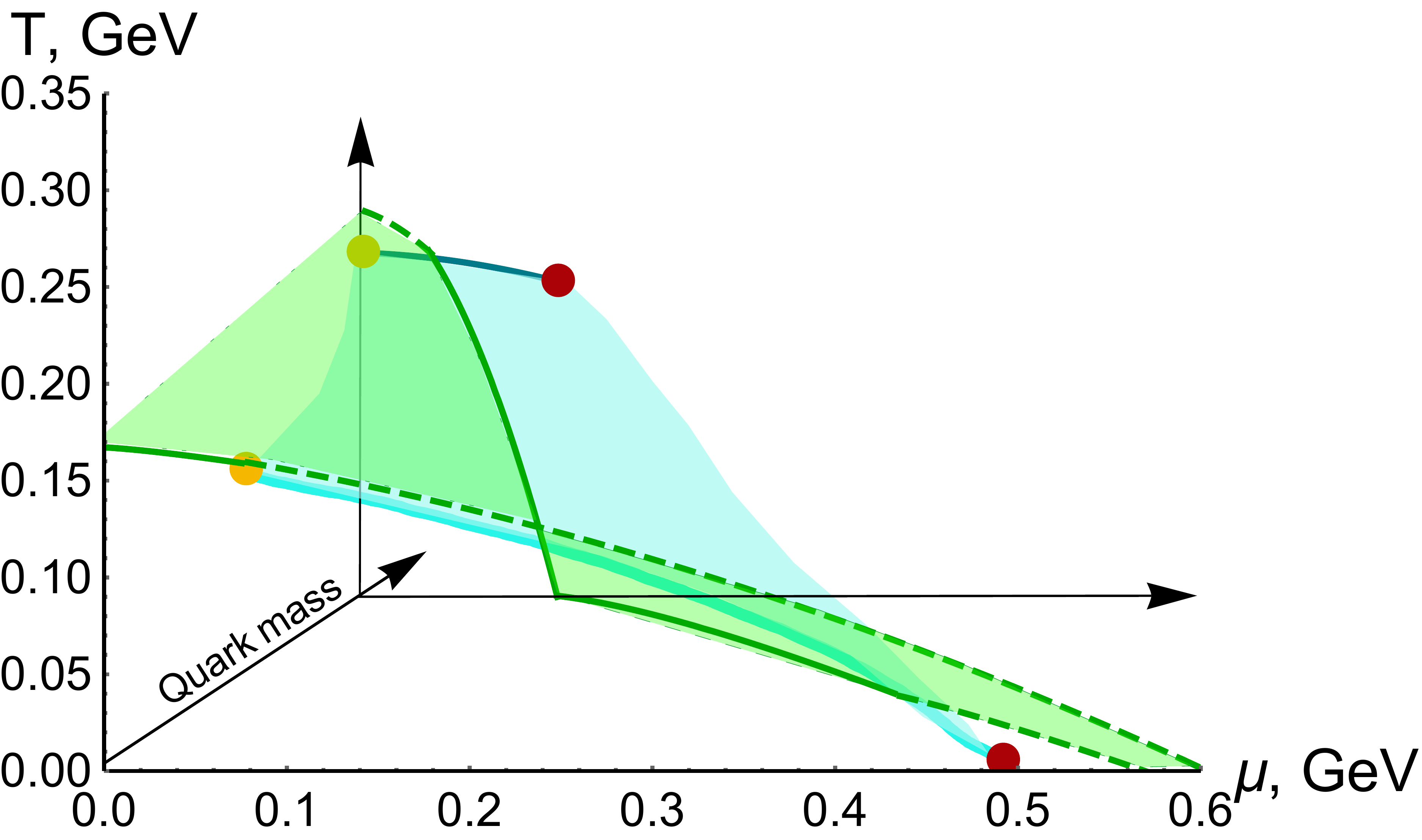} \\
A \hspace{200pt} B
\caption{Schematic pictures of phase transitions  dependence on quark mass. The cyan surface shows the first order phase transition (A)    and green surface shows the confinement/deconfinement  transition surface  (A, B). The red and yellow points are critical end points.}
\label{Fig:BPT}
\end{figure}

The structure of phase transitions strongly depends on the chosen
model -- models corresponding to light/heavy quarks have different
structures. In fact there is finer division depending on  quark type
taken into account, see for example
\cite{Brown:1990ev,1602.06129,1912.04827}. In the holographic approach
light/heavy quarks models have different structures of the background
phase transition \cite{1301.0385,1703.09184}. In most of holographic
models  the background  phase transition, see Fig.\ref{Fig:BPT}.A, is
related to chiral symmetry breaking  \cite{1206.2824, 1511.02721,
  1512.06493, 1810.12525, 1810.07019, 1910.02269, 2002.00075,
  2005.00500, 2009.05694, 2010.04578,
  2010.06762}. Confinement/deconfinement phase transition may be
related to background phase transition and may be not,
Fig.\ref{Fig:BPT}.B. This depends on the relative position of the
dynamic wall and the background phase transition, see more detail
discussion in Sect.\ref{WL}. An expected phase structure for the
intermediate version of the model between heavy and light quarks (the
so-called ``hybrid'' model) is shown on Fig.\ref{Fig:BPT}.A and
Fig.\ref{Fig:BPT}.B. The area of the assumed location of the
first-order phase transition line is depicted in cyan.\\

\begin{figure}[t!]
  \centering
  \includegraphics[scale=0.2]{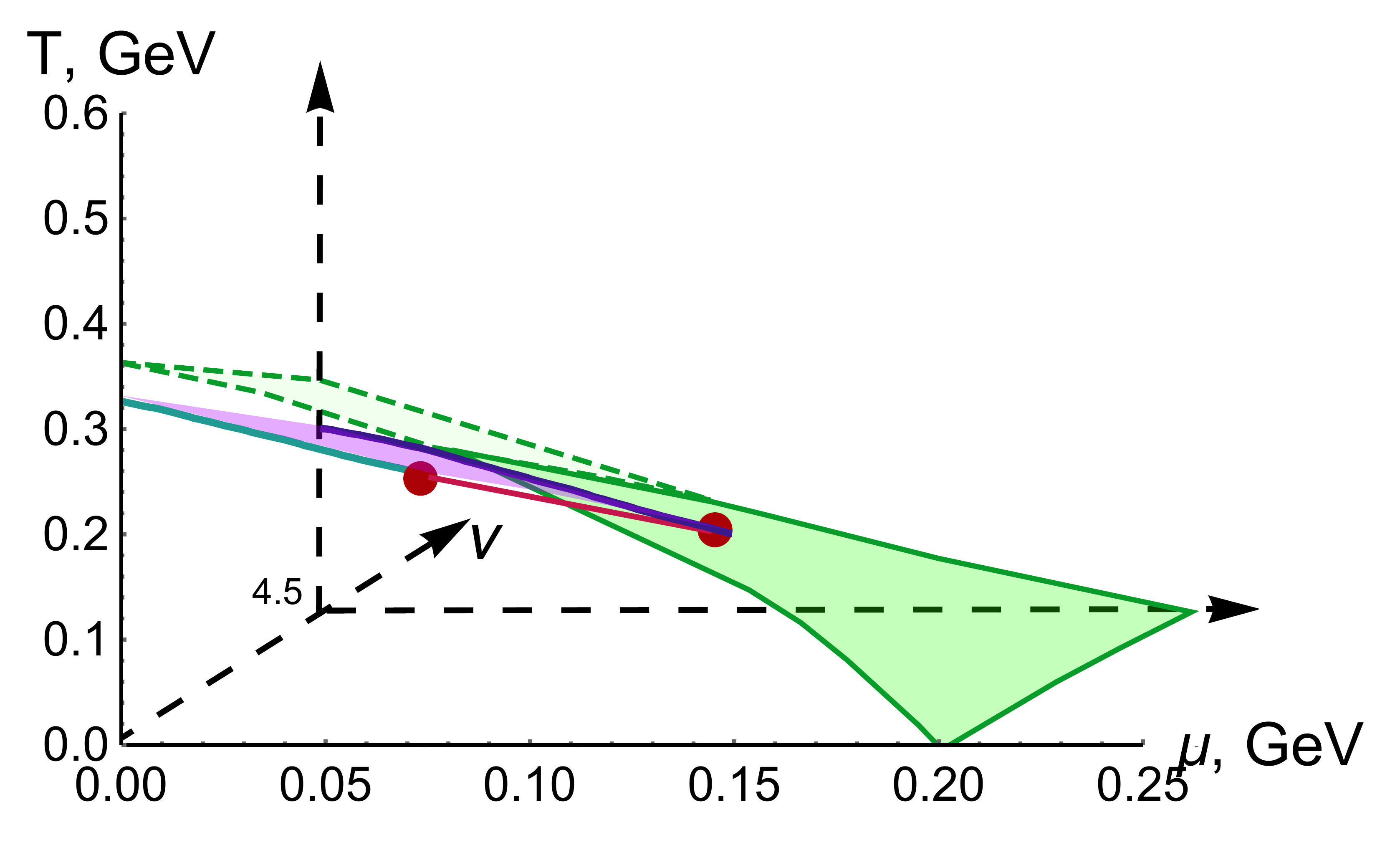} \quad
  \includegraphics[scale=0.2]{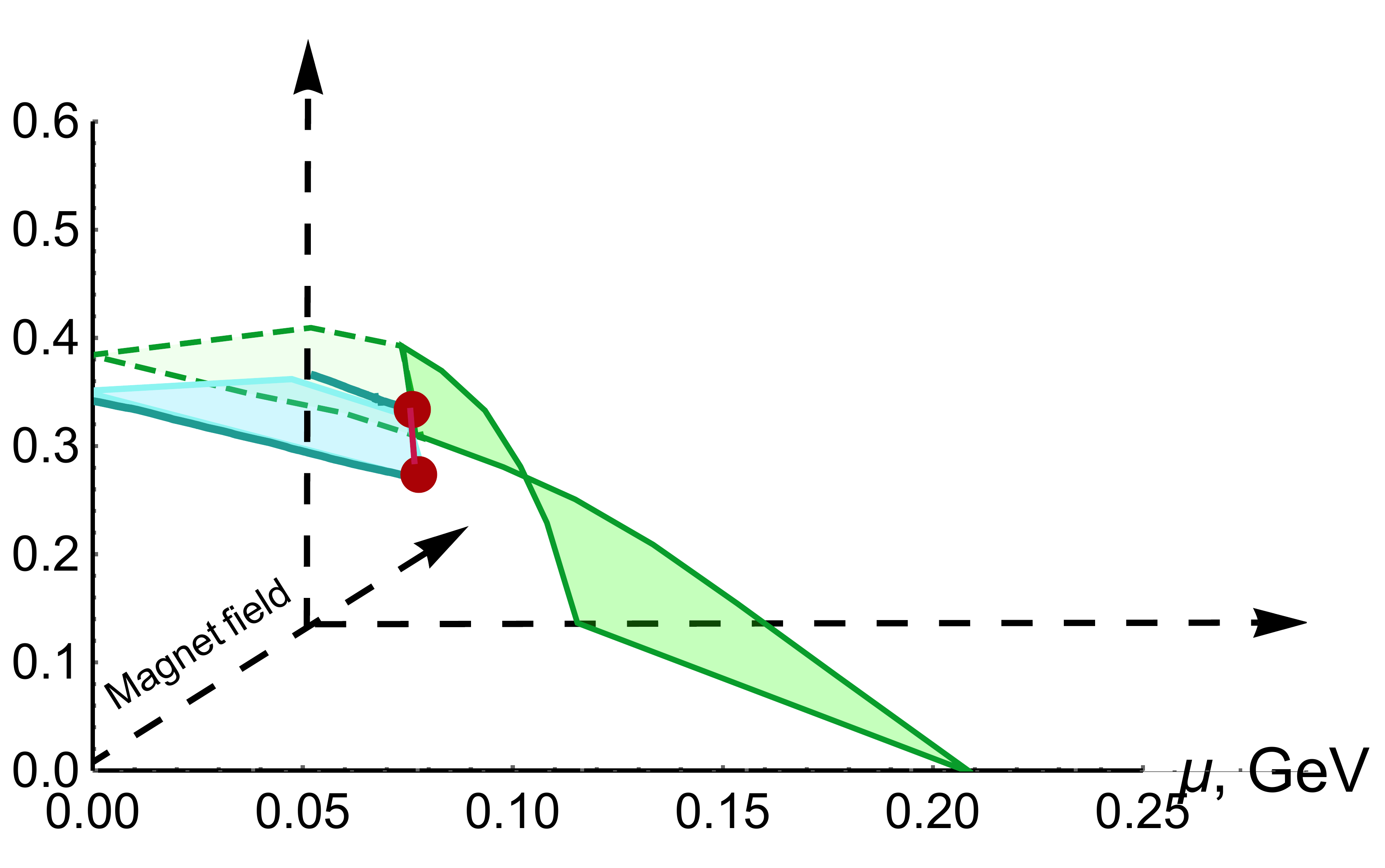} \\
  A \hspace{200pt} B
  \caption{Schematic representations of the structure of phase
    transitions of a model describing heavy quarks depending on the
    anisotropy caused by:  parameter $ \nu $ (A), magnetic field (B).}
  \label{Fig:mag-nu}
\end{figure}

In this paper we investigated the dependence of the phase transition
structure on the anisotropy, characterized by parameter $\nu$ (see
eq.\eqref{eq:2.03}), and magnetic fields. Our results are
schematically shown in the 3D diagrams $(\mu,\nu,T)$ and $(\mu,
B,T)$ on Fig.\ref{Fig:mag-nu}. The WL phase transition is indicated by
green lines, the BH-BH phase transition is indicated by cyan, blue and
magenta lines. The solid lines of BH-BH and WL phase transitions
indicate the confinement/deconfinement phase transition. In the figure
Fig.\ref{Fig:mag-nu}.A we see that while increasing anisotropy, the
BH-BH phase transition and WL phase transition occur at lower
temperatures for the same values of the chemical potential. With the
increase of $\nu$, the first-order phase transition line elongates. In
the figure Fig.\ref{Fig:mag-nu}.B, we see qualitatively the same
situation with increasing magnetic field, the phase transition occurs
at lower temperature values for the same chemical potential values,
but the first-order phase transition line shortens.\\

To
complete the confinement/de\-con\-fine\-ment picture we also
investigated string properties of the solution -- string tension
depending of orientation of the quark pair. For this purpose we
investigated behavior of differently oriented temporal Wilson loops,
that determines location of confinement/deconfinement phase
transition. Note that unlike the dynamic walls' positions the string
tension strongly depends on the integration boundary for the scalar
field \cite{2009.05562}. Here we performed calculations for the
simplest case $z_0 = \epsilon \ll 1$ only. Choice of the function $z_0
= f(z_h)$ will be consider in further investigations of the model. \\

In the model considered in this paper we have observed only IMC. Such
a behavior distinguishes this model from more complicated heavy
quarks models with higher order polynomials in the
warp-factor. Namely, the model considered in \cite{2004.01965}
demonstrates MC for small magnetic fields and IMC for large ones. This
phenomena takes place for small values of the chemical potential,
while for large admissible chemical potentials the model exhibits MC
only. Similar to  the simplest heavy quarks models \cite{1506.05930,1802.05652}
the model has CEPs, those position depend on values
of magnetic field. It seems, that this phenomena will take place also
for the case of the warp factor considered in \cite{1907.01852} (see
Table 2  in Appendix \ref{appendixB2}).\\
 
Note, that to get more realistic description of the phase structure we
have to deal with so-called hybrid warped factor that is a combination
of the exponent of polynomials with logarithmic
corrections. Constructing of such a superposition model is a must-have
item on the list of future research.

\section{Acknowledgments}

This work is supported by Russian Science Foundation grant \textnumero
20-12-00200.

$$\,$$
\newpage
\appendix 

\section*{Appendix}

\section{EOM}\label{appendixA}

For $F_{x y_1}^{(B)} = q_B$ the corresponding EOM are the following:
\begin{gather}
  \begin{split}
    \phi'' &+ \phi' \left( \cfrac{g'}{g} + \cfrac{3 \fb'}{2 \fb} -
      \cfrac{\nu + 2}{\nu z} + c_B z \right)
    + \left( \cfrac{z}{L} \right)^2 \cfrac{\partial f_1}{\partial
      \phi} \ \cfrac{(A_t')^2}{2 \fb g} \ - \\
    &\quad - \left( \cfrac{L}{z} \right)^{2-\frac{4}{\nu}}
    \cfrac{\partial f_2}{\partial \phi} \ \cfrac{q^2 \ e^{-c_Bz^2}}{2
      \fb g} \
    - \left( \cfrac{z}{L} \right)^{\frac{2}{\nu}} \cfrac{\partial
      f_B}{\partial \phi} \ \cfrac{q_B^2}{2 \fb g} \
    - \left( \cfrac{L}{z} \right)^2 \cfrac{\fb}{g} \ \cfrac{\partial
      V}{\partial \phi} = 0,
  \end{split}\label{eq:A2.05} \\
  A_t'' + A_t' \left( \cfrac{\fb'}{2 \fb} + \cfrac{f_1'}{f_1} +
    \cfrac{\nu - 2}{\nu z} + c_B z \right) = 0, \label{eq:A2.06} \\
  g'' + g' \left(\cfrac{3 \fb'}{2 \fb} - \cfrac{\nu + 2}{\nu z} + c_B
    z \right)
  - \left( \cfrac{z}{L} \right)^2 \cfrac{f_1 (A_t')^2}{\fb}
  - \left( \cfrac{z}{L} \right)^{\frac{2}{\nu}} \cfrac{q_B^2 \
    f_B}{\fb} = 0, \label{eq:A2.07} \\
  \fb'' - \cfrac{3 (\fb')^2}{2 \fb} + \cfrac{2 \fb'}{z}
  - \cfrac{4 \fb}{3 \nu z^2} \left( 1 - \cfrac{1}{\nu}
    + \left( 1 - \cfrac{3 \nu}{2} \right) c_B z^2
    - \cfrac{\nu c_B^2 z^4}{2} \right)
  + \cfrac{\fb \, (\phi')^2}{3} = 0, \label{eq:A2.08} \\
  2  g' \ \cfrac{\nu - 1}{\nu}
  + 3  g \ \cfrac{\nu - 1}{\nu} \left(
    \cfrac{\fb'}{\fb} - \cfrac{4 \left( \nu + 1 \right)}{3 \nu z}
    + \cfrac{2 c_B z}{3} \right)
  + \left( \cfrac{L}{z} \right)^{1-\frac{4}{\nu}} \cfrac{L \, q^2 \,
    e^{-c_Bz^2} f_2}{\fb} = 0, \label{eq:A2.09} \\
  \begin{split}
    \cfrac{\fb''}{\fb} &+ \cfrac{(\fb')^2}{2 \fb^2}
    + \cfrac{3 \fb'}{\fb} \left( \cfrac{g'}{2 g}
      - \cfrac{\nu + 1}{\nu z}
      + \cfrac{2 c_B z}{3} \right)
    - \cfrac{g'}{3 z g}  \left( 5 + \cfrac{4}{\nu} - 3 c_B z^2
    \right) + \\
    &+ \cfrac{8}{3 z^2} \left( 1 + \cfrac{3}{2 \nu} + \cfrac{1}{2
        \nu^2} \right) 
    - \cfrac{4 c_B}{3} \left( 1 + \cfrac{3}{2 \nu} - \cfrac{c_B
        z^2}{2} \right)
    + \cfrac{g''}{3 g} + \cfrac{2}{3} \left( \cfrac{L}{z} \right)^2
    \cfrac{\fb V}{g} = 0,
  \end{split}\label{eq:A2.10}
\end{gather}
where $'= \partial/\partial z$.


For $F_{x y_2}^{(B)} = q_B$ the corresponding EOM are the following:
\begin{gather}
  \begin{split}
    \phi'' &+ \phi' \left( \cfrac{g'}{g} + \cfrac{3 \fb'}{2 \fb} -
      \cfrac{\nu + 2}{\nu z} + c_B z \right)
    + \left( \cfrac{z}{L} \right)^2 \cfrac{\partial f_1}{\partial
      \phi} \ \cfrac{(A_t')^2}{2 \fb g} \ - \\
    &\quad - \left( \cfrac{L}{z} \right)^{2-\frac{4}{\nu}}
    \cfrac{\partial f_2}{\partial \phi} \ \cfrac{q^2 \, e^{-c_Bz^2}}{2
      \fb g} \
    - \left( \cfrac{z}{L} \right)^{\frac{2}{\nu}} \cfrac{\partial
      f_B}{\partial \phi} \ \cfrac{q_B^2 \, e^{-c_Bz^2}}{2 \fb g} \
    - \left( \cfrac{L}{z} \right)^2 \cfrac{\fb}{g} \ \cfrac{\partial
      V}{\partial \phi} = 0,
  \end{split}\label{eq:A2.11} \\
  A_t'' + A_t' \left( \cfrac{\fb'}{2 \fb} + \cfrac{f_1'}{f_1} +
    \cfrac{\nu - 2}{\nu z} + c_B z \right) = 0, \label{eq:A2.12} \\
  g'' + g' \left(\cfrac{3 \fb'}{2 \fb} - \cfrac{\nu + 2}{\nu z} + c_B
    z \right)
  - \left( \cfrac{z}{L} \right)^2 \cfrac{f_1 (A_t')^2}{\fb}
  - \left( \cfrac{z}{L} \right)^{\frac{2}{\nu}} \ \cfrac{q_B^2 \,
    e^{-c_Bz^2} f_B}{\fb} = 0, \label{eq:A2.13} \\
  \fb'' - \cfrac{3 (\fb')^2}{2 \fb} + \cfrac{2 \fb'}{z}
  - \cfrac{4 \fb}{3 \nu z^2} \left( 1 - \cfrac{1}{\nu}
    + \left( 1 - \cfrac{3 \nu}{2} \right) c_B z^2
    - \cfrac{\nu c_B^2 z^4}{2} \right)
  + \cfrac{\fb \, (\phi')^2}{3} = 0, \label{eq:A2.14} \\
  \begin{split}
    2  g' \ \cfrac{\nu - 1}{\nu}
    + 3  g \ \cfrac{\nu - 1}{\nu} \left(
      \cfrac{\fb'}{\fb} - \cfrac{4 \left( \nu + 1 \right)}{3 \nu z} +
      \cfrac{2 c_B z}{3} \right)
    &+ \left( \cfrac{L}{z} \right)^{1-\frac{4}{\nu}} \cfrac{L \, q^2
      \, e^{-c_Bz^2} f_2}{\fb} \ - \\
    &- \left( \cfrac{z}{L} \right)^{\frac{2}{\nu}} \cfrac{q_B^2 \,
      e^{-c_Bz^2} z \, f_B}{\fb} = 0,
  \end{split}\label{eq:A2.15} \\
  \begin{split}
    \cfrac{\fb''}{\fb} &+ \cfrac{(\fb')^2}{2 \fb^2}
    + \cfrac{3 \fb'}{\fb} \left( \cfrac{g'}{2 g}
      - \cfrac{\nu + 1}{\nu z}
      + \cfrac{2 c_B z}{3} \right)
    - \cfrac{g'}{3 z g}  \left( 5 + \cfrac{4}{\nu} - 3 c_B z^2
    \right)
    + \cfrac{8}{3 z^2} \left( 1 + \cfrac{3}{2 \nu} + \cfrac{1}{2
        \nu^2} \right) - \\
    &- \cfrac{4 c_B}{3} \left( 1 + \cfrac{3}{2 \nu} - \cfrac{c_B
        z^2}{2} \right)
    + \cfrac{g''}{3 g}
    + \left( \cfrac{z}{L} \right)^{\frac{2}{\nu}} \cfrac{q_B^2 \,
      e^{-c_Bz^2} f_B}{3 \fb g}
    + \cfrac{2}{3} \left( \cfrac{L}{z} \right)^2 \cfrac{\fb V}{g} =
    0.
  \end{split}\label{eq:A2.16}
\end{gather}

Note, that the form of the external magnetic field $F_{\mu\nu}^{(B)}$
has influence on the penultimate terms in (\ref{eq:A2.05},
\ref{eq:A2.11}), last terms in  (\ref{eq:A2.07}, \ref{eq:A2.13}),
equations (\ref{eq:A2.15}, \ref{eq:A2.16}) have an extra term in
comparison with (\ref{eq:A2.09}, \ref{eq:A2.10})
correspondingly. Equations (\ref{eq:A2.06}, \ref{eq:A2.12}) on electric
field $A_t$ and equations (\ref{eq:A2.08}, \ref{eq:A2.14}) on scalar
field $\phi$ are identical, thus these quantities are invariants with
respect to the orientation of the external magnetic field.

The important feature of the systems (\ref{eq:A2.05}--\ref{eq:A2.10})
and (\ref{eq:A2.11}--\ref{eq:A2.16}) is that we can't get a solution
for the blackening function from (\ref{eq:A2.07}) or (\ref{eq:A2.13})
immediately after the electric field $A_t$, as we also need to find or
exclude the 3-rd Maxwell field's coupling function $f_B$ first. 

It is possible to set $f_B$ manually, but we can avoid such an
arbitrariness. Excluding higher derivatives from equations
(\ref{eq:A2.06}--\ref{eq:A2.10}) and substituting them into
(\ref{eq:A2.05}), we get the following expression for the coupling
function $f_B$:
\begin{gather}
  f_{B} = 2 \left( \cfrac{z}{L} \right)^{-\frac{2}{\nu}} \fb g \
  \cfrac{c_B z}{q_B^2} \left( \cfrac{3 \fb'}{2 \fb} -
    \cfrac{2}{\nu z} + c_B z + \cfrac{g'}{g} \right). \label{eq:A3.02}
\end{gather}

Similarly excluding higher derivatives from equations
(\ref{eq:A2.12}--\ref{eq:A2.16}) and substituting them into
(\ref{eq:A2.10}), we get the following expression for the coupling
function $f_B$:
\begin{gather}
  f_{B} = - \, 2 \left( \cfrac{z}{L} \right)^{-\frac{2}{\nu}}
  e^{c_Bz^2} \ \fb g \ \cfrac{c_B z}{q_B^2} \left( \cfrac{3 \fb'}{2 \fb} -
    \cfrac{2}{\nu z} + c_B z + \cfrac{g'}{g} \right). \label{eq:A3.04}
\end{gather}


\section{Tables}\label{appendixB}

Here we present phase diagrams for various holographic heavy quarks
models. In Table 1 we collected known results for zero magnetic field
and in Table 2 the results for $B \neq 0$.

HQCD heavy quarks models show different behavior for different
polynomials in warped factor and function $f(z)$: the plot in the 
cell (11) of Table 2, that corresponds to the simplest warp factor,
shows the IMC, i.e. the critical temperature decreases with increasing
$|c_B|$, that corresponds to increasing the magnetic field. This is in
the opposite with lattice simulations for heavy quarks that reveal the
MC phenomenon \cite{1005.5365, 1203.3360, 1209.0374}.

The deformation of the warp-factor by adding $d z^5$ changes IMC to MC
for $d$ large enough, see plot in the cell (3,3) of Table 2. If one
considers a deformation by quadratic factor $pz^4$, then for
particular $p$, suitable to fit Regge spectrum, one gets MC for small and medium $B$, meanwhile
one observes IMC for large $B$, see plot in cell (3,2).

We also see the change of regime from IMC to MC for large enough
$\mu$, see plots in the cell (1,3). 

\newpage

\subsection{\ }\label{appendixB1}




\begin{table}[h!]\label{table1}
\centering
\caption{Holographic models for heavy quarks without magnetic
  field.} \ \\
\begin{tabular}{|c|c|c|c|}
\hline
$B=0$                                                                                              & \multicolumn{3}{c|}{Heavy quarks}                                                                    \\ \hline
${\cal A}$                                                                                         & \multicolumn{1}{l|}{$-cz^2$} & \multicolumn{1}{l|}{$-cz^2+pz^4$} & \multicolumn{1}{l|}{$-cz^2+dz^5$} \\ \hline
\multirow{2}{*}{\begin{tabular}[c]{@{}c@{}}$\nu =1$\\ $f_1(z)=1$\end{tabular}}                     & \multicolumn{1}{l|}{ \includegraphics[width=0.25\textwidth]{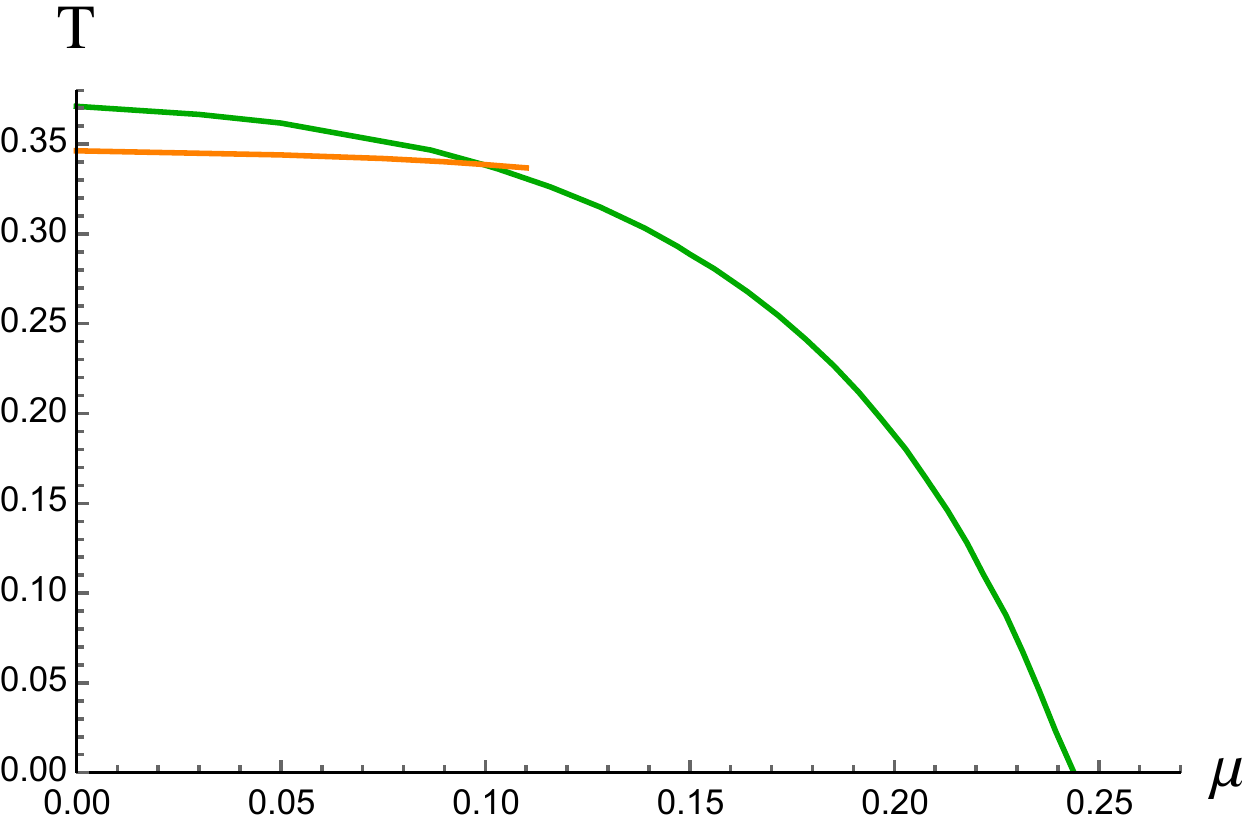} }      & \multicolumn{1}{l|}{\includegraphics[width=0.25\textwidth]{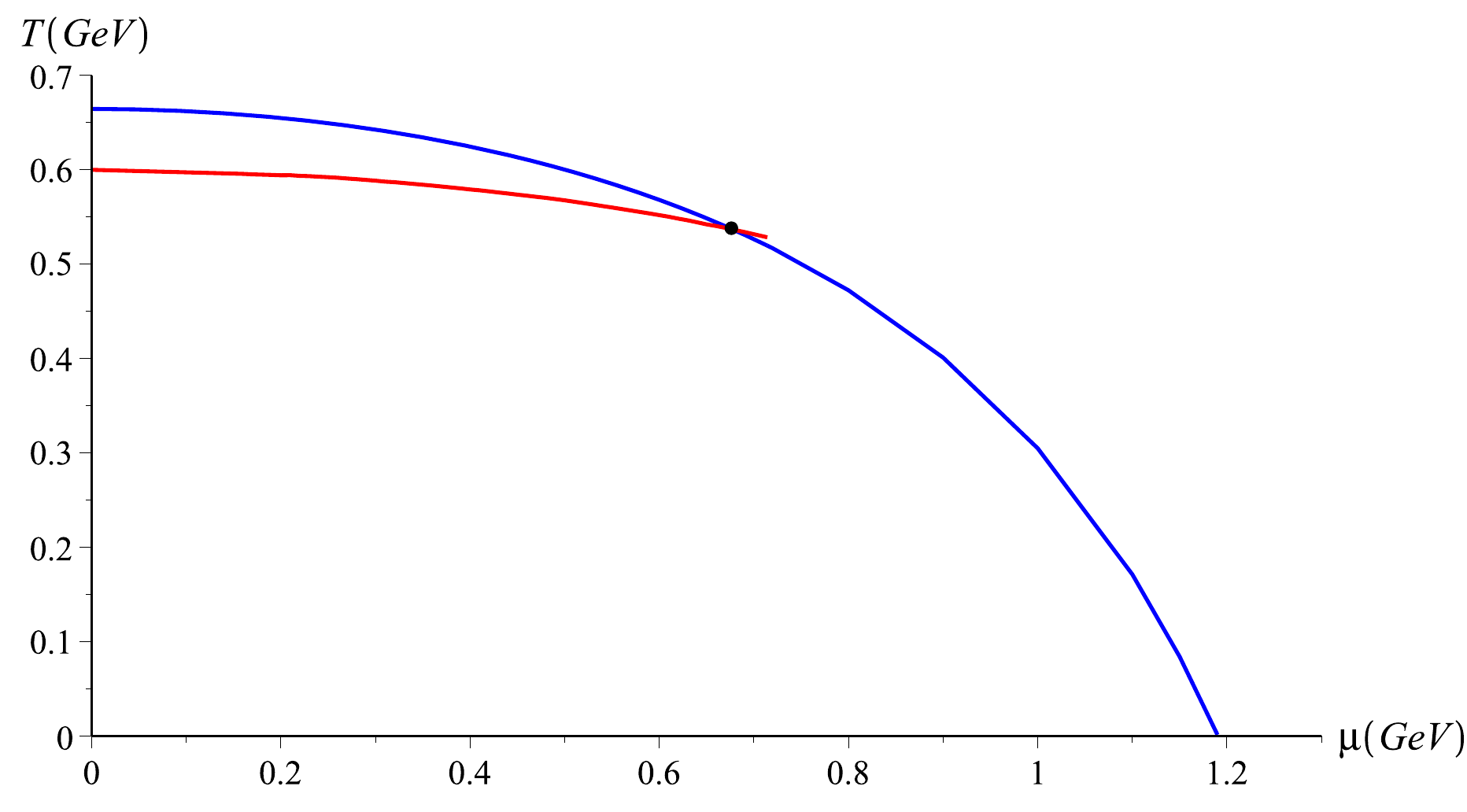}}           & \multirow{2}{*}{}                \\ \cline{2-3}
                                                                                                   & \cite{1802.05652}                           & \cite{1506.05930}                               &                                   \\ \hline
\multirow{2}{*}{\begin{tabular}[c]{@{}c@{}}$\nu=4.5$\\ $f_1=z^{2-2/\nu}$\end{tabular}}             & \multicolumn{1}{l|}{ \includegraphics[width=0.25\textwidth]{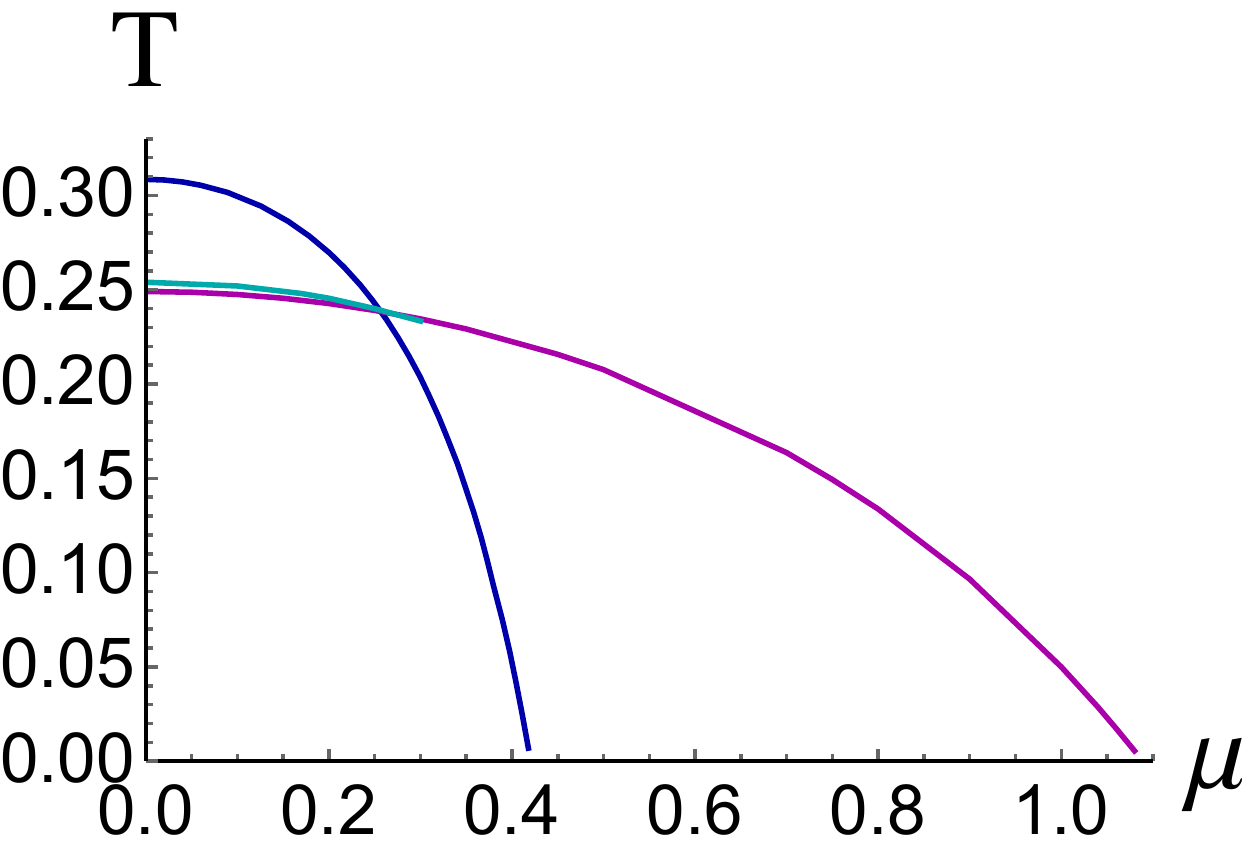} }      & \multirow{2}{*}{-}                & \multirow{2}{*}{-}                \\ \cline{2-2}
                                                                                                   & \cite{1802.05652,2009.05562}                           &                                   &                                   \\ \hline
\multirow{2}{*}{\begin{tabular}[c]{@{}c@{}}$\nu=1$\\ $f_1=e^{c_1z^2}$\end{tabular}}                & \multicolumn{1}{l|}{\includegraphics[width=0.25\textwidth]{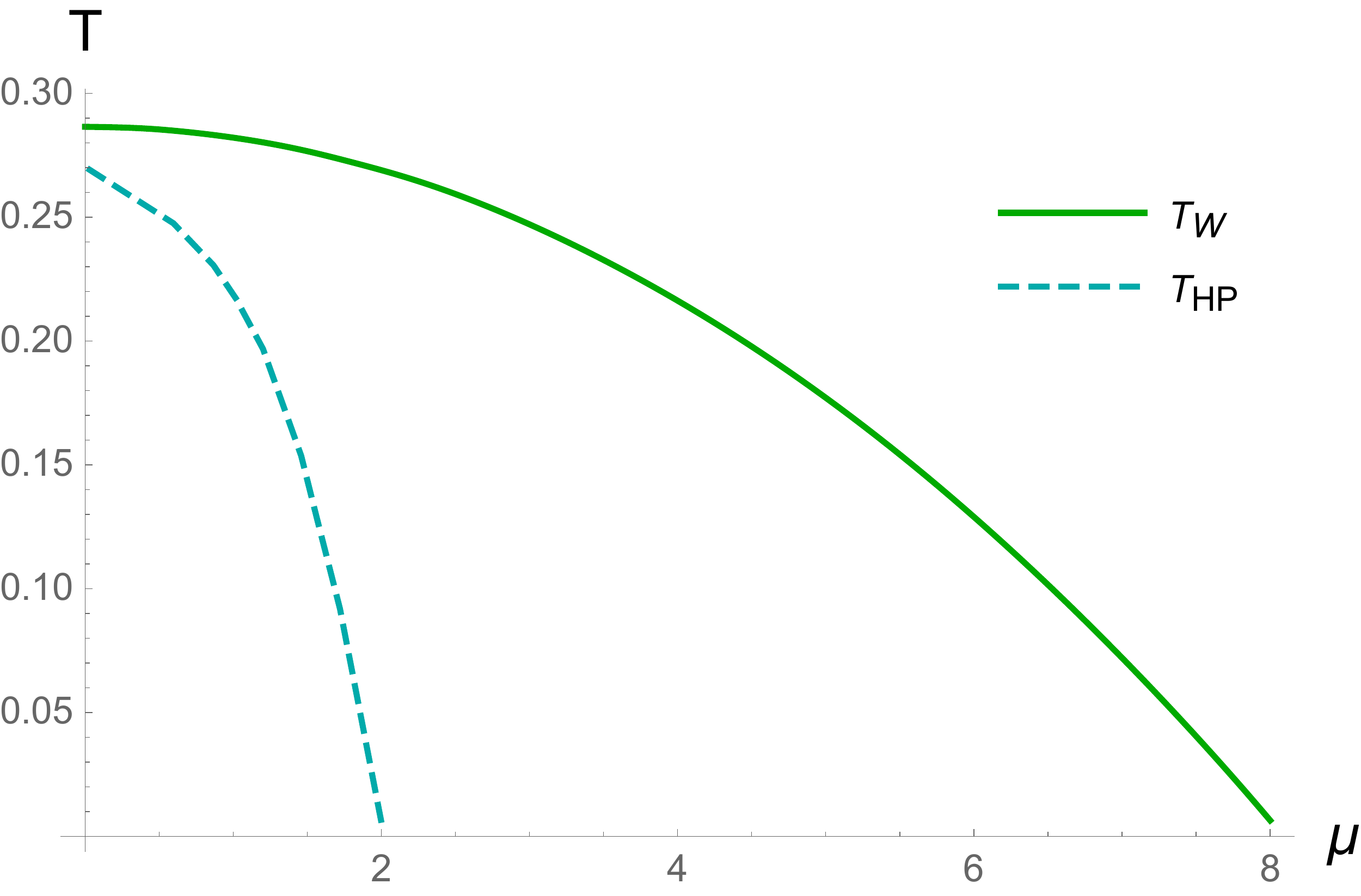}  }      & \multicolumn{1}{l|}{\includegraphics[width=0.25\textwidth]{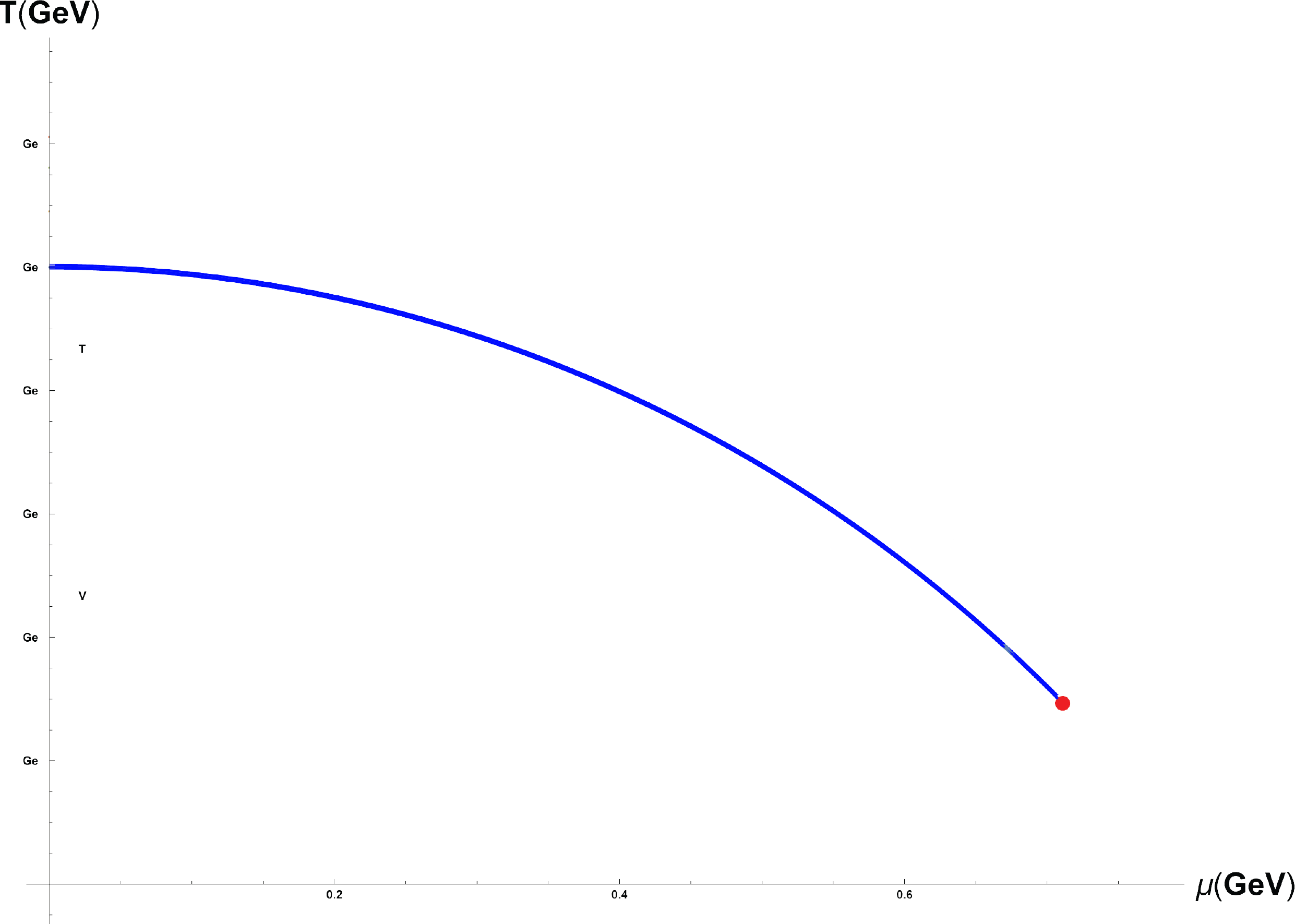}}           & \multirow{2}{*}{-}                \\ \cline{2-3}
                                                                                                   &  \cite{1907.01852}                         & \cite{1506.05930}                              &                                   \\ \hline
\multirow{2}{*}{\begin{tabular}[c]{@{}c@{}}$\nu\neq 1$\\ $f_1=z^{2-2/\nu}e^{c_1z^2}$\end{tabular}} & \multirow{2}{*}{-}           & \multirow{2}{*}{-}                & \multirow{2}{*}{-}                \\
                                                                                                   &                              &                                   &                                   \\ \hline
\end{tabular}
\end{table}

\newpage

\subsection{\ }\label{appendixB2}

\begin{table}[h!]\label{table2}
\centering
\caption{Holographic models for heavy quarks in magnetic field.} \ \\
\begin{tabular}{|c|c|c|c|}
\hline
$B\neq 0$                                                                                          & \multicolumn{3}{c|}{Heavy quarks}                                                                    \\ \hline
${\cal A}$                                                                                         & \multicolumn{1}{l|}{$-cz^2$} & \multicolumn{1}{l|}{$-cz^2+pz^4$} & \multicolumn{1}{l|}{$-cz^2+dz^5$} \\ \hline
\multirow{3}{*}{\begin{tabular}[c]{@{}c@{}}$\nu =1$\\ $f_1(z)=1$\end{tabular}}                     & \multicolumn{1}{l|}{\includegraphics[width=0.25\textwidth]{Plots/BBnu1.pdf} }      & -                                 & -                                 \\ \cline{2-4} 
                                                                                                   & \multicolumn{1}{l|}{\includegraphics[width=0.25\textwidth]
                                                                                                   {
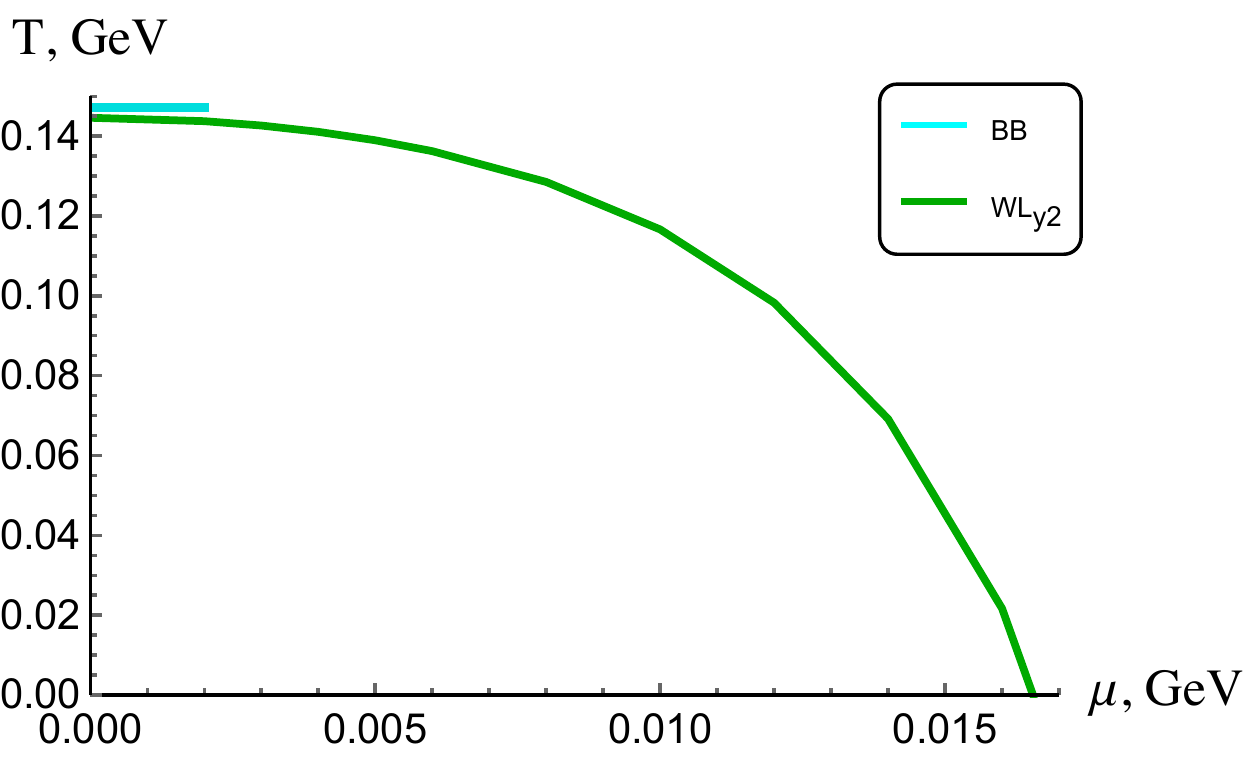} }      & -                                 & -                                 \\ \cline{2-4} 
                                                                                                   & see Fig.\ref{Fig:Tmuy2}                          & -                                & -                                 \\ \hline
\multirow{2}{*}{\begin{tabular}[c]{@{}c@{}}$\nu=4.5$\\ $f_1=z^{2-2/\nu}$\end{tabular}}             & \multicolumn{1}{l|}{\includegraphics[width=0.25\textwidth]{Plots/BBnu45.pdf}}      & \multirow{2}{*}{-}                & \multirow{2}{*}{-}                \\ \cline{2-2}
                                                                                                   &  \includegraphics[width=0.25\textwidth]
                                                                                                   {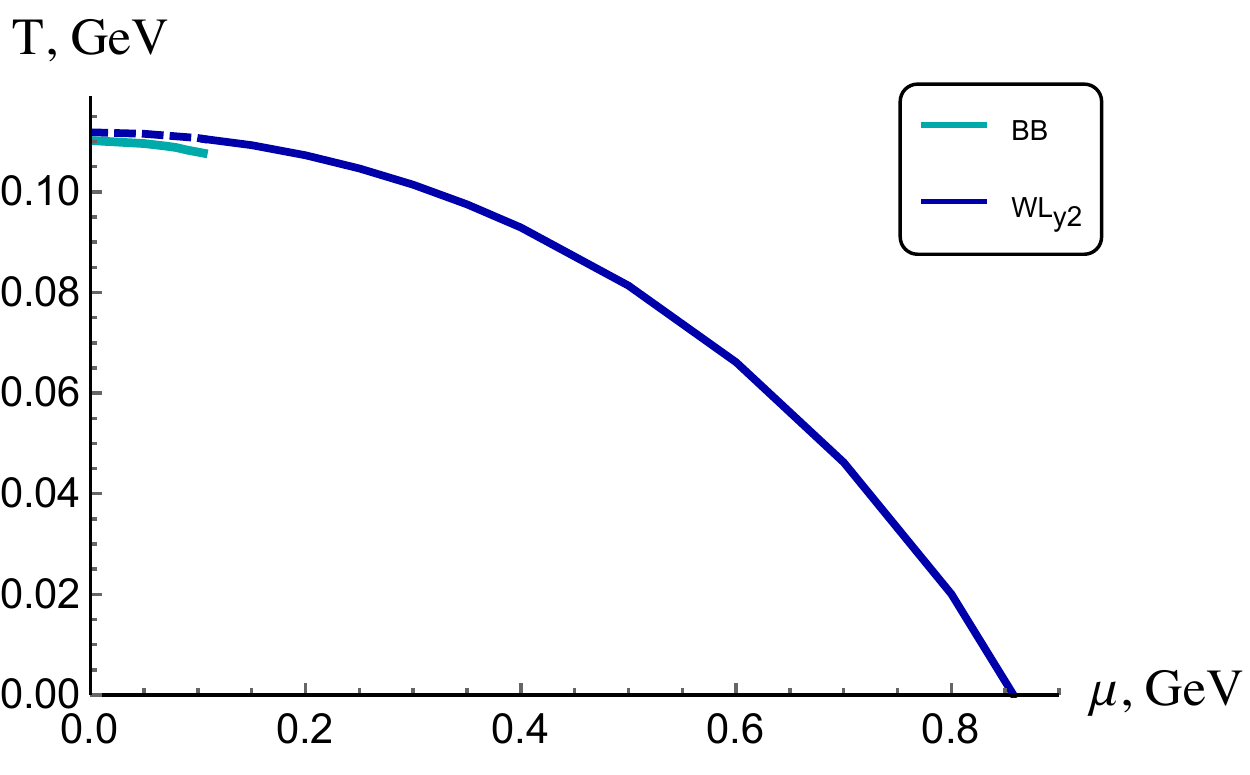}                            &                                   &                                   \\ \hline
\multirow{2}{*}{\begin{tabular}[c]{@{}c@{}}$\nu=1$\\ $f_1=e^{c_1z^2}$\end{tabular}}                & \multicolumn{1}{l|}{ \includegraphics[width=0.25\textwidth]{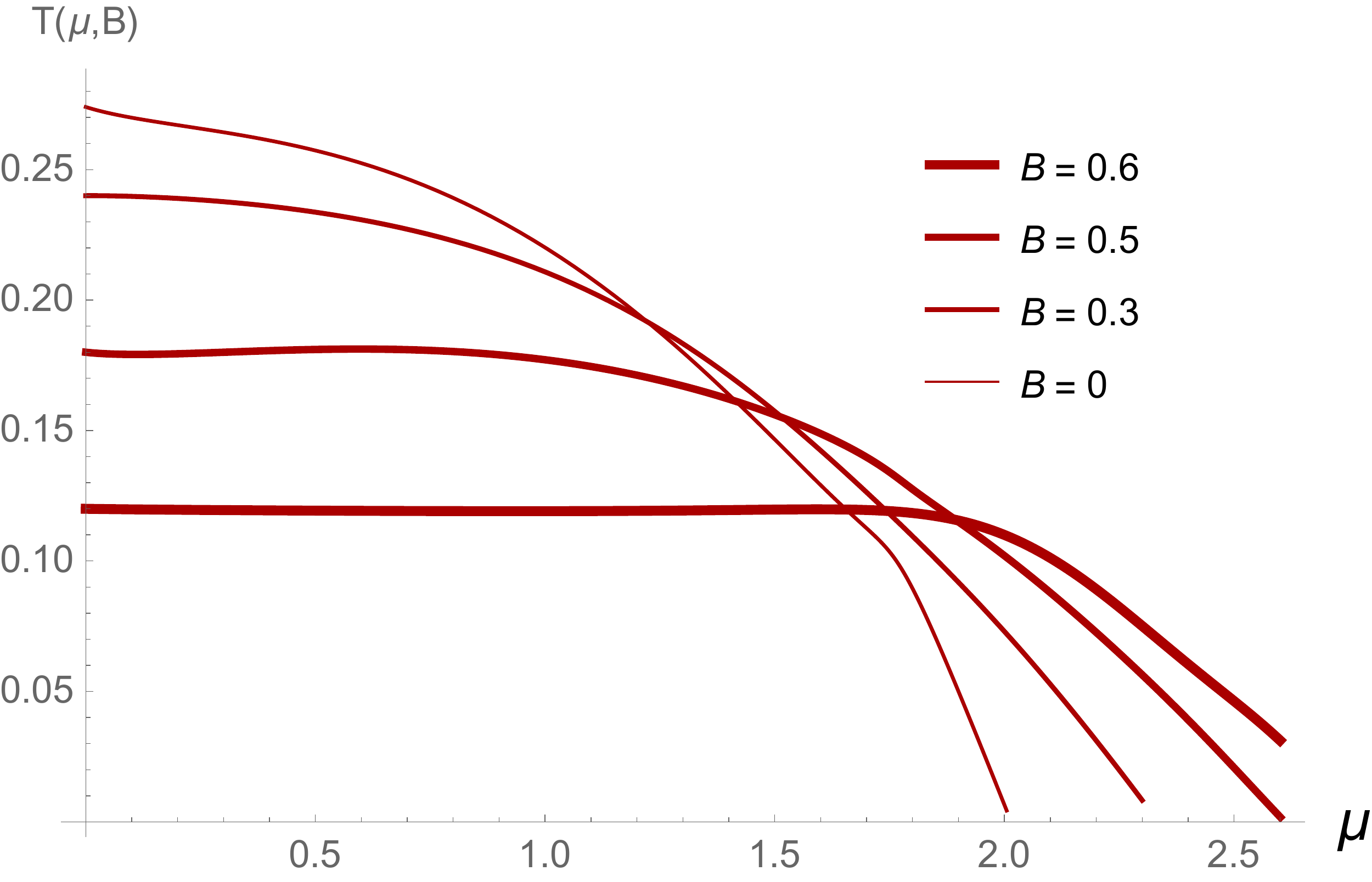}}      & \multicolumn{1}{l|}{\includegraphics[width=0.25\textwidth]{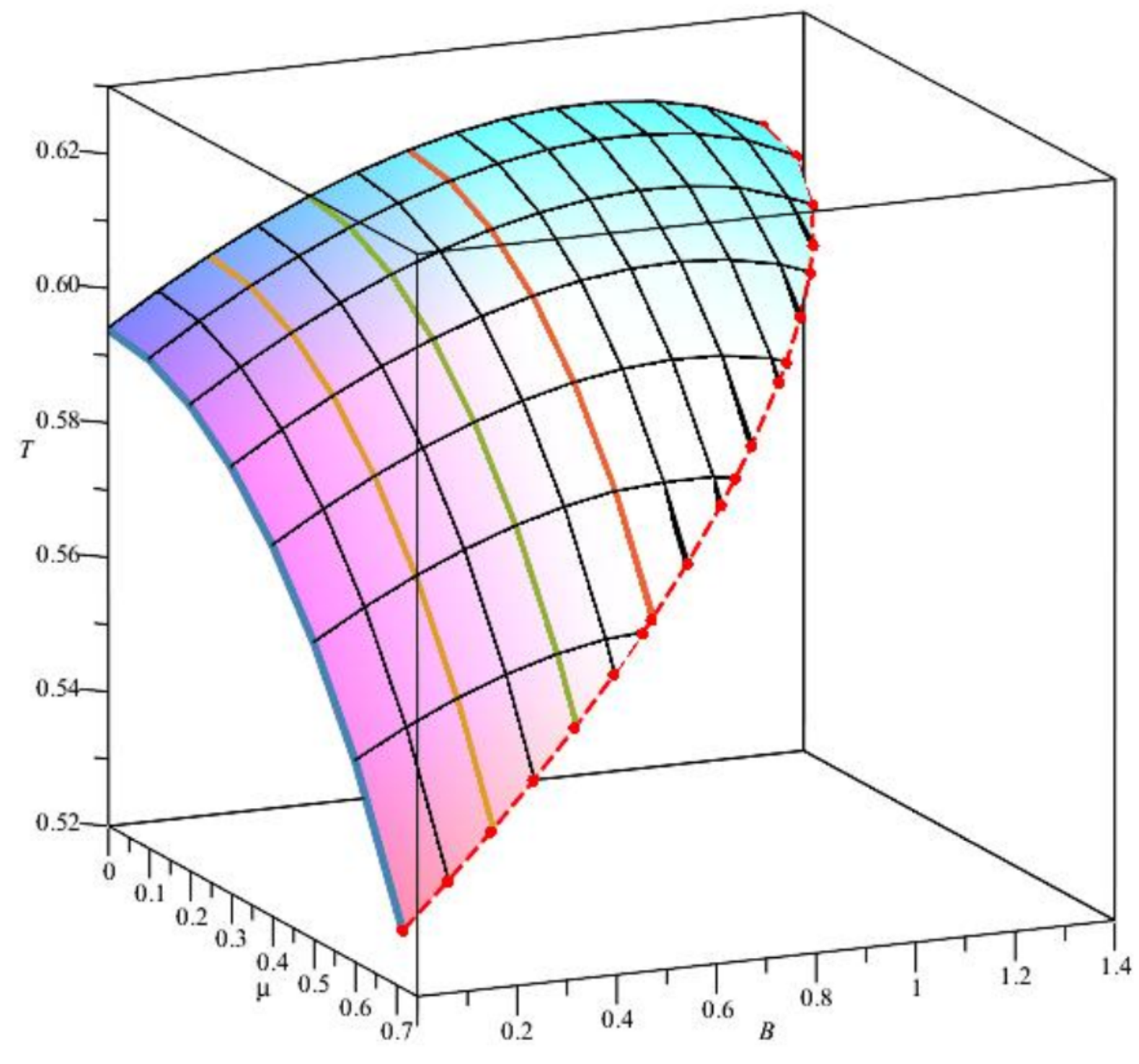}\cite{2004.01965}}            & \includegraphics[width=0.25\textwidth]{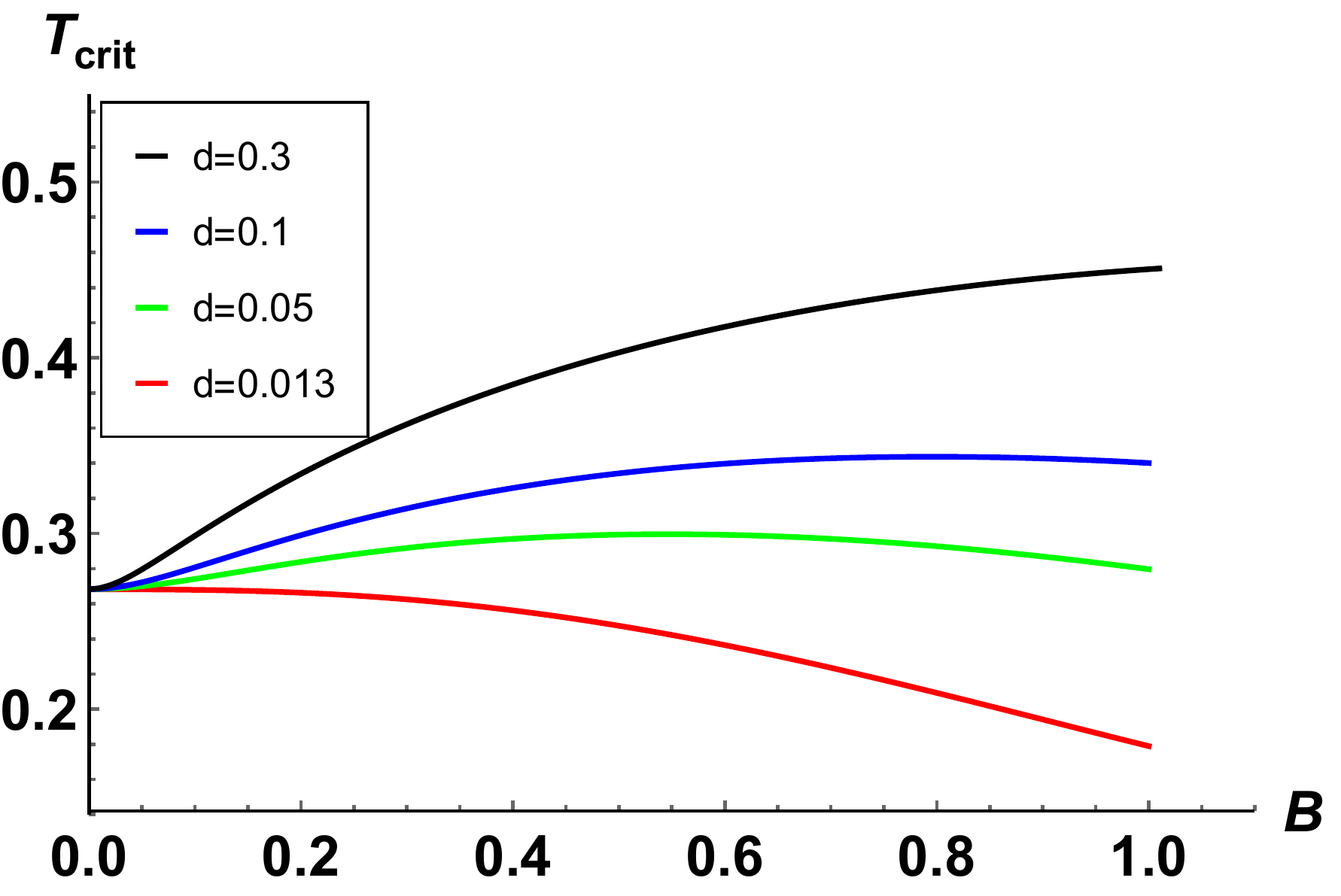}   \cite{2010.04578}                                \\ \cline{2-4} 
                                                                                                   &  \includegraphics[width=0.25\textwidth]{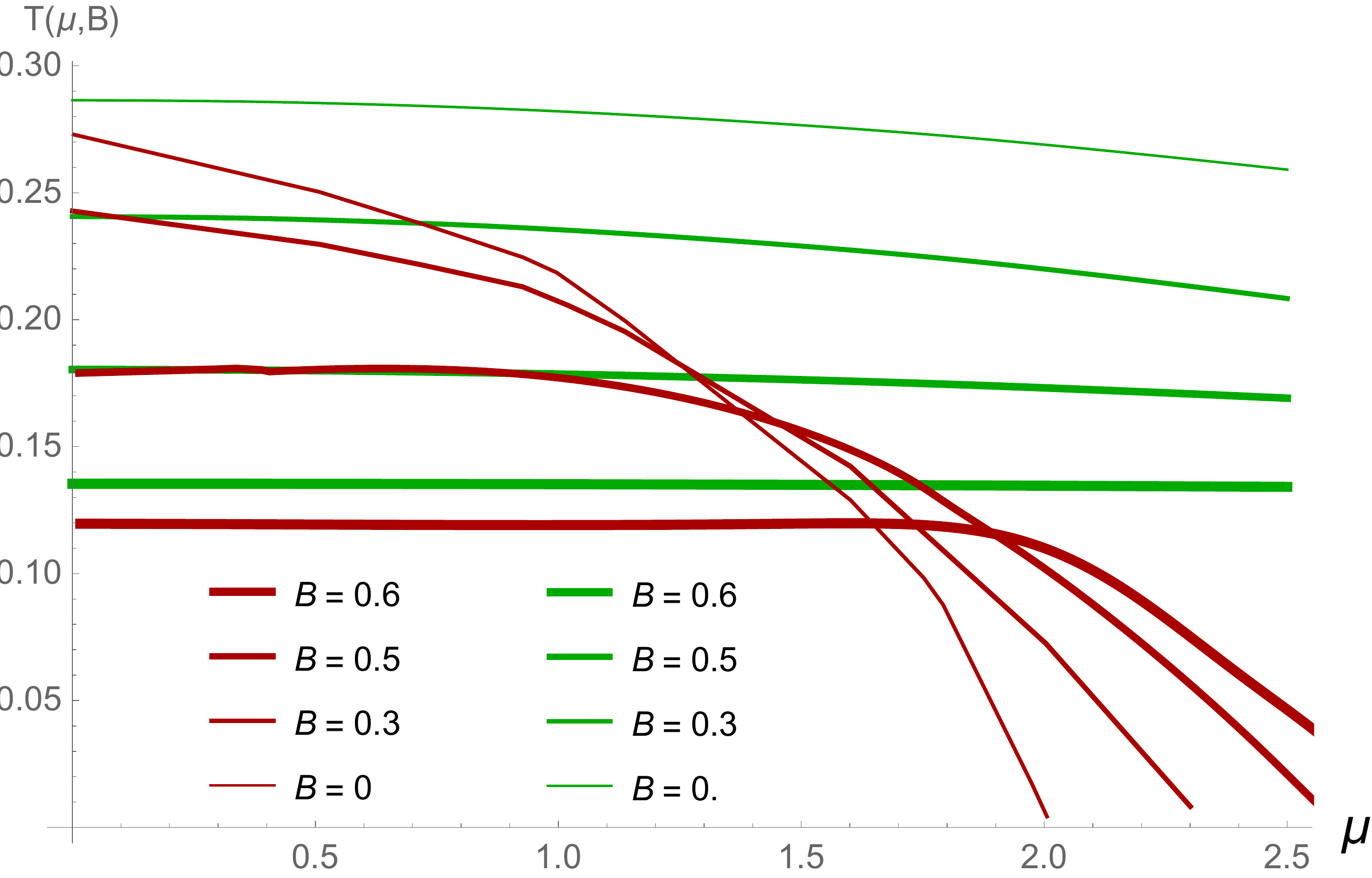}                          &               -                 & -                               \\ \hline
\multirow{2}{*}{\begin{tabular}[c]{@{}c@{}}$\nu\neq 1$\\ $f_1=z^{2-2/\nu}e^{c_1z^2}$\end{tabular}} & \multirow{2}{*}{-}           & \multirow{2}{*}{-}                & \multirow{2}{*}{-}                \\
                                                                                                   &                              &                                   &                                   \\ \hline
\end{tabular}
\end{table}

\end{document}